\documentclass[journal=chreay,manuscript=review,layout=twocolumn]{achemso}
\usepackage{chemformula} 
\usepackage[T1]{fontenc} 

\usepackage{amsmath}
\usepackage{hyperref}
\author{Julia Westermayr}
\affiliation[Vienna]{Institute of Theoretical Chemistry, Faculty of Chemistry, University of Vienna, W\"{a}hringer Str. 17, 1090 Vienna}
\author{Philipp Marquetand}
\affiliation[Vienna]{Institute of Theoretical Chemistry, Faculty of Chemistry, University of Vienna, W\"{a}hringer Str. 17, 1090 Vienna}
\alsoaffiliation[ViRAPID]{Vienna Research Platform on Accelerating Photoreaction Discovery, University of Vienna, W\"ahringer Str. 17, 1090 Vienna, Austria.}
\alsoaffiliation[DS]{Data Science @ Uni Vienna, University of Vienna, W\"ahringer Str. 29, 1090 Vienna, Austria.}
\email{philipp.marquetand@univie.ac.at}

\title[Title]
  {Machine learning for electronically excited states of molecules}

\abbreviations{ML,QC,SchNarc}

\begin{document}


\begin{abstract}
Electronically excited states of molecules are at the heart of photochemistry, photophysics, as well as photobiology and also play a role in material science. Their theoretical description requires highly accurate quantum chemical calculations, which are computationally expensive. In this review, we focus on how machine learning is employed not only to speed up such excited-state simulations but also how this branch of artificial intelligence can be used to advance this exciting research field in all its aspects. Discussed applications of machine learning for excited states include excited-state dynamics simulations, static calculations of absorption spectra, as well as many others. In order to put these studies into context, we discuss the promises and pitfalls of the involved machine learning techniques. Since the latter are mostly based on quantum chemistry calculations, we also provide a short introduction into excited-state electronic structure methods, approaches for nonadiabatic dynamics simulations and describe tricks and problems when using them in machine learning for excited states of molecules.
\end{abstract}

\includegraphics[angle=90,width=5cm]{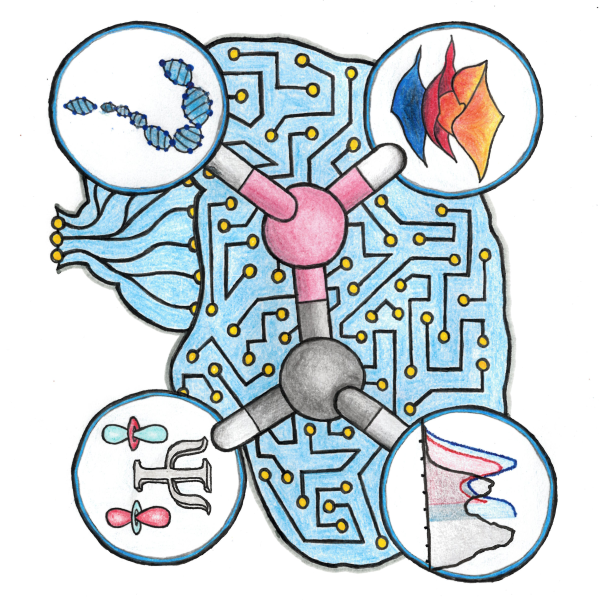}

\newpage
\tableofcontents

\section{Introduction}
\subsection{From Foundations to Applications}
In recent years, machine learning (ML) has become a pioneering field of research and has an increasing influence on our daily lives. Today it is a component of almost all applications we use. For example, when we talk to Siri or Alexa, we interact with a voice assistant and make use of natural language processing~\cite{Kepuska2018,Hoy2018MRSQ}. ML is applied for refugee integration~\cite{Silver2016N}, for playing board games~\cite{Bansak2018S}, in medicine~\cite{Leung2016PIEEE}, for example, for image recognition~\cite{Shen2017ARBE} or for autonomous driving~\cite{Chen2015}. A short historical overview over general ML is provided in ref~\citenum{Yang2019CR}.

Recently, ML has also gained increasing interest in the field of quantum chemistry~\cite{Goodfellow2016,Gomez-Bombarelli2018}. 
The power of (big) data-driven science is even seen as the "fourth paradigm of science"~\cite{Agrawal2016APLM}, which has the potential to accelerate and enable quantum chemical simulations that were considered unfeasible just a few years ago. In general, the field of ML in quantum chemistry is progressing faster and faster. In this review, we focus on an emerging part of this field, namely ML for electronically excited states. In doing so, we concentrate on singlet and triplet states of molecular systems, since almost all existing approaches of ML for the excited states focus on singlet states and only a few studies consider triplet states.~\cite{Schwilk2020arXiv,Westermayr2020JPCL,Guan2020JCTC,Taylor2020JPCA}
We note that electron detachment or uptake further leads to doublet and quartet states, and even higher spin multiplicities, such as quintets, sextets, etc. are common in transition metal complexes, where an important task is to identify which multiplicity yields the lowest energy and is thus the ground state~\cite{Taylor2020JPCA}. refs~\citenum{Kulik2019WCMS,Power2012CR,Bousseksou2004EJIC,Li2015O}
give a good overview of such processes.

The theoretical study of the excited states of molecules is crucial to complement experiments and to shed light on many fundamental processes of life and nature~\cite{Matsika2018CR}. For example, photosynthesis, human vision, photovoltaics or photodamage of biologically relevant molecules are a results of light-induced reactions~\cite{Cohen2004,Levine2007ARPC,Turro2009,Yarkony2012CR,Barbatti2014,Ibele2019MP,Nelson2020CR,Mai2020ACIE,Matsika2018CR,Lischka2018CR,Ghosh2018CR,Norman2018CR,Casanova2018CR,Hestand2018CR,Penfold2018CR,Vacher2018CR,Crespo-Otero2018CR,Gonzalez2020}. Experimental techniques like UV/visible spectroscopy or photoionization spectroscopy~\cite{Harris1989,Cheuk-Yiu1991,Zewail1994,Brixner2005,Iqbal2010JPCL,Kowalewski2017CR,Soorkia2019CR,Liu2020PRX} lack the ability to directly describe the exact electronic mechanisms of photo-induced reactions. The theoretical simulation of the corresponding experiments can go hand-in-hand with experimental results and can provide the missing details of photodamage and -stability of molecules~\cite{Matsika2018CR,Martinez2006ACR,Barbatti2008,Subotnik2016ARPC,Curchod2018CR,Ashfold2017JPCL,Tajti2009C,Barbatti2011JCP,Lu2014,Ruckenbauer2016JCP,Manathunga2016JCTC,Nogueira2017CS,Mai2018M,Rauer2018MC,Zobel2018JCTC,Nelson2020CR,Vacher2018CR,Soorkia2019CR,Pathak2020arXiv,Neves-Peterson2012,Cadet2014,Segatta2019CR,Landry2014JCTC,Schultz2003JACS,Toniolo2004FD,Domcke2011}.
However, the computation of the excited states is highly complex, costly, and often necessitates expert knowledge~\cite{Serrano-Andres2005JMST}.
As ML models have only recently been applied in the field of photochemistry, keeping track of the approaches is still possible and this field is still in its initial stage. 

\begin{figure*}[ht]
    \centering
    \includegraphics[width=0.8\textwidth]{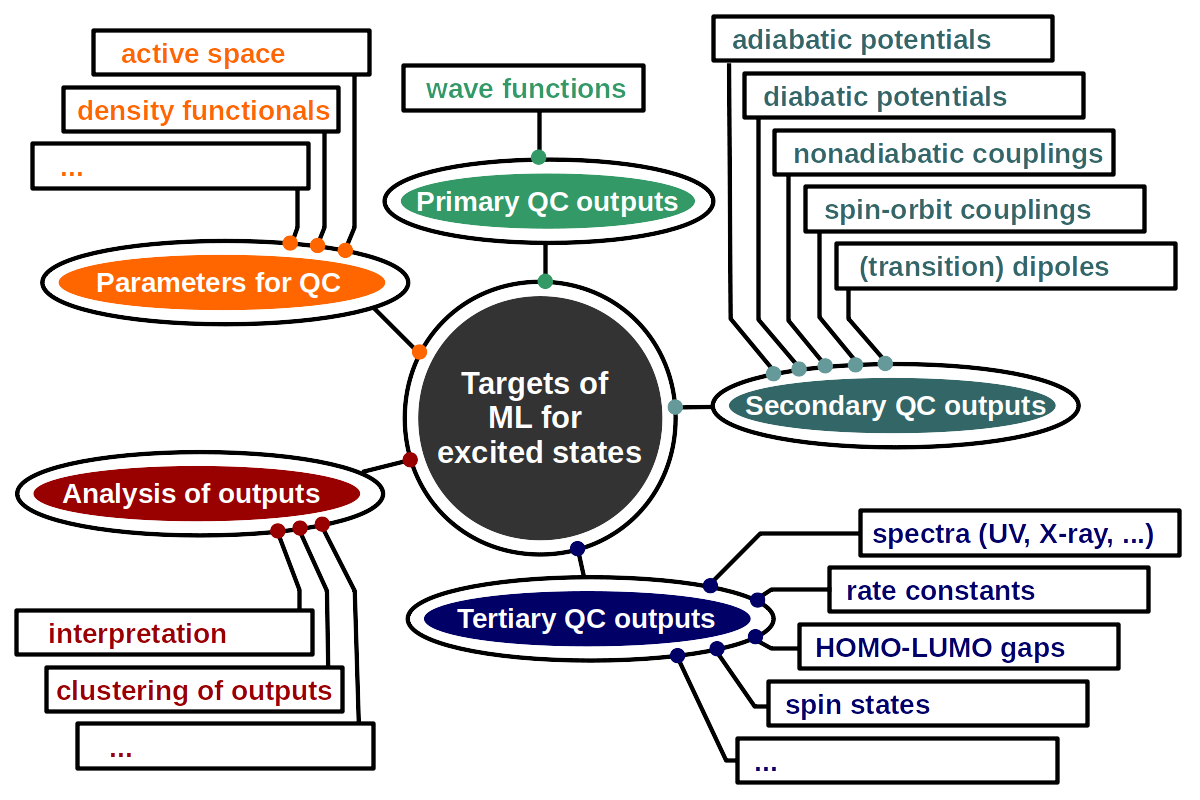}
    \caption{Targets of ML for the excited states of molecules. All areas of excited-state quantum chemistry (QC) calculations can be enhanced with ML, ranging from input to primary outputs that are used in the computation of secondary outputs, which in turn are employed to calculate tertiary outputs. Analysis can be carried out at all stages. This classification is inspired by the one in Ref.~\citenum{Chandrasekaran2019nCM}.}
    \label{fig:targets}
\end{figure*}{}
Due to the multi-faceted photochemistry of molecular systems, ML models can target this research field in many different ways, which are summarized in Figure~\ref{fig:targets}. For example, the choice of relevant molecular orbitals for active space selections can be assisted with ML.~\cite{Jeong2020JCTC} The fundamentals of quantum chemistry, e.g., to obtain an optimal solution to the Schr\"{o}dinger equation or Density Functional Theory, can be central ML applications. For the ground state, ML approximations to the molecular wave function~\cite{Carleo2017S,Saito2017JPCJ,Nomura2017PRB,Han2018JCP,Townsend2019JPCL,Schuett2019NC,Pfau2019arXiv,Hermann2019arXiv,Gastegger2020arXiv} or the density (functional) of a system exist.~\cite{Hegde2017SR,Brockherde2017NC,Gastegger2019MC,Nelson2019PRB,Chandrasekaran2019nCM,Cheng2019JCP,Lei2019PRM,Zhou2019JPCL,Kolb2017SR,Gastegger2020arXiv,Willat2019JCP} Obtaining a molecular wave function from ML can be seen as the most powerful approach in many perspectives, as any property we wish to know could be derived from it. Unfortunately, such models for the excited states are lacking and have yet only been investigated for a one-dimensional system~\cite{Choo2018PRL}, leaving much room for improvement. 

Most ML studies instead focus on predicting the output of a quantum chemical calculation, the so-called "secondary-output"~\cite{Chandrasekaran2019nCM}. Hence they fit a manifold of energetic states of different spin multiplicities, their derivatives and properties thereof. With respect to different spin states of molecular systems only a few studies exist, which predict spins of transition metal complexes~\cite{Taylor2020JPCA} or singlet and triplet energies of carbenes~\cite{Schwilk2020arXiv} of different composition or focus on the conformational changes within one molecular system~\cite{Westermayr2020JPCL,Guan2020JPCL,Carbogno2010PRB} for the sake of improving molecular dynamics (MD) simulations. The energies of a system in combination with its properties, i.e., the derivatives, the coupling values between them, and the permanent and transition dipole moments~\cite{Westermayr2019CS,Westermayr2020JPCL,Westermayr2020MLST,Guan2020JCTC,Guan2020JPCL,Shen2020JPCA,Zhang2020arXiv,Carbogno2008PRL,Carbogno2010PRB,Polyak2019JCP}, can be used for MD simulations to study the temporal evolution of a system in the ground-state ~\cite{Hobday1999MSMSE,Bartok2010PRL,Rupp2012PRL,Li2015PRL,vonLilienfeld2015IJQC,Gastegger2015JCTC,Rupp2015JPCL,Behler2016JCP,Urban2016CMS,Gastegger2016JCP,Artrith2017PRB,Gastegger2017CS,Deringer2017PRB,Botu2017JPCC,Glielmo2017PRB,Smith2017CS,Fujikake2018JCP,Behler2017ACIE,Zong2018npjCM,Wood2018JCP,Chen2018JCTC,Bartok2018PRX,Chmiela2018NC,Imbalzano2018JCP,Zhang2018NIPS,Zhang2018PRL,Chan2019JPCC,Faber2018JCP,Wang2019JCTC,Gerrits2019JPCL,Chmiela2019CPC,Carleo2019RMP,Krems2019PCCP,Deringer2019AM,Ward2019MRSC,Noe2020ARPC,Alborzpour2016JCP,Chen2020JPCA_Dynamics} and in the excited states.~\cite{Behler2008PRB,Carbogno2010PRB,Hu2018JPCL,Dral2018JPCL,Chen2018JPCL,Williams2018JCP,Xie2018JCP,Guan2019PCCP,Westermayr2019CS,Guan2020JCTC,Krems2019PCCP,Richings2018JCP,Richings2019JCTC,Guan2019JCP,Wang2019JPCA,Guan2020JPCL,Richings2018JCP,Richings2017CPL,Richings2017CPL,Netzloff2006JCP,Bettens1999JCP,Westermayr2020JPCL}

With energies and different properties, tertiary outputs can be computed, such as absorption, ionization or X-ray spectra,~\cite{Ghosh2019AS,Kananenka2019JCTC,Roch2020ACSN,Rankine2020JPCA} gaps between HOMO (highest occupied molecular orbital) and LUMO (lowest occupied MO) or vertical excitation energies.~\cite{Pereira2017JCIM,Isayev2017NC,Pronobis2018EPJB,Stuke2019JCP} 

In addition, quantum chemical outputs can also be analyzed or fitted in a direct way, e.g., reaction kinetics as results of dynamics simulations can be mapped to a set of molecular geometries and can be predicted with ML models.~\cite{Haese2019CS} Excitation energy transfer properties can be learned,~\cite{Haese2017CS,Haese2016CS}, and structure-property correlations can be explored to design materials with specific properties.~\cite{Schuett2019NC,Boyle2011JPCC,Teunissen2017JCTC,Liu2018ACSP,Elton2019MSDE,Sanchez-Lengeling2018S,Deringer2019AM,Goldsmith2018AJ,Davies2018FD,Lilienfeld2020NRC,Freeze2019CR,Lee2020OE,Kulik2019WCMS,Roch2020ACSN}

\subsection{Scope and Philosophy of this Review}
ML for the excited states is developing at a slower pace than the exploding field of ML for the electronic ground state~\cite{Cartwright2020,Gastegger2020,Schuett2020,Lilienfeld2020NRC}. The reason is in our opinion mainly a result of the complexity and high expenses of the underlying reference calculations and the associated complexity of the corresponding ML models. Simulation techniques to understand the excited-state processes are not yet viable for many applications at an acceptable cost and accuracy. Therefore, within this review we also want to highlight the existing problems of quantum chemical approaches that might be solvable with ML and put emphasis on identifying challenges and limitations that hamper the application of ML for the excited states.
The young age of this research field leaves much room for improvement and new methods.

This review is structured as follows: \\
 \indent (1) Throughout this review, we will start (non-exhaustively) discussing ground state processes, since they are inherently linked to the excited state processes and should also be considered here. 
 We will therefore start by discussing the differences between the ground-state potential energy hypersurfaces (PESs) and the excited-state PESs and will also emphasize the difference in their properties in section~\ref{sec:gs_ex}. \\
 \indent (2) Section~\ref{sec:exc} gives an overview of the theoretical methods that can be used to describe the excited states of molecules. In the forthcoming discussion, we will describe different reference methods with a view to their application in time-dependent simulations, namely MD simulations~\cite{Gastegger2020,Nelson2020CR}. It is worth mentioning, that unlike for the ground state, where a lot of different methods can provide reliable reference computations for training, choosing a proper quantum chemistry method for the treatment of excited states is a challenge on its own. Many methods require expert knowledge, prohibiting their use further~\cite{Gonzalez2020,Park2020CR}. In addition, not any method can provide the necessary properties for any type of application. 
 Subsequently, we aim to review the different flavours of excited-state MD simulations with focus on those methods that have been enhanced with ML models lately. \\
 \indent (3) After having provided the basic theoretical background, we will discuss how to generate a comprehensive, yet full-fledged training set for the excited states from the quantum chemistry data. We will summarize the existing approaches that are applied to create a comprehensive training set and put emphasis on the bottlenecks of existing methods that can limit also the application of ML. This will provide the reader with the knowledge about starting points for future research questions and clarify where method development is needed. 
 It further provides the basis for the discussion of ML models for the excited states of molecular systems. \\
 \indent (4) A summary of state-of-the-art ML methods for photochemistry follows. We will differentiate between single-state and multi-state ML models and single-property and multi-property ML models~\cite{Westermayr2020MLST}. As mentioned before, ML models can tackle a quantum chemical calculation in many different ways, see Figure~\ref{fig:targets}. 
The different ML models will be classified in the ways they enhance quantum chemical simulations. Most approaches aim at providing an ML-based force field for the excited states, so most focus will be put on this topic.
At last, the prospects of ML models to revolutionize this field of research and future avenues for ML will be highlighted. \\

Noteworthy, we focus on the excited states of molecules, as the excited electronic states in the condensed phase are challenging to fit and are thus often not explicitly considered in conventional approaches~\cite{Akimov2015CR,Frutos2007PNAS,Menger2018JCTC,Dou2020JPCA,Dou2015JCP,Tavernelli2006PRB}.
In solid state physics for example, the electronic states are usually treated as continua. The density of states at the Fermi level,~\cite{Schuett2014PRB} band gaps,~\cite{Lee2016PRB,Zhuo2018JPCL,Pilania2017CMS} and electronic friction tensors~\cite{Zhang2018PRL,Spiering2019JPCL,Zhang2020JPCC} have been described with ML models up to date and especially the electronic friction tensor is useful to study the indirect effects of electronic excitations in materials.~\cite{Zhang2019CS,Head-Gordon1995JCP,Douglas-Gallardo2019N,Yin2019JPCL,Rittmeyer2018APX,Therrien2019FD} Electron transfer processes as a result of electron-hole-pair excitations can be further investigated along with multi-quantum vibrational transitions by discretizing the continuum of electronic states and fitting them (often manually) to reproduce experimental or quantum chemical data in a model Hamiltonian.~\cite{wodtke2004IRPC,Park2019RPP,Jiang2019JCP,Dou2020JPCA,Shenvi2009JCP,Shenvi2009S,Dou2018JCP} Yet, to the best of our knowledge, the excited electronic states in the condensed phase have not been fitted with ML. A recent review on reactive and inelastic scattering processes and the use of ML for quantum dynamics reactions in the gas phase and at a gas-phase interface can be found in ref~\citenum{Jiang2020JPCL}.

Besides the electronic excitations that take place in molecules after light excitation, ML models have successfully entered research fields, which focus on other types of excitations as well. Those are for example vibrational or rotational excitations giving rise to Raman spectra or Infrared spectra,~\cite{Kowalewski2017CR,Buhrke2019CR,Raimbault2019NJP,Hu2019JPCL,Gastegger2017CS,Lussier2020TRAC,Fu2018JPCA} nuclear magnetic resonance,~\cite{Aires-de-Sousa2002AC} or magnetism~\cite{Taguchi2019JPCL,Cobas2020MRC}, which we will also not consider in this review.

\section{General Background: From the Ground State to the Excited States}\label{sec:gs_ex}

\begin{figure}[ht]
    \centering
    \includegraphics[scale=0.22]{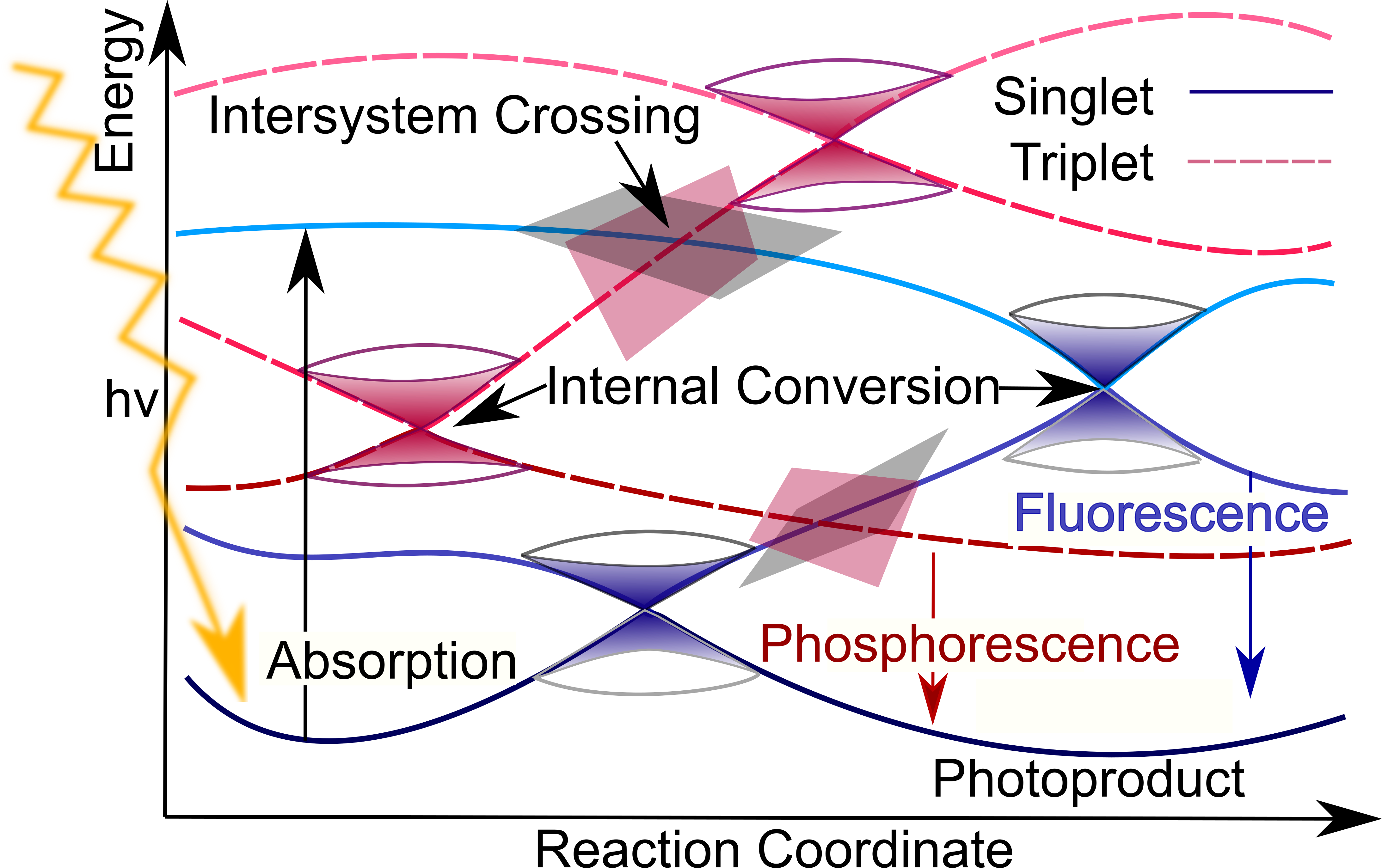}
    \caption{Excited-state processes that can take place after excitation of a molecule by light. Absorption of light can make the molecule enter a higher electronic singlet state. Intersystem crossing to a triplet state or internal conversion to another state of same spin-multiplicity can take place. Radiative emission, i.e., fluorescence and phosphorescence, are possible reactions from an excited singlet and triplet state, respectively.}
    \label{fig:jablonski}
\end{figure}{}
Figure~\ref{fig:jablonski} gives an overview of the excited state processes that will be discussed within this review.
It shows a schematic one-dimensional representation of the potential energy curves for the ground and excited states as a function of molecular coordinates. Figure \ref{fig:jablonski} illustrates that the ground state potential energy curve, given by a dark-blue solid line, is mostly a smooth function of the reaction coordinate and gives information about several local minima.
In the ground state, many methods exist to describe the physico-chemical properties of molecules and materials reasonably well, ranging from small systems up to proteins, DNA or nanoparticles. For small system sizes, highly accurate ab-initio methods can be applied, while more crude approximations have to be used for larger systems. The unfavorable scaling of many quantum chemical methods with the size of system under investigation requires this compromise between accuracy and system size. Crude approximations for systems that are larger than several 100s of atoms become inevitable~\cite{Gonzalez2020,Yarkony2012CR,Matsika2018CR}.

The chemistry we are interested in, however, is not static, but rather depends to a large extent on the changes that matter undergoes. 
In this regard, it is more intuitive to study the temporal evolution of a system. Much effort has been devoted to develop methods to study the temporal evolution of matter in the ground state potential. As an example, physical functions can be obtained with conventional force fields, such as AMBER~\cite{Salomon-Ferrer2013WCMS}, CHARMM~\cite{Brooks2009JCC} or GROMOS~\cite{Eichenberger2011JCTC,Reif2012JCTC}. 
The first ones already date back to the 1940s-1950s. Such force fields enable the study of large and complex systems, protein dynamics or binding-free energies on time scales up to a couple of nanoseconds~\cite{Perthold2020JCIM,Oelknecht2020JCC,Michlits2020ACSC,Akimov2015CR,Brunk2015CR,Bedrov2019CR,Sosso2016CR,Venable2019CR,Marrink2019CR,Groenhof2013}.
However, their applicability is restricted by the limited accuracy and inability to describe bond formation and breaking. Novel approaches, such as reactive force fields exist, but have not yet entered the mainstream and still face the problem of generally low accuracy.~\cite{Senftle2016npjCM} 

The accuracy of ab-initio methods can be combined with the efficiency of conventional force fields with ML models. The latter have shown to advance simulations in the ground state considerably and allow for the fitting of almost any input-output relation.~\cite{Hobday1999MSMSE,Bartok2010PRL,Rupp2012PRL,Li2015PRL,vonLilienfeld2015IJQC,Gastegger2015JCTC,Rupp2015JPCL,Behler2016JCP,Urban2016CMS,Gastegger2016JCP,Artrith2017PRB,Gastegger2017CS,Deringer2017PRB,Botu2017JPCC,Glielmo2017PRB,Smith2017CS,Fujikake2018JCP,Behler2017ACIE,Zong2018npjCM,Wood2018JCP,Chen2018JCTC,Bartok2018PRX,Chmiela2018NC,Imbalzano2018JCP,Zhang2018NIPS,Zhang2018PRL,Chan2019JPCC,Faber2018JCP,Wang2019JCTC,Gerrits2019JPCL,Chmiela2019CPC,Carleo2019RMP,Krems2019PCCP,Deringer2019AM,Ward2019MRSC,Sauceda2020,Gastegger2020,Chen2020JPCA_Dynamics} Accurate and reactive PESs of molecules in the ground state can be obtained with a comprehensive reference data set, which contains the energies, forces and ground-state properties of a system under investigation. Proper training of an ML model then guarantees that the accuracy of the reference method is retained, while inferences can be made much faster. In this way, they allow for a description of reactions and can overcome the limitations of existing force fields.~\cite{Noe2020ARPC,Noe2020,Glielmo2020,Abbott2019JCTC,Hellstroem2020,Vargas-Hernandez2020,Cartwright2020}

Regarding the excited states, processes become much more complex and the computation of excited state PESs is far more difficult than the computation of the ground state PESs. As can be seen in Figure~\ref{fig:jablonski}, a lot of different classes of excited states, e.g. singlet states as shown by continuous blueish lines or triplet states as shown by dashed reddish lines, have to be accounted for, which are characterized by several transition states, local minima, and crossing points. This complexity makes a separate treatment of each electronically excited state inaccurate and leads to further challenges that prohibit the straight-forward and large-scale use of many existing quantum chemical methods and consequently also existing ML models for the ground state. 

Additionally, computations of the excited states suffer from being generally less efficient. To name only one central problem: The larger the system becomes, the closer the electronic states lie in energy, and the more excited-state processes can usually take place. The necessary consideration of an increasing number of excited states increases the already substantial computational expenses even more and restricts the use of accurate methods to systems containing only a few dozens of atoms in a reasonable amount of time with current computers. This increasing complexity makes not only the reference computations, but also the application of ML models for the excited states more complicated than for the ground state. At the same time, the application of ML models for the excited states might also be more promising, because higher speed-ups can be achieved.

For the excited states, methods similar to force fields, like the linear vibronic coupling (LVC) approach~\cite{Koeppel04,Plasser2019PCCP}, are usually limited to small regions of conformational space and restricted to a single molecule. General force fields that are valid for different molecules in the excited states do not exist. Also the ML analogue, so-called transferable ML models, to fit the excited state PESs of molecules throughout chemical compound space are unavailable up to date. Nevertheless, it is out of question that an ML model, which is capable of describing the photochemistry of several different molecular systems, e.g., different amino acids or DNA bases of different sizes, is highly desirable. A lot remains to be done in order to achieve this goal and yet, to the best of our knowledge, no more than a maximum of about 20 atoms and 3 electronic states with a distinct multiplicity have been fitted accurately with ML models~\cite{Shen2020JPCA,Behler2008PRB,Carbogno2010PRB,Hu2018JPCL,Dral2018JPCL,Chen2018JPCL,Williams2018JCP,Xie2018JCP,Guan2019PCCP,Westermayr2019CS,Guan2020JCTC,Krems2019PCCP,Richings2018JCP,Richings2019JCTC,Richings2019JCTC,Guan2019JCP,Wang2019JPCA,Guan2020JPCL,Richings2018JCP,Richings2017CPL,Richings2017CPL,Netzloff2006JCP,Bettens1999JCP,Westermayr2020JPCL}.

Whether or not the excited states of a molecular system become populated depends on the ability of a molecule to absorb energy in the form of light, or more generally, electromagnetic radiation of a given wavelength.
Usually, the so-called resonance condition has to be fulfilled, i.e., the energy gap between two electronic states has to be equivalent to the photon energy of the incident light. Note however that also multi-photon processes can occur, where several photons have to be absorbed at once to bridge the energy difference between two electronic states~\cite{He2008CR,Marquetand2014,Tagliamonti2016PRA}
Further, the absorption of light does not only provide access to one, but most often to a manifold of energetically close-lying states. The number of states that can be excited is related to the range of photon energies that is contained in the electromagnetic radiation. This energy range is inversely proportional to the duration of the electric field, e.g., of a laser pulse, due to the Fourier relation of energy and time~\cite{Wollenhaupt2007}.
However, the energy range of the photons and the energy difference between the electronic states are not the only factors influencing the absorption of light, which gives rise to questions like:
Is the molecule able to absorb light of a considered wavelength? Which of the excited states is populated with the highest probability?

An answer to these questions can be obtained from an analysis of the oscillator strength. In order to make an electronic transition possible, an oscillating dipole must be induced as a result of the interaction of the molecule with light. The oscillator strength, $f^{osc}_{ij}$, between two electronic states, i and j, is proportional in atomic units (a.u.) to the respective transition dipole moment, $\mu{ij}$, and the respective energy difference, $\Delta E_{ij}$:~\cite{Hilborn1982AJP}
\begin{equation}\label{eq:osc}
    f^{osc}_{ij}=\frac{2}{3} \Delta E_{ij}\mid\mu_{ij}\mid^2.
\end{equation} If the transition dipole moment between two states is zero, no transition is allowed. The reasons can be that a change of the electronic spin would be required, and the transition is thus spin forbidden. Another reason can be the molecular symmetry, leading to symmetry forbidden transitions. The latter are common in molecules that carry an inversion centre and transitions that conserve parity are forbidden~\cite{Andrews2014}.
An energetic state is called dark, if the transition dipole moment is very small or zero. In contrast, a state is called bright, if the transition dipole moment is large. Most often, studies that target the photochemistry of molecules focus on excitation to the lowest brightest singlet state, i.e., the state that absorbs most of the incident energy.

After an excitation process, the molecule is considered to move on the excited-state PESs and is expected to undergo further conversions. The excess of energy a molecule carries -- as a result of the initial absorption of energy -- is most often converted into heat, light, such as fluorescence or phosphorescence, or into chemical energy. If the molecule returns to its original state, then the molecule is photostable. Otherwise, either photodamage, such as decomposition, or useful photochemical reactions including bond breaking/formation occur. In all cases, heat or light can be emitted, which can also be harnessed in light-emission applications~\cite{Nelson2020CR,Silva2007JACS,Hartschuh2003S,Terenziani2008AM}.
With respect to photo-stability, ultrafast transitions, in the range of femto- to picoseconds (10$^{-15}$--10$^{-12}$ seconds) take place and lead the molecule back to the ground state. This means, the electronic energy is converted into vibrations of the molecule and the molecule is termed hot. This heat is usually dissipated into the environment, a procedure that is often neglected in excited-state simulations due to the cost of describing surrounding molecules. 

Radiationless transitions from one electronic state to another take place in so-called critical regions of the PESs. As the name already suggests, critical regions are crucial for the dynamics of a molecule, but are also challenging to model accurately. A transition from one state to another that conserves the spin-multiplicity is called internal conversion. Furthermore, states of different spin-multiplicities may be accessible via intersystem crossing. 
The critical points, where transitions are most likely to occur, are called conical intersections and are illustrated in Figure \ref{fig:jablonski}.
At these crossing points, PESs computed with quantum chemistry can show  discontinuities. These discontinuities can occur also in other excited-state properties and pose an additional challenge for an ML model when fitting excited-state quantities. 

In addition to the aforementioned complications of treating a manifold of excited states, also the probability of a radiationless transition between them has to be computed somehow. This probability is usually determined by couplings between two approaching PESs. Between states of the same spin multiplicity, nonadiabatic couplings (NACs) arise, and spin-orbit couplings (SOCs) give rise to the transition probability between states of different spin multiplicities. These couplings are intimately linked to the excited-state PESs and therefore should also be considered with ML. However, only a handful of publications describe couplings with ML,~\cite{Dral2018JPCL,Shen2020JPCA,Westermayr2019CS,Westermayr2020JPCL,Westermayr2020MLST,Guan2020JPCL,Richings2017CPL,Richings2018JCP,Richings2019JCTC,Richings2020JCP} which highlights the difficulty of providing the necessary reference data as well as the challenges of accurately fitting them. New methods are constantly needed to further enhance this exciting research field.

\section{Quantum Chemical Theory and Methods}

In this section, we present some key aspects of quantum theory for excited states because (i) the outcome of the corresponding calculations serve as training data for ML and (ii) to clarify the employed nomenclature. We put special emphasis on describing the differences of excited-state computations to computations in the ground state and the challenges that arise due to the treatment of a manifold of excited states. These challenges also point at issues that are problematic for ML. These explanations will provide the groundwork to evaluate different quantum chemical methods for their use to generate a training set for ML and to use it for different types of applications, such as excited-state MD simulations. Naturally, we can only provide a general idea of this field and refer the interested reader to pertinent textbooks and reviews, such as Refs.~\cite{Gonzalez2020,Lischka2018CR,Tannor2006,Weinacht2019,Mai2018WCMS,Crespo-Otero2018CR,Ibele2019MP,Yonehara2012CR,Casida2012ARPC,Maitra2016JCP,Szalay2012CR,Helgaker2014,Roos2016}.

In order to follow a consistent notation within this review, we try to explain all basic concepts with notations that are frequently used in literature. Currently, a zoo of different notations for the same property can be found. For example, the NACs, or derivative couplings, are sometimes referred to as so-called interstate couplings, i.e. couplings between two states multiplied with the corresponding energy gap between those two states~\cite{Guan2019PCCP}, while in other works interstate couplings refer to off-diagonal elements of the Hamiltonian in another basis, where the potential energies are no eigenvalues of the electronic Schr\"{o}dinger equation. We want to avoid a confusion of the different notations and thus provide a consistent definition below. For the excited states, a number of different electronic states is required. Throughout this review, we adopt the following labelling convention for different electronic states: The lower case Latin letters, $i$, $j$, etc. will be used to denote different electronic states. The abbreviations $N_S$, $N_M$, and $N_A$ will indicate the number of states, molecules and atoms, respectively.

The foundation for the following sections is a separation of electronic and nuclear degrees of freedom, which is based on the work of Born and Oppenheimer~\cite{Born1927AP}. However, the famous Born-Oppenheimer approximation is later on (partly) lifted and the coupling between electrons and nuclei is taken into account in nonadiabatic dynamics simulations.

\subsection{Electronic Structure Theory for Excited States}\label{sec:exc}
The main goal when carrying out an electronic structure calculation is usually to compute the potential energy and other physico-chemical properties of a compound. 
We distinguish between two overarching theories to achieve this goal: Wave Function Theory (WFT) and Density Functional Theory (DFT) -- as outlined, e.g., by Kohn in his Nobel lecture~\cite{Kohn1999RMP}.

The basis of WFT, as for any electronic structure calculation, is the electronic Schr\"{o}dinger equation~\cite{Schroedinger1926PR,Schroedinger1933} with the electronic Hamilton operator, $\hat{H}_{el}$, and the N-electron wave function $\Psi_i(\mathbf{R},\mathbf{r})$ of electronic state i, which is dependent on the electronic coordinates $\mathbf{r}$ and parametrically dependent on the nuclear coordinates, $\mathbf{R}$: 
\begin{equation}\label{eq:1}
    \hat{H_{el}}(\mathbf{R},\mathbf{r})\mid\Psi_i(\mathbf{R},\mathbf{r})\rangle = E_i\mid\Psi_i(\mathbf{R},\mathbf{r})\rangle.
\end{equation}
From the wave function, the eigenvector of this eigenvalue equation, any property of the system under investigation can be derived.
How to solve the electronic Schr\"{o}dinger equation exactly to obtain the potential energy of an electronic state i, $E_i$, is known in theory. However, from a practical point of view, the computation is infeasible for molecules that are more complex than for example H$_2$, He$_2^{+}$, and similar systems.~\cite{Yu2016JCP} In order to make the computation of larger and more complex systems viable, approximated wave functions are introduced.

In contrast to WFT, DFT reformulates the energy of a system in terms of the ground state electron density rather than the N-electron wave function and the energy is expressed as a functional thereof. The advantage of DFT over WFT is a rather high accuracy for a rather low computational cost.  If DFT is applied properly, it is considered as one of the most efficient ways to obtain reliable and reasonably accurate results of molecules up to 100s of atoms. In solid state physics, DFT is even the workhorse of most studies aiming to describe ground state properties~\cite{Maurer2019ARMR}. However, the problem is that the equations to be solved are unknown. The missing piece is the exact exchange-correlation functional of a system. Up to date, researchers have come up with many different approximations to this functional that can be used to treat specific problems, but a universal functional capable of describing different problems equally accurately has not yet been found. Moreover, there is no systematic way to improve a density functional. The results obtained with DFT therefore critically depend on the choice of the functional.~\cite{Benavides-Riveros2017PCCP,Yu2016JCP}

In the following sections, we will describe both theories in the light of excited states of molecules. We will start to cover ab-initio methods, which means that they are derived from first principles without parametrization. 

\subsubsection{Wave Function Theory (WFT)}
The basis of all discussed ab-initio methods is the Hartree-Fock method. The N-electron wave function is represented by a single Slater Determinant, $\phi_0$, 
which makes N coupled one-electron problems out of the N-body problem. This Slater determinant is the anti-symmetric product of one-electron wave functions, the spin orbitals, which can be atomic, molecular or crystal orbitals, depending on the system. In the case of molecular (or also crystal) orbitals, they are usually expanded as a linear combination of atomic orbitals, where the expansion coefficients are optimized during the calculation. In order to do so efficiently, the atomic orbitals are themselves expanded with the help of a basis set. The N-electron wave function is therefore obtained as a double expansion. Two approximations are applied, which is the use of a finite basis set to represent the atomic orbitals and in turn also the molecular orbitals on the one hand and the use of a single Slater Determinant on the other hand. This usually gives a poor description of a system under investigation, due to a lack of electronic correlation. 

Electronic correlation describes how much the motion of an electron is influenced by all other electrons. Since the Hartree-Fock method can be seen as a mean-field theory, where an electron ''feels'' only the average of the other electrons, correlation is quantified by the correlation energy, which is the difference between the Hartree-Fock energy and the exact energy of a system. 

Unsurprisingly, all further discussed quantum chemical methods aim at improving the Hartree-Fock method. They can be seen as different flavors of the same solution to the problem: They all include more determinants in one way or another. Accordingly, the wave function is expanded as a linear combination of determinants, where a determinant consists of molecular orbitals, which are expanded in atomic orbitals. This ansatz contains two types of coefficients that can be optimized, the ones for the determinants and the ones yielding the molecular orbitals. If the latter are kept the same for different determinants, we speak of a single-reference wave function. If both types of coefficients are adapted, we speak of a multi-reference wave function. Similarly, the electron correlation is also divided into two parts, termed dynamic correlation and static correlation. Single-reference methods improve on the dynamic correlation, while a multi-reference wave function allows for static correlation. However, the separation is not so strict, as can be seen by the following fact: Both the aforementioned single-reference variant and the multi-reference variant become equivalent when including an infinite number of terms and deliver the exact solution to the Schr\"odinger equation if also an infinite basis set is used.

\begin{figure}[ht]
    \centering
    \includegraphics[scale=0.5]{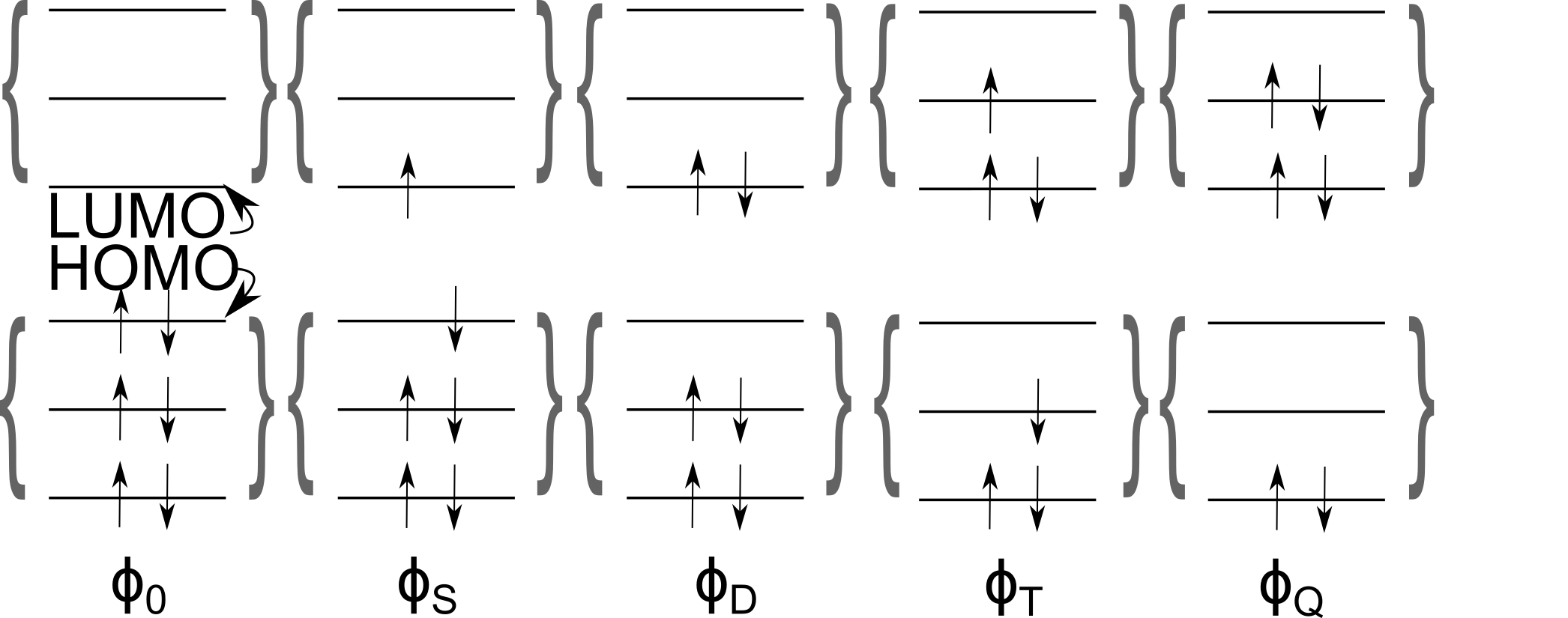}
    \caption{Different arrangements of electrons in molecular orbitals giving rise to the configuration interaction (CI) method.  Inclusion of excited configurations in addition to the ground-state, reference determinant, $\phi_0$, allows to go beyond the Hartree-Fock method. Electrons are excited into higher electronic orbitals and Slater Determinants are indicated using the letters S, D, T, and Q, which refer to single, double, triplet, and quadruple excitations.}
    \label{fig:ci}
\end{figure}

\paragraph{Configuration Interaction}
In the case of single-reference methods, the orbitals obtained from the reference calculation (usually Hartree-Fock) are kept fixed. Since usually more orbitals than the number of electrons in the system are calculated, the possibility of constructing different Slater determinants from these orbitals exist, which can be used for expanding the actual wave function:~\cite{Szabo2012,Helgaker2014HF}
\begin{equation}\label{eq:CI}
    \mid \Psi_{i}\rangle = \sum_I c_I \mid \phi_I\rangle 
\end{equation}{}
Each Slater Determinant is weighted by a coefficient, $c_I$. These coefficients can be obtained variationally by minimizing the total energy under the constraint of fixed orbitals, ending up in the Configuration Interaction (CI) methods. $\Psi_0$ is the reference, Hartree-Fock, wave function. In principle, the exact solution can be obtained by considering all possible Slater Determinants in combination with a complete basis set. The use of all possible configurations is called Full-CI and represents the case, when all electrons are arranged in all possible ways. This approach is infeasible for almost all molecular systems, more complex than e.g. He, and truncated methods are needed. Those are for example, CIS (CI Singles) or CISD (CIS and Doubles), where only single excitations or additionally double excitations are accounted for, respectively. Figure \ref{fig:ci} gives a schematic overview of the improvements of CI that one can apply. A huge advantage of these methods is, that how to obtain the exact solution is known, and that they are systematically improvable. However, truncated CI does not scale correctly with the system size and is therefore not size-extensive and also not size-consistent (i.e., the energy of two fragments A and B at large distance computed together, $E(A+B)$,  is not equal to the sum of the energies of the fragments from separate calculations, $\neq E(A) + E(B)$).~\cite{Helgaker2014CI}

The CI scheme can be employed to improve the ground-state wave function by mixing the Hartree-Fock determinant and determinants of different electron configurations. In the same way, also wave functions of excited states can be computed. Then, the coefficients $C_I$ are optimized for higher eigenvalues of the electronic Hamiltonian instead of the first one. Beginners in the field then often get confused by terms like single excitation in comparison to first excited state. A single excitation determinant (see Fig.~\ref{fig:ci}) can be part of the wave function for the first excited state but can also be a part of the ground-state wave function.

\paragraph{Electron Propagator Methods}
Another class of methods that we shortly want to mention here are electron propagator methods, that are based on one electron Green's function and are another variant of perturbation theory schemes.
One popular method that is based on Green's function one electron propagator approach is the algebraic diagrammatic construction scheme to second order perturbation theory (ADC(2)).~\cite{Dreuw2015WIREs} 
ADC(2) is a single-reference method and can be used to efficiently compute excited states of molecules. It offers a good compromise between computational efficiency and accuracy, while being systematically improvable (higher order variants like ADC(2)-x or ADC(3) exist). The time evolution of a systems polarizability is obtained by applying the polarization propagation, which contains information on a system's excited states.~\cite{Szabo2012,Niessen1984,Lindenberg2005,Melin2005JCS,Corzo2017} 
The ground-state energy of ADC(2) is based on M{\o}ller-Plesset perturbation theory of second order,~\cite{Moeller1934PR,Bartlett1981ARPC} MP2, where the latter can formally be shown to include double excitations for the improvement of Hartree-Fock, see Ref.~\citenum{Szabo2012}. The dependence of ADC(2) on MP2 gives rise to instabilities in regions, where excited states come close to the ground state or homolytic dissociation takes place.
The excited states of bound molecules are described with reasonable accuracy. Compared to multi-reference CI methods (see below), the black box behaviour of ADC(2) is a clear advantage.~\cite{Dreuw2015WIREs}

\paragraph{Coupled Cluster}
The gold standard of ab-initio methods for the ground state is the family of Coupled Cluster (CC) methods. CC is often referred to as the size-extensive and size-consistent version of CI. The different electronic configurations accounting for single or double excitations (such as in CIS and CISD for example) are obtained by applying an excitation operator, $\hat{T}$:~\cite{Helgaker2014CC}
\begin{equation}
\begin{split}
    \mid\phi_{CC}\rangle=e^{\hat{T}}\mid\Psi_0\rangle =& \\
    (\hat{T}=1+\hat{T}+\frac{1}{2!}\hat{T}^2 + \frac{1}{3!}\hat{T}^3 + ...)\mid\Psi_0\rangle.&
    \end{split}
\end{equation}
Similarly to CI, this operator can be truncated. If $\hat{T}=\hat{T}_1+\hat{T}_2$, single and double excitations are accounted for. 

Excited states can be computed in a single-reference approach by equation-of-motion-CC (EOM-CC), where the excited-state wave function is written as an excitation operator times the ground-state wave function. For further details, see, e.g., the reviews ~\citenum{Izsak2020WCMS,Krylov2008ARPC}.

\paragraph{CASSCF}
The problem of missing static correlation in the Hartree-Fock approach is tackled by a multi-reference ansatz for the wave function.~\cite{Benavides-Riveros2017PCCP}
This treatment is important for many excited-state problems, but also some transition metal complexes in their ground state, transition states or homolytic bond-breaking with the dissociation of the N$_2$ molecule being a notoriously difficult example.~\cite{Parrill2018,Pacifici2013}

\begin{figure}[ht]
    \centering
    \includegraphics[scale=0.55]{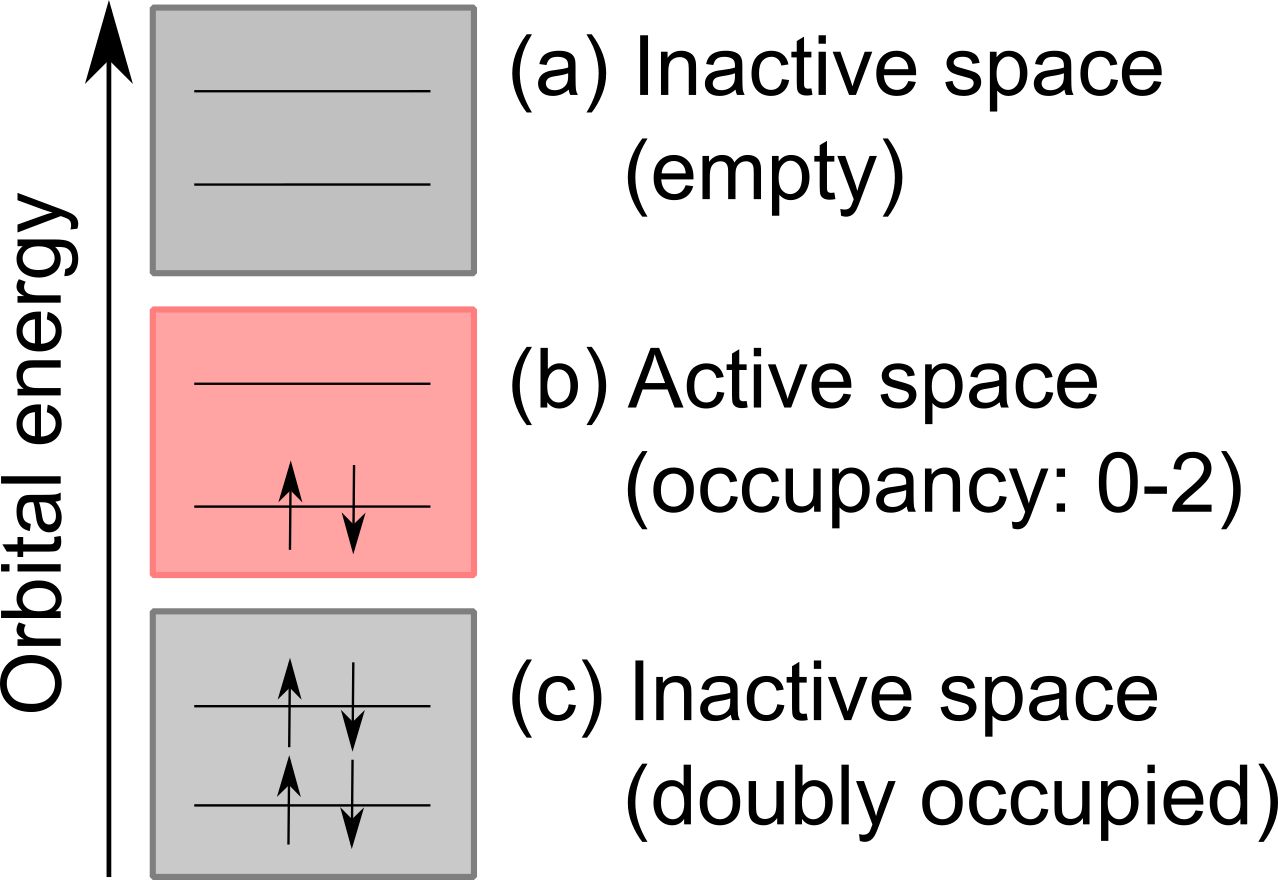}
    \caption{Electrons and orbitals of an arbitrary system to exemplify the active space needed for many multi-reference methods. (a) The highest, not considered, molecular orbitals are inactive and always empty. (c) The lowest, not considered, molecular orbitals are always doubly occupied. (b) The active space is shown with two active electrons in two active orbitals. The occupancy of the orbitals is between zero and two.}
    \label{fig:casscf}
\end{figure}

The multi-configurational self-consistent field (MCSCF) method can be seen as the multi-reference counterpart to the Hartree-Fock method.~\cite{Helgaker2014MCSCF}
One of the most popular variants of MCSCF methods is the Complete Active Space SCF (CASSCF),~\cite{Roos1980CP,Roos1980IJQC} where important atomic orbitals and electrons are selected giving rise to an active space. An example is shown in Figure~\ref{fig:casscf}.
{}
According to this scheme, the orbitals are split into an inactive, doubly occupied part, an active part and an inactive, empty part. Within the active space a FCI computation is carried out. The active space has to be chosen manually by selecting a number of active electrons and active orbitals. CASSCF is no black box method and a meaningful active space selection is the full responsibility of the user. As an advantage, CASSCF can describe static correlation well, which is necessary in systems with nearly degenerated configurations with respect to the reference Slater determinant. For completeness, state-averaging (i.e. SA-CASSCF) is most often applied, where states belonging to the same symmetry are averaged. Another variant of MCSCF methods is restricted active space SCF (RASSCF), which is very similar to CASSCF, but within RASSCF the active space is restricted and no FCI computation is carried out.~\cite{Szabo2012}

\paragraph{MR-CI}
Even higher accuracy can be obtained with multi-reference CI methods~\cite{Lischka2004JCP,Lischka2018CR,Lischka2020JCP}, such as MR-CISD, that additionally add single and double excitations out of the active space and are therefore based on CASSCF wave functions. With this approach electronic correlation, i.e. static and dynamic correlation, can be treated. 

\paragraph{CASPT2}
Alternatively, complete-active-space perturbation theory of second order, CASPT2,~\cite{Andersson1990JPC,Andersson1992JCP,Finley1998CPL} can correct electronic correlation effects via treating multi-reference problems with perturbation theory. This variant of multi-reference perturbation theory methods uses the CASSCF wave function as the zeroth order wave function. CASPT2 can be applied to each state separately (single-state (SS)-CASPT2) or correlated states can be mixed at second order resulting in a multi-state perturbation treatment (MS-CASPT2).~\cite{Andersson1990JPC,Andersson1992JCP,Finley1998CPL} Other perturbation approaches for multi-reference problems exist, like the n-electron valence state perturbation theory (NEVPT2).\cite{Angeli2001JCP,Roemelt2016JCP,Guo2017JCP}

\paragraph{MRCC}
In addition to multi-reference methods based on CI, multi-reference variants of CC approaches exist. A relatively efficient implementation is for example the Mk-MRCC approach of Mukherjee and co-workers~\cite{Maitra2012JCP} or the Brillouin-Wigner approach~\cite{Masik1998}, which is however not size extensive. Noticeably, the development of multi-reference CC approaches is a rather young research field compared to other excited-state methods and the computation of properties and forces is not well explored. Many studies therefore focus on the simulation of energies of low-lying states with MRCC methods. Additionally, such methods suffer from algebraic complexity and numerical instabilities. Interested readers that seek for a more extensive summary of existing MRCC methods are referred to Refs.~\citenum{Musial2011JCP,Evangelista2018JCP,Lischka2018CR}.

\begin{figure}[ht]
    \centering
    \includegraphics[scale=0.16]{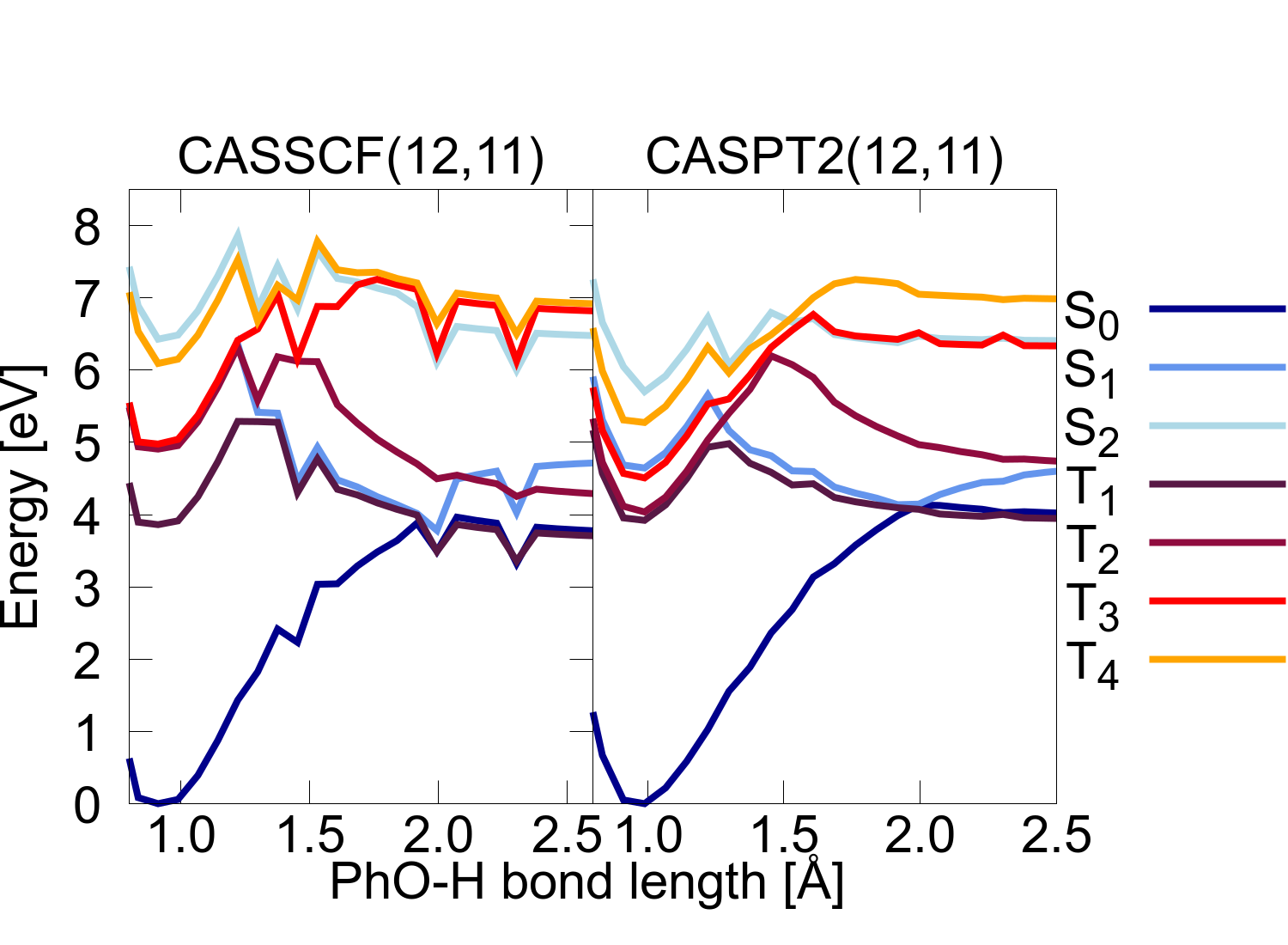}
    \caption{Potential energy curves of the three lowest singlet (S$_0$-S$_2$) and the four lowest triplet state (T$_1$-T$_4$) of the amino acid tyrosine along the O-H bond length of the hydroxy group located at the phenyl ring (Ph-OH) computed with CASSCF(12,11)/ano-rcc-pVDZ and CASPT2(12,11)/ano-rcc-pVDZ~\cite{Roos2004JPCA}. 
    }
    \label{fig:tyrosinecurves}
\end{figure}

\paragraph{Challenges}
The probably biggest drawback of the aforementioned multi-reference methods is that their protocols are very demanding. Finding a proper active space is a tedious task that often requires expert knowledge. Too small active spaces can lead to inaccurate energies and problems with so-called intruder states are common. Those are electronic states, that are high in energy at a reference molecular geometry, but become very low in energy at another molecular geometry, that is visited along a reaction coordinate. The active space then changes along this path. This behavior can result in inconsistent potential energies. In case of CASPT2, the configurations of intruder states can lead to large contributions in the second-order energy, making the assumption of small perturbations invalid.  Especially for describing molecular systems with many energetically close-lying states and for the generation of a training set for ML, such inconsistencies are problematic. Figure~\ref{fig:tyrosinecurves} shows an example of potential energy curves of 3 singlet states and 4 triplet states of tyrosine computed with (a) CASSCF(12,11) and (b) CASPT2(12,11),
where 12 refers to the number of active electrons and 11 to the number of active orbitals. We used OpenMolcas\cite{Fdez-Galvan2019JCTC} to compute an unrelaxed scan along the reaction coordinate, which is a stretching of the O-H bond located at the phenyl-ring of tyrosine. 
{}

Intruder states are no exception. Actually, they are quite common in small to medium sized organic molecules. A large enough reference space can mitigate this problem, but makes computations almost infeasible.
The computational costs increase exponentially with the number of active orbitals. In many cases, the improved accuracy due to a larger active space cannot justify the considerably higher expenses. At its best and with massively parallel simulations, an active space of about 20 electrons in 20 orbitals can be treated,~\cite{Vogiatzis2017JCP} which is impracticable for many applications, such as dynamics simulations. For medium-sized molecules, the active space that would be required for a given simulation might even be way to large to be feasible for calculations in a static picture.

Worth mentioning at this point are also Rydberg states, that often need to be considered in small to medium sized molecules. Rydberg states can be strongly interlaced with valence excited states. In such cases, the active space needs to be large enough to treat both, the valence and Rydberg molecular orbitals. Additionally, the one electron basis set should be flexible enough to describe both types of orbitals. This increases the computational costs additionally. More details on the inclusion of Rydberg states in simulations can be found in refs~\citenum{Kato1995CR,Merer1969CR,Ashfold1999,Merkt1997ARPC}.

A promising tool to eliminate the complex choice of active orbitals is autoCAS~\cite{Stein2016JCTC,Stein2017MP,Stein2017CHIMIA}. It provides a measure of the entanglement of molecular orbitals that is based on the density matrix renormalization group (DMRG). A DMRG-SCF calculation is similar to a CASSCF calculation, but instead of a FCI solution of the active space, an approximated solution with DMRG is obtained to avoid the exponential scaling of the computational costs with the number of active orbitals~\cite{Chan2006JCP,Zgid2008JCP,Keller2015JCP,Knecht2016JCTC,Freitag2017JCTC,Freitag2019JCTC}. As an alternative, ML can be used to determine an active space.\cite{Jeong2020JCTC}

\subsubsection{Density Functional Theory}
A complementary view on how to obtain the energy of a system is provided by DFT. DFT dates back to 1964, when it was formulated by Hohenberg and Kohn~\cite{Hohenberg1964PR} entirely in terms of the electron density, $\eta(\vec{r})$. A one-to-one correspondence between this density and an external potential, $v(\vec{r})$, exists and the potential acts on the electron density. The energy can be formulated in terms of a universal functional, $F[\eta(\vec{r})]$, of the electron density, which is independent of the external potential. In this way, the energy of a system's ground state can be computed with the following equation:
\begin{equation}
    E[\eta(\vec{r})]=\int v(\vec{r}) \eta(\vec{r})d\vec{r} + F[\eta(\vec{r})]
\end{equation}{}
The most widely used implementations of DFT rely on the Kohn-Sham approach.~\cite{Kohn1965PR} In fact, Kohn-Sham DFT is so successful that it is often simply referred to as DFT. In this approach, an auxiliary wave function in the form of a Slater determinant is employed. Since a single Slater determinant is the exact solution for a system of noninteracting electrons, this DFT approach can be seen as describing a system of noninteracting electrons that are forced to behave as if they were interacting. The latter effect can be achieved only by an unknown modification of the Hamiltonian or rather of the aforementioned functional. In other words, a Slater determinant as wave function ansatz is exact but the Hamiltonian can only be approximated, in contrast to Hartree-Fock, where the true electronic Hamiltonian is used but the Slater determinant is only an approximate wave function. 

The functional $F[\eta(\vec{r})]$ can be separated into Coulombic interactions and a non-Coulombic part. The latter can further be divided into two terms: the kinetic energy of the noninteracting electrons and the exchange-correlation part, which describes the interaction of electrons and thus also corrects the kinetic energy by the difference of the real kinetic energy and the kinetic energy of the fictitious system of noninteracting electrons. The exchange-correlation functional is the part of DFT that is unknown and finding the exchange-correlation functional remains the holy grail of DFT. 

In principle, if the exact functional was known, the exact ground-state energy of a system could be computed. Unfortunately, it is not known and the success of a DFT calculation critically depends on the approximation that is used to the unknown exchange-correlation functional. 
For completeness, KS-DFT is often used for closed-shell systems. In case of open-shell systems, two spin densities are distinguished, resulting in spin-polarized KS theory.~\cite{Casida2009JMST}

As explained above, the electron density is computed from a single reference Kohn-Sham wave function, i.e., the one of noninteracting electrons with the density of the real system. This single-reference wave function makes DFT a single-reference method. In fact, most failures of DFT are a consequence of an improper description of static correlation.~\cite{Benavides-Riveros2017PCCP}
In order to describe excited states, the time-dependent (TD) version of DFT, namely TDDFT, can be used. The foundation of this theory was laid in the 1980s with the Runge-Gross theorems~\cite{Runge1984PRL}, which can be regarded as analogies to the Hohenberg-Kohn theorems. They are based on the assumption that a one-to-one correspondence exists also between a time-dependent potential and a time-dependent electron density in this potential. A system can therefore be completely described by its time-dependent density. Also in the time-dependent case, the variational principle for the density is proposed. 

The most widely used approach of TDDFT is linear response TDDFT (LR-TDDFT). Again, often TDDFT is used synonymously with LR-TDDFT due to its extensive use. Within this theory and the KS approximation, no time dependent density is necessary to compute excitation energies and excited state properties. Linear response theory can be directly applied to the ground state density~\cite{Zangwill1980PRA,Chong1995}. Casida's formulation of this theory is the most popular one and gives rise to random-phase approximation pseudo-eigenvalue equations, which are also known as the Casida equations. Within the adiabatic approximation, they are implemented efficiently in many existing electronic structure programs. The Tamm-Dancoff approximation~\cite{Tamm1945JP,Dancoff1950PR} further simplifies the equations to an eigenvalue problem, resulting in the counterpart to CIS~\cite{Hirata1999CPL}. Especially in cases, when the time evolution of a system is studied, the Tamm-Dancoff approximation is beneficial, since it leads to more stable computations close to critical regions of the PESs~\cite{Yu2016JCP,Casida2009JMST}. 

The advantage of LR-TDDFT is its computational efficiency. The reasonable accuracy if a proper functional is chosen makes this approach often the method of choice to study the photochemistry of medium-sized to large and complex systems, which are not feasible to treat with costly multi-reference WFT based methods.~\cite{Cordova2007JCP,Lischka2018CR,Goerigk2020AJC}
Shortcomings of LR-TDDFT are the incorrect dimensionality of conical intersections, which are, however, one of the most important regions during nonadiabatic MD simulations~\cite{Worth2004ARPC,Doltsinis2006NIC,Jacquemin2016}. 
The incorrect dimensionality of conical intersections with standard TDDFT implementations leads to a qualitatively incorrect description of such critical regions. The missing couplings can be corrected for example with the CI-corrected Tamm-Dancoff approximation~\cite{Li2014JPCL} or the hole-hole Tamm-Dancoff approximation,\cite{Bannwarth2020chemrxiv}
which can recover the missing couplings and provide correct dimensionality at conical intersections. 

In addition, one should be aware that by definition, double excitations cannot be accounted for with LR-TDDFT. The computation of double excitations can be achieved by using a frequency dependent exchange kernel, which is known as dressed TDDFT~\cite{Maitra2004JCP,Elliot2011CP}. Alternatively, spin-flip TDDFT~\cite{Katriel2001IJQC,Shao2003JCP} can be used, where a triplet state is taken as a reference state and single excitations are treated with a flip in the electron's spin.
However, spin-contamination is quite common within these methods. In general, the description of double excitations from a multi-reference state would be more favorable, although spin-flip TDDFT is often considered to be a multi-reference method. In order to compute specific orbital occupations and consequently excitations and charge-transfer states, an alternative approximation exists, which is known as the $\Delta$-SCF approach. In this theory, the electrons are forced into specific KS orbitals. The SCF is applied to converge the energy with respect to this configuration~\cite{Gavnholt2008PRB,Maurer2011JCP,Maurer2013JCP}.
Other multi-reference variants of TDDFT exist too. However, their description is beyond the scope of this review and we refer the reader to a review covering this topic in much more detail.~\cite{Lischka2018CR}

Last but not least, we shortly want to discuss the most critical part of a DFT calculation, which is the proper choice of the exchange-correlation functional. In case of excited states, the treatment of valence excitations, Rydberg states and long-range charge transfer excitations on the same footing is highly problematic. While hybrid (meta-) generalized gradient approximation (GGA) or range-separated hybrid functionals~\cite{Chai2008JCP} are for example well suited for vertical excitations and the latter also for Rydberg states, global hybrid meta GGA or range-separated hybrid GGA functionals are better to describe charge transfer~\cite{Tozer2000PCCP,Yu2016JCP}. Most often, functionals are accurate for one specific problem, but they fail to describe others. Although much effort has been devoted to develop functionals, finding a universal functional for DFT is still far from being achieved~\cite{Casida2009JMST,Yu2016JCP,Akimov2015CR,Lischka2018CR}.
\\~
\\
In summary, it should be stressed that, in general, there is not only one single solution to a particular problem, but that many possible ways can be considered which lead to an equivalent description of a particular problem. Considering the excited states of molecules, it should be mentioned that it is of utmost importance to think carefully about the photochemical processes that may occur in order to find the most appropriate method for most of the assumed reactions. It often happens that within the same molecular system, one method can describe a certain photochemical reaction quite well, while another reaction can be described better with another method. However, the mixing of methods is not practicable for standard applications. Recently, studies on ML models have emerged that combine the different strengths of several methods, e.g. $\Delta$-learning techniques~\cite{Ramakrishnan2015JCP,Dral2020JCP} or transfer learning~\cite{Smith2019NC}. 
These methods could be well-suited solutions for many future applications to overcome the current limitations of existing quantum mechanical methods for the excited states. Even more than for ground state properties, the quality of the excited states depends critically on the ability of a method to describe the different possible reactions - as a consequence of the larger accessible configuration space of a molecular system. Even for medium-sized systems it should be clear that a suitable method may already be computationally impracticable and a balance between accuracy and computational effort has to be found. 

\subsection{Bases}
\begin{figure*}[ht]
    \centering
    \includegraphics[scale=0.4]{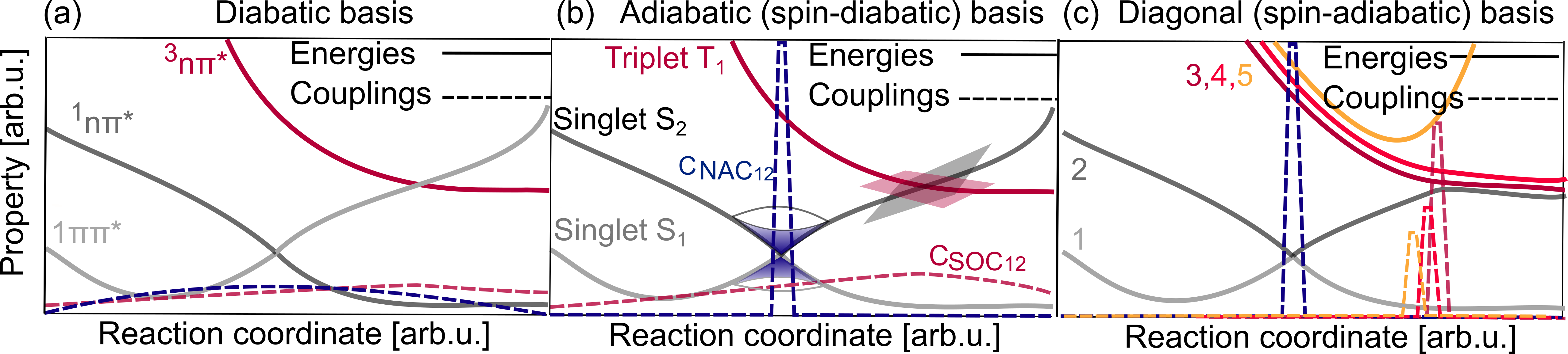}
    \caption{(a) Example of three potential energy curves ordered by their character along with respective potential couplings between different states shown by dashed lines. (b) Two singlets (E$_i$ and E$_j$) and one triplet state (E$_k$) including coupling values (with vectorial properties, C$_{ij}^{NAC}$, shown by their norm) in the the adiabatic basis, in which the triplet state crosses singlet states. (c) The diagonal, or spin-adiabatic, basis, in which all states are ordered by their energy and are spin-mixed. Kinetic couplings are shown by their norm. Note that the ground state is not shown.}
    \label{fig:diab_adiab}
\end{figure*}

The potentials computed with the aforementioned methods for different nuclear geometries can be represented in different bases, which are connected by unitary transformations. An example of five states in different bases are given in Figure~\ref{fig:diab_adiab}. Note that often a system in a certain basis is also referred to as being in a certain picture or representation; here we will not use the term representation in order to not confuse the reader with molecular representations used in ML. As it is visible in the figure, we focus on three types of bases: (a) the diabatic basis, (b) the adiabatic (spin-diabatic) basis, i.e., the direct output of standard electronic structure programs, (c) the diagonalized version of (a) and (b), i.e., the spin-adiabatic basis. Throughout literature, different names are given to these bases, which are summarized in Table~\ref{tab:bases}. 
They stem from a partition of the total wave function into a sum of electronic and nuclear contributions, which can be written for all bases as:
\begin{equation}\label{eq:wfbasis}
    \Psi(r,\mathbf{R},t)=\sum_i\psi_i^{\text{basis}}(r,\mathbf{R})\chi_i^{\text{basis}}(\mathbf{R},t).
\end{equation}
In a similar way as the number 20 can be factored into $4\cdot 5$ or $4.5\cdot 4.\bar{4}$, the total wave function can be expanded in the different bases. Here, $\psi_i^{\text{basis}}(r,\mathbf{R})$ corresponds to the eigenfunctions of the electronic Hamiltonian only for one of the bases (namely the one from column B of table~\ref{tab:bases}). Associated with these functions are the corresponding potentials, depicted for a model system in Fig.~\ref{fig:diab_adiab}.
Note that a different approach is taken in the exact factorization method,\cite{Abedi2010PRL} 
where the total wave function is expanded only in a single product, i.e., without the sum in eq.~\ref{eq:wfbasis}, giving rise to only one (time-dependent) potential.

\begin{table}[ht]
    \centering
    \begin{tabular}{ccc}
    \hline
a &b &c\\
     \hline\hline
        \parbox{2.4cm}{diabatic}& \parbox{2.4cm}{adiabatic} & \parbox{2.4cm}{diagonal}  \\\parbox{2.4cm}{crude adiabatic}&
         \parbox{2.4cm}{spin-diabatic}  &\parbox{2.4cm}{spin-adiabatic} \\
        \parbox{2.4cm}{spectroscopic}& \parbox{2.4cm}{MCH} &\parbox{2.4cm}{field-adiabatic}\\ 
         \parbox{2.4cm}{quasi-diabatic}&\parbox{2.4cm}{field-free} &\parbox{2.4cm}{field-dressed} \\
         \end{tabular}
    \caption{Commonly used names of bases for the excited-state potential energy surfaces based on refs~\citenum{Koeppel04,Thachuk1996JCP,Mitric2009PRA,Mitric2011PCCP,Granucci2012JCP,Mai2018WCMS}. The labels a, b, and c are consistent with Fig.~\ref{fig:diab_adiab}}.
    \label{tab:bases}
\end{table}
{}

\subsubsection{Adiabatic (Spin-Diabatic) Basis}
The direct output of an electronic structure calculation usually provides the eigenenergies and eigenfunctions of the electronic Hamiltonian. In many cases, only one spin multiplicity is calculated. If this procedure is repeated along a nuclear coordinate, potential curves result that are termed adiabatic. Adiabatic means ''not going through'' (from greek a=not, dia=through, batos=passable) and, indeed, the potentials never cross when considering one multiplicity. This situation is schematically illustrated in Figure~\ref{fig:diab_adiab}(b) for singlet S$_i$ and singlet S$_j$.

Within one multiplicity, 3N$_A$-dimensional adiabatic PESs are obtained that are strictly ordered by energy. Hence, the states are usually denominated with the first letter of the multiplicity and a number as subscript, e.g., S$_0$, S$_1$, etc. For states of the same multiplicity, critical points and seams exist. These regions of the PESs are referred to as conical intersection (seams), in which the corresponding states become degenerate.
Such features make adiabatic PESs non-smooth functions of the atomic coordinates, which make them difficult to predict with the intrinsically smooth regressors of ML. At a conical intersection, the approaching potential energy curves form a cone and the NACs, denoted as $C_{\text{NAC}_{ij}}$, between them show singularities as a result of the inverse proportionality to the vanishing energy gap~\cite{Baer2002PR,Lischka2004JCP}:
\begin{equation}
\begin{array}{lr}
    \label{eq:nac5}
        C_{\text{NAC}_{ij}} \approx \langle  \Psi_i \mid \frac{\partial}{\partial \mathbf{R}}\Psi_j \rangle =\\
        \frac{1}{E_i-E_j}{\langle \Psi_i \mid \frac{\partial H_{el}}{\partial \mathbf{R}}\mid \Psi_j \rangle }   ~~~\text{for}~  i \neq j,
\end{array}{}
\end{equation}
Second order derivatives are neglected here, as is done in many quantum chemistry programs that compute NAC vectors. The blue dashed curve in panel (b) of Fig.~\ref{fig:diab_adiab} illustrates the norm of the NAC vector, C$_{ij}^{NAC}$, that couples the states S$_i$ and S$_j$. At the avoided crossing points of the states, the NAC norm shows a sharp spike, but is almost vanishing elsewhere. 
If more than one multiplicity is considered, the term adiabatic is not adequate anymore, because potentials of different multiplicity might cross through each other. This situation is then called diabatic with respect to the spin multiplicities, or spin-diabatic in short. For example, singlets are adiabatic among each other, triplets are adiabatic among each other but singlets are diabatic with respect to triplets. However, also the diabatic basis (see Fig.~\ref{fig:diab_adiab}(a) and also below) qualifies as spin-diabatic. Due to this nomenclature issue, which even gets experts confused sometimes, we refer to this basis as MCH (Molecular Coulomb Hamiltonian) because it is obtained from the eigenfunctions and eigenvalues of the non-relativistic electronic Hamiltonian, where only Coulomb interactions are considered.

As an example, a crossing of a singlet state and a triplet state is shown in Fig.~\ref{fig:diab_adiab}(b). As it is visible, the triplet components, which are defined by different magnetic quantum numbers, are degenerate. The states are coupled by SOCs (denoted as $C_{\text{SOC}_{ij}}$), which are usually obtained as smooth potential couplings
with standard quantum chemistry programs~\cite{Mai2018WCMS,Granucci2012JCP,Mai2020ACIE}: \begin{equation}\label{eq:soc}
C_{\text{SOC}_{jk}}=\langle \Psi_j \mid \hat{H}^{SO}\mid \Psi_k \rangle .
\end{equation}
These couplings are single real-valued or complex-valued properties~\cite{Penfold2018CR,Kryachko2000IJQC}.  Whether they are complex or not depends on the electronic structure program employed, but they can be converted into each other.~\cite{Marian2012WIRE,Mai2018WCMS,Crespo-Otero2018CR,Penfold2018CR}

$\hat{H}^{SO}$ in eq. \ref{eq:soc} is the spin-orbit Hamilton operator, which describes the relativistic effect due to interactions of the electron-spin with the orbital angular momentum, allowing states of different spin-multiplicities to couple.~\cite{Dyall2007,Reiher2009,Marian2012WIRE}
Note that also SOCs between different states of the same multiplicity exist except for singlets.
No exact expression on how to include relativistic effects into the many-body equations has been found, yet. Among the most popular approximations used is the Breit equation~\cite{Bethe1957}, applying an adapted Hamiltonian instead of the electronic Hamiltonian, which comprises, among other terms, a relativistic part. This additional part of the Hamiltonian accounts for spin-orbit effects and is proportional to the atomic charge,~\cite{Bethe1957,Reiher2009,Marian2012WIRE,Crespo-Otero2018CR,Penfold2018CR} leading to the belief that SOCs would only be relevant in systems with heavy atoms.~\cite{Pyykko1988CR,Neese2007CCR} Today it is known, that spin-orbit effects also play a crucial role in many other molecular systems and are important for intersystem crossing between states of different spin multiplicities.~\cite{Neese2005JCP,Richter2011JCTC,Mai2015IJQC,Gonzalez2020}

The states in the MCH basis can also be coupled via external electric-magnetic fields, e.g., by sunlight or a laser. The corresponding couplings stem from the transition dipole moments multiplied with the electric field. Since the effect of the field is not included in the potentials but as off-diagonal potential couplings, the MCH basis is also called field-free.~\cite{Thachuk1996JCP,Mitric2009PRA,Mitric2011PCCP,Mai2018} However, also the diabatic basis qualifies as field-free.

\subsubsection{Diabatic Basis}\label{sec:diabatic}
In the diabatic basis, the electronic wave function is not parametrically dependent on the nuclear coordinates. Note that such a strictly diabatic basis for polyatomic systems does not exist in practice and only approximated, so called, quasi-diabatic, PESs can be fit. In literature, quasi-diabatic PESs are most often referred to as diabatic ones, so we will also use this notation here. Further, diabatic potentials usually need to be determined from adiabatic potentials and are not unique, i.e., they rely on the method and the reference point, which is chosen in the adiabatic basis to fit diabatic potentials.~\cite{Koeppel04,Mai2018WCMS}

An example of a system in the diabatic basis as given in panel (a) of Figure~\ref{fig:diab_adiab} and commonly used notations can be found in Table~\ref{tab:bases} in the first column.
In regions, where an avoided crossing is present in the adiabatic basis, the coupled diabatic potential energy curves cross. Since the electronic wave function of a state is ideally independent of the nuclear coordinates, its character is conserved. Consequently the states are labeled according to their character and multiplicity, e.g., as $^1\pi\pi\ast$ or according to symmetry labels. Similar to the character, also spectroscopically important quantities like the dipole moment are mostly conserved or vary smoothly along the nuclear coordinates. Therefore, spectroscopic experiments can easily be interpreted when using the diabatic basis, which is thus sometimes also called spectroscopic basis. Note that sometimes labels like S$_1$, etc. are used also when referring to the diabatic basis, especially in experimental papers when an identification of the wave function's character has not been carried out and only one geometry is considered. However, at a different geometry, the energetic order of the states might have changed such that a state previously labeled as S$_2$ might now be lower in energy than a state previously labeled as S$_1$. Furthermore, this labeling scheme in the diabatic basis can lead to confusion with the labels from the MCH basis, and we suggest to reserve it only for the MCH basis.

Due to the mostly conserved characters and the crossing of states, diabatic potentials are smooth functions of the nuclear coordinates, in contrast to adiabatic potentials. A diabatic PES is thus highly favorable for several numerical applications including ML.

The MCH and diabatic bases can be interconverted by a unitary transformation
\begin{equation}\label{eq:atd}
\Psi^{MCH}(\mathbf{r},\mathbf{R})=\mathbf{U}(\mathbf{R})\Psi^{diab}(\mathbf{r},\mathbf{R})
\end{equation}
with a unitary matrix, $\mathbf{U}$, that is determined up to an arbitrary sign (as a result of the arbitrary sign of the wave function, which will be discussed in detail in section~\ref{sec:phase}). 
In the case of two states, $\mathbf{U}$, is a rotation matrix:
\begin{equation}\label{eq:rotation}
    \mathbf{U}=\left( 
 {\begin{array}{cc}
   cos\theta(\mathbf{R}) & -sin\theta(\mathbf{R}) \\
   sin\theta(\mathbf{R}) & cos\theta(\mathbf{R}) \\
  \end{array} }
    \right)
\end{equation}{}
and is dependent on the rotation angle, $\theta$.
Accordingly, the peaky NACs, which are obtained as derivative couplings (also called kinetic couplings) in the MCH basis, are converted to smooth potential couplings in the diabatic basis. The smooth SOCs from the MCH basis become even smoother (ideally constant) in the diabatic basis. 

While one can straightforwardly apply diagonalization to convert diabatic PESs to adiabatic PESs (and similarly adiabatic PESs to diagonal PESs), a dilemma arises when one wants to take the inverse way to obtain diabatic PESs from adiabatic ones (and similarly adiabatic PESs from diagonal ones). In fact, finding diabatic PESs is highly complex and most often requires expert knowledge. Up to date, only small molecules could be represented with accurate diabatic potentials and developing a method to automatically generate diabatic PESs remains an active field of research. Existing methods to obtain diabatic potentials require human input and are mostly applicable to small systems and certain reaction coordinates. Early pioneering works can be found in refs~\cite{Koeppel2001JCP,Koeppel04}. Today, a lot more variants exist. Examples are the propagation diabatization~procedure~\cite{Richings2015JPCA}, diabatization by localization~\cite{Accomasso2019CPC}, Procrustes diabatization~\cite{Richings2020JCP} or diabatization by ansatz~\cite{Lenzen2017JCP,Williams2018JCP}. Further, methods can be based on couplings or other properties~\cite{Subotnik2008JCP,Hoyer2016JCP,Wittenbrink2013JPCA,Varga2018PCCP}, configuration uniformity~\cite{Nakamura2002JCP}, block-diagonalization~\cite{Cave2014JCP,Venghaus2016JCP}, CI vectors~\cite{Robertson2019JCC} or (partly) on ML~\cite{Li2013JCP,Jiang2013JCP,Jiang2014JCP,Jiang2016IRPC,Lenzen2017JCP,Williams2018JCP,Xie2018JCP}.

\subsubsection{Diagonal Basis}\label{sec:diagonal}
As the name indicates, the diagonal basis can be obtained by a diagonalization from the MCH or diabatic bases. In this case, a strictly adiabatic picture is obtained, where states never cross.~\cite{Mai2018WCMS} Accordingly, the concept of multiplicity for a single state is lost because the state might be of singlet character in one region and of triplet character in another region. Therefore, the basis is also called spin-mixed or spin-adiabatic.~\cite{Kryachko2000IJQC,Crespo-Otero2018CR,Mai2014} The states are strictly ordered by energy and can be labeled simply with numbers (see Fig.~\ref{fig:diab_adiab}(c)). The resulting wave functions are eigenfunctions of the relativistic electronic Hamiltonian.\cite{Mai2018WCMS,Richter2011JCTC,Crespo-Otero2018CR} These eigenfunctions as well as the eigenenergies can be also obtained directly with e.g. relativistic two-component or four-component calculations,\cite{Liu2020JCP} instead of via diagonalization. 

In this basis, the effect of the SOCs are incorporated into the PESs to a large extent. What remains are localized kinetic couplings, which are similar in nature to the NACs in the MCH basis.
An example is given in Fig.~\ref{fig:diab_adiab}(c). The parts of the potentials that correspond to the different triplet components in the MCH basis are split energetically in the diagonal basis. In the case of small SOCs, the diagonal potentials look similar to the MCH potentials. However, if the SOCs are strong, potentials that are degenerate in the MCH basis can be easily shifted apart by 1 eV in the diagonal basis. Such splittings are then also experimentally observable, and the diagonal basis yields a more intuitive interpretation of these experiments.~\cite{Horton2019JCP,Liu2020PRX,Horton2018PRA}

As mentioned above, the states in the MCH basis can also be coupled via electromagnetic fields. A diagonalization of the potential matrix then yields so-called field-dressed states or light-induced potentials, which can also be termed field-adiabatic.\cite{Sussman2006S,Thachuk1996JCP,Marquetand2011FD,Bajo2012JPCA,Mai2018} Since the fields are usually time-dependent, the most important axis along which the potentials in this field-dressed basis need to be plotted is time.~\cite{Mai2018} 

In principle, all these bases are equivalent but only if an infinite number of terms is considered in eq.~(\ref{eq:wfbasis}). In practice, potentials represented in different bases have different advantages for dynamics simulations, especially in combination with different approximations made in the different dynamics methods as outlined below.

\subsection{Excited-State Dynamics Simulations}\label{sec:MD}
In order to investigate the temporal evolution of an isolated molecular system in the excited states, the time-dependent Schr\"odinger equation has to be solved:\cite{Tannor2006}
\begin{equation}
    i\hbar \frac{\partial \Psi(\mathbf{r},\mathbf{R},t)}{\partial t} = \hat{H}_{el}(\mathbf{r},\mathbf{R})\Psi(\mathbf{r},\mathbf{R},t).
\end{equation}{}
From a technical point of view, a sequence of time steps is computed, where in every step the electronic problem is solved to yield potentials, which determine the forces acting on the nuclei such that the nuclear equations of motion can be solved for the current time step.

\begin{figure}
    \centering
    \includegraphics[scale=0.5]{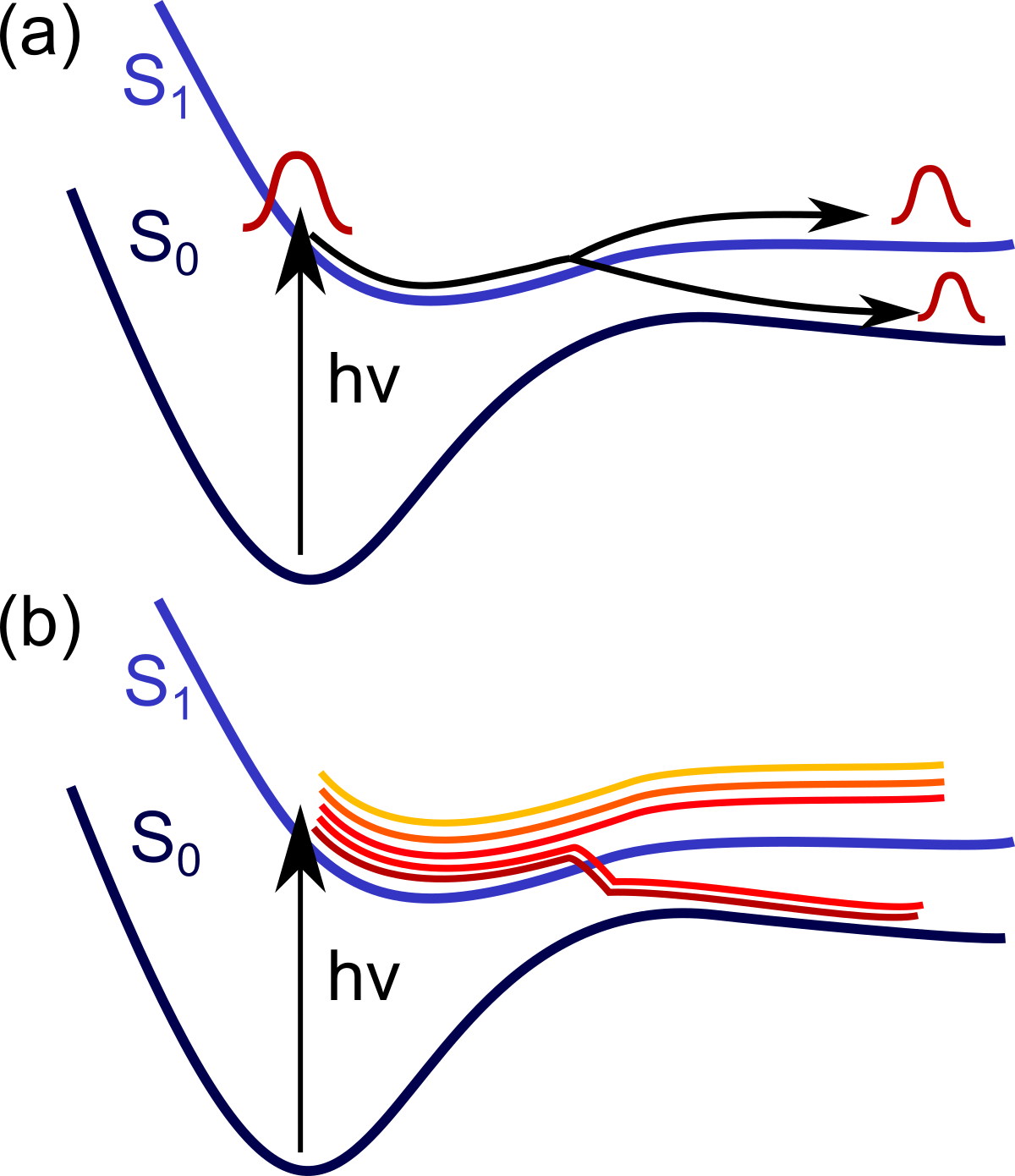}
    \caption{Excited-state dynamics can be treated with (a) quantum approaches, where wave functions are used for the nuclei, or (b) classical approaches, based on trajectories.}
    \label{fig:MQCD}
\end{figure}

Ideally, the nuclei are treated quantum mechanically. In this case, the PESs are usually computed in advance and either interpolated or stored on a grid for later use. The hope is that ML can improve the interpolation of potentials drastically. Such global PESs are needed because a wave function is employed for the nuclei, which extends over a range of nuclear coordinates at the same time (see Fig.~\ref{fig:MQCD}(a)). An overview over corresponding dynamics methods is given in section~\ref{sec:qd}.

The nuclear dynamics can also be approximated classically while quantum potentials are used, i.e., mixed quantum classical dynamics (MQCD) simulations are carried out. Such methods is discussed in section~\ref{sec:MQCD}. Since the classical nuclear trajectories are defined only at one nuclear geometry at a time (see Fig.~\ref{fig:MQCD}(b)), on-the-fly calculations of the potential energies are possible. An on-the-fly scheme is computationally advantageous, if the number of visited geometries during the dynamics is smaller than the number of points needed to represent the conformational space on a grid or via interpolation.~\cite{Doltsinis2006NIC,Koeppel84ACP,Worth2004ARPC,Richter2011JCTC,Mai2018WCMS,Ibele2019MP,Mai2020ACIE} 
No fitting of PESs is necessary in an on-the-fly approach but fitted PESs can still be used as an alternative. Since ML approaches provide such interpolated potentials, the amount of training points generated with quantum chemistry must be less than the number of points needed in an on-the-fly approach in order to be advantageous. This demand is satisfied, e.g., for long time scales or if many trajectories are necessary.

In the following, we will shortly discuss the different types of nuclear motion and the opportunities of ML models to enhance the respective dynamics simulations. 
{}

\subsubsection{Quantum Nuclear Dynamics}\label{sec:qd}
The computational cost of an exact nuclear dynamics simulation scales exponentially with the nuclear degrees of freedom. Hence simulations are limited to small systems, typically containing less than 5 atoms~\cite{Penfold2018CR,Nelson2020CR,Ben-Nun2002}. 
Still, the calculation of the PESs of the molecule can be a rather expensive part of the whole scheme and the use of ML algorithms is advisable even for such small systems. 

To treat larger systems, approximations have to be invoked. A prominent approach that can be converged to the exact solution is the 
multi-configurational time-dependent Hartree (MCTDH) approach~\cite{Beck2000PR,Curchod2018CR,Yeager1979CPL,Manthe2017JPCM}.
Its high efficiency stems from the use of time-dependent basis functions to represent the nuclear wave functions. Nonetheless, the computations are computationally costly and the nuclear degrees of freedom are often reduced to only a few important key coordinates,\cite{Koeppel04,Eng2015ACR} 
where classical simulations can help identifying the latter.\cite{Gomez2019JPCA}
Whether quantum dynamics of such reduced-dimensionality models are better than using classical dynamics of a full-dimensional system is still under debate and probably depends on the system. The potentials need to be presented to the algorithm in the diabatic basis, mostly due to numerical stability (e.g., smooth couplings are easier to integrate than singular ones). Since more than 20 years, (modified) Shepard interpolation is used to fit diabatic potentials~\cite{Ischtwan1994JCP,Bettens1999JCP,Evenhuis2004JCP,Evenhuis2011JCP,Mukherjee2013JPCA}. Notably, the grow algorithm~\cite{Bettens1999JCP} can be used to efficiently generate the database of points upon which the interpolation is based. However, it is clearly desirable to treat larger systems, and ML models like neural networks (NNs) promise higher performance or more flexibility in such cases.~\cite{Li2013JCP,Jiang2013JCP,Jiang2014JCP,Richings2015JPCA,Jiang2016IRPC,Richings2017CPL,Xie2018JCP,Guan2019JCP,Richings2019JCTC}

More recently, on-the-fly methods addressing quantum dynamics have been developed.~\cite{Worth2008MP,Persico2014TCA,Komarova2018PCL,Richings2018JCP} They mostly rely on a combination of Gaussians to represent the nuclear wave function.\cite{Ibele2019MP} For example, the variational multi-configuration Gaussian method (dd-vMCG)~\cite{Lasorne2007PCCP} offers a variational and thus accurate solution for the equations of motion. Also full multiple spawning\cite{Martinez2006ACR,Ben-Nun2002,Ben-Nun2000CP} can be regarded as fully quantum mechanically by describing the wave function with a number of time-dependent Gaussian functions, that follow classical trajectories with quantum mechanically determined time-dependent coefficients. In its more affordable ab-initio multiple spawning variant, more approximations are introduced such that the results sometimes draw near the classical solutions.\cite{Curchod2016JCP,Mignolet2018JCP}
Further related methods exist, like the ab-initio multiple cloning method,\cite{Freixas2018PCCP} or the thawed Gaussian approximation.\cite{Begusic2018CP}

Another class of dynamics methods are semi-classical approaches, which allow the inclusion of quantum effects in the classical dynamics of nuclei, such as quantum mechanical tunnelling or coherence.~\cite{Markland2018NRC} Note that these methods, where the nuclear dynamics is treated semi-classically, should not be confused with the MQCD approaches (see below) that are also often termed semi-classical (because the nuclei are treated classically and the electrons quantum-mechanically). The semi-classical dynamics methods range from the initial value representation,~\cite{Miller1970JCP,Ceotto2009PCCP}
adapted with the Zhu-Nakamura approach leading to the Zhu-Nakamura-Herman-Kluk initial value representation,~\cite{Nakamura2016PCCP} to path integral approaches.\cite{Gao2020JCTC}

The path integral formalism is especially interesting when the quantum and classical degrees of freedom should be coupled in a dynamically consistent manner. By using so-called ring-polymers, i.e., replica of the original classical system, a deviation of the nuclear dynamics from the classical path can be obtained and the time evolution of a system including nuclear quantum effects can be investigated. However, ring-polymer dynamics suffer from high computational efforts as a consequence of the large number of replica required. Accelerated formalism exist, which are for example implemented in the Python wrapper i-PI,~\cite{Ceriotti2014CPC,Kapil2019CPC}
which allows to interface path-integral methods with programs that provide PESs, but are mostly dedicated to the electronic ground state. Up to date, only a few implementations of semi-classical methods in atomistic simulation software are available. Compared to classical mechanics, the computational costs increase by a factor of about 10 to 100.~\cite{Markland2018NRC,Thoss2000JCP,Lee2016ARPC}

\subsubsection{Mixed Quantum-Classical Molecular Dynamics}\label{sec:MQCD}
While semi-classical methods are promising to simulate the dynamics of molecular systems containing up to tens of atoms highly accurately, the study of larger systems is still dominated by computationally cheaper MQCD methods, where the nuclear motion is treated fully classically.~\cite{Nelson2020CR,Stock1997PRL,Thoss2000JCP,Lee2016ARPC} In  contrast to quantum dynamics, the motion of the nuclei can be computed very fast using classical mechanics, and the computation of the PESs, on which the nuclei are assumed to move, remains the time limiting step. In this sense, ML models have a huge potential to enhance MQCD simulations by providing the electronic PESs and enabling the investigation of reactions that are not feasible with conventional approaches.~\cite{Westermayr2020MLST_Perspective,Westermayr2020JPCL,Weinreich2020JPCC,Lin2020JCP}
In fact, most studies that describe photochemistry with ML up to date aim to replace the quantum chemical calculation of the PESs in MQCD approaches.

The most popular MQCD method is trajectory surface hopping,~\cite{Tully1990JCP,Tully1991IJQC,Tully1998FD} schematically represented in Figure~\ref{fig:MQCD}(b). 
A manifold of independent trajectories is required to obtain statistically relevant results and to mimic the extended nuclear wave functions. For a single trajectory, the nuclei move classically on one of the quantum potentials, hence only one state is considered to be active, but transitions between different states are allowed.\cite{Mai2020}

Different approaches exist to determine the probability of such a transition, also called hop or jump in surface hopping methods. To this aim, different quantities are needed that are commonly provided in the MCH basis, as it is the direct outcome of a quantum chemical simulation. One of the first implementations to compute the hopping probability is based on the Landau-Zener formalism.~\cite{Zener1932,Wittig2005JPCB}  Based on the Landau-Zener formula, the potential energy differences are used to determine the hopping probability. No information about couplings is required, which implies that the approach must fail for  states that do not couple but lie close in energy. Very similar to this approach is the Zhu-Nakamura theory~\cite{Zhu2001JCP,Zhu2002JCP,Oloyede2006JCP,Ishida2017IRPC}. Also here, the computation of couplings is omitted and only information about PESs is used. Among the mostly used hopping algorithm is Tully's fewest switches algorithm~\cite{Tully1990JCP}, which is valid for many cases and based on the NACs between different PESs. An extension to other couplings is provided e.g. in the SHARC (surface hopping including arbitrary couplings) method.\cite{Richter2011JCTC} When couplings are considered, an internal transformation from the MCH basis to the diagonal basis is most advantageous because the localized couplings of the diagonal picture precisely indicate, where the few switches of the fewest switches approach should take place. In cases, where the PESs are fit in advance, either with ML models or other types of analytical functions, the use of a diabatic basis is favorable (because of the Berry phase, see below) but should be transformed to the diagonal picture for the calculation of hopping probabilities. Other flavors to account for transitions exist. However, they have not been applied in simulations with ML algorithms yet. Interested readers are therefore referred to refs~\cite{Tully1990JCP,Zhu1995PRL,Zhu2002JCP,Doltsinis2006NIC,Granucci2007JCP,Fabiano2008CP,Richter2011JCTC,Malhado2014FC,Wang2016JPCL,Subotnik2016ARPC,Mai2018WCMS,Crespo-Otero2018CR} for further information.

The bottleneck of approaches that require NACs is that the computation of the couplings remains one of the most expensive part of a quantum chemical calculation. The computational effort to compute a NAC vector is comparable to that of a force calculation. However, more NACs are present than there are forces, i.e. $N_S \times (N_S-1)/2$ NACs need to be computed, whereas $N_S$ forces are needed (respectively with entries for the Cartesian coordinates of each nucleus). Note that in case of fitted PESs with ML, all of these vectors have to be computed for each data point. Conventional approaches with an ab-initio on-the-fly evaluation of the PESs can make use of the fact, that only one active state needs to be considered at a certain time step. Many MD programs therefore only require a computation of the forces of the active state and the respective couplings arising from this state. 

Note that despite the benefits of MQCD simulations, they obey micro-reversibility only approximately\cite{Subotnik2015JPCA}
and effects due to coherences or tunneling necessitate additional considerations as a consequence of the classical treatment of nuclear motion.\cite{Hammes-Schiffer1994JCP}

A more approximate approach is the Ehrenfest dynamics method, also referred to as mean-field trajectory method. It is often used for large systems and also frequently in material science.~\cite{Dou2020JPCA,Douglas-Gallardo2019N} The Ehrenfest method is based on the approximation that nuclei move classically on an average potential, rather than switching from one specific state to another.\cite{Sawada1985PRB,Doltsinis2006NIC,Li2005JCP} Due to the treatment of each electronic state separately, surface hopping methods allow the accurate bifurcation into different reaction channels, while such effects are neglected in a mean-field treatment of PESs.

The main limitation of MQCD approaches are the expensive evaluation of ab-initio potentials, which allows dynamics simulations only for up to a couple of picoseconds. In addition, rare reaction channels are hardly explored as a result of usually bad statistics~\cite{Mai2016JPCL,Mai2017CP,Crespo-Otero2018CR}. In this sense, MQCD simulations offer a perfect place for ML to enter this field of research and advance it significantly. The fast evaluation of the ML PESs can help to explore different reaction channels and to obtain accurate reaction kinetics. Observables and macroscopic properties can be computed directly or with post-processing as well as analysis runs, and offer another fulcrum for ML. The computed observables should then be directly compared to experiments.

\subsection{Dipole Moments and Spectra}
An important property for comparing experiment and theory is the dipole moment. The permanent dipole moment of the ground state is a frequent target of studies with ML~\cite{Ramakrishnan2014SD,Artrith2011PRB,Huang2016JCP,Gastegger2017CS,Yao2018CS,Schuett2018JCP,Nebgen2018JCTC,Sifain2018JPCL,Schuett2019,Schuett2019JCTC,Christensen2019JCP,Veit2020arXiv,Gastegger2018arXiv}. The permanent dipole moment, $\mu_i$ (or $\mu_{ii}$) , of a state $i$ can be obtained via the dipole moment operator (see eq. (\ref{eq:dipole}) below) or as the sum of partial charges, $q_{a,i}$ of atom $a$ in state $i$, and the vector that describes the distance of the position of atom $a$ to the center of mass of the molecule, $r_\alpha$: $\mu_i = \sum_a^{N_A} = q_{a,i} r_\alpha$.
It can be used for the computation of infrared spectra with MD simulations. The spectrum is then obtained as the Fourier transform of the time auto-correlation function of the time derivative of the dipole moment~\cite{Thomas2013PCCP}.

In contrast to the ground state, excited-state simulations often make use of the transition dipole moments, which are computed from the dipole moment operator within many quantum chemistry programs:
\begin{equation}\label{eq:dipole}
    \mu_{ij}= \langle \Psi_i\mid\hat{\mu}\mid\Psi_j\rangle.
\end{equation}{} The ground state dipole moment can differ strongly from those in the excited states, due to a frequency shift and altered electron distribution upon light-excitation~\cite{Wilke2016JCP}. 

Transition and permanent dipole moments can be fit with the charge model of ref.~\citenum{Gastegger2017CS}, where point charges are never learned directly, but instead are inferred as latent variables by an NN dipole model making use of r$_\alpha$.\cite{Gastegger2018arXiv}
Noticeably, the computation of absolute values of permanent and transition dipole moments is very challenging even when highly accurate quantum chemistry methods are employed and experimental values are hardly reproduced~\cite{Tennyson2016JCP,Westermayr2020MLST}. However, also experimental studies provide absolute values only in few cases. Most computational studies therefore do not aim to reproduce the absolute values of transition dipole moments but rather use relative values to obtain reasonably accurate absorption spectra, which can be compared to experiments.~\cite{Tannor2006,Marquetand2017M,Nogueira2017CS,Nogueira2018ARPC,Mai2018WCMS,Barbatti2007JPPA} Since many molecules absorb in the UV, the terms UV spectra and absorption spectra are often used interchangeably. However, absorption can take place in many regions of the electromagnetic spectrum, including, e.g., X-rays, where rather core electrons than valence electrons are excited.\cite{Norman2018CR}

As already mentioned shortly, absorption spectra can be obtained from a calculation of excited-state energies and oscillator strengths, which are proportional to the squared transition dipole moments. Noticeably, the transition dipole moment is only defined up to an arbitrary sign as a result of the arbitrary phase of the wave function (see section~\ref{sec:phase}).  To circumvent this ill-definition, oscillator strengths or the lengths of dipole vectors can be fitted with ML. However, this workaround can be problematic if explicit field-dipole interactions should be considered with ML models. 

\section{Data Sets for Excited States}
The basis of any successful ML model is a comprehensive and accurate training set that can describe the required conformational space of a molecule comprehensively and accurately with as little noise as possible~\cite{Dral2020JPCL}. 
While electronic structure theory for ground state problems is almost free of noise, the same cannot be said so easily for problems in the excited states. "Bad points with abrupt changes"~\cite{Guan2020JCTC} within ab-initio calculations for the excited states are frequently observed, which can occur even far away from any critical point of the PESs and are difficult to detect~\cite{Westermayr2019CS,Westermayr2020JPCL,Guan2020JCTC}. 
The amount of noise in the reference data does not only depend on the chosen method (and in case of multi-reference methods on the selected active space), but also on the number of electronic states considered and the photochemistry of the molecule under investigation. 

\subsection{Choosing the Right Reference Method for Excited-State Data}
Many existing training sets for ML in quantum chemistry are based on DFT~\cite{Gastegger2015JCTC,Chen2019JPCL,Chen2020JPCA_Hessian,Behler2008PRL,Li2015PRL,Gastegger2018JCP,Deringer2017PRB}.
The ease of use and low computational costs of DFT-based methods make them suitable to treat large systems with acceptable accuracy. In fact, DFT is the workhorse of many studies solving ground-state problems. In contrast, TDDFT has not yet managed to equal DFT for the treatment of excited-state problems. Consequently, training sets for the excited states are less frequently computed with TDDFT~\cite{Carbogno2008PRL,Carbogno2010PRB,Ramakrishnan2015JCP,Haese2016CS,Zhang2020arXiv,Chen2020arXiv} and rely most often on multi-reference methods. Examples of applied methods are CASSCF~\cite{Richings2017CPL,Hu2018JPCL,Chen2018JPCL,Dral2018JPCL,Richings2018JCP,Richings2019JCTC,Haese2019CS,Westermayr2020JPCL} or MR-CI schemes~\cite{Koch2014JCP,He2016SR,Guan2017JCP,Williams2018JCP,Wang2018SR,Yuan2018PCCP,Yin2019PCCP,Guan2019PCCP,Westermayr2019CS,Westermayr2020MLST,Schwilk2020arXiv,Guan2020JPCL,Guan2020JCTC,Westermayr2020JPCL,Shen2020JPCA}, where the latter method is more expensive than the former and therefore limited to describe small systems. 

In general, the computation of excited-state PESs is much more expensive than the computation of the ground state potential of the same molecule. Not only highly accurate ab-initio methods have to be applied for many systems, but also forces and couplings are required for the considered states. A high density of electronic states present in a molecular system can thus increase the costs of a calculation considerably. In this regard, an active, efficient and meaningful training set generation is indispensable, especially when photodynamics simulations are the target of a study. 

Keeping in mind, that the quality of the reference data confines the quality of an ML model, several key questions can be identified when designing a study based on ML potentials. 
We believe the following questions to be important for the selection of a suitable reference method: 
\\
\indent 1) What is the goal of an ML model and what properties must it predict in order to benefit from the advantages that ML can offer? Are only energy gaps of different electronic states to the electronic ground state necessary or are gaps between other states and couplings between them also relevant? Especially the description of couplings requires further consideration, as they cannot be calculated with all quantum chemistry methods and additionally face the problem of random sign jumps along different reaction coordinates~\cite{Akimov2018JPCL,Westermayr2019CS,Guan2020JPCL}. 
\\
\indent 2) How many excited states are relevant and which method is computationally affordable to treat the amount of states required? 
A comparison with experiment and the computation of vertical excitation spectra with reference methods can help to obtain an answer to this question. 
\\
\indent 3) How large is the system under investigation and how complex are the excited state processes that are considered to be important?
This question is important in order to identify if single reference methods like LR-TDDFT or ADC(2) make sense for certain reactions that might occur. While large and flexible molecules with a lot of energetically close-lying states can give rise to a multifaceted photochemistry including dissociation, homolytic bond-breaking, and bond-formation, the dynamics of rigid molecules might only be dominated by one main reaction channel and lose the additional energy in form of molecular vibrations. The complexity of the excited-state processes can help to estimate the number of necessary data points to describe the relevant configurational space of the molecule. 

In case multi-reference methods are necessary to describe many different excited-state processes of a molecule, the training set generation can become infeasible. For example, 356 data points were computed for the 15-atom cyclopentoxy molecule with MR-CISD(5,3)/cc-pVD(T)Z.~\cite{Shen2020JPCA} Respective calculations comprised 19,302,445 configuration state functions and one reaction coordinate could be fitted in the diabatic basis. We also ran into a similar problem when fitting the excited states of the amino acid tyrosine containing 24 atoms, which also requires a multi-reference treatment. The size of the active space and the number of states needed for an accurate description made multi-reference methods like CASSCF or CASPT2 computationally too expensive, see Fig.~\ref{fig:tyrosinecurves}.
In these cases, the computation of an ample training set is far too expensive with multi-reference methods and the quantum chemistry calculations remain the bottleneck even when using ML.

In addition to the aforementioned intricacies to build up a meaningful, yet accurate training set for the excited-states, the process is further complicated by the arbitrary phase of the wave function. As a consequence, excited-state properties resulting from two different electronic states, such as transition dipole moments or couplings between different electronic states~\cite{Akimov2018JPCL,Westermayr2019CS,Westermayr2020MLST,Westermayr2020JPCL,Guan2020JPCL,Guan2020JCTC}, are not uniquely defined and cannot simply be fitted with conventional ML models. Either an additional data preprocessing or an adaption of the learning algorithm has to be incorporated to render data learnable with ML models. 

\subsection{Phase of the Wave Function}\label{sec:phase}
In contrast to ground state properties, excited-state properties such as transition dipole moments, NACs or SOCs arise from two different electronic states. As a consequence of the arbitrary phase of the wave function of each electronic state, properties resulting from two different states carry an arbitrary sign, which makes them generally double-valued. 
In case of vectorial properties, such as dipole moments or coupling vectors, the whole vector can be multiplied by +1 or -1 and is still a valid solution. Similarly, single valued properties, such as SOCs obtained from electronic structure programs, can be multiplied by +1 or -1 and are equally correct. This additional complexity prohibits that conventional ML algorithms learn such raw data of quantum chemistry and hampers the training process to find a proper relation between a molecular geometry and the excited-state property.~\cite{Akimov2018JPCL,Westermayr2019CS}

\begin{figure}[ht]
    \centering
    \includegraphics[scale=0.45]{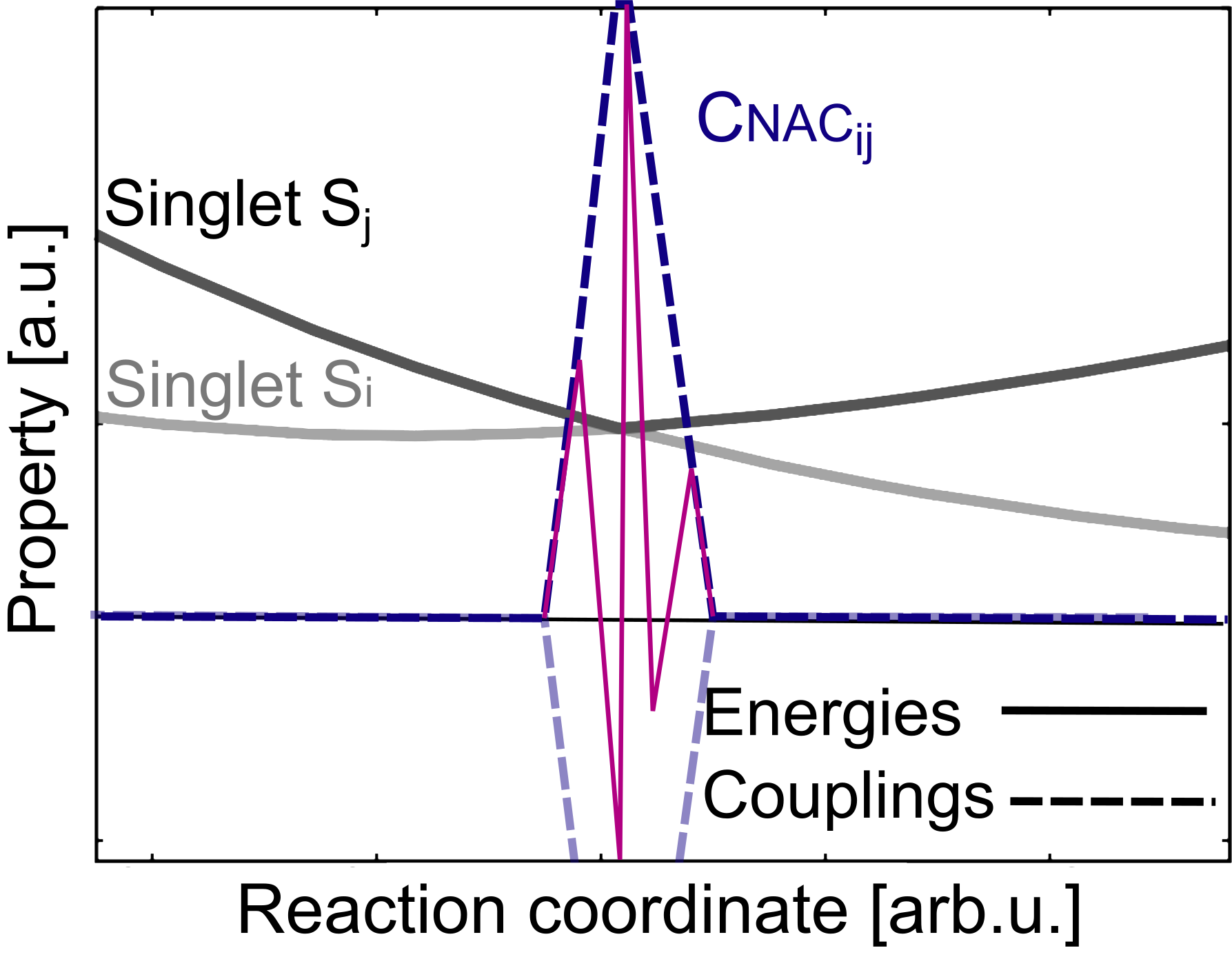}
    \caption{NAC value between singlet state S$_i$ and S$_j$ in the MCH basis. A consistent sign along the reaction path of couplings is shown by blue dashed lines. The direct output of a quantum chemical calculation is shown by a magenta line.}\label{fig:phase}
\end{figure}

A one-dimensional example of this problem is illustrated for the NAC (exemplified using one single value along the reaction coordinate) that couples an excited singlet state, S$_i$, and a second excited singlet state, S$_j$, in Figure \ref{fig:phase}. A positively signed function of atomic coordinates is shown by dashed blue lines with a cusp at the point at which the two singlet states are degenerate. Such a smooth function (besides the sharp spike at the conical intersection) is highly desirable when fitting with ML models is aimed for. It is worth mentioning that a consistent negative sign (light-blue dashed line) along this reaction coordinate is equally correct and that it is desirable to seek for one global sign. However, the direct output of a quantum chemistry program along this reaction coordinate looks more similar to the dashed magenta line in-between the blue curves. As one can imagine, no proper training can be guaranteed with these inconsistent data. Note that existing MD programs for the excited states usually track such phase jumps within electronic wave functions in order to account for nonadiabatic transitions correctly.\cite{Mai2018WCMS}

The idea of phase tracking can also be applied in ML in order to thwart the problems due to the arbitrariness within coupling or dipole elements. Some algorithms have been developed to remove the arbitrary sign jumps and provide smooth functions of atomic coordinates~\cite{Westermayr2019CS,Westermayr2020JPCL,Guan2020JCTC,Shu2019JCP}. Noticeably, the properties obtained after a transformation to the diabatic basis are already smoothly varying functions of atomic coordinates~\cite{Kryachko2000IJQC}. However, the challenges arising due to the arbitrary phase of the wave function still persist, because the inconsistencies within adiabatic properties have to be removed in order to make the diabatization process feasible.~\cite{Guan2020JCTC,Guan2020JPCL} 

It is worth mentioning at this point that also another kind of phase exists that cannot be eliminated in the aforementioned way. It is called Berry phase or geometric phase. After performing a loop in space around a conical intersection and returning to the original point, a change in the phase of the wave function of $\pi$ can be observed, i.e., the same point is only reached after two loops around the conical intersection. Neglecting this effect can lead to false transition probabilities, depending on the dynamics method and the system. While in most cases in MQCD the Berry phase can be safely neglected, this is not possible in quantum dynamics simulations. A diabatic basis is advantageous in this case, because the Berry phase is absent in this picture. However, the Berry phase has to be kept in mind, when  fitting diabatic potentials~\cite{Yarkony1996JCPa,Yarkony1996JCPb,Ryabinkin2014JCP,Gherib2015JCTC,Ryabinkin2017ACR}.

\subsubsection{Phase Correction of Adiabatic Data}
First ML studies on dynamics in the adiabatic basis omitted a preprocessing and were unable to reproduce reference results based on ML alone,\cite{Dral2018JPCL} or avoided the phase problem by using the Zhu-Nakamura method.\cite{Hu2018JPCL,Chen2018JPCL}
Evidently, potentials and forces can be learned with conventional ML approaches but adaptations or a preprocessing of data is necessary to learn coupling elements or transition dipole moments.
Independent of the purpose -- the fitting of adiabatic quantities~\cite{Akimov2018JPCL,Westermayr2019CS} or the diabatization of adiabatic data with property-dependent diabatization schemes~\cite{Guan2020JCTC} -- the adiabatic data has to be corrected to remove the arbitrary sign jumps that are due to the arbitrary phase of the wave function. Several ways for these corrections exist, which have been shown to work well for different excited-state problems.

One possibility is to preprocess data according to the wave function overlap -- betweem the wave functions from a geometry of interest and a reference geometry -- for each electronic state. This process is termed phase correction~\cite{Akimov2018JPCL,Mai2018WCMS} and has been applied by us in order to generate a training set for three singlet states of CH$_2$NH$_2^+$~\cite{Westermayr2019CS} and 2 singlet and 2 triplet states of CSH$_2$. SOCs~\cite{Westermayr2020JPCL}, NACs~\cite{Westermayr2019CS,Westermayr2020MLST,Westermayr2020JPCL}, and transition dipole moments~\cite{Westermayr2019CS,Westermayr2020MLST} could be fitted in the adiabatic basis with deep NNs and kernel ridge regression (KRR)~\cite{Westermayr2019CS,Westermayr2020JPCL,Westermayr2020MLST}. Very recently, Zhang et al.~\cite{Zhang2020arXiv} applied this procedure to describe transition dipole moments of N-methylacetamide.

The wave function overlap matrix, $\mathbf{S}$, with size N$_S\times$N$_S$, is computed between two molecular geometries $\alpha$ and $\beta$~\cite{Plasser2016JCTC}: 
\begin{equation}
    \boldsymbol{S}=\langle \Psi_\alpha \mid \Psi_\beta \rangle.
\end{equation}
In many cases along a given reaction path, the off-diagonal elements of the overlap matrix are very close to zero and the diagonal elements are very close to +1 or -1, indicating whether the phase of a state has changed along this path or not. Whenever a new state enters along the reaction path or adiabatic states switch their character, which is common after passing through a conical intersection for example, the off-diagonal elements provide the relevant phase information instead of the diagonal elements. Taking all these effects into account, a phase vector, $\mathbf{p}$, can be derived for each given molecular geometry. A property resulting from electronic state i and j has to be multiplied by the corresponding phase factors of these states~\cite{Westermayr2019CS}. 

An advantage of this algorithm is that it does not require any manual fitting of data. However, this procedure has to be carried out for every data point included in the training set with respect to one pre-defined reference wave function. This reference wave function can be for example the wave function of the ground-state equilibrium structure of the molecule and needs to be identified to guarantee an almost globally consistent sign of elements. 
During a photo-initiated simulation, it is common that geometries quickly start to differ from the reference geometry. The wave function overlap then tends to zero and cannot provide information about the correct sign of a certain electronic state. In this case, the phase must be propagated from the reference geometry on with $n$ interpolation steps. The phase vector applicable for the correction of the data point to be included in the training set is then obtained by multiplication with all previously obtained phase vectors, $p_0$ to $p_{n-1}$:
\begin{equation}
    \boldsymbol{p}=\prod_{\alpha=0}^{n-1} p_\alpha.
\end{equation}{}

Intruder states prohibit a proper tracking because their wave function is absent at the earlier geometries. Hence, a phase correction may be rendered infeasible for systems with a high density of states. 

In order to obtain the correct phase, more states can be included in the simulations, which however increases the computational cost. A solution is to take many electronic states into account only close to the reference geometry. The amount of states can then be reduced along a given reaction coordinate and relevant states can be disentangled from irrelevant ones. Further, it makes sense to save the already phase-corrected wave functions of several geometries in addition to the reference geometry. Whenever a new data point should be included into the training set, the distance to each saved data point can be computed in order to find the closest available structure and reduce the amount of interpolation steps~\cite{Westermayr2019CS,Westermayr2020MLST_Perspective}. 

This problem has also been recognized by Robertson et al.~\cite{Robertson2019JCC} for a diabatization process, where a sufficiently large vector space of the CAS wave function is required for proper diabatization. The overlaps of electronic states can be maximized by rotation of CI vectors of CAS wave function states. A similar version to use the information of CI vectors for diabatization was applied by Williams et al.~\cite{Williams2018JCP}, who used NNs to assist the diabatization process of adiabatic NO$_3$ potentials.

Another way to correct the sign of data points was carried out by Guan et al.~\cite{Guan2020JCTC}, who fitted diabatic 1,2$^1$A PESs and dipole moment surfaces of NH$_3$ from MR-CISD/aug-cc-pVTZ 
data with NNs. 
The diabatic PESs were taken from a previous study and obtained with the Zhu-Yarkony diabatization procedure~\cite{Zhu2010JCP,Zhu2012JCP,Zhu2012JCPb}. By diagonalization, the rotation matrix defined in eq~\ref{eq:rotation} could be obtained, which connects the diabatic and the adiabatic basis (see eq.~(\ref{eq:atd})). The adiabatic dipole moments, $\mu^{MCH}$, could then be transformed into the diabatic basis using the unitary matrix, $\mathbf{U}$:
\begin{equation}\label{eq:atd_dipole}
\mu^{diab} = \mathbf{U}\mu^{MCH} \mathbf{U}^{\dagger}.
\end{equation}
As the unitary matrix $\mathbf{U}$ is only defined up to an arbitrary sign, the signs of the diabatic dipole moments have to be corrected in order to provide a consistent diabatic dipole moment surface. This correction has been done with a so-called cluster growing algorithm~\cite{Shu2019JCP}.

The cluster growing algorithm requires an initial set of phase corrected data points. In this work, 347 data points were adjusted manually for this purpose. Subsequently, a Gaussian process regression (GPR) model~\cite{Rasmussen2004} was fitted to these data points. The signs of the rest of the data points to be corrected were then adjusted with the GPR model. Several iterations were carried out, where each iteration aims for the inclusion of close-lying points to the cluster, leading to the name "cluster growing" algorithm~\cite{Wang2019JPCA}. 

The singularities in regions close to conical intersections can make this algorithm fail. Therefore, data points in such regions have been removed by setting a threshold. Data points with energy gaps lower than this threshold were excluded from the cluster. The regions around conical intersections could not be fitted as comprehensively as other regions of the PESs. 
As another drawback, the authors note that the initial manual fitting of the signs is a tedious task, especially when larger systems and more dimensions are described. 

Two of the authors also fitted diabatic PESs of two singlet states and one triplet state as well as the SOCs between singlets and triplets of formaldehyde, CH$_2$O, with NNs.~\cite{Guan2020JPCL} The electronic structure reference method was MR-CISD/cc-pVTZ. The diabatic potentials were obtained using an adapted version of the Boys localization~\cite{Subotnik2008JCP}. The energy differences between two states are incorporated in the equations in order to remove earlier identified diabolic singularities~\cite{Wang2019JPCA}. The range of $\pi$, which the rotation angle for the diabatization covers, guarantees a proper treatment of the Berry phase. The diabatization procedure further requires consistent transition dipole moments, which were adjusted manually for this purpose. The diabatic SOCs were then obtained as a linear combination of the adiabatic SOCs by applying the same rotation matrix as for the energies. One separate NN function was used to fit each coupling value and electronic state separately.

It becomes clear that only a small number of works on this topic exist. At the moment, many problems remain unsolved for generating a training set that properly accounts for both types of phases, the arbitrary phase and the Berry phase, and is applicable for large systems with many states. An automatic phase correction procedure without the need of manual input would be very advantageous, especially when larger and more flexible systems are treated. Further developments are needed.

\subsubsection{ML-Based Internal Phase Correction}\label{sec:phaseless}

One step towards a routine application of ML for photochemical studies and an easier training set generation with quantum chemistry is an ML-based internal phase correction, which has been implemented by us into the SchNarc approach for photodynamics simulations~\cite{Westermayr2020JPCL}. In contrast to the phase correction algorithm to correct the training data, this procedure renders the learning of inconsistent quantum chemical data possible. A modification of the training process, termed phase-free training, is required for this purpose.~\cite{Westermayr2020JPCL} 

We implemented this training algorithm in a combination of the deep continuous-filter convolutional-layer NN SchNet~\cite{Schuett2018JCP,Schuett2019JCTC}, adapted for excited states, and the MD program SHARC~\cite{Richter2011JCTC,Mai2018WCMS,sharc-md2}

Similar to standard training algorithms, parameters of an ML model are optimized in order to minimize a cost function. Most frequently, the L$_1$ or L$_2$ loss functions are applied, which take the mean absolute error or mean squared error between predicted and reference data into account. The phase-free training algorithm uses a phase-less loss function, which includes all trained properties at once and additionally removes the influence of the random phase switches. In this way, the computational costs for the training set generation can be reduced.

Compared to the previously reported ML models for photochemistry, where each state was fitted independently~\cite{Chen2018JPCL,Guan2020JPCL,Guan2020JCTC}, SchNarc is capable of describing all PESs at once, including the elements resulting from different pairs of states. This results in an overall loss function with several terms, where each term is weighted with a different trade-off value, $t$, that can be defined manually:
\begin{equation}
\begin{array}{ll}
       L_{ph} =\\ t_E\mid\mid E^{QC} - E^{ML} \mid\mid^2 \\
       + t_F \mid\mid F^{QC}-F^{ML}\mid\mid^2 \\
       + t_{SOC} \cdot L_{SOC} \\
       + t_{NAC} \cdot L_{NAC}
       \end{array}
\end{equation}
If only energies (E) and forces (F) are fitted, then the loss function is equal to a linear combination of L$_2$ loss functions for energies and forces~\cite{Gastegger2017CS,Westermayr2020JPCL}. The parts of the SOCs and NACs are
\begin{equation}
    \begin{array}{ll} 
    L_{SOC}=min(\mid \varepsilon_{SOC}^{\kappa}\mid )\\
    \text{with}~ 0 \leq \kappa \leq 2^{N_S-1}
    \end{array}
\end{equation}{}
and
\begin{equation}
    \begin{array}{ll} 
    L_{NAC}=min(\mid \varepsilon_{NAC}^{\kappa}\mid )\\
    \text{with}~ 0 \leq \kappa \leq 2^{N_S-1},
    \end{array}
\end{equation}{}
respectively. The error for SOCs and NACs that enters the loss function is the minimum error that can be achieved when trying out all possible combinations of phases for each pair of states, i.e., 2$^{N_S-1}$ possible solutions. The algorithm takes into account that the signs of SOCs and NACs coupling different pairs of states depend on each other. 

The error function containing all possible solutions for SOCs, $\varepsilon_{SOC}^{phase}$, and NACs, $\varepsilon_{NAC}^{phase}$, can be obtained as follows:
\begin{equation}
\begin{array}{ll}
     \varepsilon_{SOC}^{\kappa} =\\
     \frac{1}{N_S^2}\sum_{i=1}^{N_S}\sum_{i\neq j}^{N_S} \mid\mid C^{QC}_{SOC_{ij}} - C^{ML}_{SOC_{ij}} \cdot p_i^{\kappa} \cdot p_j^{\kappa} \mid\mid ^2 \\
     \text{with}~ 0 \leq \kappa \leq 2^{N_S-1}
\end{array}
\end{equation}
     
\begin{equation}
\begin{array}{ll}
     \varepsilon_{NAC}^{\kappa} =\\
     \frac{1}{N_S^2}\sum_{i=1}^{N_S}\sum_{i\neq j}^{N_S} \frac{1}{N_A}\sum_{a=1}^{N_A} \\
     \mid\mid C^{QC}_{NAC_{ij,a}} - C^{ML}_{NAC_{ij,a}} \cdot p_i^{\kappa} \cdot p_j^{\kappa} \mid\mid ^2 \\
     \text{with}~ 0 \leq \kappa \leq 2^{N_S-1}
\end{array}
\end{equation}

This phase-less loss procedure does not require any preprocessing of training data. Quantum chemistry calculations can be directly fitted with this adaption of the loss function. The power of this approach is that, once a given phase vector for a data point has been found, it can be directly applied to correct the arbitrary signs of other properties, such as transition dipole moments. If other properties are targeted, the loss function applied for NACs can be similarly used for other vectorial properties, and the loss function applied for SOCs can be used for any other single- or complex-valued element of arbitrary sign~\cite{Westermayr2020JPCL}.
However, as a consequence of the higher complexity of the loss function, the training process is generally more expensive. The computational effort required for training can be reduced if only one type of coupling is treated within MD simulations. In these cases, a simpler adaption of the phase-free loss is also applicable.~\cite{Westermayr2020JPCL} 

\subsection{Training Set Generation}\label{sec:trainingsetgeneration}

The requirements and desirable specifications for a training set can vary strongly, dependent on the type of application: When the focus of a study is the investigation of the huge chemical space and the search for certain patterns thereof or the design of new molecules with targeted properties, usually the training set should be as large as possible to cover as many molecules as possible. In the best case, the data points are computed with high accuracy and this reference method is accurate for the excited states of many different types of systems. In terms of accuracy and general applicability, ab-initio methods are more suitable, as they do not require the selection of a density functional, which might be accurate for some cases, but fail for others. However, the costs and complexity of highly accurate multi-reference ab-initio methods limit their applicability, so that TDDFT remains the method of choice when making predictions throughout chemical compound space~\cite{Ramakrishnan2015JCP,Ghosh2019AS,Beard2019SD}. The most widely applied approach to generate a training set for this purpose is to start from an existing (ground-state) data base that already covers a large chemical space of certain types of molecules. In this way, not much effort has to be devoted into the exploration of chemical space and structure optimizations to get the most stable conformations of different molecules.

For the purpose of ML-based excited-state dynamics simulations, things look quite different. Note that for photodynamics simulations, only molecule-specific ML model exist until now, which can potentially develop into a universal excited-state force field, but much remains to be done to achieve this goal. Indeed, the generalization of the excited state PESs and corresponding couplings is expected to be a highly complex task, especially due to the problematic generalization of excited states~\cite{Westermayr2019CS}. A comparison of the isoelectronic molecules CH$_2$NH$_2^+$ and C$_2$H$_4$ can serve as an example. Their conical intersection between the first excited singlet state and the ground state is accompanied by a rotation along the dihedral angle, which could lead to very similar photo-initiated processes. However, higher-lying excited states are ordered completely different in both molecules and excitation leads to completely different photodynamics.~\cite{Barbatti2005JCP,Tapavicza2007PRL,Tavernelli2009JCP,Tavernelli2009JMS,Barbatti2006MP,Tao2011JCP,Allison2012JCP,Mori2012JPCA,Sellner2013MP,Barbatti2012JCP,Hollas2018JCTC,Westermayr2019CS} 

As it stands, existing ML models for photodynamics simulations are developed to investigate the photo-initiated processes of one specific molecule.

Overall, we arrive at the following wish list for the training set, which has been identified also for MD in the ground state~\cite{Botu2015IJQC,Li2015PRL,Behler2015IJQC,Behler2017ACIE}: 1) The training set should be as small as possible to keep the number of reference calculations at a minimum. 2) At the same time, the relevant conformational space of the molecule that is required for the reaction under investigation should be sampled comprehensively~\cite{Behler2015IJQC,Behler2017ACIE,Westermayr2019CS,Dral2020JPCL,Dral2020JCP}.\\
Keeping this in mind, an efficient procedure to obtain relevant molecular structures has to be applied. A large number of schemes to achieve this goal have been proposed, which are mainly based on two different strategies: One approach is to simulate MD in the ground and excited states with the reference method and putting much effort into covering critical regions of the PESs comprehensively~\cite{Hu2018JPCL,Dral2018JPCL,Chen2018JPCL}. Structure-based sampling or subsequent clustering is beneficial in this case~\cite{Dral2018JPCL,Hu2018JPCL,Ceriotti2013JCTC,Dral2017JCP,Sobol2011W}. The other strategy is to use an active learning approach, which decreases the number of necessary reference calculations considerably, but is usually more time-consuming~\cite{Behler2015IJQC}. Noticeable, within ML for quantum chemistry, active learning often refers to an approach, where an initial training set is used to fit an ML model and this previously learned information is applied to expand the training set~\cite{Lin2020JCP}. 
The latter approach is often carried out with the help of MD simulations, but has also recently been adapted in a trajectory-free way~\cite{Uteva2018JCP,Lin2020JCP}.

\subsubsection{Basic Sampling Techniques and Existing Databases}
To find patterns within certain groups of molecules, to explore chemical space and to develop new methods that can fit for example different properties of molecules, such as the valence density used in DFT~\cite{Brockherde2017NC}, or large molecules from small building blocks~\cite{Gastegger2016JCP}, a good starting point is often considered to be an already existing data base. Prominent examples are the QM data bases, namely QM7, QM7b, QM8, and QM9~\cite{Ramakrishnan2014SD}, which have been used in a large number of publications up to date and provide a benchmark for many ML studies~\cite{Hansen2013JCTC,Christensen2019JCP,Chmiela2017SA,Gastegger2018JCP,Schuett2018JCP,Christensen2020JCP,Kim2019SD,Glavatskikh2019JC,Schwilk2020arXiv,Ghosh2019AS,Veit2020arXiv}. 
Especially the QM9~\cite{Ramakrishnan2014SD} data set containing more than 133k small organic molecular structures and corresponding DFT energies, enthalpies, harmonic frequencies, and dipole moments (to name only a few properties) is very popular among the scientific community and has also been used in challenges on kaggle, where researcher and layperson all over the world can compete against each other to find the most suitable solution to a given task. Prices up to several thousand dollar are quite common~\cite{kaggle1}. 
In a similar spirit, the QM9 IPAM ML 2016 challenge requires to predict the energies of QM9 from only 100 training points within chemical accuracy (error of $\approx$0.05 eV).\cite{vonLilienfeld2018twitter} 

All aforementioned data bases originate from GDB data bases~\cite{Fink2005ACIE,Fink2007JCIM,Blum2009JACS}, and are often a subset thereof. The chemical universe GDB data bases have been designed using molecular graphs to sample a comprehensive space of molecular structures for the search of new lead compounds in drug design~\cite{Blum2009JACS}.

One of the first data bases available for the scientific community to treat the excited states of molecules is most probably the QM7b~\cite{Montavon2013NJP} data set, that contains the excitation energies computed with TDDFT for a total amount of $>$14k molecules with atoms C, N, O, H, S, and Cl. This data set is based on the molecular geometries of the QM7~\cite{Blum2009JACS,Rupp2012PRL} data set plus an additional amount of 7211 molecules containing a chlorine atom. The excitation energies of the first singlet state and other properties were recomputed for each optimized molecular geometry. Very similar, the QM8~\cite{Ramakrishnan2015JCP} data base was developed, based on the GDB-17 data base~\cite{Ruddigkeit2012JCIM}. This data set can be used for the computation of vertical excitation spectra. It hence includes not only the vertical excitation energies of the first excited singlet state, but also the corresponding oscillator strengths. Oscillator strengths are also reported in an auto-generated data set for optoelectronic materials with DFT.~\cite{Beard2019SD} Note that the oscillator strength is computed from the squared transition dipole moment, hence an arbitrary phase factor cancels out and the data does not have to be preprocessed. In addition to the TDDFT energies, CCSD energies are reported, having enabled the development of the so-called $\Delta$-learning approach - a powerful way to obtain the accuracy of highly accurate ab-initio methods with only a small amount of respective reference calculations. Two ML models are trained in this approach, one on a less accurate method and another one on the difference between the less accurate and higher sophisticated method~\cite{Ramakrishnan2015JCTC}. This scheme can also be applied multiple times to achieve increasing accuracy with little additional computational effort~\cite{Dral2020JCP} and has been adapted for spectroscopy in the condensed phase as well~\cite{Kananenka2019JCTC}.

The QM9 data set has further been the basis of a very recently constructed data set for singlet and triplet states of $>$13k carbene structures, termed QMspin~\cite{Schwilk2020arXiv}. 4,000 geometries from the QM9 data set were randomly selected, hydrogen atoms were subtracted and singlet and triplet states were optimized using CASSCF(2,2)/cc-pVDZ-F12 and open-shell restricted KS-DFT with the B3LYP~\cite{Lee1988PRB,Becke1988PRA} functional, respectively. The MR-CI method was subsequently used to compute the electronic energies of singlet and triplet states. This data set has been used to investigate structural and electronic relationships in carbenes, which are important intermediates in many organic reaction networks~\cite{Schwilk2020arXiv}.

The OE62~\cite{Stuke2020SD} data base, a benchmark data set applicable for spectroscopy, is another descent of several existing data sets, such as the QM8 and QM9 data sets. It consists of $>$61k organic molecules able to form crystals including up to 174 non-hydrogen atoms. Reported are the orbital energies of molecules computed with DFT/PBE~\cite{Perdew1996PRL}.

Another database, which also contains excited state data, is the PubChemQC data base.~\cite{Nakata2017JCIM} It contains over three million molecules, whose structures are reported along with the energies at DFT/B3LYP/6-31G* level of theory. In addition, the excitation energies of at least three million structures are reported for the 10 energetically lowest-lying singlet states at TDDFT/B3LYP/6-31G* level of theory.

A simple strategy was carried out by Kolb et al.,~\cite{Kolb2016JCP} who used an existing analytical PES to create an ML potential: They randomly sampled data points, trained an ML model and added more points in regions with deviations from the original PES. Other strategies have been carried out mainly for the fitting of ground state potentials and for materials, which are however also relevant to consider for the excited states. One novel, suitable strategy is for example "de novo exploration" of PESs using a similarity measure provided by ML models.~\cite{Bernstein2019npjCM} At least for material discovery, this method can be used to omit any additional active learning procedure to converge PESs. Another way to build a training set is to employ molecule-generating ML models,\cite{Elton2019MSDE,Yao2020arXiv,Krenn2019arXiv}
such as the recently developed Gschnet.~\cite{Gebauer2019arXiv} Alternatively, MD simulations with the reference method can provide a good starting point for training.~\cite{Chmiela2017SA,Schuett2017NC,Chmiela2018NC}
Ye et al.~\cite{Ye2019PNAS} sampled 70k conformations for \textit{N}-methylacetamide via MD simulations with the OPLS force field\cite{Jorgensen1996JACS}
within GROMACS~\cite{VanDerSpoel2005JCC} for subsequent UV spectra calculations.
We have applied a similar scheme to generate a training set of SO$_2$ based on an LVC model~\cite{Plasser2019PCCP}. Surface hopping MD simulations with the SHARC method~\cite{Richter2011JCTC,Mai2018WCMS,sharc-md2} were carried out with the reference method LVC(MR-CISD) ending up in $>$200k data points of different conformations of SO$_2$~\cite{Westermayr2020JPCL}. Due to the crude sampling and low cost of the reference method, no emphasis was put on clustering the training set into a smaller, still comprehensive set. 

90k data points were required in an ML-based surface hopping study of CH$_2$NH with the Zhu-Nakamura method. Reference data for the ground and first excited singlet state, S$_0$ and S$_1$, were generated with CASSCF(2,2)/6-31G via ground-state and surface hopping MD simulations. The latter method was applied to sample the regions around conical intersections between the S$_0$ and S$_1$ state.~\cite{Chen2018JPCL}

Similarly, Hu et al.~\cite{Hu2018JPCL} sampled 200k data points of 6-aminopyrimidine using ground-state and surface hopping MD with CASSCF(10,8)/6-31G*. State-averaging over three singlet states was applied. In addition, structures that led to hops between different states were used as starting points to find minimum energy conical intersections and clustering was carried out to reduce the amount of data for training. 

One way to select data points more efficiently is a structure-based sampling scheme, as proposed for instance by Ceriotti et al. with sketch map,~\cite{Ceriotti2011PNAS,Tribello2012PNAS,Ceriotti2013JCTC} an algorithm for dimensionality reduction of atomistic MD simulations or enhanced sampling simulations. 
Likewise, Dral et al.~\cite{Dral2018JPCL} applied a grid-based sampling method to construct PESs of a model spin-boson Hamiltonian to execute surface hopping MD with KRR. The energetically low-lying regions of the PESs were first sampled via an inexpensive method and subsequently the distances between the molecular structures were computed. In this way, 10,000 data points were obtained~\cite{Dral2017JCP,Dral2018JPCL}. ML models trained on only 1,000 data points were accurate enough to reproduce reference dynamics. This approach was compared with random sampling for the methyl chloride molecule and was shown to reduce the amount of training data needed up to 90\% for static calculations~\cite{Dral2017JCP,Sobol2011W}.

\subsubsection{Active Learning}
As shown in the previous section, training sets with the respective equilibrium structure of a large number of molecules are very powerful for investigating the huge chemical space or for the design of new molecules. However, the usefulness of such training sets for photodynamics is rather questionable. The reason for this deficiency is that, especially in MD simulations in the excited states, the excess of energy carried by a molecule very quickly leads to conformations that are far beyond the equilibrium structure and most likely far away from originally sampled structures. The formation and breaking of bonds is quite common in photodynamics simulations and is usually only accessible from an excited, dissociative state. The use of photodynamics simulations with the reference method could solve this problem, but are not feasible if specific reactions occur on a rather slow time scale or if many different processes take place.~\cite{Mai2020ACIE,Nelson2020CR,Crespo-Otero2018CR,Gonzalez2020,Cartwright2020,Westermayr2020MLST_Perspective} As previous studies have shown, inefficient sampling techniques lead to a huge amount of data, which still does not guarantee that the training set is comprehensive enough for excited-state MLMD simulations. In fact, ML models fail dramatically in under-sampled and extrapolative regions of the PESs. 
A smarter sampling technique is advantageous in these cases in order to efficiently identify such under-sampled regions and build trustworthy ML models. 

Active learning, where ML ''asks'' for its training data, is one solution to create a data set more efficiently. An example from chemistry is the adaption of an initially generated training set due to an uncertainty measure for ML models trained on this initial training set. This concept has already been introduced in 1992 as query by committee~\cite{Seung1992} and has been adapted for quantum chemistry quite fast due to the required fitting and interpolation of PESs for grid-based quantum dynamics simulations.
Pioneering works by Collins and co-workers~\cite{Ischtwan1994JCP,Bettens1999JCP,Netzloff2006JCP,Collins2002TCA} applied modified Shepard interpolation to fit PESs and iteratively adapt them in out-of-confidence regions using the GROW algorithm~\cite{Collins2002TCA,Godsi2010JCP}. Since then, several sampling techniques have been developed that are based on MD and an extension of data bases using interpolation moving least squares~\cite{Dawes2007JCP,Dawes2008JCP}, permutation invariant polynomial fitting~\cite{Braams2009IRPC,Qu2018ARPC}, and different ML models for the ground state~\cite{Lorenz2004CPL,Raff2005JCP,Behler2007PRL,Bartok2010PRL,Chen2013JCP,Jiang2015PRL,Shen2015JCP,Botu2015IJQC,Li2015PRL,Behler2015IJQC,Shao2016JCP,Cui2016JPB,Gastegger2017CS,Kolb2017JPCA,Kolb2017JPCL,Huang2017CS,Zhou2017PCCP,Zhang2019JPCL} and also excited states~\cite{Dral2018JPCL,Westermayr2019CS,Westermayr2020JPCL}.

As active learning starts from already trained ML models, an initial training set has to be provided. Some strategies to provide this initial reference data set will be discussed, following strategies applied to adapt this initial training set. Note that all previously discussed methods can be similarly applied to generate an initial training set. 

\paragraph{Initial training set}
In general, an initial training set can be obtained in many different ways. As photo-initiated MD simulations usually start from vertical excitation of the ground state equilibrium geometry, this structure is commonly used as the starting point and reference geometry for the training set generation.
In principle, any technique can be applied to then add conformations to obtain a preliminary training set. 
A good starting guess is to use normal modes of a molecule, as they are generally important for dynamics. In two recent works, we carried out scans along different normal modes and combinations thereof to sample conformations of small molecules.~\cite{Westermayr2019CS,Westermayr2020JPCL} Normal modes are also sampled for generating ANI-1 NN PESs.~\cite{Smith2017CS} For the excited states, it is favorable to include critical regions of the molecule in the initial training set by carrying out optimization of these geometries and including the calculations into the training set~\cite{Westermayr2019CS,Hu2018JPCL}. 

When small molecules are targeted, this initial training set can already be comprehensive to start the training of ML models and adapt the training set based on an uncertainty measure provided by the ML models~\cite{Westermayr2019CS}.
In case more flexible and larger molecules are studied that give rise to a complex photochemistry and a high density of states including different spin multiplicities, a small initial training set might not be sufficient and a larger conformational space of the molecule needs to be sampled. This can be done for example via Wigner sampling~\cite{Wigner1932PR} and also with MD simulations in the ground state~\cite{Bruccoleri1990B,Maximova2016PLOSCB}. Suitable methods are for example umbrella sampling~\cite{Kaestner2011WIRCMS}, trajectory-guided sampling~\cite{Tao2019TCA}, enhanced sampling~\cite{Yang2019JCP} or metadynamics~\cite{Herr2018JCP} in combination with a cheap electronic structure method like the semi-empirical tight-binding based quantum chemistry method GFN2-xTB~\cite{Grimme2019JCTC} or existing ground-state force fields. A large amount of different geometries can be created very fast and inexpensively, which then can be clustered to exclude similar conformations of the molecule to keep the number of reference simulations at a minimum. The selected data points for the training set can then be computed with the chosen reference method, whose accuracy is targeted with ML. 
Additionally, if certain reaction coordinates have been shown to be important in experiments or previous studies, then it is favorable to include data from scans along these reaction coordinates.~\cite{Westermayr2020MLST_Perspective,Shen2020JPCA}

As soon as meaningful ML models can be obtained from the initial training set, active learning techniques can be applied to enlarge the set. What number of data points turns out to be sufficient for the initial training set is dependent on a lot of different factors, such as the size and flexibility of the molecule under investigation, the number of excited electronic states described, and the ML model and descriptor applied~\cite{Westermayr2019CS,Westermayr2020MLST}. In order to give a ballpark figure, we note that we used approximately 1000 data points as initial training set for small molecules in recent studies using deep multi-layer feed-forward NNs.~\cite{Westermayr2019CS,Westermayr2020JPCL}

\paragraph{Strategies for actively expanding the training set}
The next step in active learning is to expand the initial training set by adding points from out-of-confidence regions.
The detection of these undersampled regions can be done in many different ways, whereby most approaches rely on MD simulations. 

Among the most popular strategy is the iterative sampling scheme of Behler~\cite{Behler2015IJQC}, originally developed for fitting ground-state PESs. Today, it is widely used, see for example refs~\citenum{Botu2015IJQC,Li2015PRL,Smith2018JCP}, and has been modified as a so-called adaptive sampling approach.\cite{Gastegger2017CS} The latter has been adapted by us for the generation of a training set for the excited state PESs of molecules including couplings~\cite{Westermayr2019CS}. 
The basis of almost any iterative or adaptive sampling scheme is a similarity measure to judge whether a molecular geometry can be predicted reliably with ML models or not. While kernel methods intrinsically provide a measure of similarity for each molecular geometry, NNs do not. Therefore, adaptive sampling with NNs requires at least two ML models. In case of KRR or GPR, two ML models can be used as well, but are not necessarily needed. Indeed, the statistical uncertainty estimate of the predictions remains a huge advantage of GPR models.~\cite{Malbon2020PCCP,Dral2020JPCL} 

\begin{figure*}[ht]
    \centering
    \includegraphics[scale=0.45]{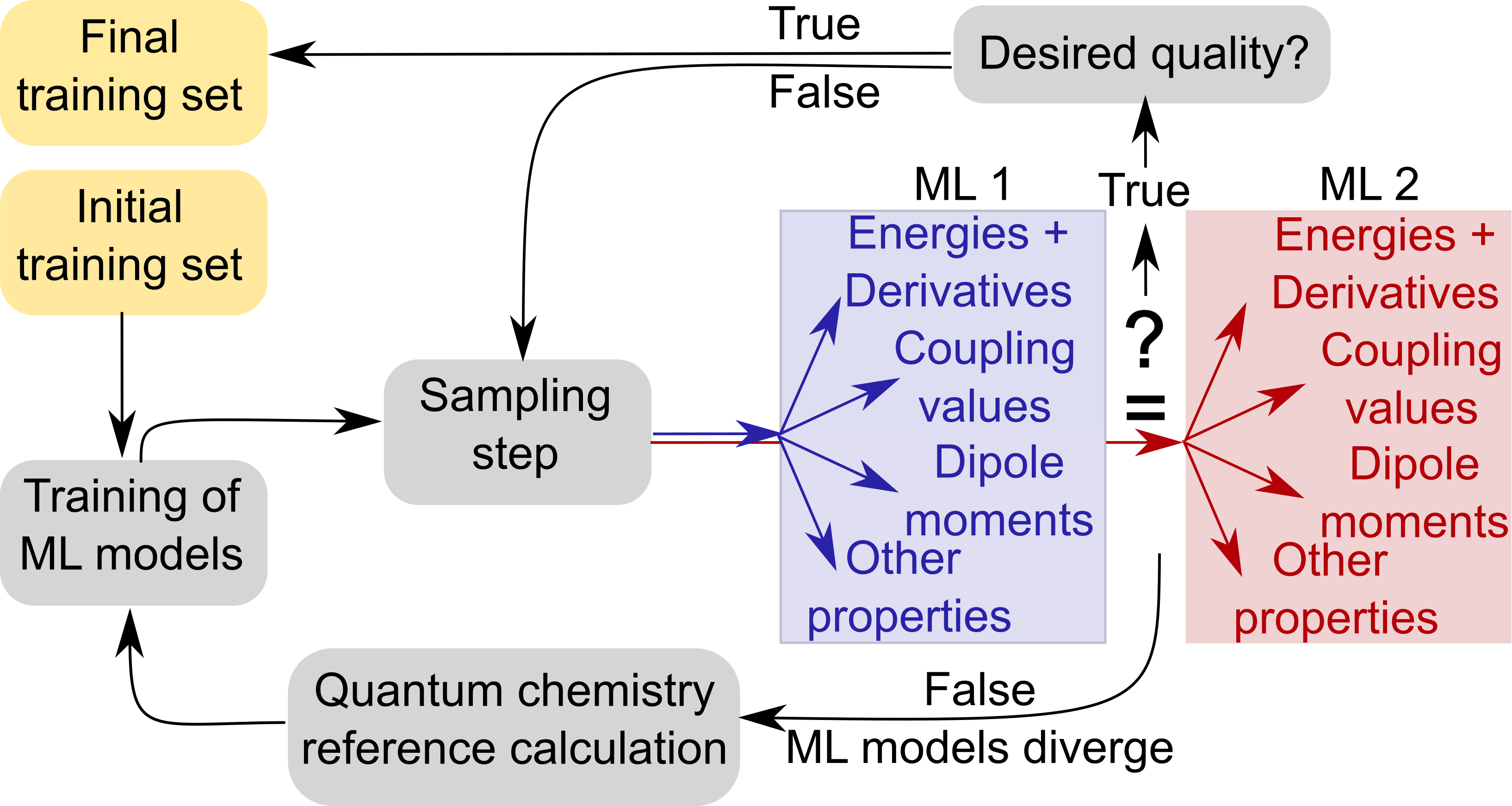}
    \caption{Adaptive sampling scheme illustrated using two ML models (blue and red blocks). The active learning procedure starts from an initial, preliminary training set (yellow), which is used to train ML models. A sampling step, e.g. a time step of an MD simulation, is executed: The ML models take the molecular geometry of the sampling step as an input and predict the energies of the considered excited states, their derivatives, and additional required photo-chemical properties. In case the predictions of the ML models are deemed to be different, quantum chemical reference calculations are carried out, ML models are retrained and the serial steps are carried out again. This procedure is executed until the desired quality of the ML PESs is attained in order to sufficiently describe the chemical problem under investigation. }
    \label{fig:adaptive}
\end{figure*}

The adaptive sampling scheme for the excited states is illustrated in Figure~\ref{fig:adaptive} and exemplified with two ML models. The whole process starts with an initial training set, which is used to train the two (or more) preliminary ML models. These models differ in their initial weights or model parameters. The resulting dissimilar ML architectures guarantee that the ML models do not predict the exact same number for a given molecular input. The hypothesis underlying this scheme is that inferences of different ML models trained on the same training set will be similar to each other as long as an interpolative regime is given. The inferences of the ML models are inaccurate and should differ from each other to a much larger extent if a molecular input lies in an unknown or under-sampled region of the PESs.
{}

In order to find such regions, sampling steps are carried out, e.g., by running (excited-state) MD simulations based on 
the mean of the inferences made by the different ML models for energies, $\overline{E}^{ML}$, forces, $\overline{F}^{ML}$, and if required also couplings, $\overline{C}^{ML}$. In each sampling step, the variances for each predicted property are computed. In the present example, energies and forces are treated together as $\sigma_{E+F}^{ML}$ (but can also be used separately), separately from variance of the couplings $\sigma_{C}^{ML}$. If a variance exceeds a pre-defined threshold, the ML models diverge and the predictions are deemed untrustworthy. $N_{ML}$ refers to the number of different ML models, $\zeta$, used for adaptive sampling:
\begin{equation}
\begin{array}{ll}
 \sigma_{E+F}^{ML} =\\
 \frac{1}{N_S}\sum_i^{N_S}\left(\sqrt{\frac{1}{N_{ML}-1}\sum_{\zeta=1}^{N_{ML}}  \left( E^{ML}_\zeta-\overline{E}^{ML} \right) ^2 }\color{white}\right)+ \\ 
\color{white}\left(
 \color{black}\sqrt{\frac{1}{N_{ML}-1}\sum_{\zeta=1}^{N_{ML}}\left(\frac{1}{3N_{A}}\sum_{a}^{3N_{A}} \left( F^{ML}_{\zeta,a}-\overline{F}^{ML}_a\right)^2\right) }\right)
 \end{array}
\end{equation}
\begin{equation}
\begin{array}{lr}
 \sigma_{C}^{ML} =\\
 \frac{1}{2N_S^2}\sum_i^{N_S}\sum_j^{N_S}\sqrt{\frac{1}{N_{ML}-1}\sum_{\zeta=1}^{N_{ML}} \left( C^{ML}_\zeta-\overline{C}^{ML}\right)^2}
\end{array}
\end{equation}
Note that the variance is averaged over all states for energies and forces and over all pairs of states for couplings, that are described with the ML models. As a variant, each state could also be treated separately. However, as the different electronic states are not independent of each other, a mean-treatment is assumed to be advantageous~\cite{Westermayr2020MLST}.

Each data point that is predicted with a variance larger than the pre-defined threshold for a given property, is recomputed with the reference quantum chemistry method and added to the training set. In this way, undersampled or generally unknown regions of the PESs are identified. Whenever the variance of each property is within the range that is thought to be reliable, the mean of the inferences is forwarded to the MD program to propagate the nuclei and continue MLMD simulations. 
The name adaptive sampling is based on the recommendation to choose a rather large threshold in the beginning of the adaptive sampling procedure and to adapt this threshold to smaller values as the ML models become more accurate and robust~\cite{Gastegger2017CS}. A first estimate for the initial value of a threshold can be obtained from the MAE of the corresponding ML model on the initial training set.

In principle, adaptive sampling can be carried out for every property, that should be represented with ML potentials, and is not restricted to energies, forces, and couplings. Similarly, it does not need to be executed with excited-state dynamics, but could also be done with ground-state MD or any sampling method that is considered to be suitable.

As a negative side effect, this procedure is generally more time-consuming than many other sampling techniques, because ML models have to be trained each time a new data point is added to the training set. To apply adaptive sampling in a more efficient way, it is advantageous to execute not only one ML trajectory, but many hundred trajectories in parallel, as it is usually done in MD simulations. The ML models should then only be retrained, when all ML-based trajectories have reached an undersampled conformational region~\cite{Behler2015IJQC,Gastegger2017CS,Westermayr2019CS}. 
Despite the higher complexity of adaptive sampling compared to random sampling, it can reduce the number of required data points for MLMD simulations substantially. In this regard, also the computational costs for the training set generation can be kept at a minimum.

Adaptive sampling was carried out successfully to generate a training set of 4,000 data points of CH$_2$NH$_2^+$ containing three singlet states and couplings. ML-based surface hopping MD simulation could be carried out on long time scales using the average of two deep NNs. The concept of iterative sampling also proved beneficial for the long MD simulation to guarantee accurate ML potentials throughout the production run. Here, the threshold was not adapted anymore and the MD was continued from the current geometry after a training cycle was completed~\cite{Westermayr2019CS}. In addition, the average of more NNs turned out to be more accurate than the prediction of only one NN, which was also shown in Ref.~\citenum{Gastegger2017CS}. 

Another quality control besides the property-based one proposed by Behler can be obtained by comparing the molecular structures at each time step as done by Dral et al.~\cite{Dral2018JPCL,Dral2017JCP} and Ceriotti et al.~\cite{Ceriotti2013JCTC}. 
A combination of a structure-based and property-based detection of sparsely sampled regions of the PESs has been done by Zhang et al. and Guo et al.~\cite{Chen2013JCP,Jiang2016IRPC,Xu2014CJCP,Li2015JCP,Jiang2014PCCP} Very recently, an alternative approach has been applied with NNs by Lin et al.~\cite{Lin2020JCP} that does not require MD simulations. It is based on the finding that the negative of the squared difference surface obtained from NNs approaches zero in regions, where no data points are available.~\cite{Dawes2008JCP}
Therefore, new points can be computed at the minima of the negative squared difference surfaces of at least two NNs (or, equivalently, at local maxima of the squared difference surface). This method is supposed to be very efficient in cases, where different conformations are separated by large energy barriers or strongly stabilized local minima are common. MD simulations would take a long time to overcome the potential barriers and reach the region of unknown molecular structures.~\cite{Lin2020JCP}

The idea behind this technique is similar to previous works with GPR. A measure of confidence can be provided with GPR models that enables the search of regions with large variance in ML predictions. In these regions, data points can be added to build up a training set~.\cite{Toyoura2016PRB,Guan2018MP,Uteva2018JCP,Vargas-Hernandez2019NJP} Similarly, Bayesian Optimisation Structure Search (BOSS) has been proposed for constructing energy landscapes of organic and inorganic interfaces.~\cite{Todorovic2018npjCM}
A combination of different approaches has also been applied by H\"ase et al.,~\cite{Haese2016CS} who fitted TDDFT excited-state energies of a light-harvesting system. Given a large enough, error-free, and comprehensive data set, ML has the potential to determine known and unknown (un)physical laws within the data.~\cite{Butler2018N}

\section{ML Models}
Besides the training set, which defines the highest possible accuracy an ML model can attain, the type of regressor and the descriptor to represent a molecule to the ML model play also important roles.\cite{Haghighatlari2020C} Improper choices of regressors and descriptors can result in inaccurate ML models.
\subsection{ML Models: Type of Regressor}\label{sec:mlreg}
Given the vast number of ML algorithms applied in the field of computational chemistry, one might ask which one to use or adapt for photochemistry. As recent studies applying ML for quantum chemistry have shown, many possible choices of ML approaches exist and there is no single solution. Nevertheless, a trend can be observed: Many studies that use ML in the research field of quantum chemistry employ labelled data sets, i.e., supervised learning techniques. Within supervised learning, one can distinguish between regression and classification. Classification aims at finding patterns and at grouping data into certain clusters.~\cite{Bishop2006} Those types of ML models are often used e.g. in spam filters, in medicine to diagnose diseases~\cite{Halama2019BJC,Bychkov2018SR}, or in food research, e.g. to guarantee a certain wine quality or origin.~\cite{Gomez2014FRI} Examples of applied classification models in the field of computational chemistry are for example support vector machines, random forests or decision trees used, e.g., to classify enzymes~\cite{Watanabe2020JCIM} or for the selection of an active space~\cite{Jeong2020JCTC,Chen2016}.

More often than classification models, regression models are applied to assist the search for a solution of a quantum chemical problem. Regression is used to fit functions that can relate a molecular input, $X$, to a quantum chemical output, $Y$. 
The simplest relation that can be assumed is linear. Although many quantum chemical problems cannot be accurately described with a linear function as given in eq.~\ref{eq:linear}, it can serve as a baseline model to evaluate the minimum accuracy one can obtain.~\cite{Bishop2006,Cartwright2020,Ahneman2018S,Atahan-Evrenk2019JPCA,Westermayr2019CS}
\begin{equation}\label{eq:linear}
    Y = b + w \cdot X
\end{equation}
The regression coefficients, also known as weights, $w$, and biases, $b$, are tailored for a given problem under investigation. In case of linear regression, ordinary least squares regression can be applied to find these coefficients. The process of finding the optimal relation between $X$ and $Y$ is termed training. The coefficients are optimized by minimizing a so-called loss function, $L$, which monitors the error between the original property, $Y^{QC}$, and the predicted property by the ML model, $Y^{ML}$, with respect to the training instances. Most often, the L$_1$ loss or the L$_2$ loss is used as an indicator for the training convergence. The L$_1$ monitors the mean average error (MAE) and the L$_2$ loss the mean squared error (MSE) of predictions:
\begin{equation}\label{eq:loss}
    L_2 = \frac{1}{N_M} \sum_\beta^{N_M} \left( Y_\beta^{ML} - Y_\beta^{QC} \right).  
\end{equation}
The Greek letter $\beta$ runs over all molecules, N$_M$, inside the training set. In principle, any error estimate can be used to train an ML model and find suitable regression coefficients.

An example specifically developed for excited-state problems is the aforementioned phase-less loss (see section~\ref{sec:phaseless}).\cite{Westermayr2020JPCL}
Such adapted loss functions and also conventional ones are employed in different types of ML models. In the following, we focus on the two most widely used models for the description of the excited states: Kernel methods and NNs.

\paragraph{Kernel methods}
Kernel methods~\cite{Hofmann2008} are based on a similarity measure between data points. Examples are KRR or GPR, which go beyond linear regression by applying the kernel trick and ridge regression. Ridge regression is used to find the weights, which differs from linear regression by a regularization term, $\lambda$:
\begin{equation}
    w = (K+\lambda \mathbf{1})^{-1} Y^{QC}
\end{equation}
Y$^{QC}$ refers to the training data and $K$ to the kernel matrix.

The kernel trick makes it possible to apply ridge regression to non-linearly separable data by mapping them into a higher-dimensional feature space, in which the data points are linearly separable.
Therefore, a kernel function, $k$, e.g. a Gaussian or Laplacian, is placed on each compound to measure the distance to all of the other compounds in the training set. The kernel function defines the non-linearity of the model. A property of a query compound, $\alpha$, can be obtained as the weighted sum of regression coefficients and kernel instances:
\begin{equation}\label{krr}
    Y^{ML}(X_\alpha) = \sum_\beta^{N_M} w_\beta K(X_\alpha, X_\beta).
\end{equation}
The size of the kernel matrix is dependent on the number of training points and hence the depth of the model is inherently linked to the size of the training set, which is why they are called ''non-parametric''.\cite{Bishop2006,Raschka2019}

An advantage of kernel methods is that they mainly contain two hyperparameters, i.e., internal model parameters, which need to be optimized for proper training.
Most important are the width of the non-linear kernel function, $\sigma$, and the regularization. The latter is used to prevent the model from overfitting -- the case when the model fits training data including noise almost exactly and fails to accurately predict data points not included in the training set but stemming from an interpolative regime. As quantum chemical data is most often noise-free, the regularization term is usually small.

As the optimization of hyperparameters is often a tedious task, kernel methods with their few hyperparameters are easier to use than, e.g., NNs with many hyperparameters. Nonetheless, kernel methods can provide almost exact solutions of problems under investigation.~\cite{Faber2018JCP} A drawback is, however, that the inversion of the kernel matrix can become expensive and even be rendered infeasible on current computers due to increasing memory requirements with increasing training set size.~\cite{Westermayr2020MLST}

Further, kernel methods are usually defined to only map an input to a single output. Therefore, they can treat only one electronic state at a time in standard implementations and, thus, can be referred to as single-state models.  A single-state treatment requires a separate ML model for each electronic state or for each property resulting of a pair of states, whereas a multi-state ML model describes all electronic states and properties resulting from different pairs of states at once.~\cite{Westermayr2020MLST_Perspective,Westermayr2020MLST} Hence in their standard implementation, the treatment of several excited states necessitates the use of several kernel models, which is commonly done in the research field of quantum chemistry.~\cite{Xue2020chemrxiv,Ramakrishnan2015,Richings2017CPL,Dral2018JPCL,Hu2018JPCL} The description of forces is possible for the ground state or a single excited state and is implemented, e.g., in the QML toolkit using KRR and the Faber-Christensen-Huang-Lilienfeld (FCHL) representation,~\cite{Christensen2019JCP} in the symmetric gradient domain ML (sGDML)~\cite{Chmiela2018NC,Chmiela2018NC} method or with smooth overlaps of atomic positions (SOAP)~\cite{Bartok2013PRB} for GPR.~\cite{Bartok2018PRX} 

\begin{figure}[ht]
    \centering
    \includegraphics[scale=0.5]{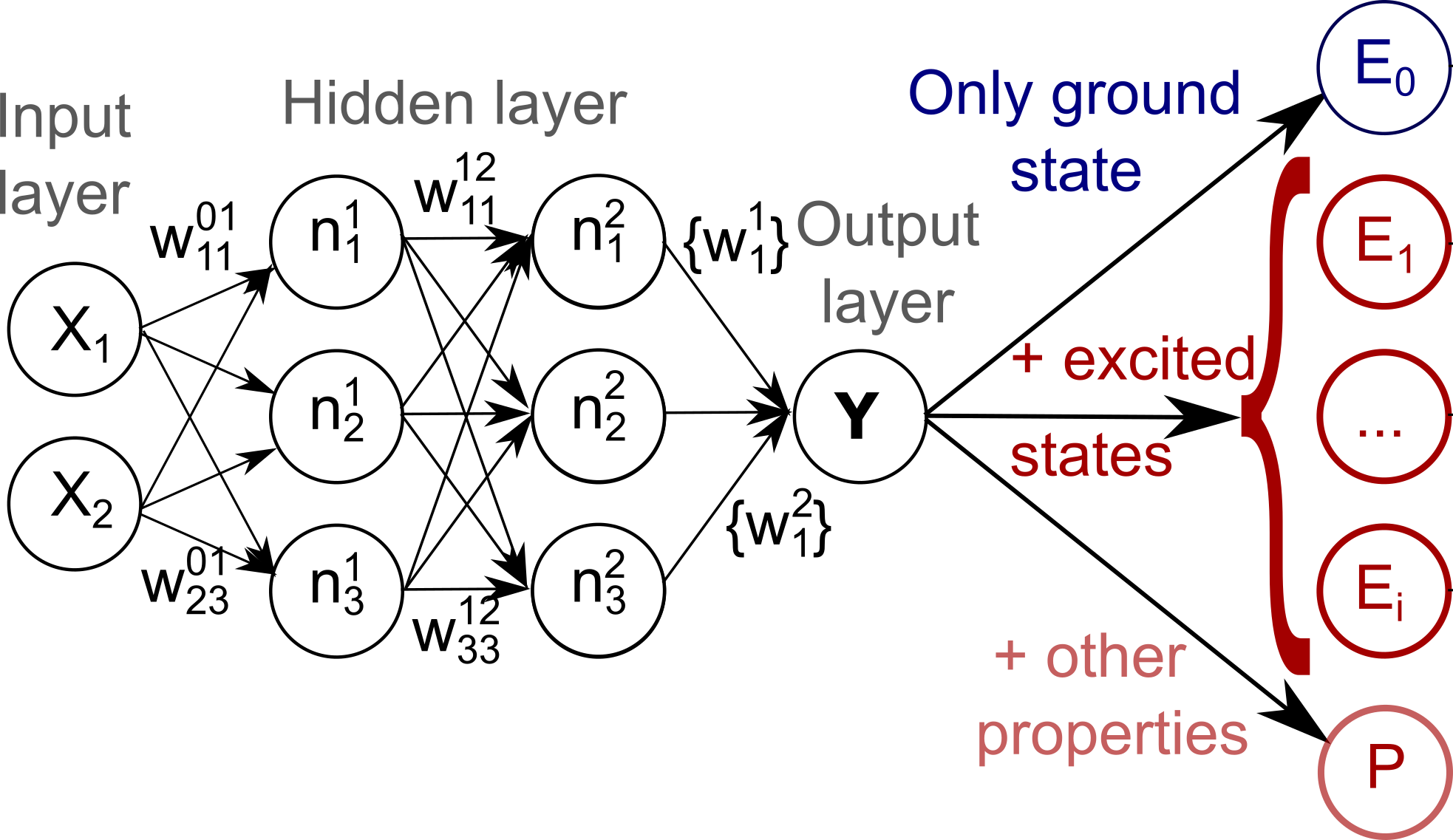}
    \caption{Schematic representation of a multi-layer feed-forward NN with inputs, ${X}$, nodes, $n$, and outputs, $Y$. In the usual implementation for the fitting of PESs, the NN maps a molecular geometry to the ground state, which could be similarly done for any other single state. In case a manifold of excited states is described, one molecular input can also be mapped to a vector of different excited states and additionally, other properties can be included. The forces are treated as derivatives of the NN potentials with respect to Cartesian coordinates. 
    }
    \label{fig:NN}
\end{figure}

\paragraph{Neural Networks} Another prominent approach in ML is the use of NNs as highly flexible parametric functions, which can fit huge amounts of data and can map a molecular input to many quantum chemical outputs.~\cite{Westermayr2020MLST}
The simplest form of NNs are multi-layer feed-forward NNs, which are schematically represented in Fig.~\ref{fig:NN}.
As it is visible in Fig.~\ref{fig:NN}, the width of the model is dependent on the number of nodes, $n_r^t$, which are connected to each other using weights, $w_{rs}^{tu}$. The indices refer to a connection between node r and node s from layer t and layer u, respectively. The number of nodes and hidden layers can be chosen independently of the training set size. 

Due to the highly flexible functional form of NNs, highly complex relationships can be fit, but an analytical solution to find the weights is not available (in contrast to KRR). A numerical solution can be obtained with stochastic gradient algorithms, which are frequently applied to obtain a step-wise update of the weights:
\begin{equation}
    \label{eq:sgd}
    w_{k+1}=w_k - l_r \nabla L_2(w).
\end{equation}
The gradient of the loss function as given in eq.~(\ref{eq:loss}) with respect to the weights is multiplied with a so-called learning rate, $l_r$. This hyperparameter is deemed one of the most important hyperparameters used for training.~\cite{Goodfellow2016,Glorot2010} In order to obtain an optimal solution, the learning rate needs to be chosen properly.
Algorithms such as AdaGrad~\cite{Duchi2011JMLR} or Adam~\cite{Adam2014} can automatically adapt the learning rate during training. Further, the second-order derivatives can be included into algorithms, which is for instance done in the global extended Kalman filter~\cite{Puskorius1991}, in its parallel variant,\cite{Singraber2019JCTCa}
or the element-decoupled variant~\cite{Gastegger2015JCTC}. The loss function can be adapted so that more than only one property can be trained at once. This is often done to include the forces in the training process. 

In general, NNs possess various hyperparameters like the learning rate, regularizers, number of nodes, etc. As a consequence, an extensive hyperparameter search complicates the use of NNs and makes them more complex to apply than kernel methods.

Besides simple multi-layer feed-forward NNs, high-dimensional variants exist. These networks comprise several atomic NNs, which represent atoms in their chemical and structural environment and are thus also called atomistic NNs. 
Each local atomic contribution, $E_a$, can be summed up to provide the energy of the whole system, $E$, which is well known to work for the ground state PESs:
\begin{equation}
    \label{eq:hdnn}
    E=\sum_{a=1}^{N_A} E_a,
\end{equation}
and was originally implemented by Behler to construct high-dimensional NN potentials.~\cite{Behler2011JCP}
Embedded-atom NNs~\cite{Zhang2019JPCL} are similar to high-dimensional NNs in their way of constructing the energy of a system. They differ in the underlying descriptors to the ones of Behler. Atomic contributions to the energy are dependent on the embedded density of atoms and are summed up according to eq~\ref{eq:hdnn}. These embedded density-like descriptors are approximated from atomic orbitals.\\

Independent of a simple or an atomistic architecture, the model can be used to fit a single output or a vector of many outputs at the same time. For ground state problems, a single-state model is usually used, which maps an input to a single output, e.g. the PES of the ground state. Oftentimes, this single-state fashion is adapted to fit different excited states with different NN models.~\cite{Guan2020JCTC,Jiang2020JPCL,Chen2018JPCL,Ye2019PNAS} However, it has been shown that including more excited-states in one model can be advantageous~\cite{Westermayr2020MLST}, as the excited-states are inherently linked to each other and so are the excited-state properties.~\cite{Gonzalez2020} Treating many excited states can be referred to as multi-state model and the inclusion of more properties can result in a multi-property model.~\cite{Westermayr2020MLST,Westermayr2020MLST_Perspective,Schuett2019NC,Zhang2020arXiv,Zhang2020JPCC} The different properties can be weighted with respect to their magnitudes or importance for a given chemical problem under investigation, such that the best possible accuracy can be obtained.~\cite{Westermayr2020JPCL}

Another type of networks are convolutional NNs, which are most often applied in image or speech recognition~\cite{LeCun1995,Krizhevsky2012NIPS,Sainath2015NN}, but can also be adapted to process a molecular input and identify an optimal molecular descriptor. This type of network can be combined in an end-to-end fashion with an architecture, which fits this generated molecular representation to a query output.~\cite{Gilmer2017,Schuett2017NC,phdschuett,Schuett2018JCP,Schuett2019JCTC} 

An important ingredient of all these ML models is the descriptor, which is mapped to the output. In most studies, the descriptor is one of many different possibilities to represent a molecule, which will be discussed in the next section.

\subsection{Descriptors and Features}
Electronic structure methods can process and uniquely identify molecules using e.g. Cartesian coordinates. In contrast, such types of inputs are not optimal for ML models as the same molecular geometry, but translated or rotated, could only be mapped to the same output with great effort and unnecessary computational cost. Hence, a molecular descriptor should fulfill the following requirements: It should be translationally, rotationally, and permutationally invariant as well as differentiable.~\cite{vonLilienfeld2015IJQC} It should also be unique with respect to the relative spatial arrangement of atoms, universally applicable for any kind of system, and computationally efficient.\cite{Haghighatlari2020C}
However, a descriptor can be more than that; it can already include a part of the mapping, e.g., from a molecular structure to an energy. It can thus ease the task of the regressor and help to attain the best possible accuracy for a given training set. 

The ways to represent a molecule to an ML model can be classified roughly into two categories: molecule-wise descriptors, which represent the molecule as a whole to the ML model, and atom-wise descriptors, which represent atoms in their chemical and structural environment and build up a property using local contributions.~\cite{Behler2015IJQC,vonLilienfeld2015IJQC} Both ways in describing a molecular system have their merits and pitfalls and will be discussed along with their applications in recent studies for the excited states in the following.

\paragraph{Molecule-wise descriptors}
The distance matrix is one of the simplest descriptors that preserves rotational and translational invariance. Most often it is used in its inverse form with distances between atoms a and b,
\begin{equation}\label{eq:invD}
    D_{ab}=\frac{1}{\mid\mid r_a-r_b\mid\mid},
\end{equation}
giving rise to the symmetric inverse distance matrix, $\mathbf{D}$. Due to the ill-definition of diagonal elements, which are not differentiable, the diagonal elements are excluded and only the upper or lower triangular matrix is used to represent a molecule to an ML model.~\cite{Ramakrishnan2015} Since the Hamiltonian contains distances rather in the denominator, it makes sense to also use the matrix of inverse distances.\cite{Westermayr2019CS} The matrix of inverse distances is very similar to the Coulomb Matrix, $\mathbf{C}$:~\cite{Rupp2012PRL}
\begin{equation}
\label{eq:coulomb}
    C_{ab} =\begin{cases} 0.5Z_a^{2.4}~~\text{if} ~~ a=b  \\ 
   \frac{Z_a Z_b}{\mid\mid r_a-r_b\mid\mid}\end{cases}
\end{equation}
but the Coulomb matrix additionally considers the atomic charges, Z. These types of descriptors are frequently used in ML studies for the excited states. For example, MLMD simulations in the excited states could be advanced using these simple descriptors~\cite{Dral2018JPCL,Hu2018JPCL,Westermayr2019CS,Westermayr2020MLST} and were also accurate enough to fit NNs and KRR models for excited-state properties~\cite{Ye2019PNAS,Westermayr2020MLST,Westermayr2019CS,Ramakrishnan2015JCP,Haese2016CS,Xue2020chemrxiv}. Distance based descriptors are further implemented in several program packages that have been used for photodynamics simulations with KRR. For example, MLAtom~\cite{Dral2019JCC} contains the Coulomb Matrix and a representation that includes all nuclear pairs in form of normalized inverted internuclear distances~\cite{Dral2017JCP}. The QML toolkit~\cite{QML} includes the Coulomb matrix in addition to other representations, such as bag of bonds~\cite{Hansen2015JPCL}. Another variant are polynomials formed from inverse distances~\cite{Westermayr2019CS}.

These molecule-wise  descriptors have the advantage of being easy to use and implement. Especially for small molecular systems and with regard to the training of an ML model, they are cheap. However, they might miss some important information based on angular distributions. Currently, it is also investigated, whether representations based on two-body or three-body terms are accurate enough to uniquely identify a molecule.~\cite{Pozdnyakov2020arXiv}

A problematic issue of the aforementioned types of distance-based molecular descriptors is that they are not permutationally invariant.~\cite{Behler2015IJQC,vonLilienfeld2015IJQC,phdschuett,Westermayr2020MLST_Perspective} This problem can be mitigated by data augmentation, i.e., randomly permutation of atoms by mixing of matrix rows, which results in more data points for the same molecular input. The additional amount of data increases rapidly with the system size and could lead to long training times.~\cite{Behler2015IJQC,phdschuett} Alternatively, another metric than the commonly used L$_1$ or L$_2$ norms can be employed, the so-called Wasserstein metric, which was tested with the Coulomb matrix.~\cite{Caylak2020arXiv}

Permutation invariant polynomials (PIPs), introduced by Bowman and co-workers,~\cite{Braams2009IRPC,Bowman2011PCCP,Qu2018ARPC} are frequently applied in a PIP-NN approach by Guo and coworkers to investigate photochemical problems~\cite{Li2013JCP,Jiang2013JCP,Jiang2014JCP,Jiang2016IRPC,Xie2018JCP,Guan2019JCP,Guan2019PCCP}. The advantage of these polynomials is that they are invariant to permutation of atoms and inversion.~\cite{Guan2019JCP} They comprise single-valued functions, $p_{ab}$, such as logarithmic or Morse like functions, which incorporate internuclear distances, $r_{ab}$. The PIP vector, $\mathbf{G}$ is obtained applying a symmetrization operator, $\hat{S}$, accounting for possible permutation operations:
\begin{equation}
    \mathbf{G}=\hat{S} \prod_{a<b}^{N_A}p_{ab}
\end{equation}
with an example of p$_{ab}$:
\begin{equation}
    p_{ab}= e^{-cr_{ab}}.
\end{equation}
Evidently, additional hyperparameters such as $c$ have to be optimized and the choice of PIPs is generally not unique.~\cite{Jiang2013JCP,Malbon2020PCCP}

Another negative aspect of molecule-wise descriptors is that they can only treat one molecular system, because the input size is fixed. The input dimension could, in principle, be defined according to the largest system included in the training set, but this would lead to unnecessarily large input  vectors for smaller systems, which would then contain many zero values.~\cite{phdschuett,Behler2011JCP} The training of more ML models, each for one specific system size, is one possible solution~\cite{Haese2016CS}, but obviously necessitates the training and evaluation of more than one ML model.

\paragraph{Atom-wise descriptors}
In contrast, atom-wise representations allow for a fitting of molecules of arbitrary size and composition. Such descriptors are state-of-the-art for ground-state problems with commonly used examples being the SOAP~\cite{Bartok2013PRB}, atom-centered symmetry functions (ACSF)~\cite{Behler2011JCP}, weighted ACSFs~\cite{Gastegger2018JCP,Herr2019JCP} or the FCHL representation.~\cite{Faber2018JCP,Christensen2020JCP} These representations describe atoms in their chemical and structural local environment and usually rely on a cut-off function. This cut-off function defines the sphere around an atom, which is deemed to be important and is therefore considered when modelling the atomic local environment. Radial distribution functions, so-called second-order terms, account for inter-atomic distances and are often used together with angular distribution functions, i.e., third-order terms. It is further beneficial to include first-order terms, i.e., the stoichiometry of atoms.~\cite{Faber2018JCP,Schuett2018JCP,Gastegger2018JCP,Herr2019JCP} Most often, higher order terms than third-order terms are not included due to increasing costs and little improvements in accuracy.~\cite{phdschuett} 

The description of PESs from atomic contributions is beneficial in order to treat systems of arbitrary sizes and to use systematic molecular fragmentation methods.~\cite{Gastegger2016JCP}
Admittedly, the validity of this approach is not so clear for the excited-states and consequently, such representations are less frequently used in ML studies targeting the excited states. Up to day, only small molecules have been fitted with atom-wise representations, which are too small to prove the validity of excited-state PESs, which are constructed from local atomic contributions. To the best of our knowledge, the largest molecule fitted with atom-wise descriptors contained 12 atoms and was \textit{N}-methylacetamide.~\cite{Zhang2020arXiv}
Other molecules were CH$_2$NH$_2^+$~\cite{Westermayr2020MLST,Westermayr2020JPCL}, CH$_2$NH~\cite{Chen2018JPCL}, SO$_2$~\cite{Westermayr2020JPCL} or CSH$_2$~\cite{Westermayr2020JPCL}. Further studies are needed to demonstrate whether an atom-wise construction of excited-state properties and PESs is possible or not. Nevertheless, this approach is most powerful for studies that aim to describe large and complex systems, which could potentially be described from smaller building blocks. For instance, the construction of a DNA double strand or a peptide could be, at least in principle, constructed from ML models that are trained on their smaller subsystems, i.e., DNA bases and amino acids, respectively. Unfortunately, we are far away from having achieved a description of large molecular systems for the excited states, let alone the construction of accurate PESs of medium-sized molecular systems, such as DNA bases or amino acids.

\paragraph{Other types of descriptors} Besides the benefits high-dimensional ML models offer for the fitting of PESs of molecules, descriptors are not restricted to the aforementioned examples. In general, any type of descriptor might be suitable for a given problem. Applied descriptors range from topological and binary features generated from SMILES strings~\cite{Kang2020arXiv} to normal modes, which are often used as a coordinate system and descriptors to fit diabatic PESs.~\cite{Richings2018JCP,Alborzpour2016JCP,Richings2019JCTC,Polyak2019JCP,Richings2017JCTC,Richings2018JCP,Richings2017CPL,Li2013JCP,Jiang2013JCP,Jiang2014JCP,Jiang2016IRPC,Xie2018JCP,Guan2019JCP,Guan2020JCTC} Other types of molecular features besides structure-based ones, e.g. electronegativity, bond-order, oxidation states, ...,~\cite{Jeong2020JCTC,Taylor2020JPCA} are also used.

\paragraph{Automatically generated descriptors} The selection of an optimal descriptor and the optimization of the related parameters for this descriptor is no trivial task and requires expert knowledge in many cases.~\cite{phdschuett} 
A way to circumvent an extensive parameter search is offered by the aforementioned message passing NNs~\cite{Gilmer2017}, which include the descriptor parameters in the network architecture. In this way, they automatically fit the optimal parameters of a descriptor for a given problem, i.e., training set under investigation. Such tailored descriptors can guarantee highly accurate solutions if the NN model is trained properly.  PhysNet,~\cite{Unke2019JCTC} HIP-NN~\cite{Lubbers2018JCP} or Deep Tensor NN (DTNN)~\cite{Schuett2017NC}, which forms the basis of the deep learning model SchNet,~\cite{Schuett2018JCP,Schuett2019JCTC}, which in turn is used within the SchNarc approach for excited states,\cite{Westermayr2020JPCL} are examples of such NNs. 

\section{Application of ML for Excited States}
In this chapter, we review ML studies of excited states and their properties. We aim to show how they have been employed to improve static and dynamics calculations and focus on the used type of regressor, descriptor, training set, and property. We will classify the approaches according to Figure~\ref{fig:targets}.

\subsection{Parameters for Quantum Chemistry}
At the current state of research, the user must decide whether a multi-reference method is necessary or a single reference method is sufficient to describe a chemical problem. It would be helpful if ML models could suggest a suitable reference method, e.g. based on a literature search. Unfortunately, such a tool is not yet available, but ML can help to select an active space for multi-reference methods. Jeong et. al~\cite{Jeong2020JCTC} developed an ML protocol for classification based on XGBoost~\cite{Chen2016} to allow for a ''black box'' use of many multi-reference methods by automatically selecting the relevant active space for molecular systems. The tedious selection of active orbitals and active electrons can thus be avoided. The accuracy of this approach was demonstrated for diatomic molecules in the dissociation limit and the molecules were represented via the molecular orbital bond order and the average electronegativity of the system.

\subsection{ML of Primary Outputs}
To the best of our knowledge, no ML models for providing primary outputs of quantum chemistry exist for excited states (see Figure~\ref{fig:targets}). Targeting the primary output of a quantum chemistry simulation, i.e., the N-electron wave function, or providing ML density (functionals) is far from trivial even for ground-state problems.~\cite{Carleo2017S,Saito2017JPCJ,Nomura2017PRB,Han2018JCP,Townsend2019JPCL,Schuett2019NC,Pfau2019arXiv,Hermann2019arXiv,Gastegger2020arXiv,Choo2018PRL,Zheng2019PRL,Brockherde2017NC,Fabrizio2019CS,Grisafi2019ACSCS,Fabrizio2020CHIMIA} However, such an approach for excited states could solve many problems and allow for wave function analysis, providing additional insights like the excited state characters.\cite{Mai2018CCR}
Therefore, we expect such models to appear in the near future.

\subsection{ML of Secondary Outputs}
In the following, we summarize the contributions of ML models that fit the secondary output of quantum chemical calculations, i.e., PESs, SOCs, NACs, and transition as well as permanent dipole moments in the adiabatic and diabatic basis (Figure \ref{fig:targets}). 
The prediction of the manifold quantities (see Fig.~\ref{fig:jablonski}) can be done in two ways, i.e., in a single-state fashion and in a multi-state fashion.~\cite{Westermayr2020MLST} 
The applicability of such ML models to the simulation of photodynamics will be discussed.

\subsubsection{ML in the Diabatic Basis}

Diabatic PESs are fitted with ML and related methods since more than 25 years.~\cite{Ischtwan1994JCP,Bettens1999JCP} An advantage of diabatic PESs is their smoothness, which is perfectly matched by ML models built upon smooth functions. However, the tedious procedure to generate diabatic PESs remains. Some effort is therefore devoted to develop ML-assisted diabatization procedures and eliminate this limiting step.
\paragraph{Diabatization}
Williams et. al~\cite{Williams2018JCP} incorporated NNs into diabatization by ansatz and fit diabatic NO$_3$ PESs.
Recently, Shen and Yarkony~\cite{Shen2020JPCA} fit two diabatic potentials of the cyclopentoxy radical, C$_5$H$_9$O, and one state of cyclopentoxide, C$_5$H$_9$O$^-$,  with 356 data points sampled from scans along different reaction coordinates. The diabatization was assisted with NNs. Due to the high dimensionality of the system,
the authors resort to application of regularization in the fitting algorithm and an adapted loss function to obtain an accurate representation of two-state diabatic PESs with NNs. This novel strategy is envisioned for the computation of the photoelectron spectrum of cyclopentoxide~\cite{Shen2020JPCA}. Fitting 39 degrees of freedom in the diabatic basis is a huge improvement in this research field. 
The authors further note that a comprehensive sampling of the full relevant PESs in such high dimensional space is problematic.

Due to the aforementioned problems, a description of medium-sized to large molecules with diabatic potentials is often done with more crude approximations.~\cite{Williams2018JCP,Gomez2019JPCA} An example is the LVC model~\cite{Koeppel04}, with its one-shot variant~\cite{Plasser2019PCCP}, or the exciton model.~\cite{Mennucci2009JPCA,Menger2018JCTC} For more details on this topic, the reader is referred to refs~\cite{Jasper2004,Koeppel04,Yarkony2004,Worth2004ARPC,Segatta2019CR,Cupellini2020BBAB}.
The Frenkel exciton Hamiltonian can be used to describe light-harvesting systems or charge-transfer.~\cite{Menger2018JCTC,Mennucci2009JPCA} Such a Hamiltonian was constructed for the investigation of the excited state energies of bacteriochlorophylls of the Fenna-Matthews-Olson complex. Multi-layer feed-forward NNs with the Coulomb matrix as a molecular descriptor could accelerate the construction of such Hamiltonians for the prediction of excited-state energies.~\cite{Haese2016CS} The effective Hamiltonian of the whole complex was subsequently used to predict excitation energy transfer times and efficiencies. Therefore, H\"{a}se et al. used exciton Hamiltonians as an input.~\cite{Haese2017CS}

\paragraph{Fitting diabatic potentials and properties}
Given diabatic PESs, ML models can be used to fit them. KRR models are often employed for this task, due to their ease of use and ability to provide accurate predictions, as mentioned above. 
Recent studies by Habershon and co-workers focus on interpolation of diabatic PESs and their use for grid-based quantum dynamics methods, i.e., variational Gaussian wavepackets and MCTDH. The butatriene cation has been investigated in two-dimensions comprising two electronic states.~\cite{Richings2017CPL} The description of this molecule has been recently advanced with a new diabatization scheme, namely Procrustes diabatization. The method was evaluated with two-state direct-dynamics MCTDH (DD-MCTDH) simulations of LiF and applied to four electronic states of butatriene.~\cite{Richings2020JCP}
Some of the authors also carried out DD-MCTDH 4-mode/2-state~\cite{Richings2018JCP} and subsequently 12-mode/2-state dynamics of pyrazine~\cite{Richings2019JCTC}. The investigation of the higher-dimensional space of pyrazine could be achieved by systematic tensor decomposition of KRR and advances conventional MCTDH simulations considerably with respect to accuracy and computational efficiency. Further, the method was applied to investigate the ultrafast photodynamics of mycosporine-like amino acids, which are suitable as ingredients in sunscreens due to their photochemical properties and photostability.~\cite{Richings2019FD} However, the reduced 6-dimensional and 14-dimensional DD-MCTDH simulations with KRR interpolated PESs were unable to reproduce the expected ultrafast photodynamics, which had been observed in previously performed surface hopping calculations and is typical for sunscreen ingredients. 
The authors note that the inclusion of more adiabatic states for the diabatization procedure and the consideration of additional relevant modes can lead to more accurate results. All of the reference simulations were carried out at the CASSCF level of theory with KRR fitted diabatic PESs.

In addition to KRR models, NNs were also used to describe diabatic PESs. Seminal works include PIP-based NNs by Guo, Yarkony and co-workers. Absorption spectra and the dynamics of excited states of NH$_3$ and H$_2$O could be studied by fitting potential energy matrix elements.~\cite{Li2013JCP,Jiang2013JCP,Jiang2014JCP,Jiang2016IRPC,Xie2018JCP,Guan2019JCP,Malbon2020PCCP} Subsequently, some of the authors fit the dipole moments corresponding to the diabatic 1,2$^1$A surface of NH$_3$.~\cite{Guan2020JCTC}
SOCs of formaldehyde were learned with NNs in the diabatic picture.~\cite{Guan2020JPCL} 341 data points were used for training of SOCs. A singlet and a triplet state in the adiabatic basis were transformed to diabatic states using Boys localization~\cite{Subotnik2008JCP}. Since this diabatization is based on transition dipole moments, the respective properties of the excited states had to be phase corrected. The authors proved the accuracy of their fitted PESs and emphasized the usability of the ML models to describe full-dimensional quantum dynamics.~\cite{Guan2020JCTC,Guan2020JPCL,Malbon2020PCCP} 
Very recently, they investigated the $OH+H_2$ reaction, i.e., the nonadiabatic quenching of the hydroxyl radical colliding with molecular hydrogen. Four diabatic potentials including forces and couplings were fitted using a least squares fitting procedure. 1345 data points of 1,2,3 $^2A$ adiabatic PESs were computed with MR-CISD.~\cite{Malbon2020PCCP}

The aforementioned ML models are single-state models. Each energetic state and each coupling or dipole moment value resulting from different pairs of states is fitted with a separate ML model. While this yields justifiable accuracy for energies and diabatic coupling values~\cite{Westermayr2020MLST}, dipole moments are vectorial properties and need to preserve rotational covariance.~\cite{Zhang2020arXiv} \\

As the aforementioned studies show, ML models are generally powerful to advance quantum dynamics simulations for the excited states and can also assist the construction of effective Hamiltonians.
However currently, diabatic PESs cannot simply be fit for systems with arbitrary size and arbitrary complexity. The diabatization remains a methodological bottleneck, where additional developments are needed.

The investigation of medium-sized to larger molecular systems, especially the investigation of their temporal evolution, is more often carried out in the adiabatic basis using on-the-fly simulations. An increasing number of recent studies focus on fitting such adiabatic PESs.  The inconsistencies in adiabatic properties make such quantities generally more challenging to fit, which is why this field of research gained a lot of attention relatively late, i.e., only in the last 3 years.

\subsubsection{ML in the Adiabatic Basis}

\paragraph{Surface hopping MD}
Probably, the first ML models for MQCD calculations date back to the year 2008.~\cite{Carbogno2008PRL} Nonadiabatic MD simulations were carried out with NN-interpolated PESs to investigate O$_2$ scattered from Al(III). Symmetry functions were used as descriptors.~\cite{Behler2007JCP} A spin-unpolarized singlet and a spin-polarized triplet state at DFT level of theory were fitted with 3768 data points.~\cite{phdbehler,Behler2007JCP} This two-state spin-diabatic problem allowed for evaluation of coupling values and singlet-triplet transitions with the fewest switches surface hopping approach.~\cite{Tully1990JCP,Tully1991IJQC} In a later study, another adiabatic spin-polarized PES was included and coupling values were computed between singlets and triplets~\cite{Jansen1999IJQC} and evaluated from constructed Hamiltonian matrices.~\cite{Carbogno2010PRB} MD simulations were executed using a manifold of ML-fitted PESs according to different spin-configurations. The studies showed that singlet-triplet transitions are highly probable during the scattering event of O$_2$ on Au(III).\cite{Carbogno2008PRL,Carbogno2010PRB}

After these two seminal studies, the interest in advancing MQC photodynamics simulations in the adiabatic basis increased mainly in the last three years. One of the first works during this time was conducted by Hu et. al~\cite{Hu2018JPCL}, who investigated the nonadiabatic dynamics of 6-aminopyrimidine with KRR and the Coulomb matrix.
Due to the many degrees of freedom of the molecule and including three singlet states, a large amount of training data was required ($>$ 65k data points). Coupling values were not fitted but, instead, the Zhu-Nakamura approach was used to compute hopping probabilities.

Later, Dral et al.~\cite{Dral2018JPCL} applied KRR models to accurately fit a two-state spin-Boson Hamiltonian and reproduce reference dynamics using 1,000 and 10,000 data points. NAC vectors were fit in a single-state fashion. 
During dynamics simulations, conformations close to critical regions were computed with the reference method instead of the ML model in order to allow for accurate transitions. 

In another study, Chen et al.\cite{Chen2018JPCL} used two separate deep NNs to fit the energies and forces of two adiabatic singlet states of CH$_2$NH. About 90k data points were used to generate these single-state models. Using the Zhu-Nakamura approach to account for hopping probabilities, the reference dynamics could be reproduced and quantum chemical calculations were replaced completely during the dynamics.

Cui and coworkers~\cite{Chen2019PCCP} further developed a multi-layer energy-based fragmentation method to study the excited-state dynamics and photochemistry of larger systems. This scheme composes a molecular system into a photochemically active (inner) region and a photochemically inert (outer) region. In the original scheme, the active region and the interactions with the outer region are described with the multi-reference method CASSCF, whereas the outer region is treated with DFT. This decomposition of the total energy of a system allows to treat larger systems, which cannot be described fully with CASSCF. The approach is similar to QM/MM (quantum mechanics/molecular mechanics) schemes in the mechanical embedding framework. The authors simulated two-state photodynamics of CH$_3$N=NCH$_3$ (inner region) including five water molecules (outer region) without the use of ML. The Zhu-Nakamura approximation to model hopping probabilities in nonadiabatic MD simulations was applied.~\cite{Chen2019PCCP} In order to make the simulations more efficient, the authors replaced the DFT calculations with deep multi-layer feed-forward NNs using a distance-based descriptor~\cite{Zhang2018PRL}, hence they describe the ground state energies and forces of the photochemically inert region with ML and describe the S$_1$ and S$_0$ state of the inner region with CASSCF. The hybrid ML multi-layer energy-based fragmentation method can reproduce the photodynamics of the system.~\cite{Chen2019JPCL} Subsequently, the deep NNs were replaced with embedded-atom NNs~\cite{Zhang2019JPCL} and accurate second derivatives could be computed efficiently.~\cite{Chen2020JPCA_Hessian}

Recently, we sought to fit NACs and transition and permanent dipole moments in addition to energies and forces of three singlet states of the methylenimmonium cation, CH$_2$NH$_2^+$, using deep NNs and the matrix of inverse distances as a molecular descriptor.~\cite{Westermayr2019CS} We were able to perform ML-enhanced excited-state MD simulations with hopping probabilities based on ML-fitted NACs. NNs could replace the reference method MR-CISD completely during the dynamics. Long time scale photodynamics simulations for 1 ns were achieved using the mean of 2 NN models in approximately two months, whereas the reference method would have taken an estimated 19 years to compute the dynamics for 1 ns on the same computer. This study demonstrated the possibility of MLMD simulations to go beyond time scales of conventional methods. As another benefit of the ML models, it was shown that a large ensemble of trajectories could be calculated, still at lower cost than a few trajectories with the reference method.~\cite{Westermayr2019CS} 

With the same training set, we further assessed the performance of KRR together with von Lilienfeld and co-workers.~\cite{Westermayr2020MLST} The operator formalism~\cite{Christensen2019CHIMIA} and the FCHL representation~\cite{Faber2018JCP,Christensen2020JCP} were used to fit the three singlet states of CH$_2$NH$_2^+$. A single-state treatment and a multi-state treatment for predicting energies were compared. To this aim, a multi-state KRR approach as developed with an additional kernel that encodes the quantum energy levels. The accuracy of KRR models could be improved using this extended approach.~\cite{Westermayr2020MLST}
The KRR models were further compared to deep NN models regarding their ability to predict dipole moments and NACs.  
While NNs yielded slightly higher accuracy at the largest available training set size, KRR models exhibited a steeper learning curve, hence more efficient learning. 
The different performance of NNs and KRR models was proposed to be a result of the parametric dependence of the depth of NNs and the non-parametric dependence of the depth of KRR models.  
Results further suggested that small differences between the reference method and ML models, especially in critical regions of the PESs, can lead to completely wrong photodynamics simulations.~\cite{Westermayr2020MLST}
Nevertheless, multi-reference quantum chemical potential energy curves could be faithfully reproduced with KRR models and NN models for the three singlet energies of CH$_2$NH$_2^+$. 

In order to omit the extensive hyperparameter search of the descriptor and regressor, we further developed the SchNarc approach for photodynamics~\cite{Westermayr2020JPCL}, which is based on SchNet.~\cite{Schuett2018JCP,Schuett2019JCTC} SchNarc allows for (1) a description of SOCs, (2) an NAC approximation based on ML-fitted PESs, their first and second derivatives with respect to Cartesian coordinates, and (3) a phase-free training algorithm to enable a training of raw quantum chemical data. The SchNarc approach is based on the message passing NN SchNet~\cite{Schuett2018JCP,Schuett2019JCTC}, which was adapted by us for the treatment of a manifold of excited electronic states. Additionally, this model can describe dipole moments using the charge model of ref~\cite{Gastegger2017CS}, also adapted for excited-states.
All excited-state properties can be described in one ML model in a multi-state fashion. The performance of SchNarc was evaluated with surface hopping dynamics: Three singlet and three triplet states of SO$_2$ were computed with ML models for 700 fs and the underlying PESs were based on an "one-shot" LVC(MR-CISD) model~\cite{Plasser2019PCCP}. CSH$_2$ was investigated using 2 singlets and 2 triplet states for 3 ps at CASSCF level of theory representing slow population transfer, and the performance of SchNarc to reproduce ultrafast transitions during dynamics was assessed using CH$_2$NH$_2^+$ with the aforementioned training set. The hopping probabilities were computed according to ML-fitted SOCs and NACs -- the latter being fitted in a rotationally covariant way as derivatives of virtual ML properties and approximated from ML PESs. 
In all cases, excellent agreement with the reference method could be achieved.
Noticeably, all the aforementioned photodynamics studies with ML models~\cite{Hu2018JPCL,Dral2018JPCL,Chen2018JPCL,Westermayr2020JPCL,Westermayr2020MLST,Westermayr2019CS} make use of Tully's fewest switches surface hopping approach with hopping probabilities based on coupling values or approximated schemes.~\cite{Tully1990JCP,Tully1991IJQC}

\paragraph{Exemplary timings for MLMD, LVC dynamics, and MQCD} The speed-up of simulations is one of the main arguments employed for promoting ML in quantum chemistry. In order to get an idea about the computational time used in different calculations, we provide an example here. The timings of surface hopping MD with analytical PESs (from LVC), quantum chemical PESs, and ML-fitted PESs based on fitted and approximated NACs from Hessians can be found for three exemplary molecules in Table~\ref{tab:timing}. 

Obviously, crude excited-state force fields like the LVC model are faster than ML models, e.g., for SO$_2$. We note that even such force field implementations can probably still be streamlined for speed but will always be more expensive than ground-state MD simulations, where it would take approximately 0.005 seconds to simulate 100 fs for the gas-phase methylenimmonium cation, CH$_2$NH$_2^+$, using a state-of-the-art program like Amber.\cite{Salomon-Ferrer2013WCMS}

However, dynamics based on highly accurate quantum chemical calculations can be accelerated significantly with ML-fitted PESs, e.g., SchNarc models for CH$_2$NH$_2^+$ based on MR-CISD/aug-cc-pVDZ.\cite{Westermayr2020JPCL} The speedup is higher if NACs are learned directly (MLMD1) compared to when they are approximated from Hessians (MLMD2). A lot of Hessian evaluations are required  in this example because ultrafast transitions occur in CH$_2$NH$_2^+$. The second-order derivatives reduce the efficiency by a factor of about ten. Nevertheless, Hessian calculations of ML-PESs can be accelerated by a factor of about 5-10 using a GPU (dependent on the molecule and GPU used).

Table~\ref{tab:timing} further shows that a cheaper underlying reference method, such as CASSCF(6,5)/def2-SVP used for CSH$_2$, does not allow for such a significant speed-up. In this example however, the difference between simulations with learned NACs and approximated NACs is small because the dynamics of CSH$_2$ is characterized by slow population transfer. Hence, less Hessian evaluations are required to estimate the hopping probabilities.

The time required to train a SchNarc model on a GeForce GTX 1080 Ti GPU is approximately 11 hours for energies and forces of 3 singlet states with 3,000 data points of CH$_2$NH$_2^+$, about 13 hours for energies, forces, and SOCs of 2 singlet and 2 triplet states using 4,000 data points of CSH$_2$ and about 4 hours for energies and forces of 3 singlet states of SO$_2$ using 5,000 data points.

\begin{table}[ht]
 \caption{Comparison of the timings to compute 100 fs with the surface hopping including arbitrary couplings (SHARC)~\cite{sharc-md2,Mai2018WCMS,Richter2011JCTC} method. For SO$_2$ and CH$_2$NH$_2^+$, three singlet states are described and for CSH$_2$ two singlet and two triplet states. The molecule SO$_2$ is approximated using a highly efficient LVC model~\cite{Plasser2019PCCP}, while the underlying reference method to describe the excited states of CH$_2$NH$_2^+$ is MR-CISD/aug-cc-pVDZ and of CSH$_2$ is CASSCF(6,5)/def2-SVP. SchNarc is used for the MLMD simulations. Once, energies, forces, and NACs are trained and predicted (MLMD1) and once, NACs are approximated from first- and second-order derivatives of ML PESs (MLMD2). 2x Intel Xeon E5-2650 v3 CPUs are used.\cite{Westermayr2020JPCL} }
    \centering
    \begin{tabular}{c|c|c|c }
         &\multicolumn{3}{c}{100 fs dynamics [s/CPU]}\\
         & MLMD1 &MLMD2& Reference\\
         \hline
         \hline
         SO$_2$&10 &12 & 2-3\\
         CH$_2$NH$_2^+$ &24 &250& 74,224\\
         CSH$_2$ & 14 & 16 &104\\
    \end{tabular}
    \label{tab:timing}
\end{table}{}

\paragraph{Dipole Moments} In addition to the investigation of the temporal evolution of some systems in the excited states, permanent and transition dipole moments have been computed with ML models. As mentioned before, in our earlier approaches, we fitted permanent and transition dipole moments as single values with NNs and KRR -- strictly speaking we were neglecting the rotational covariance of the vectors (since rotations were negligible in these simulations).~\cite{Westermayr2019CS,Westermayr2020MLST} The SchNarc model improved on this description by treating dipole moments as vectorial properties.
The NN and KRR models for dipole moments have been evaluated and compared to quantum chemical reference dipole moments using learning curves and MAEs. Their potential to compute UV spectra was emphasized.

The use of dipole moments to actually simulate UV spectra was demonstrated by Jiang, Mukamel, and co-workers using
\textit{N}-methylacetamide, a model system to investigate peptide bonds.~\cite{Ye2019PNAS,Zhang2020arXiv} They evaluated the ability of ML to describe transition dipole moments at TDDFT level of theory. In a first attempt,\cite{Ye2019PNAS} the authors predicted dipole vectors as independent values. 14 internal coordinates in combination with multi-layer feed-forward NNs were used to predict transition energies of \textit{N}-methylacetamide. Xyz representations served as an input for fitting ground state dipole moments. The Coulomb matrix was employed to fit transition dipole moments for the n$\pi^\ast$ and $\pi\pi^\ast$ transitions, but did not lead to sufficiently accurate results. Higher accuracy was obtained by replacing the atomic charges in the Coulomb matrix (eq~\ref{eq:coulomb}) with charges from natural population analysis. The choice of descriptors was justified by screening different types of descriptors for prediction of different properties. 
In a later work, some of the authors used embedded-atom NNs to predict transition dipole moments from atomic contributions in a rotationally covariant way. The dipole moment vector between two states i and j was obtained as a linear combination of three contributions:
\begin{equation}
    \mu_{ij} = \mu_T^i + \mu_T^j + \mu_T^3
\end{equation}
$\mu_T^i$ and $\mu_T^j$ were modeled using the charge model of ref~\citenum{Gastegger2017CS}. A third contribution, $\mu_T^3$, was obtained as the cross product of $\mu_T^i$ and $\mu_T^j$:
\begin{equation}
    \mu_T^3 = \sum_a^{N_A}q_a^3(\mu_T^i \times \mu_T^j)
\end{equation}
$\mu_T^i$, $\mu_T^j$ and $q_a^3$ were outputs of the same embedded-atom NN.

\subsection{ML of Tertiary Outputs}
The secondary outputs, such as dipole moments or excited state energies can be used to calculate oscillator strengths (eq~\ref{eq:osc}) and energy gaps (Fig.~\ref{fig:targets}(d)). These properties can serve for the modelling of UV absorption spectra. UV spectra were computed in the previously described studies of \textit{N}-methylacetamid with the ML fitted transition dipole moments. Jiang, Mukamel and co-workers~\cite{Ye2019PNAS} applied the transition dipole moment and additionally fitted n$\pi^\ast$ and $\pi\pi^\ast$ excitation energies to compute UV spectra this molecule with NNs. Subsequently, some of the authors~\cite{Zhang2020arXiv} used these excitation energies and the transition dipole moments to model a Frenkel exciton Hamiltonian for proteins using amino acid residues and peptide bonds. This effective Hamiltonian could subsequently be used to approximate UV spectra of proteins. The interaction between amino acid residues and peptides was neglected so only the isolated peptide excitation energies, i.e., those of \textit{N}-methylacetamid, and the respective transition dipole moments were needed to construct the Hamiltonian. The authors made use of the dipole-dipole approximation~\cite{Kasha1965} and applied embedded-atom NNs. 

Ramakrishnan et. al~\cite{Ramakrishnan2015JCP} predicted excitation energies of the lowest-lying two excited singlet states, S$_1$ and S$_2$, as well as corresponding oscillator strengths obtained from TDDFT calculations with KRR. The QM8~\cite{Ruddigkeit2012JCIM} data base was used consisting of 20k organic molecules. With the $\Delta$-learning approach, CC2 accuracy could be obtained. 
Very recently, Xue et al.~\cite{Xue2020chemrxiv} assessed the performance of KRR models with the normalized inverse distances as a molecular descriptor to predict absorption spectra of benzene and a derivative of acridine containing 38 atoms. Therefore, the authors learned the excited-state energy gaps of several states and the corresponding oscillator strengths in a single-state fashion. Applying Gaussian broadening, the absorption cross sections could be computed at TDDFT accuracy.

Pronobis et al.~\cite{Pronobis2018EPJB} compared 2-body, 3-body and automatically designed descriptors to learn TDDFT HOMO-LUMO gaps as well as first and second vertical excitation energies. 
More than 20k molecules of the QM9 data base~\cite{Ruddigkeit2012JCIM,Ramakrishnan2014SD} were selected for this purpose and learning curves were used to evaluate the learning behaviour of different ML models.
While atom-wise descriptors worked well for HOMO-LUMO gaps, the authors concluded that the accuracy of predicted transition energies is not sufficiently accurate and suggested that advanced non-local descriptors might be necessary to achieve higher accuracy. They further proposed the idea of encoding information about the electronic state in the ML model.~\cite{Pronobis2018EPJB} Indeed, our recent study, in which we compared the performance of KRR and NN models with atom-wise and molecule-wise descriptors demonstrated that encoding of the energy level is advantageous.~\cite{Westermayr2020MLST}

Recently, Kang et. al~\cite{Kang2020arXiv} used 500,000 molecules of the PubChemQC~\cite{Nakata2017JCIM} data base to train a random forest model on the excitation energy and the oscillator strength corresponding to the electronic state with the highest oscillator strength. 10 singlet states, as available in the PubChemQC data base, were evaluated for that purpose. The authors used SMILES (simplified molecular-input line-entry system) strings and converted them into descriptors. The descriptors comprised several topological~\cite{Rogers2010JCIM} and binary~\cite{Durant2002JCICS} fingerprints, which were calculated with the help of the RDkit library~\cite{Landrum2016}.
The authors compared the prediction accuracy to the aforementioned models and stated that their model outperformed previous ML models in the task of predicting accurate oscillator strengths and excitation energies for the most probable transition in organic molecules. Analysis of important features led the authors identify that nitrogen-containing heterocycles are important for high oscillator strengths in molecules. The authors concluded that their study could serve the design of new fluorophores with high oscillator strengths.~\cite{Kang2020arXiv}

Ghosh et. al~\cite{Ghosh2019AS} used multi-layer feed-forward NNs, convolutional NNs and DTNNs to fit 16 highest occupied orbital energies from DFT, i.e., the respective eigenvalues, for the computation of molecular spectra with a full width at half maximum of 0.5 eV for Gaussian broadening. 
Geometries from the QM7b~\cite{Blum2009JACS,Montavon2013NJP} and QM9~\cite{Ruddigkeit2012JCIM,Ramakrishnan2014SD} data base were used for training and molecular spectra were tested using 10k additional diastereomers, which were also used by Ramakrishnan et. al~\cite{Ramakrishnan2015JCP} to evaluate the $\Delta$-learning approach. The convolutional NNs with the Coulomb matrix and DTNNs with an automatically generated representation outperformed the simpler NNs. Overall, good agreement to reference DFT spectra could be achieved.~\cite{Ghosh2019AS}

Markland and co-workers~\cite{Chen2020arXiv} trained NNs with atom-centered Chebyshev polynomial descriptors~\cite{Artrith2017PRB} on the TDDFT/CAM-B3LYP/6-31+G* S$_0$-S$_1$ energy gap of the deprotonated trans-thiophenyl-p-coumarate (chromophore of yellow protein) in water and Nile red chromophore in water and benzene. Farthest point sampling~\cite{Imbalzano2018JCP} was used to select about 2,000 data points from a larger set of 36,000 data points and was compared to random sampling.  
The authors assessed the performance of three different ML approaches to compute absorption spectra, spectral densities and 2-dimensional electronic spectra. One model (hidden solvation) completely ignored any environmental effects and only described the chromophore, another model (indirect solvation) incorporated environmental effects within a 5{\AA} cutoff of the atomistic descriptor for the chromophore and a third model (direct solvation) treated the whole system, i.e., the chromophore and the atoms of the solvent, explicitly. As expected, the hidden solvation model turned out to be insufficiently accurate for systems with strong solvent-chromophore interactions, but was comparable to the hidden solvation model when describing Nile red chromophore in benzene. The indirect solvation and direct solvation models were comparable to each other, but with respect to the computational efficiency, the indirect solvation model was beneficial. This model could reproduce reference linear absorption spectra, spectral densities, and could capture spectral diffusion of 2-dimensional electronic spectra of all treated chromophores.~\cite{Chen2020arXiv}

Penfold and co-workers~\cite{Rankine2020JPCA} applied deep multi-layer feed-forward NNs to proof the ability of ML to predict X-ray absorption spectra (XAS), which provide a wealth of information on the geometry and electronic structure of chemical systems, especially in the near-edge structure region. Note that X-Ray free-electron laser spectroscopy can further be used to generate ultrashort X-ray pulses to investigate photodynamics simulations in real-time. The training set for the prediction of Fe K-edge X-ray near-edge structure spectra contained 9040 data points. The inputs for NNs were generated using local radial distributions around the Fe absorption site of arbitrary systems taken from the Materials Project Database.~\cite{Jain2013APLM} Qualitatively accurate peak positions and intensities could be obtained computationally efficient and the structural refinement of nitrosylmyoglobin and [Fe(bpy)$_3$]$^{2+}$ was assessed with NNs.
The authors noted that future development is needed to accurately capture structures far from equilibrium as well as irregularities in the bulk. 

Another study was executed by Aarva et al.,~\cite{Aarva2019CMb} who focused on XAS and X-ray photoelectron spectra of functionalized amorphous carbonaceous materials. By clustering of DFT data with unsupervised ML techniques average fingerprint spectra of distinct functionalized surfaces could be obtained. The authors use GPR. Similarly to the aforementioned state encoding,~\cite{Westermayr2020MLST} the authors encoded the electronic structure, i.e., the $\Delta$-Kohn Sham values (core-electron binding energies), in a Gaussian kernel. This kernel was then linearly combined with a structure-based kernel based on the SOAP~\cite{Bartok2017SA} descriptor. The spectra computed from the different clusters were used to fit experimental spectra allowing for an approximation to the composition of experimental samples on a semi-quantitative level. The so-called fingerprint spectra, which enabled the differentiation of the spectral signatures, were assessed in a previous study using different models for amorphous carbon~\cite{Aarva2019CMa}, among them an ML fitted PES using GPR~\cite{Deringer2017PRB,Deringer2018CM}.

Kulik and co-workers~\cite{Taylor2020JPCA} used deep NNs to predict the spin-state ordering in transition metal complexes to determine the spin of the lowest lying energetic state in open-shell systems. The determination of spin states is important to evaluate catalytic and material properties of metal complexes. Descriptors based on a selection of empirical features were used to capture the bonding in inorganic molecular systems. The performance of descriptors including different features was assessed for a set of octahedral complexes with first-row transition metals. The most important features were identified to be the atom, which connects the ligand to the metal, its environment and its electronegativity, the metal identity and its oxidation state, as well as the formal charge and denticity of the ligand.~\cite{Janet2017CS} The ML models were tested on spin-crossover complexes and could assign the correct spin in most cases. Additionally, ML models were applied for the discovery of inorganic complexes~\cite{Janet2019CS,Janet2017IECR,Janet2018JPCL,Janet2020ACSCS}

The inverse design of molecules with specific properties was further targeted by Sch\"{u}tt et. al,~\cite{Schuett2019NC} who developed SchNOrb, a deep NN model based on SchNet. The automatically generated descriptor was extended with a description of atom pairs in their chemical and structural environment. An analytic representation of the electronic structure of a molecular system was obtained in a local atomic orbital representation. The analytic derivatives of the electronic structure allowed for optimization of electronic properties. This was demonstrated by minimizing and maximizing the HOMO-LUMO gap of malonaldehyde~\cite{Chmiela2017SA}. Besides, the ML method was used to predict the lowest 20 molecular orbitals of ethanol at DFT level of theory, to investigate proton transfer in malonaldehyde using ground-state dynamics and to analyze bond order and partial charges of uracil. 

Bayesian NN models were applied by H\"{a}se et. al~\cite{Haese2019CS} to relate molecular geometries to the outcome of nonadiabatic MD simulations obtained with CASSCF. Normal modes with and without velocities of initial conditions served as an input for NN models. 
Velocities in addition to normal modes as descriptors improved the accuracy of ML models slightly, pointing out that normal modes contain already enough information for the sake of their study. The dissociation times of 1,2-dioxetane obtained from nonadiabatic MD simulations was the targeted output. The NNs could faithfully reproduce dissociation times and further provided a measure of uncertainty. The authors noted that their method could be particularly interesting for analysis of MLMD simulations. 

\subsection{ML-Assisted Analysis}
The aforementioned studies have shown that ML enables the simulation of MD simulations and spectra predictions at low computational costs. The computational efficiency allows for enhanced statistics, i.e., in case of MD simulations a huge number of trajectories and the simulations on long time scales.~\cite{Westermayr2019CS,Westermayr2020JPCL} Therefore, subsequent analyses of production runs can become a time limiting step of studies. This problem was identified in the aforementioned study on the dissociation times of 1,2-dioxetane by H\"ase et. al.~\cite{Haese2019CS} Therefore, the authors further used their method to interpret the outcomes of nonadiabatic MD simulations. 
1,2-dioxetane is the target of their study as it is the smallest molecule known to show chemilumiescence after nonadiabatic transitions from an excited state to the ground state. The chemiluminescent properties of this compound were related to its decomposition rate into two formaldehyde molecules. By analysis of the ML models that fit the dissociation times, correlations could be observed between the normal modes and the dissociation times. For example, the modes corresponding to C-C bond stretching and C-O bond stretching were relevant for the accurate prediction of dissociation times.
It was further emphasized by the authors that although the findings of NNs were expected and obey physical laws, ML models were helpful to extract relevant information of large amount of data and could potentially serve as an inspiration to humans. 

Time-resolved experimental photoluminescence spectra could be analyzed with the LumiML software developed by {\DH}or{\dj}evi\'{c} et. al,~\cite{Dordevic2018ACSP} who applied linear regression models to learn from computer-generated photoluminescence data. The software was employed to predict decay rate distributions~\cite{Abramavicius2010JACS} of perovskite nanocrystals from data generated with femtosecond broadband fluorescence upconversion spectroscopy.~\cite{Zhang2011RSI} The authors highlighted the applicability of their method to enhance studies on the optimization and design of optical devices and further noted that their approach can also be used to analyze transient absorption spectra. 
Aspuru-Guzik and co-workers~\cite{Roch2020ACSN} applied Bayesian NNs to find correlations of nanoaggregates with electronic coupling in semiconducting materials using absorption spectra. In general, the analysis of experimental spectra and the inverse design of compounds is most frequently applied in the research field of material science. Their description goes beyond the scope of this review and the reader is referred to Refs~\citenum{Freeze2019CR,Liu2018ACSP,Elton2019MSDE,Sanchez-Lengeling2018S,Goldsmith2018AJ,Davies2018FD}.

\section{Conclusion and Future Perspectives}
In the last few years, machine learning (ML) has started to slowly enter the research field of photochemistry, especially the photochemistry of molecular systems. Although this field of research is rather young compared to ML for the electronic ground-state, some groundbreaking works have already shown the potential of ML models to significantly accelerate and improve existing simulation techniques. So far, most studies provide a proof of concept using small molecular systems or model systems. Different applications are targeted and will also be aimed at in the future, ranging from dynamics with excited-state ML potentials via absorption spectra to the interpretation of data, see Fig.~\ref{fig:targets}.

\begin{figure}[ht]
    \centering
    \includegraphics[scale=0.3]{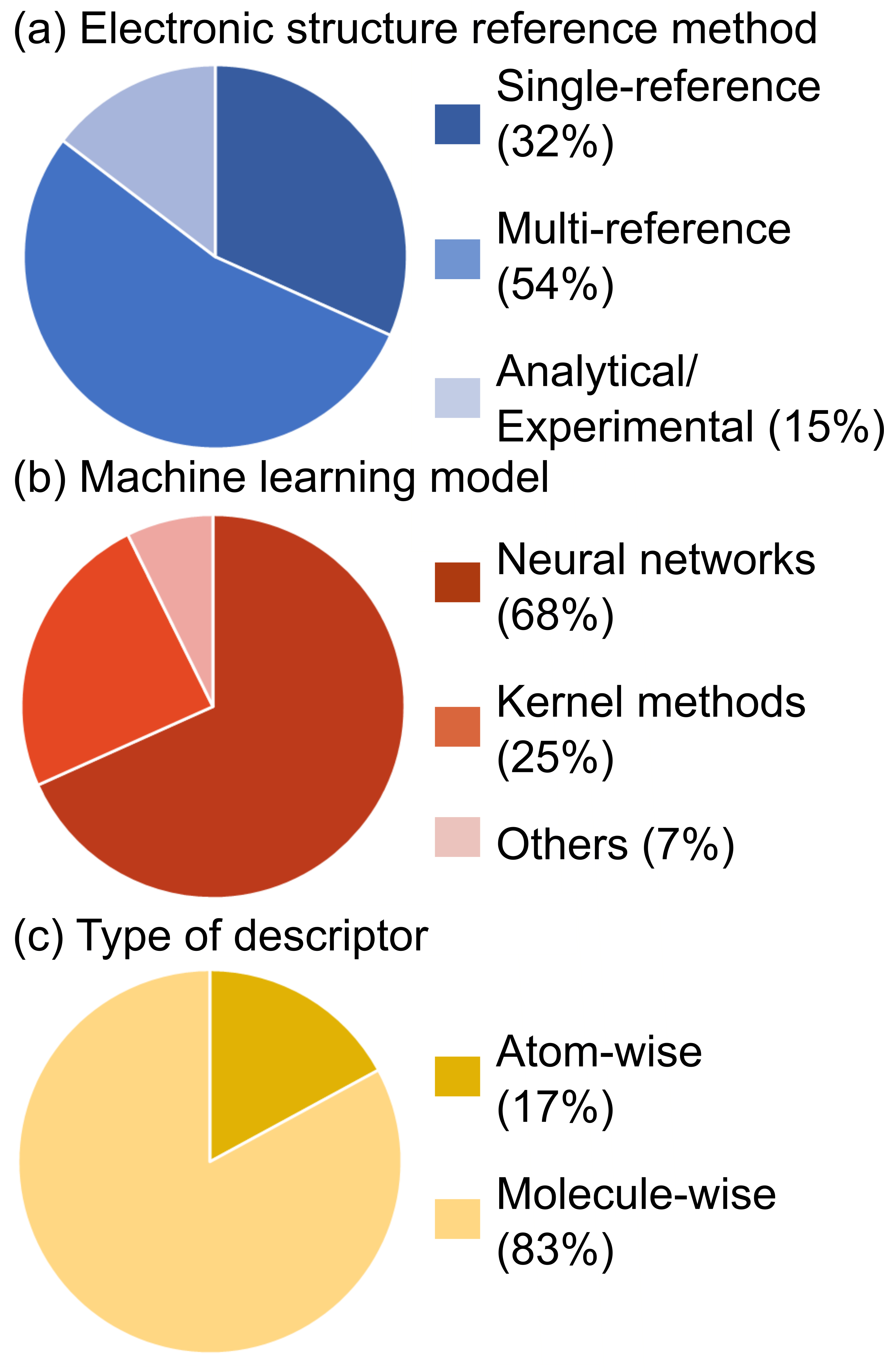}
    \caption{Pie diagrams summarizing the reference methods used for the training set generation, the chosen ML models and the type of descriptors for the description of the excited states with ML.}
    \label{fig:pie}
\end{figure}

Analysing the different studies reviewed here, some trends in the choice of reference methods, ML models, and descriptors can be observed. These trends are illustrated in Figure~\ref{fig:pie}.

The pie chart in panel~\ref{fig:pie}(a) shows the used reference methods for the computation of a training set to describe the excited states or excited-state properties of molecules. As can be seen, about half of the training sets are computed with multi-reference methods.~\cite{Schwilk2020arXiv,Williams2018JCP,Haese2019CS,Jeong2020JCTC,Guan2020JCTC,Guan2020JPCL,Shen2020JPCA,Westermayr2019CS,Westermayr2020JPCL,Westermayr2020MLST,Hu2018JPCL,Chen2018JPCL,Richings2020JCP,Richings2019JCTC,Richings2018JCP,Richings2017CPL,Jiang2016IRPC,Xie2018JCP,Li2013JCP,Guan2019PCCP,Jiang2013JCP,Guan2019JCP,Richings2019FD} The employed single-reference approaches are exclusively based on DFT.~\cite{Ghosh2019AS,Zhang2020arXiv,Ye2019PNAS,Schuett2019NC,Pronobis2018EPJB,Taylor2020JPCA,Beard2019SD,Ramakrishnan2015JCP,Carbogno2008PRL,Carbogno2010PRB,Kang2020arXiv,Chen2020arXiv,Aarva2019CMa} 
Analytical methods or experimental data are also applied.~\cite{Dordevic2018ACSP,Dral2018JPCL,Roch2020ACSN,Haese2016CS,Haese2017CS} 

When restricting the analysis to studies targeting dynamics, the fraction that employs multi-reference methods even increases. About 70\% of all dynamics studies use multi-reference methods to compute the training data for ML models. 15\% of the studies use single-reference methods and an equally large portion apply model Hamiltonians or analytical potentials. This shows that most chemical problems for the investigation of the excited states of molecules require multi-reference accuracy. 

Recent studies of ML-based photodynamics simulations have shown that many thousands of data points are necessary to describe a few excited-state potentials of small molecular systems. To the best of our knowledge, the dynamics in the excited states with ML for molecules with more than 12 atoms in full dimensions has not yet been investigated.~\cite{Hu2018JPCL,Richings2019JCTC,Richings2018JCP} Especially the huge number of data points is concerning in this case, as larger molecules with more energetic states and a complex photochemistry could require many more data points. A meaningful training set generation, which can be achieved with active learning, adaptive sampling and structure-based sampling techniques, is thus essential for dynamics simulations.~\cite{Behler2015IJQC,Gastegger2017CS,Westermayr2019CS,Botu2015IJQC} Clustering of molecular geometries obtained from dynamics simulations with a cheap method further is beneficial for selecting important reference geometries.~\cite{Dral2018JPCL,Hu2018JPCL,Ceriotti2013JCTC,Dral2017JCP,Sobol2011W}
Still, the high costs and the complexity of multi-reference methods to compute an ample training set for ML also hampers the application of ML models to fit the excited states of larger polyatomic systems, whose accurate photochemical description is often additionally complicated by a high density of electronic states. 

Single reference methods, such as time-dependent DFT, are advantageous with respect to the computational costs of the training set, but suffer from qualitatively incorrect PESs in some conformational regions of molecules, such as dissociative regions. In principle, these conformational regions could be excluded from the training set and the remaining conformational space could be interpolated using ML, but the training set would then remain incomplete and so would the dynamics. Schemes like the $\Delta-learning$ approach~\cite{Ramakrishnan2015JCP} or transfer learning~\cite{Smith2019NC} could be helpful in this regard. These approaches might be useful to let ML models learn from single-reference data and adjust their accuracy according to multi-reference methods. The direct use of approximated methods, such as time-dependent DFT-based tight binding, is most likely not suitable for photodynamics on long time scales, because such approaches might easily be quantitatively incorrect. Of particular concern is then the accumulation of quantitatively tiny errors in the underlying potentials toward wrong dynamics trends. At the current stage of research, it is not clear whether such approximate potentials can provide qualitatively correct trends for reaction dynamics.~\cite{Westermayr2020MLST_Perspective}

In addition to the aforementioned problems, the training set generation is complicated by the arbitrariness of the signs of coupling values and properties resulting from two different electronic states.~\cite{Guan2020JCTC,Guan2020JPCL,Westermayr2019CS,Westermayr2020MLST,Westermayr2020JPCL,Zhang2020arXiv} This arbitrariness has to be removed in order to make data learnable with conventional methods. Such a correction scheme is termed phase correction and has been applied to correct coupling values and dipole moments.~\cite{Westermayr2019CS,Akimov2018JPCL,Guan2020JCTC,Guan2020JPCL,Zhang2020arXiv} An alternative phase correction training algorithm has been shown to be beneficial with respect to the costs of the training set generation and has enabled the learning of raw quantum chemical data.~\cite{Westermayr2020JPCL} 
\\

Figure \ref{fig:pie}(b) shows which ML models are applied in the discussed studies. About two thirds rely on NNs, whereby simple multi-layer feed-forward NNs are most often employed. Several research fields were advanced with NN-fitted functions: photodynamics simulations~\cite{Westermayr2019CS,Westermayr2020JPCL,Westermayr2020MLST,Dral2018JPCL,Chen2018JPCL,Jiang2016IRPC,Xie2018JCP,Li2013JCP,Guan2019PCCP,Jiang2013JCP,Guan2019JCP,Carbogno2008PRL,Carbogno2010PRB}, spectra predictions and analysis,~\cite{Xue2020chemrxiv,Rankine2020JPCA,Ghosh2019AS,Chen2020arXiv,Ye2019PNAS,Zhang2020arXiv} 
excited-state properties,~\cite{Guan2020JCTC,Guan2020JPCL,Westermayr2020MLST,Westermayr2020JPCL,Taylor2020JPCA,Zhang2020arXiv,Ye2019PNAS} diabatization procedures,~\cite{Williams2018JCP,Shen2020JPCA} interpretation of reaction outcomes,~\cite{Haese2019CS,Dordevic2018ACSP} and the prediction of HOMO-LUMO gaps or gaps between energetic states.~\cite{Pronobis2018EPJB,Ghosh2019AS,Schuett2019NC} KRR methods were mainly applied to interpolate diabatic potentials~\cite{Richings2020JCP,Richings2019JCTC,Richings2018JCP,Richings2017CPL,Richings2019FD} and in studies focusing on more than one molecular systems.~\cite{Ramakrishnan2015JCP} In general, only a few studies focused on extrapolation throughout chemical compound space in the excited states. Yet only the energies, HOMO-LUMO gaps or spectra based on fitted oscillator strengths could be predicted using a single ML model for different molecules.~\cite{Taylor2020JPCA,Rankine2020JPCA,Ramakrishnan2015JCP,Pronobis2018EPJB}
Decision trees were used to select an active space for diatomic molecules.~\cite{Jeong2020JCTC} 
 
One drawback of recently developed ML models is that they are molecule-specific and thus not universal. In part, this issue is related to the used molecular descriptors. As can be seen in panel (c) in Figure~\ref{fig:pie}, most studies apply descriptors that capture molecules as a whole. The few studies, which describe PESs and properties of molecular systems from atomic contributions, either treat small molecular systems~\cite{Westermayr2020JPCL,Westermayr2020MLST,Zhang2020arXiv} or predict properties related to the ground-state equilibrium structure of a molecular system or to electronic ground state calculations, e.g. the HOMO-LUMO gaps.~\cite{Pronobis2018EPJB,Ghosh2019AS,Schuett2019NC} Due to the limited transferability of existing ML models to predict the excited state PESs and properties of different molecular systems, an extrapolation throughout chemical compound space is hindered in many cases.

In order to fully exploit the advantages that ML models offer and to achieve the aforementioned goal of a transferable ML model for the excited states, a highly versatile descriptor is required, which can describe atoms in their chemical and structural environment and enables an ML model to treat molecules of arbitrary size and composition. It would be highly desirable, if an ML model could then describe the photochemistry of large systems, which are too expensive to compute with precise multi-reference methods, using only small building blocks, i.e., small enough ones to describe their electronic structure accurately. For example, the excited states of proteins or DNA strands could potentially be predicted from contributions of amino acids or DNA bases, respectively, which is most often done using effective model Hamiltonians up to date.\cite{Nogueira2017CS} A local description of the excited-state PESs and their properties derived from the ML-fitted PESs, could further provide a way toward excited-state ML/MM simulations alike QM/MM (quantum mechanics/molecular mechanics) techniques.~\cite{Westermayr2020MLST_Perspective,Chen2019PCCP,Chen2019JPCL} Unfortunately, it is not yet known whether the excited-state PESs and properties can be constructed from atomic contributions or not.~\cite{Westermayr2020MLST_Perspective}  

In studies comparing different ML models, it was even suggested that non-local descriptors might be needed or that the electronic state has to be encoded explicitly in the molecular representation to enable a transferable description of the excited states with ML~\cite{Pronobis2018EPJB,Westermayr2020MLST}. 

To conclude, the reviewed studies focus on almost all aspects of excited-state quantum chemistry and improve them successfully: ML models can help to choose a proper active space for multi-reference methods, they predict secondary and tertiary outputs of quantum chemical calculations and help in the interpretation of theoretical studies. ML models push the boundaries of computed time scales~\cite{Westermayr2019CS} and are used to investigate and analyze the huge amount of data we produce every day in experiments or with high-performance computers.~\cite{Dordevic2018ACSP,Haese2019CS} 

It should be emphasized once more that the recent studies show that the goal of ML is not to replace existing methods completely, but to provide a way to improve them. In fact, ML models for the excited states at their current stage are far from replacing existing quantum chemical methods, and they are also far from being routine. Without human intervention, ML cannot solve existing problems and much remains to be done to describe systems beyond single, isolated molecules.

To the best of our knowledge, what is still missing is the proof that ML can provide an approximation to the multi-reference wave function of a molecular system. Such an achievement would be a great advancement in the research field of photochemistry, as any property we wish to know could possibly be derived from the ML wave function. An ML representation of the electronic structure would further be beneficial to allow for an inverse design of molecules with specific properties, which has been shown to be feasible for the ground state of a molecular system~\cite{Schuett2019NC}. The optimization of photochemical properties with respect to molecular geometries would be useful for many exciting research fields, e.g. photocatalysis~\cite{Sanchez-Lengeling2018S}, photosensitive drug design~\cite{Ahmad2016IJP} or photovoltaics~\cite{Mathew2014NC,Bartok2017SA}.

The multi-faceted photochemistry offers a perfect playground for ML models. 
It may be important to highlight that, despite the negative image ML has suffered in some research communities, it cannot be denied that it opens up many new ways and possibilities to improve simulations and make studies feasible that were considered unattainable only a few years, if not only months ago.~\cite{Dral2020JPCL} The computational efficiency and high flexibility of deep learning models can lead this research field toward simulations of long time and large length scales. The possibilities ML models offer are far from being being exhausted.
Considering the enormous chemical space, estimated to consist of more than $10^{60}$ molecules~\cite{Dobson2004N}, and the desire to develop methods, which could develop into a universal approximator, make ML models perfectly suited to advance this research field.
The possibility of deep ML models to process a huge amount of data can even assist the interpretation and analysis~\cite{Haese2019CS,Dordevic2018ACSP} of many photochemical studies and can help to explore unknown physical relations and be a source of potential human inspiration.

\begin{acknowledgement}
This work was financially supported by the Austrian Science Fund, W 1232 (MolTag) and the uni:docs program of the University of Vienna (J.W.). P. M. thanks the University of Vienna for continuous support, also in the frame of the research platform ViRAPID. We thank P. A. S\'anchez-Murcia for help in setting up the quick Amber simulation for MD timings.
\end{acknowledgement}

\begin{mcitethebibliography}{623}
\providecommand*\natexlab[1]{#1}
\providecommand*\mciteSetBstSublistMode[1]{}
\providecommand*\mciteSetBstMaxWidthForm[2]{}
\providecommand*\mciteBstWouldAddEndPuncttrue
  {\def\EndOfBibitem{\unskip.}}
\providecommand*\mciteBstWouldAddEndPunctfalse
  {\let\EndOfBibitem\relax}
\providecommand*\mciteSetBstMidEndSepPunct[3]{}
\providecommand*\mciteSetBstSublistLabelBeginEnd[3]{}
\providecommand*\EndOfBibitem{}
\mciteSetBstSublistMode{f}
\mciteSetBstMaxWidthForm{subitem}{(\alph{mcitesubitemcount})}
\mciteSetBstSublistLabelBeginEnd
  {\mcitemaxwidthsubitemform\space}
  {\relax}
  {\relax}

\bibitem[{K\"epuska} and {Bohouta}(2018){K\"epuska}, and
  {Bohouta}]{Kepuska2018}
{K\"epuska},~V.; {Bohouta},~G. {Next-Generation of Virtual Personal Assistants
  (Microsoft Cortana, Apple Siri, Amazon Alexa and Google Home)}. {2018 IEEE
  8th Annual Computing and Communication Workshop and Conference (CCWC)}. 2018;
  pp 99--103\relax
\mciteBstWouldAddEndPuncttrue
\mciteSetBstMidEndSepPunct{\mcitedefaultmidpunct}
{\mcitedefaultendpunct}{\mcitedefaultseppunct}\relax
\EndOfBibitem
\bibitem[Hoy(2018)]{Hoy2018MRSQ}
Hoy,~M.~B. {Alexa, Siri, Cortana, and More: An Introduction to Voice
  Assistants}. \emph{Med. Ref. Serv. Q.} \textbf{2018}, \emph{37}, 81--88\relax
\mciteBstWouldAddEndPuncttrue
\mciteSetBstMidEndSepPunct{\mcitedefaultmidpunct}
{\mcitedefaultendpunct}{\mcitedefaultseppunct}\relax
\EndOfBibitem
\bibitem[Silver \latin{et~al.}(2016)Silver, Huang, Maddison, Guez, Sifre,
  van~den Driessche, Schrittwieser, Antonoglou, Panneershelvam, Lanctot,
  Dieleman, Grewe, Nham, Kalchbrenner, Sutskever, Lillicrap, Leach,
  Kavukcuoglu, Graepel, and Hassabis]{Silver2016N}
Silver,~D. \latin{et~al.}  Mastering the Game of Go with Deep Neural Networks
  and Tree Search. \emph{Nature} \textbf{2016}, \emph{529}, 484--489\relax
\mciteBstWouldAddEndPuncttrue
\mciteSetBstMidEndSepPunct{\mcitedefaultmidpunct}
{\mcitedefaultendpunct}{\mcitedefaultseppunct}\relax
\EndOfBibitem
\bibitem[Bansak \latin{et~al.}(2018)Bansak, Ferwerda, Hainmueller, Dillon,
  Hangartner, Lawrence, and Weinstein]{Bansak2018S}
Bansak,~K.; Ferwerda,~J.; Hainmueller,~J.; Dillon,~A.; Hangartner,~D.;
  Lawrence,~D.; Weinstein,~J. {Improving Refugee Integration through
  Data-Driven Algorithmic Assignment}. \emph{Science} \textbf{2018},
  \emph{359}, 325--329\relax
\mciteBstWouldAddEndPuncttrue
\mciteSetBstMidEndSepPunct{\mcitedefaultmidpunct}
{\mcitedefaultendpunct}{\mcitedefaultseppunct}\relax
\EndOfBibitem
\bibitem[{Leung} \latin{et~al.}(2016){Leung}, {Delong}, {Alipanahi}, and
  {Frey}]{Leung2016PIEEE}
{Leung},~M. K.~K.; {Delong},~A.; {Alipanahi},~B.; {Frey},~B.~J. Machine
  Learning in Genomic Medicine: A Review of Computational Problems and Data
  Sets. \emph{Proc. IEEE} \textbf{2016}, \emph{104}, 176--197\relax
\mciteBstWouldAddEndPuncttrue
\mciteSetBstMidEndSepPunct{\mcitedefaultmidpunct}
{\mcitedefaultendpunct}{\mcitedefaultseppunct}\relax
\EndOfBibitem
\bibitem[Shen \latin{et~al.}(2017)Shen, Wu, and Suk]{Shen2017ARBE}
Shen,~D.; Wu,~G.; Suk,~H.-I. Deep Learning in Medical Image Analysis.
  \emph{Annu. Rev. Biomed. Eng.} \textbf{2017}, \emph{19}, 221--248\relax
\mciteBstWouldAddEndPuncttrue
\mciteSetBstMidEndSepPunct{\mcitedefaultmidpunct}
{\mcitedefaultendpunct}{\mcitedefaultseppunct}\relax
\EndOfBibitem
\bibitem[Chen \latin{et~al.}(2015)Chen, Seff, Kornhauser, and Xiao]{Chen2015}
Chen,~C.; Seff,~A.; Kornhauser,~A.; Xiao,~J. DeepDriving: Learning Affordance
  for Direct Perception in Autonomous Driving. The IEEE International
  Conference on Computer Vision (ICCV). 2015\relax
\mciteBstWouldAddEndPuncttrue
\mciteSetBstMidEndSepPunct{\mcitedefaultmidpunct}
{\mcitedefaultendpunct}{\mcitedefaultseppunct}\relax
\EndOfBibitem
\bibitem[Yang \latin{et~al.}(2019)Yang, Wang, Byrne, Schneider, and
  Yang]{Yang2019CR}
Yang,~X.; Wang,~Y.; Byrne,~R.; Schneider,~G.; Yang,~S. Concepts of Artificial
  Intelligence for Computer-Assisted Drug Discovery. \emph{Chem. Rev.}
  \textbf{2019}, \emph{119}, 10520--10594\relax
\mciteBstWouldAddEndPuncttrue
\mciteSetBstMidEndSepPunct{\mcitedefaultmidpunct}
{\mcitedefaultendpunct}{\mcitedefaultseppunct}\relax
\EndOfBibitem
\bibitem[Goodfellow \latin{et~al.}(2016)Goodfellow, Bengio, and
  Courville]{Goodfellow2016}
Goodfellow,~I.; Bengio,~Y.; Courville,~A. \emph{Deep Learning}; MIT Press,
  2016\relax
\mciteBstWouldAddEndPuncttrue
\mciteSetBstMidEndSepPunct{\mcitedefaultmidpunct}
{\mcitedefaultendpunct}{\mcitedefaultseppunct}\relax
\EndOfBibitem
\bibitem[G{\'o}mez-Bombarelli and Aspuru-Guzik(2018)G{\'o}mez-Bombarelli, and
  Aspuru-Guzik]{Gomez-Bombarelli2018}
G{\'o}mez-Bombarelli,~R.; Aspuru-Guzik,~A. In \emph{Handbook of Materials
  Modeling : Methods: Theory and Modeling}; Andreoni,~W., Yip,~S., Eds.;
  Springer International Publishing: Cham, 2018; pp 1--24\relax
\mciteBstWouldAddEndPuncttrue
\mciteSetBstMidEndSepPunct{\mcitedefaultmidpunct}
{\mcitedefaultendpunct}{\mcitedefaultseppunct}\relax
\EndOfBibitem
\bibitem[Agrawal and Choudhary(2016)Agrawal, and Choudhary]{Agrawal2016APLM}
Agrawal,~A.; Choudhary,~A. {Perspective: Materials Informatics and Big Data:
  Realization of the “Fourth Paradigm” of Science in Materials Science}.
  \emph{APL Mat.} \textbf{2016}, \emph{4}, 053208\relax
\mciteBstWouldAddEndPuncttrue
\mciteSetBstMidEndSepPunct{\mcitedefaultmidpunct}
{\mcitedefaultendpunct}{\mcitedefaultseppunct}\relax
\EndOfBibitem
\bibitem[Schwilk \latin{et~al.}(2020)Schwilk, Tahchieva, and von
  Lilienfeld]{Schwilk2020arXiv}
Schwilk,~M.; Tahchieva,~D.~N.; von Lilienfeld,~O.~A. Large yet Bounded: Spin
  Gap Ranges in Carbenes. \emph{arXiv} \textbf{2020}, \emph{2004.10600}\relax
\mciteBstWouldAddEndPuncttrue
\mciteSetBstMidEndSepPunct{\mcitedefaultmidpunct}
{\mcitedefaultendpunct}{\mcitedefaultseppunct}\relax
\EndOfBibitem
\bibitem[Westermayr \latin{et~al.}(2020)Westermayr, Gastegger, and
  Marquetand]{Westermayr2020JPCL}
Westermayr,~J.; Gastegger,~M.; Marquetand,~P. {Combining SchNet and SHARC: The
  SchNarc Machine Learning Approach for Excited-State Dynamics}. \emph{J. Phys.
  Chem. Lett.} \textbf{2020}, \emph{11}, 3828--3834\relax
\mciteBstWouldAddEndPuncttrue
\mciteSetBstMidEndSepPunct{\mcitedefaultmidpunct}
{\mcitedefaultendpunct}{\mcitedefaultseppunct}\relax
\EndOfBibitem
\bibitem[Guan \latin{et~al.}(2020)Guan, Guo, and Yarkony]{Guan2020JCTC}
Guan,~Y.; Guo,~H.; Yarkony,~D.~R. Extending the Representation of Multistate
  Coupled Potential Energy Surfaces to Include Properties Operators using
  Neural Networks: Application to the 1,2$^1$A States of Ammonia. \emph{J.
  Chem. Theory Comput.} \textbf{2020}, \emph{16}, 302--313\relax
\mciteBstWouldAddEndPuncttrue
\mciteSetBstMidEndSepPunct{\mcitedefaultmidpunct}
{\mcitedefaultendpunct}{\mcitedefaultseppunct}\relax
\EndOfBibitem
\bibitem[Taylor \latin{et~al.}(2020)Taylor, Yang, Lin, Nandy, Janet, Duan, and
  Kulik]{Taylor2020JPCA}
Taylor,~M.~G.; Yang,~T.; Lin,~S.; Nandy,~A.; Janet,~J.~P.; Duan,~C.;
  Kulik,~H.~J. Seeing Is Believing: Experimental Spin States from Machine
  Learning Model Structure Predictions. \emph{J. Phys. Chem. A} \textbf{2020},
  \emph{124}, 3286--3299\relax
\mciteBstWouldAddEndPuncttrue
\mciteSetBstMidEndSepPunct{\mcitedefaultmidpunct}
{\mcitedefaultendpunct}{\mcitedefaultseppunct}\relax
\EndOfBibitem
\bibitem[Kulik(2020)]{Kulik2019WCMS}
Kulik,~H.~J. Making Machine Learning a Useful Tool in the Accelerated Discovery
  of Transition Metal Complexes. \emph{WIREs Comput. Mol. Sci.} \textbf{2020},
  \emph{10}, e1439\relax
\mciteBstWouldAddEndPuncttrue
\mciteSetBstMidEndSepPunct{\mcitedefaultmidpunct}
{\mcitedefaultendpunct}{\mcitedefaultseppunct}\relax
\EndOfBibitem
\bibitem[Power(2012)]{Power2012CR}
Power,~P.~P. {Stable Two-Coordinate, Open-Shell (d1–d9) Transition Metal
  Complexes}. \emph{Chem. Rev.} \textbf{2012}, \emph{112}, 3482--3507\relax
\mciteBstWouldAddEndPuncttrue
\mciteSetBstMidEndSepPunct{\mcitedefaultmidpunct}
{\mcitedefaultendpunct}{\mcitedefaultseppunct}\relax
\EndOfBibitem
\bibitem[Bousseksou \latin{et~al.}(2004)Bousseksou, Molnár, and
  Matouzenko]{Bousseksou2004EJIC}
Bousseksou,~A.; Molnár,~G.; Matouzenko,~G. Switching of Molecular Spin States
  in Inorganic Complexes by Temperature, Pressure, Magnetic Field and Light:
  Towards Molecular Devices. \emph{Eur. J. Inorg. chem.} \textbf{2004},
  \emph{2004}, 4353--4369\relax
\mciteBstWouldAddEndPuncttrue
\mciteSetBstMidEndSepPunct{\mcitedefaultmidpunct}
{\mcitedefaultendpunct}{\mcitedefaultseppunct}\relax
\EndOfBibitem
\bibitem[Li \latin{et~al.}(2014)Li, Feng, Sun, Fan, King, and
  Schaefer]{Li2015O}
Li,~H.; Feng,~H.; Sun,~W.; Fan,~Q.; King,~R.~B.; Schaefer,~H.~F. {First-Row
  Transition Metals in Binuclear Cyclopentadienylmetal Derivatives of
  Tetramethyleneethane: $\eta^3,\eta^3$ versus $\eta^4,\eta^4$ Ligand–Metal
  Bonding Related to Spin State and Metal–Metal Bonds}.
  \emph{Organometallics} \textbf{2014}, \emph{33}, 3489--3499\relax
\mciteBstWouldAddEndPuncttrue
\mciteSetBstMidEndSepPunct{\mcitedefaultmidpunct}
{\mcitedefaultendpunct}{\mcitedefaultseppunct}\relax
\EndOfBibitem
\bibitem[Matsika and Krylov(2018)Matsika, and Krylov]{Matsika2018CR}
Matsika,~S.; Krylov,~A.~I. Introduction: Theoretical Modeling of Excited State
  Processes. \emph{Chem. Rev.} \textbf{2018}, \emph{118}, 6925--6926\relax
\mciteBstWouldAddEndPuncttrue
\mciteSetBstMidEndSepPunct{\mcitedefaultmidpunct}
{\mcitedefaultendpunct}{\mcitedefaultseppunct}\relax
\EndOfBibitem
\bibitem[Cohen \latin{et~al.}(2004)Cohen, Crespo-Hern\'{a}ndez, Hare, and
  Kohler]{Cohen2004}
Cohen,~B.; Crespo-Hern\'{a}ndez,~C.~E.; Hare,~P.~M.; Kohler,~B. \emph{Ultrafast
  Excited-State Dynamics in DNA and RNA Polymers}; Elsevier: Amsterdam, 2004;
  Chapter Ultrafast Excited-State Dynamics in DNA and RNA Polymers, pp
  463--470\relax
\mciteBstWouldAddEndPuncttrue
\mciteSetBstMidEndSepPunct{\mcitedefaultmidpunct}
{\mcitedefaultendpunct}{\mcitedefaultseppunct}\relax
\EndOfBibitem
\bibitem[Levine and Mart{{\'i}}nez(2007)Levine, and
  Mart{{\'i}}nez]{Levine2007ARPC}
Levine,~B.~G.; Mart{{\'i}}nez,~T.~J. Isomerization Through Conical
  Intersections. \emph{Annu. Rev. Phys. Chem.} \textbf{2007}, \emph{58},
  613--634\relax
\mciteBstWouldAddEndPuncttrue
\mciteSetBstMidEndSepPunct{\mcitedefaultmidpunct}
{\mcitedefaultendpunct}{\mcitedefaultseppunct}\relax
\EndOfBibitem
\bibitem[Turro \latin{et~al.}(2009)Turro, Ramamurthy, and Scaiano]{Turro2009}
Turro,~N.~J.; Ramamurthy,~V.; Scaiano,~J.~C. Principles of Molecular
  Photochemistry: An Introduction. 2009\relax
\mciteBstWouldAddEndPuncttrue
\mciteSetBstMidEndSepPunct{\mcitedefaultmidpunct}
{\mcitedefaultendpunct}{\mcitedefaultseppunct}\relax
\EndOfBibitem
\bibitem[Yarkony(2012)]{Yarkony2012CR}
Yarkony,~D.~R. Nonadiabatic Quantum Chemistry - Past, Present, and Future.
  \emph{Chem. Rev.} \textbf{2012}, \emph{112}, 481--498\relax
\mciteBstWouldAddEndPuncttrue
\mciteSetBstMidEndSepPunct{\mcitedefaultmidpunct}
{\mcitedefaultendpunct}{\mcitedefaultseppunct}\relax
\EndOfBibitem
\bibitem[Barbatti \latin{et~al.}(2014)Barbatti, Borin, and
  Ullrich]{Barbatti2014}
Barbatti,~M.; Borin,~A.~C.; Ullrich,~S. \emph{Photoinduced Phenomena in Nucleic
  Acids I}; Topics in Current Chemistry; Springer Berlin Heidelberg, 2014; Vol.
  355; pp 1--32\relax
\mciteBstWouldAddEndPuncttrue
\mciteSetBstMidEndSepPunct{\mcitedefaultmidpunct}
{\mcitedefaultendpunct}{\mcitedefaultseppunct}\relax
\EndOfBibitem
\bibitem[Ibele \latin{et~al.}(2020)Ibele, Nicolson, and Curchod]{Ibele2019MP}
Ibele,~L.~M.; Nicolson,~A.; Curchod,~B. F.~E. Excited-State Dynamics of
  molecules with classically driven trajectories and Gaussians. \emph{Mol.
  Phys.} \textbf{2020}, \emph{118}, e1665199\relax
\mciteBstWouldAddEndPuncttrue
\mciteSetBstMidEndSepPunct{\mcitedefaultmidpunct}
{\mcitedefaultendpunct}{\mcitedefaultseppunct}\relax
\EndOfBibitem
\bibitem[Nelson \latin{et~al.}(2020)Nelson, White, Bjorgaard, Sifain, Zhang,
  Nebgen, Fernandez-Alberti, Mozyrsky, Roitberg, and Tretiak]{Nelson2020CR}
Nelson,~T.~R.; White,~A.~J.; Bjorgaard,~J.~A.; Sifain,~A.~E.; Zhang,~Y.;
  Nebgen,~B.; Fernandez-Alberti,~S.; Mozyrsky,~D.; Roitberg,~A.~E.; Tretiak,~S.
  Non-adiabatic Excited-State Molecular Dynamics: Theory and Applications for
  Modeling Photophysics in Extended Molecular Materials. \emph{Chem. Rev.}
  \textbf{2020}, \emph{120}, 2215--2287\relax
\mciteBstWouldAddEndPuncttrue
\mciteSetBstMidEndSepPunct{\mcitedefaultmidpunct}
{\mcitedefaultendpunct}{\mcitedefaultseppunct}\relax
\EndOfBibitem
\bibitem[Mai and González(2020)Mai, and González]{Mai2020ACIE}
Mai,~S.; González,~L. Molecular Photochemistry: Recent Developments in Theory.
  \emph{Angew. Chem. Int. Ed.} \textbf{2020}, \emph{n/a}\relax
\mciteBstWouldAddEndPuncttrue
\mciteSetBstMidEndSepPunct{\mcitedefaultmidpunct}
{\mcitedefaultendpunct}{\mcitedefaultseppunct}\relax
\EndOfBibitem
\bibitem[Lischka \latin{et~al.}(2018)Lischka, Nachtigallov{{\'a}}, Aquino,
  Szalay, Plasser, Machado, and Barbatti]{Lischka2018CR}
Lischka,~H.; Nachtigallov{{\'a}},~D.; Aquino,~A. J.~A.; Szalay,~P.~G.;
  Plasser,~F.; Machado,~F. B.~C.; Barbatti,~M. Multireference Approaches for
  Excited States of Molecules. \emph{Chem. Rev.} \textbf{2018}, \emph{118},
  7293--7361\relax
\mciteBstWouldAddEndPuncttrue
\mciteSetBstMidEndSepPunct{\mcitedefaultmidpunct}
{\mcitedefaultendpunct}{\mcitedefaultseppunct}\relax
\EndOfBibitem
\bibitem[Ghosh \latin{et~al.}(2018)Ghosh, Verma, Cramer, Gagliardi, and
  Truhlar]{Ghosh2018CR}
Ghosh,~S.; Verma,~P.; Cramer,~C.~J.; Gagliardi,~L.; Truhlar,~D.~G. Combining
  Wave Function Methods with Density Functional Theory for Excited States.
  \emph{Chem. Rev.} \textbf{2018}, \emph{118}, 7249--7292\relax
\mciteBstWouldAddEndPuncttrue
\mciteSetBstMidEndSepPunct{\mcitedefaultmidpunct}
{\mcitedefaultendpunct}{\mcitedefaultseppunct}\relax
\EndOfBibitem
\bibitem[Norman and Dreuw(2018)Norman, and Dreuw]{Norman2018CR}
Norman,~P.; Dreuw,~A. Simulating X-Ray Spectroscopies and Calculating
  Core-Excited States of Molecules. \emph{Chem. Rev.} \textbf{2018},
  \emph{118}, 7208--7248\relax
\mciteBstWouldAddEndPuncttrue
\mciteSetBstMidEndSepPunct{\mcitedefaultmidpunct}
{\mcitedefaultendpunct}{\mcitedefaultseppunct}\relax
\EndOfBibitem
\bibitem[Casanova(2018)]{Casanova2018CR}
Casanova,~D. Theoretical Modeling of Singlet Fission. \emph{Chem. Rev.}
  \textbf{2018}, \emph{118}, 7164--7207\relax
\mciteBstWouldAddEndPuncttrue
\mciteSetBstMidEndSepPunct{\mcitedefaultmidpunct}
{\mcitedefaultendpunct}{\mcitedefaultseppunct}\relax
\EndOfBibitem
\bibitem[Hestand and Spano(2018)Hestand, and Spano]{Hestand2018CR}
Hestand,~N.~J.; Spano,~F.~C. Expanded Theory of H- and J-Molecular Aggregates:
  The Effects of Vibronic coupling and Intermolecular Charge Transfer.
  \emph{Chem. Rev.} \textbf{2018}, \emph{118}, 7069--7163\relax
\mciteBstWouldAddEndPuncttrue
\mciteSetBstMidEndSepPunct{\mcitedefaultmidpunct}
{\mcitedefaultendpunct}{\mcitedefaultseppunct}\relax
\EndOfBibitem
\bibitem[Penfold \latin{et~al.}(2018)Penfold, Gindensperger, Daniel, and
  Marian]{Penfold2018CR}
Penfold,~T.~J.; Gindensperger,~E.; Daniel,~C.; Marian,~C.~M. Spin-Vibronic
  Mechanism for Intersystem Crossing. \emph{Chem. Rev.} \textbf{2018},
  \emph{118}, 6975--7025\relax
\mciteBstWouldAddEndPuncttrue
\mciteSetBstMidEndSepPunct{\mcitedefaultmidpunct}
{\mcitedefaultendpunct}{\mcitedefaultseppunct}\relax
\EndOfBibitem
\bibitem[Vacher \latin{et~al.}(2018)Vacher, Fdez.~Galv{{\'a}}n, Ding, Schramm,
  Berraud-Pache, Naumov, Ferr{\'e}, Liu, Navizet, Roca-Sanju\'{a}n, Baader, and
  Lindh]{Vacher2018CR}
Vacher,~M.; Fdez.~Galv{{\'a}}n,~I.; Ding,~B.-W.; Schramm,~S.;
  Berraud-Pache,~R.; Naumov,~P.; Ferr{\'e},~N.; Liu,~Y.-J.; Navizet,~I.;
  Roca-Sanju\'{a}n,~D.; Baader,~W.~J.; Lindh,~R. Chemi- and Bioluminescence of
  Cyclic Peroxides. \emph{Chem. Rev.} \textbf{2018}, \emph{118},
  6927--6974\relax
\mciteBstWouldAddEndPuncttrue
\mciteSetBstMidEndSepPunct{\mcitedefaultmidpunct}
{\mcitedefaultendpunct}{\mcitedefaultseppunct}\relax
\EndOfBibitem
\bibitem[Crespo-Otero and Barbatti(2018)Crespo-Otero, and
  Barbatti]{Crespo-Otero2018CR}
Crespo-Otero,~R.; Barbatti,~M. Recent Advances and Perspectives on Nonadiabatic
  Mixed Quantum–Classical Dynamics. \emph{Chem. Rev.} \textbf{2018},
  \emph{118}, 7026--7068\relax
\mciteBstWouldAddEndPuncttrue
\mciteSetBstMidEndSepPunct{\mcitedefaultmidpunct}
{\mcitedefaultendpunct}{\mcitedefaultseppunct}\relax
\EndOfBibitem
\bibitem[Gonz{{\'a}}lez and Lindh(2020)Gonz{{\'a}}lez, and Lindh]{Gonzalez2020}
Gonz{{\'a}}lez,~L.; Lindh,~R. \emph{Quantum Chemistry and Dynamics of Excited
  States : Methods and Applications}; John Wiley and Sons Ltd, 2020\relax
\mciteBstWouldAddEndPuncttrue
\mciteSetBstMidEndSepPunct{\mcitedefaultmidpunct}
{\mcitedefaultendpunct}{\mcitedefaultseppunct}\relax
\EndOfBibitem
\bibitem[Harris and Bertolucci(1989)Harris, and Bertolucci]{Harris1989}
Harris,~D.~C.; Bertolucci,~M.~D. \emph{Symmetry and Spectroscopy: an
  Introduction to Vibrational and Electronic Spectroscopy}; New York: Dover
  Publications, 1989\relax
\mciteBstWouldAddEndPuncttrue
\mciteSetBstMidEndSepPunct{\mcitedefaultmidpunct}
{\mcitedefaultendpunct}{\mcitedefaultseppunct}\relax
\EndOfBibitem
\bibitem[Ng(1991)]{Cheuk-Yiu1991}
Ng,~C.-Y. \emph{Vacuum Ultraviolet Photoionization and Photodissociation of
  Molecules and Clusters}; World Scientific, 1991\relax
\mciteBstWouldAddEndPuncttrue
\mciteSetBstMidEndSepPunct{\mcitedefaultmidpunct}
{\mcitedefaultendpunct}{\mcitedefaultseppunct}\relax
\EndOfBibitem
\bibitem[Zewail(1994)]{Zewail1994}
Zewail,~A.~H. \emph{Femtochemistry: Ultrafast Dynamics of the Chemical Bond};
  World Scientific, 1994; pp 3--22\relax
\mciteBstWouldAddEndPuncttrue
\mciteSetBstMidEndSepPunct{\mcitedefaultmidpunct}
{\mcitedefaultendpunct}{\mcitedefaultseppunct}\relax
\EndOfBibitem
\bibitem[Brixner \latin{et~al.}(2005)Brixner, Pfeifer, Gerber, Wollenhaupt, and
  Baumert]{Brixner2005}
Brixner,~T.; Pfeifer,~T.; Gerber,~G.; Wollenhaupt,~M.; Baumert,~T. In
  \emph{Femtosecond Laser Spectroscopy}; Hannaford,~P., Ed.; {Springer-Verlag}:
  New York, 2005; pp 225--266\relax
\mciteBstWouldAddEndPuncttrue
\mciteSetBstMidEndSepPunct{\mcitedefaultmidpunct}
{\mcitedefaultendpunct}{\mcitedefaultseppunct}\relax
\EndOfBibitem
\bibitem[Iqbal and Stavros(2010)Iqbal, and Stavros]{Iqbal2010JPCL}
Iqbal,~A.; Stavros,~V.~G. Active Participation of $1\pi\sigma${*} States in the
  Photodissociation of Tyrosine and its Subunits. \emph{J. Phys. Chem. Lett.}
  \textbf{2010}, \emph{1}, 2274--2278\relax
\mciteBstWouldAddEndPuncttrue
\mciteSetBstMidEndSepPunct{\mcitedefaultmidpunct}
{\mcitedefaultendpunct}{\mcitedefaultseppunct}\relax
\EndOfBibitem
\bibitem[Kowalewski \latin{et~al.}(2017)Kowalewski, Fingerhut, Dorfman,
  Bennett, and Mukamel]{Kowalewski2017CR}
Kowalewski,~M.; Fingerhut,~B.~P.; Dorfman,~K.~E.; Bennett,~K.; Mukamel,~S.
  Simulating Coherent Multidimensional Spectroscopy of Nonadiabatic Molecular
  Processes: From the Infrared to the X-Ray Regime. \emph{Chem. Rev.}
  \textbf{2017}, \emph{117}, 12165--12226\relax
\mciteBstWouldAddEndPuncttrue
\mciteSetBstMidEndSepPunct{\mcitedefaultmidpunct}
{\mcitedefaultendpunct}{\mcitedefaultseppunct}\relax
\EndOfBibitem
\bibitem[Soorkia \latin{et~al.}(2020)Soorkia, Jouvet, and
  Gr{\'e}goire]{Soorkia2019CR}
Soorkia,~S.; Jouvet,~C.; Gr{\'e}goire,~G. UV Photoinduced Dynamics of
  Conformer-Resolved Aromatic Peptides. \emph{Chem. Rev.} \textbf{2020},
  \emph{120}, 3296–3327\relax
\mciteBstWouldAddEndPuncttrue
\mciteSetBstMidEndSepPunct{\mcitedefaultmidpunct}
{\mcitedefaultendpunct}{\mcitedefaultseppunct}\relax
\EndOfBibitem
\bibitem[{Yusong Liu, et al.}(2020)]{Liu2020PRX}
{Yusong Liu, et al.}, Spectroscopic and Structural Probing of Excited-State
  Molecular Dynamics with Time-Resolved Photoelectron Spectroscopy and
  Ultrafast Electron Diffraction. \emph{Phys. Rev. X} \textbf{2020}, \emph{10},
  021016\relax
\mciteBstWouldAddEndPuncttrue
\mciteSetBstMidEndSepPunct{\mcitedefaultmidpunct}
{\mcitedefaultendpunct}{\mcitedefaultseppunct}\relax
\EndOfBibitem
\bibitem[Mart\'{i}nez(2006)]{Martinez2006ACR}
Mart\'{i}nez,~T.~J. Insights for Light-Driven Molecular Devices from \emph{Ab
  Initio} Multiple Spawning Excited-State Dynamics of Organic and Biological
  Chromophores. \emph{Acc. Chem. Res.} \textbf{2006}, \emph{39}, 119--126\relax
\mciteBstWouldAddEndPuncttrue
\mciteSetBstMidEndSepPunct{\mcitedefaultmidpunct}
{\mcitedefaultendpunct}{\mcitedefaultseppunct}\relax
\EndOfBibitem
\bibitem[Barbatti \latin{et~al.}(2008)Barbatti, Sellner, Aquino, and
  Lischka]{Barbatti2008}
Barbatti,~M.; Sellner,~B.; Aquino,~A. J.~A.; Lischka,~H. In \emph{Radiation
  Induced Molecular Phenomena in Nucleic Acids}; Shukla,~M., Leszczynski,~J.,
  Eds.; Challenges and Advances in Computational Chemistry and Physics;
  Springer Netherlands, 2008; Vol.~5; pp 209--235\relax
\mciteBstWouldAddEndPuncttrue
\mciteSetBstMidEndSepPunct{\mcitedefaultmidpunct}
{\mcitedefaultendpunct}{\mcitedefaultseppunct}\relax
\EndOfBibitem
\bibitem[Subotnik \latin{et~al.}(2016)Subotnik, Jain, Landry, Petit, Ouyang,
  and Bellonzi]{Subotnik2016ARPC}
Subotnik,~J.~E.; Jain,~A.; Landry,~B.; Petit,~A.; Ouyang,~W.; Bellonzi,~N.
  Understanding the Surface Hopping View of Electronic Transitions and
  Decoherence. \emph{Annu. Rev. Phys. Chem.} \textbf{2016}, \emph{67},
  387--417\relax
\mciteBstWouldAddEndPuncttrue
\mciteSetBstMidEndSepPunct{\mcitedefaultmidpunct}
{\mcitedefaultendpunct}{\mcitedefaultseppunct}\relax
\EndOfBibitem
\bibitem[Curchod and Mart\'{i}nez(2018)Curchod, and
  Mart\'{i}nez]{Curchod2018CR}
Curchod,~B. F.~E.; Mart\'{i}nez,~T.~J. \emph{Ab Initio} Nonadiabatic Quantum
  Molecular Dynamics. \emph{Chem. Rev.} \textbf{2018}, \emph{118},
  3305--3336\relax
\mciteBstWouldAddEndPuncttrue
\mciteSetBstMidEndSepPunct{\mcitedefaultmidpunct}
{\mcitedefaultendpunct}{\mcitedefaultseppunct}\relax
\EndOfBibitem
\bibitem[Ashfold \latin{et~al.}(2017)Ashfold, Bain, Hansen, Ingle, Karsili,
  Marchetti, and Murdock]{Ashfold2017JPCL}
Ashfold,~M. N.~R.; Bain,~M.; Hansen,~C.~S.; Ingle,~R.~A.; Karsili,~T. N.~V.;
  Marchetti,~B.; Murdock,~D. Exploring the Dynamics of the Photoinduced
  Ring-Opening of Heterocyclic Molecules. \emph{J. Phys. Chem. Lett.}
  \textbf{2017}, \emph{8}, 3440--3451\relax
\mciteBstWouldAddEndPuncttrue
\mciteSetBstMidEndSepPunct{\mcitedefaultmidpunct}
{\mcitedefaultendpunct}{\mcitedefaultseppunct}\relax
\EndOfBibitem
\bibitem[Tajti \latin{et~al.}(2009)Tajti, Fogarasi, and Szalay]{Tajti2009C}
Tajti,~A.; Fogarasi,~G.; Szalay,~P.~G. Reinterpretation of the {UV} Spectrum of
  Cytosine: Only Two Electronic Transitions? \emph{ChemPhysChem} \textbf{2009},
  \emph{10}, 1603--1606\relax
\mciteBstWouldAddEndPuncttrue
\mciteSetBstMidEndSepPunct{\mcitedefaultmidpunct}
{\mcitedefaultendpunct}{\mcitedefaultseppunct}\relax
\EndOfBibitem
\bibitem[Barbatti \latin{et~al.}(2011)Barbatti, Szymczak, Aquino,
  Nachtigallov{{\'a}}, and Lischka]{Barbatti2011JCP}
Barbatti,~M.; Szymczak,~J.~J.; Aquino,~A. J.~A.; Nachtigallov{{\'a}},~D.;
  Lischka,~H. The Decay Mechanism of Photoexcited Guanine -- A Nonadiabatic
  Dynamics Study. \emph{J. Chem. Phys.} \textbf{2011}, \emph{134}, 014304\relax
\mciteBstWouldAddEndPuncttrue
\mciteSetBstMidEndSepPunct{\mcitedefaultmidpunct}
{\mcitedefaultendpunct}{\mcitedefaultseppunct}\relax
\EndOfBibitem
\bibitem[Lu \latin{et~al.}(2014)Lu, Lan, and Thiel]{Lu2014}
Lu,~Y.; Lan,~Z.; Thiel,~W. \emph{Photoinduced Phenomena in Nucleic Acids II};
  Topics in Current Chemistry; Springer Berlin Heidelberg, 2014; Vol. 356; pp
  89--122\relax
\mciteBstWouldAddEndPuncttrue
\mciteSetBstMidEndSepPunct{\mcitedefaultmidpunct}
{\mcitedefaultendpunct}{\mcitedefaultseppunct}\relax
\EndOfBibitem
\bibitem[Ruckenbauer \latin{et~al.}(2016)Ruckenbauer, Mai, Marquetand, and
  Gonz\'{a}lez]{Ruckenbauer2016JCP}
Ruckenbauer,~M.; Mai,~S.; Marquetand,~P.; Gonz\'{a}lez,~L. Photoelectron
  Spectra of 2-Thiouracil, 4-Thiouracil, and 2,4-Dithiouracil. \emph{J. Chem.
  Phys.} \textbf{2016}, \emph{144}, 074303\relax
\mciteBstWouldAddEndPuncttrue
\mciteSetBstMidEndSepPunct{\mcitedefaultmidpunct}
{\mcitedefaultendpunct}{\mcitedefaultseppunct}\relax
\EndOfBibitem
\bibitem[Manathunga \latin{et~al.}(2016)Manathunga, Yang, Luk, Gozem, Frutos,
  Valentini, Ferr\`e, and Olivucci]{Manathunga2016JCTC}
Manathunga,~M.; Yang,~X.; Luk,~H.~L.; Gozem,~S.; Frutos,~L.~M.; Valentini,~A.;
  Ferr\`e,~N.; Olivucci,~M. Probing the Photodynamics of Rhodopsins with
  Reduced Retinal Chromophores. \emph{J. Chem. Theory Comput.} \textbf{2016},
  \emph{12}, 839--850\relax
\mciteBstWouldAddEndPuncttrue
\mciteSetBstMidEndSepPunct{\mcitedefaultmidpunct}
{\mcitedefaultendpunct}{\mcitedefaultseppunct}\relax
\EndOfBibitem
\bibitem[Nogueira \latin{et~al.}(2017)Nogueira, Plasser, and
  Gonz{\'a}lez]{Nogueira2017CS}
Nogueira,~J.~J.; Plasser,~F.; Gonz{\'a}lez,~L. {Electronic Delocalization{,}
  Charge Transfer and Hypochromism in the UV Absorption Spectrum of Polyadenine
  Unravelled by Multiscale Computations and Quantitative Wavefunction
  Analysis}. \emph{Chem. Sci.} \textbf{2017}, \emph{8}, 5682--5691\relax
\mciteBstWouldAddEndPuncttrue
\mciteSetBstMidEndSepPunct{\mcitedefaultmidpunct}
{\mcitedefaultendpunct}{\mcitedefaultseppunct}\relax
\EndOfBibitem
\bibitem[Mai \latin{et~al.}(2018)Mai, Mohamadzade, Marquetand, Gonz{{\'a}}lez,
  and Ullrich]{Mai2018M}
Mai,~S.; Mohamadzade,~A.; Marquetand,~P.; Gonz{{\'a}}lez,~L.; Ullrich,~S.
  Simulated and Experimental Time-Resolved Photoelectron Spectra of the
  Intersystem Crossing Dynamics in 2-Thiouracil. \emph{Molecules}
  \textbf{2018}, \emph{23}, 2836\relax
\mciteBstWouldAddEndPuncttrue
\mciteSetBstMidEndSepPunct{\mcitedefaultmidpunct}
{\mcitedefaultendpunct}{\mcitedefaultseppunct}\relax
\EndOfBibitem
\bibitem[Rauer \latin{et~al.}(2018)Rauer, Nogueira, Marquetand, and
  Gonz\'{a}lez]{Rauer2018MC}
Rauer,~C.; Nogueira,~J.~J.; Marquetand,~P.; Gonz\'{a}lez,~L. Stepwise
  photosensitized thymine dimerization mediated by an exciton intermediate.
  \emph{Monatsh. Chem.} \textbf{2018}, \emph{149}, 1–9\relax
\mciteBstWouldAddEndPuncttrue
\mciteSetBstMidEndSepPunct{\mcitedefaultmidpunct}
{\mcitedefaultendpunct}{\mcitedefaultseppunct}\relax
\EndOfBibitem
\bibitem[Zobel \latin{et~al.}(2018)Zobel, Heindl, Nogueira, and
  Gonz{{\'a}}lez]{Zobel2018JCTC}
Zobel,~J.~P.; Heindl,~M.; Nogueira,~J.~J.; Gonz{{\'a}}lez,~L. Vibrational
  Sampling and Solvent Effects on the Electronic Structure of the Absorption
  Spectrum of 2-Nitronaphthalene. \emph{J. Chem. Theory Comput.} \textbf{2018},
  \emph{14}, 3205--3217\relax
\mciteBstWouldAddEndPuncttrue
\mciteSetBstMidEndSepPunct{\mcitedefaultmidpunct}
{\mcitedefaultendpunct}{\mcitedefaultseppunct}\relax
\EndOfBibitem
\bibitem[Pathak \latin{et~al.}(2019)Pathak, Ibele, Boll, Callegari, Demidovich,
  Erk, Feifel, Forbes, Fraia, Giannessi, Hansen, Holland, Ingle, Mason, Plekan,
  Prince, Rouzée, Squibb, Tross, Ashfold, Curchod, and
  Rolles]{Pathak2020arXiv}
Pathak,~S. \latin{et~al.}  Tracking the Ultraviolet Photochemistry of
  Thiophenone During and Beyond the Initial Ultrafast Ring Opening. 2019\relax
\mciteBstWouldAddEndPuncttrue
\mciteSetBstMidEndSepPunct{\mcitedefaultmidpunct}
{\mcitedefaultendpunct}{\mcitedefaultseppunct}\relax
\EndOfBibitem
\bibitem[Maria Teresa Neves-Petersen and Gajula(2012)Maria Teresa
  Neves-Petersen, and Gajula]{Neves-Peterson2012}
Maria Teresa Neves-Petersen,~S.~P.; Gajula,~G.~P. \emph{UV Light Effects on
  Proteins: From Photochemistry to Nanomedicine, Molecular Photochemistry -
  Various Aspects}; IntechOpen, 2012; Chapter 7\relax
\mciteBstWouldAddEndPuncttrue
\mciteSetBstMidEndSepPunct{\mcitedefaultmidpunct}
{\mcitedefaultendpunct}{\mcitedefaultseppunct}\relax
\EndOfBibitem
\bibitem[Cadet \latin{et~al.}(2014)Cadet, Grand, and Douki]{Cadet2014}
Cadet,~J.; Grand,~A.; Douki,~T. \emph{Photoinduced Phenomena in Nucleic Acids
  II}; Topics in Current Chemistry; Springer Berlin Heidelberg, 2014; Vol. 356;
  pp 249--275\relax
\mciteBstWouldAddEndPuncttrue
\mciteSetBstMidEndSepPunct{\mcitedefaultmidpunct}
{\mcitedefaultendpunct}{\mcitedefaultseppunct}\relax
\EndOfBibitem
\bibitem[Segatta \latin{et~al.}(2019)Segatta, Cupellini, Garavelli, and
  Mennucci]{Segatta2019CR}
Segatta,~F.; Cupellini,~L.; Garavelli,~M.; Mennucci,~B. Quantum Chemical
  Modeling of the Photoinduced Activity of Multichromophoric Biosystems.
  \emph{Chem. Rev.} \textbf{2019}, \emph{119}, 9361--9380\relax
\mciteBstWouldAddEndPuncttrue
\mciteSetBstMidEndSepPunct{\mcitedefaultmidpunct}
{\mcitedefaultendpunct}{\mcitedefaultseppunct}\relax
\EndOfBibitem
\bibitem[Landry and Subotnik(2014)Landry, and Subotnik]{Landry2014JCTC}
Landry,~B.~R.; Subotnik,~J.~E. Quantifying the Lifetime of Triplet Energy
  Transfer Processes in Organic Chromophores: A Case Study of
  4-(2-Naphthylmethyl)benzaldehyde. \emph{J. Chem. Theory Comput.}
  \textbf{2014}, \emph{10}, 4253--4263\relax
\mciteBstWouldAddEndPuncttrue
\mciteSetBstMidEndSepPunct{\mcitedefaultmidpunct}
{\mcitedefaultendpunct}{\mcitedefaultseppunct}\relax
\EndOfBibitem
\bibitem[Schultz \latin{et~al.}(2003)Schultz, Quenneville, Levine, Toniolo,
  Mart\'{i}nez, Lochbrunner, Schmitt, Shaffer, Zgierski, and
  Stolow]{Schultz2003JACS}
Schultz,~T.; Quenneville,~J.; Levine,~B.; Toniolo,~A.; Mart\'{i}nez,~T.~J.;
  Lochbrunner,~S.; Schmitt,~M.; Shaffer,~J.~P.; Zgierski,~M.~Z.; Stolow,~A.
  Mechanism and Dynamics of Azobenzene Photoisomerization. \emph{J. Am. Chem.
  Soc.} \textbf{2003}, \emph{125}, 8098--8099\relax
\mciteBstWouldAddEndPuncttrue
\mciteSetBstMidEndSepPunct{\mcitedefaultmidpunct}
{\mcitedefaultendpunct}{\mcitedefaultseppunct}\relax
\EndOfBibitem
\bibitem[Toniolo \latin{et~al.}(2004)Toniolo, Olsen, Manohar, and
  Mart\'{i}nez]{Toniolo2004FD}
Toniolo,~A.; Olsen,~S.; Manohar,~L.; Mart\'{i}nez,~T.~J. {Conical Intersection
  Dynamics in Solution: The Chromophore of Green Fluorescent Protein}.
  \emph{Faraday Discuss.} \textbf{2004}, \emph{127}, 149--163\relax
\mciteBstWouldAddEndPuncttrue
\mciteSetBstMidEndSepPunct{\mcitedefaultmidpunct}
{\mcitedefaultendpunct}{\mcitedefaultseppunct}\relax
\EndOfBibitem
\bibitem[Domcke \latin{et~al.}(2011)Domcke, Yarkony, and
  K{\"o}ppel]{Domcke2011}
Domcke,~W.; Yarkony,~D.; K{\"o}ppel,~H. \emph{Conical Intersections: Theory,
  Computation and Experiment}; Advanced Series in Physical Chemistry; World
  Scientific Publishing Company, 2011\relax
\mciteBstWouldAddEndPuncttrue
\mciteSetBstMidEndSepPunct{\mcitedefaultmidpunct}
{\mcitedefaultendpunct}{\mcitedefaultseppunct}\relax
\EndOfBibitem
\bibitem[Serrano-Andr\'{e}s and Merch\'{a}n(2005)Serrano-Andr\'{e}s, and
  Merch\'{a}n]{Serrano-Andres2005JMST}
Serrano-Andr\'{e}s,~L.; Merch\'{a}n,~M. Quantum chemistry of the excited state:
  2005 overview. \emph{J. Mol. Struc.-THEOCHEM} \textbf{2005}, \emph{729}, 99
  -- 108\relax
\mciteBstWouldAddEndPuncttrue
\mciteSetBstMidEndSepPunct{\mcitedefaultmidpunct}
{\mcitedefaultendpunct}{\mcitedefaultseppunct}\relax
\EndOfBibitem
\bibitem[Chandrasekaran \latin{et~al.}(2019)Chandrasekaran, Kamal, Batra, Kim,
  Chen, and Ramprasad]{Chandrasekaran2019nCM}
Chandrasekaran,~A.; Kamal,~D.; Batra,~R.; Kim,~C.; Chen,~L.; Ramprasad,~R.
  Solving the Electronic Structure Problem with Machine Learning. \emph{npj
  Comput. Mater.} \textbf{2019}, \emph{5}, 22\relax
\mciteBstWouldAddEndPuncttrue
\mciteSetBstMidEndSepPunct{\mcitedefaultmidpunct}
{\mcitedefaultendpunct}{\mcitedefaultseppunct}\relax
\EndOfBibitem
\bibitem[Jeong \latin{et~al.}(2020)Jeong, Stoneburner, King, Li, Walker, Lindh,
  and Gagliardi]{Jeong2020JCTC}
Jeong,~W.; Stoneburner,~S.~J.; King,~D.; Li,~R.; Walker,~A.; Lindh,~R.;
  Gagliardi,~L. Automation of Active Space Selection for Multireference Methods
  via Machine Learning on Chemical Bond Dissociation. \emph{J. Chem. Theory
  Comput.} \textbf{2020}, \emph{16}, 2389--2399\relax
\mciteBstWouldAddEndPuncttrue
\mciteSetBstMidEndSepPunct{\mcitedefaultmidpunct}
{\mcitedefaultendpunct}{\mcitedefaultseppunct}\relax
\EndOfBibitem
\bibitem[Carleo and Troyer(2017)Carleo, and Troyer]{Carleo2017S}
Carleo,~G.; Troyer,~M. Solving the Quantum Many-Body Problem with Artificial
  Neural Networks. \emph{Science} \textbf{2017}, \emph{355}, 602--606\relax
\mciteBstWouldAddEndPuncttrue
\mciteSetBstMidEndSepPunct{\mcitedefaultmidpunct}
{\mcitedefaultendpunct}{\mcitedefaultseppunct}\relax
\EndOfBibitem
\bibitem[Saito(2017)]{Saito2017JPCJ}
Saito,~H. Solving the Bose–Hubbard Model with Machine Learning. \emph{J.
  Phys. Soc. Jpn.} \textbf{2017}, \emph{86}, 093001\relax
\mciteBstWouldAddEndPuncttrue
\mciteSetBstMidEndSepPunct{\mcitedefaultmidpunct}
{\mcitedefaultendpunct}{\mcitedefaultseppunct}\relax
\EndOfBibitem
\bibitem[Nomura \latin{et~al.}(2017)Nomura, Darmawan, Yamaji, and
  Imada]{Nomura2017PRB}
Nomura,~Y.; Darmawan,~A.~S.; Yamaji,~Y.; Imada,~M. Restricted Boltzmann Machine
  Learning for solving strongly correlated quantum systems. \emph{Phys. Rev. B}
  \textbf{2017}, \emph{96}, 205152\relax
\mciteBstWouldAddEndPuncttrue
\mciteSetBstMidEndSepPunct{\mcitedefaultmidpunct}
{\mcitedefaultendpunct}{\mcitedefaultseppunct}\relax
\EndOfBibitem
\bibitem[Han \latin{et~al.}(2019)Han, Zhang, and E]{Han2018JCP}
Han,~J.; Zhang,~L.; E,~W. {Solving Many-Electron Schr\"{o}dinger Equation using
  Deep Neural Networks}. \emph{J. Comput. Phys.} \textbf{2019}, \emph{399},
  108929\relax
\mciteBstWouldAddEndPuncttrue
\mciteSetBstMidEndSepPunct{\mcitedefaultmidpunct}
{\mcitedefaultendpunct}{\mcitedefaultseppunct}\relax
\EndOfBibitem
\bibitem[Townsend and Vogiatzis(2019)Townsend, and Vogiatzis]{Townsend2019JPCL}
Townsend,~J.; Vogiatzis,~K.~D. Data-Driven Acceleration of the coupled-Cluster
  Singles and Doubles Iterative Solver. \emph{J. Phys. Chem. Lett.}
  \textbf{2019}, \emph{10}, 4129--4135\relax
\mciteBstWouldAddEndPuncttrue
\mciteSetBstMidEndSepPunct{\mcitedefaultmidpunct}
{\mcitedefaultendpunct}{\mcitedefaultseppunct}\relax
\EndOfBibitem
\bibitem[Sch\"{u}tt \latin{et~al.}(2019)Sch\"{u}tt, Gastegger, Tkatchenko,
  M\"{u}ller, and Maurer]{Schuett2019NC}
Sch\"{u}tt,~K.~T.; Gastegger,~M.; Tkatchenko,~A.; M\"{u}ller,~K.-R.;
  Maurer,~R.~J. Unifying Machine Learning and quantum chemistry with a deep
  neural network for molecular wavefunctions. \emph{Nat. Commun.}
  \textbf{2019}, \emph{10}, 5024\relax
\mciteBstWouldAddEndPuncttrue
\mciteSetBstMidEndSepPunct{\mcitedefaultmidpunct}
{\mcitedefaultendpunct}{\mcitedefaultseppunct}\relax
\EndOfBibitem
\bibitem[Pfau \latin{et~al.}(2019)Pfau, Spencer, de~G.~Matthews, and
  Foulkes]{Pfau2019arXiv}
Pfau,~D.; Spencer,~J.~S.; de~G.~Matthews,~A.~G.; Foulkes,~W. M.~C. Ab-Initio
  Solution of the Many-Electron Schr\"{o}dinger Equation with Deep Neural
  Networks. 2019\relax
\mciteBstWouldAddEndPuncttrue
\mciteSetBstMidEndSepPunct{\mcitedefaultmidpunct}
{\mcitedefaultendpunct}{\mcitedefaultseppunct}\relax
\EndOfBibitem
\bibitem[Hermann \latin{et~al.}(2019)Hermann, Sch\"{a}tzle, and
  No\'e]{Hermann2019arXiv}
Hermann,~J.; Sch\"{a}tzle,~Z.; No\'e,~F. Deep Neural Network Solution of the
  Electronic Schr\"{o}dinger Equation. 2019\relax
\mciteBstWouldAddEndPuncttrue
\mciteSetBstMidEndSepPunct{\mcitedefaultmidpunct}
{\mcitedefaultendpunct}{\mcitedefaultseppunct}\relax
\EndOfBibitem
\bibitem[Gastegger \latin{et~al.}(2020)Gastegger, McSloy, Luya, Sch\"{u}tt, and
  Maurer]{Gastegger2020arXiv}
Gastegger,~M.; McSloy,~A.; Luya,~M.; Sch\"{u}tt,~K.~T.; Maurer,~R.~J. {A Deep
  Neural Network for Molecular Wave Functions in Quasi-Atomic Minimal Basis
  Representation}. \emph{https://arxiv.org/abs/2005.06979} \textbf{2020},
  \relax
\mciteBstWouldAddEndPunctfalse
\mciteSetBstMidEndSepPunct{\mcitedefaultmidpunct}
{}{\mcitedefaultseppunct}\relax
\EndOfBibitem
\bibitem[Hegde and Bowen(2017)Hegde, and Bowen]{Hegde2017SR}
Hegde,~G.; Bowen,~R.~C. Machine-Learned Approximations to Density Functional
  Theory {Hamiltonian}s. \emph{Sci. Rep.} \textbf{2017}, \emph{7}, 42669\relax
\mciteBstWouldAddEndPuncttrue
\mciteSetBstMidEndSepPunct{\mcitedefaultmidpunct}
{\mcitedefaultendpunct}{\mcitedefaultseppunct}\relax
\EndOfBibitem
\bibitem[Brockherde \latin{et~al.}(2017)Brockherde, Vogt, Li, Tuckerman, Burke,
  and M{\"u}ller]{Brockherde2017NC}
Brockherde,~F.; Vogt,~L.; Li,~L.; Tuckerman,~M.~E.; Burke,~K.;
  M{\"u}ller,~K.-R. Bypassing the Kohn-Sham Equations with Machine Learning.
  \emph{Nat. Commun.} \textbf{2017}, \emph{8}, 872\relax
\mciteBstWouldAddEndPuncttrue
\mciteSetBstMidEndSepPunct{\mcitedefaultmidpunct}
{\mcitedefaultendpunct}{\mcitedefaultseppunct}\relax
\EndOfBibitem
\bibitem[Gastegger \latin{et~al.}(2019)Gastegger, Gonz{\'a}lez, and
  Marquetand]{Gastegger2019MC}
Gastegger,~M.; Gonz{\'a}lez,~L.; Marquetand,~P. Exploring Density Functional
  Subspaces with Genetic Algorithms. \emph{Monatsh. Chem.} \textbf{2019},
  \emph{150}, 173--182\relax
\mciteBstWouldAddEndPuncttrue
\mciteSetBstMidEndSepPunct{\mcitedefaultmidpunct}
{\mcitedefaultendpunct}{\mcitedefaultseppunct}\relax
\EndOfBibitem
\bibitem[Nelson \latin{et~al.}(2019)Nelson, Tiwari, and Sanvito]{Nelson2019PRB}
Nelson,~J.; Tiwari,~R.; Sanvito,~S. Machine Learning Density Functional Theory
  for the Hubbard Model. \emph{Phys. Rev. B} \textbf{2019}, \emph{99},
  075132\relax
\mciteBstWouldAddEndPuncttrue
\mciteSetBstMidEndSepPunct{\mcitedefaultmidpunct}
{\mcitedefaultendpunct}{\mcitedefaultseppunct}\relax
\EndOfBibitem
\bibitem[Cheng \latin{et~al.}(2019)Cheng, Welborn, Christensen, and
  Miller]{Cheng2019JCP}
Cheng,~L.; Welborn,~M.; Christensen,~A.~S.; Miller,~T.~F. A Universal Density
  Matrix Functional from Molecular Orbital-Based Machine Learning:
  Transferability Across Organic Molecules. \emph{J. Chem. Phys.}
  \textbf{2019}, \emph{150}, 131103\relax
\mciteBstWouldAddEndPuncttrue
\mciteSetBstMidEndSepPunct{\mcitedefaultmidpunct}
{\mcitedefaultendpunct}{\mcitedefaultseppunct}\relax
\EndOfBibitem
\bibitem[Lei and Medford(2019)Lei, and Medford]{Lei2019PRM}
Lei,~X.; Medford,~A.~J. Design and Analysis of Machine Learning
  Exchange-Correlation Functionals via Rotationally Invariant Convolutional
  Descriptors. \emph{Phys. Rev. Materials} \textbf{2019}, \emph{3},
  063801\relax
\mciteBstWouldAddEndPuncttrue
\mciteSetBstMidEndSepPunct{\mcitedefaultmidpunct}
{\mcitedefaultendpunct}{\mcitedefaultseppunct}\relax
\EndOfBibitem
\bibitem[Zhou \latin{et~al.}(2019)Zhou, Wu, Chen, and Chen]{Zhou2019JPCL}
Zhou,~Y.; Wu,~J.; Chen,~S.; Chen,~G. Toward the Exact Exchange–Correlation
  Potential: A Three-Dimensional Convolutional Neural Network Construct.
  \emph{J. Phys. Chem. Lett.} \textbf{2019}, \emph{10}, 7264--7269\relax
\mciteBstWouldAddEndPuncttrue
\mciteSetBstMidEndSepPunct{\mcitedefaultmidpunct}
{\mcitedefaultendpunct}{\mcitedefaultseppunct}\relax
\EndOfBibitem
\bibitem[Kolb \latin{et~al.}(2017)Kolb, Lentz, and Kolpak]{Kolb2017SR}
Kolb,~B.; Lentz,~L.~C.; Kolpak,~A.~M. Discovering Charge Density Functionals
  and Structure-Property Relationships with PROPhet: A General Framework for
  Coupling Machine Learning and First-Principles Methods. \emph{Sci. Rep.}
  \textbf{2017}, \emph{7}\relax
\mciteBstWouldAddEndPuncttrue
\mciteSetBstMidEndSepPunct{\mcitedefaultmidpunct}
{\mcitedefaultendpunct}{\mcitedefaultseppunct}\relax
\EndOfBibitem
\bibitem[Willatt \latin{et~al.}(2019)Willatt, Musil, and
  Ceriotti]{Willat2019JCP}
Willatt,~M.~J.; Musil,~F.; Ceriotti,~M. Atom-Density Representations for
  Machine Learning. \emph{J. Chem. Phys.} \textbf{2019}, \emph{150},
  154110\relax
\mciteBstWouldAddEndPuncttrue
\mciteSetBstMidEndSepPunct{\mcitedefaultmidpunct}
{\mcitedefaultendpunct}{\mcitedefaultseppunct}\relax
\EndOfBibitem
\bibitem[Choo \latin{et~al.}(2018)Choo, Carleo, Regnault, and
  Neupert]{Choo2018PRL}
Choo,~K.; Carleo,~G.; Regnault,~N.; Neupert,~T. Symmetries and Many-Body
  Excitations with Neural-Network Quantum States. \emph{Phys. Rev. Lett.}
  \textbf{2018}, \emph{121}, 167204\relax
\mciteBstWouldAddEndPuncttrue
\mciteSetBstMidEndSepPunct{\mcitedefaultmidpunct}
{\mcitedefaultendpunct}{\mcitedefaultseppunct}\relax
\EndOfBibitem
\bibitem[Guan and Yarkony(2020)Guan, and Yarkony]{Guan2020JPCL}
Guan,~Y.; Yarkony,~D.~R. Accurate Neural Network Representation of the \emph{Ab
  Initio} Determined Spin–Orbit Interaction in the Diabatic Representation
  Including the Effects of Conical Intersections. \emph{J. Phys. Chem. Lett.}
  \textbf{2020}, \emph{11}, 1848--1858\relax
\mciteBstWouldAddEndPuncttrue
\mciteSetBstMidEndSepPunct{\mcitedefaultmidpunct}
{\mcitedefaultendpunct}{\mcitedefaultseppunct}\relax
\EndOfBibitem
\bibitem[Carbogno \latin{et~al.}(2010)Carbogno, Behler, Reuter, and
  Gro\ss{}]{Carbogno2010PRB}
Carbogno,~C.; Behler,~J.; Reuter,~K.; Gro\ss{},~A. Signatures of Nonadiabatic
  {O}$_2$ Dissociation at {Al}(111): First-Principles Fewest-Switches Study.
  \emph{Phys. Rev. B} \textbf{2010}, \emph{81}, 035410\relax
\mciteBstWouldAddEndPuncttrue
\mciteSetBstMidEndSepPunct{\mcitedefaultmidpunct}
{\mcitedefaultendpunct}{\mcitedefaultseppunct}\relax
\EndOfBibitem
\bibitem[Westermayr \latin{et~al.}(2019)Westermayr, Gastegger, Menger, Mai,
  Gonz\'{a}lez, and Marquetand]{Westermayr2019CS}
Westermayr,~J.; Gastegger,~M.; Menger,~M. F. S.~J.; Mai,~S.; Gonz\'{a}lez,~L.;
  Marquetand,~P. {Machine Learning Enables Long Time Scale Molecular
  Photodynamics Simulations}. \emph{Chem. Sci.} \textbf{2019}, \emph{10},
  8100--8107\relax
\mciteBstWouldAddEndPuncttrue
\mciteSetBstMidEndSepPunct{\mcitedefaultmidpunct}
{\mcitedefaultendpunct}{\mcitedefaultseppunct}\relax
\EndOfBibitem
\bibitem[Westermayr \latin{et~al.}(2020)Westermayr, Faber, Christensen, von
  Lilienfeld, and Marquetand]{Westermayr2020MLST}
Westermayr,~J.; Faber,~F.~A.; Christensen,~A.~S.; von Lilienfeld,~O.~A.;
  Marquetand,~P. Neural Networks and Kernel Ridge Regression for Excited States
  Dynamics of {CH}$_2${NH}$_2^+$: From Single-State to Multi-State
  Representations and Multi-Property Machine Learning Models. \emph{Mach.
  Learn.: Sci. Technol.} \textbf{2020}, \emph{1}, 025009\relax
\mciteBstWouldAddEndPuncttrue
\mciteSetBstMidEndSepPunct{\mcitedefaultmidpunct}
{\mcitedefaultendpunct}{\mcitedefaultseppunct}\relax
\EndOfBibitem
\bibitem[Shen and Yarkony(2020)Shen, and Yarkony]{Shen2020JPCA}
Shen,~Y.; Yarkony,~D.~R. Construction of Quasi-diabatic Hamiltonians That
  Accurately Represent \emph{Ab Initio} Determined Adiabatic Electronic States
  Coupled by Conical Intersections for Systems on the Order of 15 Atoms.
  Application to Cyclopentoxide Photoelectron Detachment in the Full 39 Degrees
  of Freedom. \emph{J. Phys. Chem. A} \textbf{2020}, \emph{124},
  4539--4548\relax
\mciteBstWouldAddEndPuncttrue
\mciteSetBstMidEndSepPunct{\mcitedefaultmidpunct}
{\mcitedefaultendpunct}{\mcitedefaultseppunct}\relax
\EndOfBibitem
\bibitem[Zhang \latin{et~al.}(2020)Zhang, Ye, Zhang, Jiang, and
  Jiang]{Zhang2020arXiv}
Zhang,~Y.; Ye,~S.; Zhang,~J.; Jiang,~J.; Jiang,~B. Efficient and Accurate
  Spectroscopic Simulations with Symmetry-Preserving Neural Network Models for
  Tensorial Properties. \emph{arXiv} \textbf{2020}, \emph{2004.13605}\relax
\mciteBstWouldAddEndPuncttrue
\mciteSetBstMidEndSepPunct{\mcitedefaultmidpunct}
{\mcitedefaultendpunct}{\mcitedefaultseppunct}\relax
\EndOfBibitem
\bibitem[Carbogno \latin{et~al.}(2008)Carbogno, Behler, Gro{\ss}, and
  Reuter]{Carbogno2008PRL}
Carbogno,~C.; Behler,~J.; Gro{\ss},~A.; Reuter,~K. Fingerprints for
  {Spin}-{Selection} {Rules} in the {Interaction} {Dynamics} of {O}$_2$ at
  {Al}(111). \emph{Phys. Rev. Lett.} \textbf{2008}, \emph{101}, 096104\relax
\mciteBstWouldAddEndPuncttrue
\mciteSetBstMidEndSepPunct{\mcitedefaultmidpunct}
{\mcitedefaultendpunct}{\mcitedefaultseppunct}\relax
\EndOfBibitem
\bibitem[Polyak \latin{et~al.}(2019)Polyak, Richings, Habershon, and
  Knowles]{Polyak2019JCP}
Polyak,~I.; Richings,~G.~W.; Habershon,~S.; Knowles,~P.~J. Direct Quantum
  Dynamics using Variational Gaussian Wavepackets and Gaussian Process
  Regression. \emph{J. Chem. Phys.} \textbf{2019}, \emph{150}, 041101\relax
\mciteBstWouldAddEndPuncttrue
\mciteSetBstMidEndSepPunct{\mcitedefaultmidpunct}
{\mcitedefaultendpunct}{\mcitedefaultseppunct}\relax
\EndOfBibitem
\bibitem[Hobday \latin{et~al.}(1999)Hobday, Smith, and
  Belbruno]{Hobday1999MSMSE}
Hobday,~S.; Smith,~R.; Belbruno,~J. Applications of Neural Networks to Fitting
  Interatomic Potential Functions. \emph{Modell. Simul. Mater. Sci. Eng.}
  \textbf{1999}, \emph{7}, 397\relax
\mciteBstWouldAddEndPuncttrue
\mciteSetBstMidEndSepPunct{\mcitedefaultmidpunct}
{\mcitedefaultendpunct}{\mcitedefaultseppunct}\relax
\EndOfBibitem
\bibitem[Bart{\'o}k \latin{et~al.}(2010)Bart{\'o}k, Payne, Kondor, and
  Cs{\'a}nyi]{Bartok2010PRL}
Bart{\'o}k,~A.~P.; Payne,~M.~C.; Kondor,~R.; Cs{\'a}nyi,~G. Gaussian
  Approximation Potentials: The Accuracy of Quantum Mechanics, without the
  Electrons. \emph{Phys. Rev. Lett.} \textbf{2010}, \emph{104}, 136403\relax
\mciteBstWouldAddEndPuncttrue
\mciteSetBstMidEndSepPunct{\mcitedefaultmidpunct}
{\mcitedefaultendpunct}{\mcitedefaultseppunct}\relax
\EndOfBibitem
\bibitem[Rupp \latin{et~al.}(2012)Rupp, Tkatchenko, M{\"u}ller, and von
  Lilienfeld]{Rupp2012PRL}
Rupp,~M.; Tkatchenko,~A.; M{\"u}ller,~K.-R.; von Lilienfeld,~O.~A. Fast and
  {Accurate} {Modeling} of {Molecular} {Atomization} {Energies} with {Machine}
  {Learning}. \emph{Phys. Rev. Lett.} \textbf{2012}, \emph{108}, 058301\relax
\mciteBstWouldAddEndPuncttrue
\mciteSetBstMidEndSepPunct{\mcitedefaultmidpunct}
{\mcitedefaultendpunct}{\mcitedefaultseppunct}\relax
\EndOfBibitem
\bibitem[Li \latin{et~al.}(2015)Li, Kermode, and De~Vita]{Li2015PRL}
Li,~Z.; Kermode,~J.~R.; De~Vita,~A. Molecular Dynamics with On-the-Fly Machine
  Learning of Quantum-Mechanical Forces. \emph{Phys. Rev. Lett.} \textbf{2015},
  \emph{114}, 096405\relax
\mciteBstWouldAddEndPuncttrue
\mciteSetBstMidEndSepPunct{\mcitedefaultmidpunct}
{\mcitedefaultendpunct}{\mcitedefaultseppunct}\relax
\EndOfBibitem
\bibitem[{von Lilienfeld} \latin{et~al.}(2015){von Lilienfeld}, Ramakrishnan,
  Rupp, and Knoll]{vonLilienfeld2015IJQC}
{von Lilienfeld},~O.~A.; Ramakrishnan,~R.; Rupp,~M.; Knoll,~A. Fourier Series
  of Atomic Radial Distribution Functions: A Molecular Fingerprint for Machine
  Learning Models of Quantum Chemical Properties. \emph{Int. J. Quantum Chem.}
  \textbf{2015}, \emph{115}, 1084--1093\relax
\mciteBstWouldAddEndPuncttrue
\mciteSetBstMidEndSepPunct{\mcitedefaultmidpunct}
{\mcitedefaultendpunct}{\mcitedefaultseppunct}\relax
\EndOfBibitem
\bibitem[Gastegger and Marquetand(2015)Gastegger, and
  Marquetand]{Gastegger2015JCTC}
Gastegger,~M.; Marquetand,~P. High-Dimensional Neural Network Potentials for
  Organic Reactions and an Improved Training Algorithm. \emph{J. Chem. Theory
  Comput.} \textbf{2015}, \emph{11}, 2187--2198\relax
\mciteBstWouldAddEndPuncttrue
\mciteSetBstMidEndSepPunct{\mcitedefaultmidpunct}
{\mcitedefaultendpunct}{\mcitedefaultseppunct}\relax
\EndOfBibitem
\bibitem[Rupp \latin{et~al.}(2015)Rupp, Ramakrishnan, and von
  Lilienfeld]{Rupp2015JPCL}
Rupp,~M.; Ramakrishnan,~R.; von Lilienfeld,~O.~A. Machine Learning for Quantum
  Mechanical Properties of Atoms in Molecules. \emph{J. Phys. Chem. Lett.}
  \textbf{2015}, \emph{6}, 3309--3313\relax
\mciteBstWouldAddEndPuncttrue
\mciteSetBstMidEndSepPunct{\mcitedefaultmidpunct}
{\mcitedefaultendpunct}{\mcitedefaultseppunct}\relax
\EndOfBibitem
\bibitem[Behler(2016)]{Behler2016JCP}
Behler,~J. Perspective: Machine Learning Potentials for Atomistic Simulations.
  \emph{J. Chem. Phys.} \textbf{2016}, \emph{145}, 170901\relax
\mciteBstWouldAddEndPuncttrue
\mciteSetBstMidEndSepPunct{\mcitedefaultmidpunct}
{\mcitedefaultendpunct}{\mcitedefaultseppunct}\relax
\EndOfBibitem
\bibitem[Artrith and Urban(2016)Artrith, and Urban]{Urban2016CMS}
Artrith,~N.; Urban,~A. An implementation of artificial neural-network
  potentials for atomistic materials simulations: Performance for TiO$_2$.
  \emph{computational Materials Science} \textbf{2016}, \emph{114}, 135 --
  150\relax
\mciteBstWouldAddEndPuncttrue
\mciteSetBstMidEndSepPunct{\mcitedefaultmidpunct}
{\mcitedefaultendpunct}{\mcitedefaultseppunct}\relax
\EndOfBibitem
\bibitem[Gastegger \latin{et~al.}(2016)Gastegger, Kauffmann, Behler, and
  Marquetand]{Gastegger2016JCP}
Gastegger,~M.; Kauffmann,~C.; Behler,~J.; Marquetand,~P. Comparing the Accuracy
  of High-Dimensional Neural Network Potentials and the Systematic Molecular
  Fragmentation Method: A Benchmark Study for All-Trans Alkanes. \emph{J. Chem.
  Phys.} \textbf{2016}, \emph{144}\relax
\mciteBstWouldAddEndPuncttrue
\mciteSetBstMidEndSepPunct{\mcitedefaultmidpunct}
{\mcitedefaultendpunct}{\mcitedefaultseppunct}\relax
\EndOfBibitem
\bibitem[Artrith \latin{et~al.}(2017)Artrith, Urban, and Ceder]{Artrith2017PRB}
Artrith,~N.; Urban,~A.; Ceder,~G. Efficient and Accurate Machine-Learning
  Interpolation of Atomic Energies in Compositions with Many Species.
  \emph{Phys. Rev. B} \textbf{2017}, \emph{96}, 014112\relax
\mciteBstWouldAddEndPuncttrue
\mciteSetBstMidEndSepPunct{\mcitedefaultmidpunct}
{\mcitedefaultendpunct}{\mcitedefaultseppunct}\relax
\EndOfBibitem
\bibitem[Gastegger \latin{et~al.}(2017)Gastegger, Behler, and
  Marquetand]{Gastegger2017CS}
Gastegger,~M.; Behler,~J.; Marquetand,~P. Machine Learning Molecular Dynamics
  for the Simulation of Infrared Spectra. \emph{Chem. Sci.} \textbf{2017},
  \emph{8}, 6924--6935\relax
\mciteBstWouldAddEndPuncttrue
\mciteSetBstMidEndSepPunct{\mcitedefaultmidpunct}
{\mcitedefaultendpunct}{\mcitedefaultseppunct}\relax
\EndOfBibitem
\bibitem[Deringer and Cs{{\'a}}nyi(2017)Deringer, and
  Cs{{\'a}}nyi]{Deringer2017PRB}
Deringer,~V.~L.; Cs{{\'a}}nyi,~G. Machine Learning Based Interatomic Potential
  for Amorphous Carbon. \emph{Phys. Rev. B: Condens. Matter Mater. Phys.}
  \textbf{2017}, \emph{95}, 094203\relax
\mciteBstWouldAddEndPuncttrue
\mciteSetBstMidEndSepPunct{\mcitedefaultmidpunct}
{\mcitedefaultendpunct}{\mcitedefaultseppunct}\relax
\EndOfBibitem
\bibitem[Botu \latin{et~al.}(2017)Botu, Batra, Chapman, and
  Ramprasad]{Botu2017JPCC}
Botu,~V.; Batra,~R.; Chapman,~J.; Ramprasad,~R. Machine Learning Force Fields:
  Construction, Validation, and Outlook. \emph{J. Phys. Chem. C} \textbf{2017},
  \emph{121}, 511--522\relax
\mciteBstWouldAddEndPuncttrue
\mciteSetBstMidEndSepPunct{\mcitedefaultmidpunct}
{\mcitedefaultendpunct}{\mcitedefaultseppunct}\relax
\EndOfBibitem
\bibitem[Glielmo \latin{et~al.}(2017)Glielmo, Sollich, and
  De~Vita]{Glielmo2017PRB}
Glielmo,~A.; Sollich,~P.; De~Vita,~A. Accurate Interatomic Force Fields via
  Machine Learning with Covariant Kernels. \emph{Phys. Rev. B} \textbf{2017},
  \emph{95}, 214302\relax
\mciteBstWouldAddEndPuncttrue
\mciteSetBstMidEndSepPunct{\mcitedefaultmidpunct}
{\mcitedefaultendpunct}{\mcitedefaultseppunct}\relax
\EndOfBibitem
\bibitem[Smith \latin{et~al.}(2017)Smith, Isayev, and Roitberg]{Smith2017CS}
Smith,~J.~S.; Isayev,~O.; Roitberg,~A.~E. {ANI}-1: An Extensible Neural Network
  Potential with DFT Accuracy at Force Field Computational Cost. \emph{Chem.
  Sci.} \textbf{2017}, \emph{8}, 3192--3203\relax
\mciteBstWouldAddEndPuncttrue
\mciteSetBstMidEndSepPunct{\mcitedefaultmidpunct}
{\mcitedefaultendpunct}{\mcitedefaultseppunct}\relax
\EndOfBibitem
\bibitem[Fujikake \latin{et~al.}(2018)Fujikake, Deringer, Lee, Krynski,
  Elliott, and Cs{{\'a}}nyi]{Fujikake2018JCP}
Fujikake,~S.; Deringer,~V.~L.; Lee,~T.~H.; Krynski,~M.; Elliott,~S.~R.;
  Cs{{\'a}}nyi,~G. Gaussian Approximation Potential Modeling of Lithium
  Intercalation in Carbon Nanostructures. \emph{J. Chem. Phys.} \textbf{2018},
  \emph{148}, 241714\relax
\mciteBstWouldAddEndPuncttrue
\mciteSetBstMidEndSepPunct{\mcitedefaultmidpunct}
{\mcitedefaultendpunct}{\mcitedefaultseppunct}\relax
\EndOfBibitem
\bibitem[Behler(2017)]{Behler2017ACIE}
Behler,~J. First Principles Neural Network Potentials for Reactive Simulations
  of Large Molecular and Condensed Systems. \emph{Angew. Chem. Int. Edit.}
  \textbf{2017}, \emph{56}, 12828--12840\relax
\mciteBstWouldAddEndPuncttrue
\mciteSetBstMidEndSepPunct{\mcitedefaultmidpunct}
{\mcitedefaultendpunct}{\mcitedefaultseppunct}\relax
\EndOfBibitem
\bibitem[Zong \latin{et~al.}(2018)Zong, Pilania, Ding, Ackland, and
  Lookman]{Zong2018npjCM}
Zong,~H.; Pilania,~G.; Ding,~X.; Ackland,~G.~J.; Lookman,~T. Developing an
  Interatomic Potential for Martensitic Phase Transformations in Zirconium by
  Machine Learning. \emph{npj comput Mater} \textbf{2018}, \emph{4}\relax
\mciteBstWouldAddEndPuncttrue
\mciteSetBstMidEndSepPunct{\mcitedefaultmidpunct}
{\mcitedefaultendpunct}{\mcitedefaultseppunct}\relax
\EndOfBibitem
\bibitem[Wood and Thompson(2018)Wood, and Thompson]{Wood2018JCP}
Wood,~M.~A.; Thompson,~A.~P. Extending the Accuracy of the SNAP Interatomic
  Potential Form. \emph{J. Chem. Phys.} \textbf{2018}, \emph{148}, 241721\relax
\mciteBstWouldAddEndPuncttrue
\mciteSetBstMidEndSepPunct{\mcitedefaultmidpunct}
{\mcitedefaultendpunct}{\mcitedefaultseppunct}\relax
\EndOfBibitem
\bibitem[Chen \latin{et~al.}(2018)Chen, J{\o}rgensen, Li, and
  Hammer]{Chen2018JCTC}
Chen,~X.; J{\o}rgensen,~M.~S.; Li,~J.; Hammer,~B. Atomic Energies from a
  Convolutional Neural Network. \emph{J. Chem. Theory Comput.} \textbf{2018},
  \emph{14}, 3933--3942\relax
\mciteBstWouldAddEndPuncttrue
\mciteSetBstMidEndSepPunct{\mcitedefaultmidpunct}
{\mcitedefaultendpunct}{\mcitedefaultseppunct}\relax
\EndOfBibitem
\bibitem[Bart{\'o}k \latin{et~al.}(2018)Bart{\'o}k, Kermode, Bernstein, and
  Cs{{\'a}}nyi]{Bartok2018PRX}
Bart{\'o}k,~A.~P.; Kermode,~J.; Bernstein,~N.; Cs{{\'a}}nyi,~G. Machine
  Learning a General-Purpose Interatomic Potential for Silicon. \emph{Phys.
  Rev. X} \textbf{2018}, \emph{8}, 041048\relax
\mciteBstWouldAddEndPuncttrue
\mciteSetBstMidEndSepPunct{\mcitedefaultmidpunct}
{\mcitedefaultendpunct}{\mcitedefaultseppunct}\relax
\EndOfBibitem
\bibitem[Chmiela \latin{et~al.}(2018)Chmiela, Sauceda, M\"{u}ller, and
  Tkatchenko]{Chmiela2018NC}
Chmiela,~S.; Sauceda,~H.~E.; M\"{u}ller,~K.-R.; Tkatchenko,~A. Towards Exact
  Molecular Dynamics Simulations with Machine-Learned Force Fields. \emph{Nat.
  Commun.} \textbf{2018}, \emph{9}, 3887\relax
\mciteBstWouldAddEndPuncttrue
\mciteSetBstMidEndSepPunct{\mcitedefaultmidpunct}
{\mcitedefaultendpunct}{\mcitedefaultseppunct}\relax
\EndOfBibitem
\bibitem[Imbalzano \latin{et~al.}(2018)Imbalzano, Anelli, Giofr{\'e}, Klees,
  Behler, and Ceriotti]{Imbalzano2018JCP}
Imbalzano,~G.; Anelli,~A.; Giofr{\'e},~D.; Klees,~S.; Behler,~J.; Ceriotti,~M.
  Automatic Selection of Atomic Fingerprints and Reference Configurations for
  Machine-Learning Potentials. \emph{J. Chem. Phys.} \textbf{2018}, \emph{148},
  241730\relax
\mciteBstWouldAddEndPuncttrue
\mciteSetBstMidEndSepPunct{\mcitedefaultmidpunct}
{\mcitedefaultendpunct}{\mcitedefaultseppunct}\relax
\EndOfBibitem
\bibitem[Zhang \latin{et~al.}(2018)Zhang, Han, Wang, Saidi, Car, and
  Weinan]{Zhang2018NIPS}
Zhang,~L.; Han,~J.; Wang,~H.; Saidi,~W.~A.; Car,~R.; Weinan,~E. End-to-end
  Symmetry Preserving Inter-atomic Potential Energy Model for Finite and
  Extended Systems. Proceedings of the 32Nd International conference on Neural
  Information Processing Systems. USA, 2018; pp 4441--4451\relax
\mciteBstWouldAddEndPuncttrue
\mciteSetBstMidEndSepPunct{\mcitedefaultmidpunct}
{\mcitedefaultendpunct}{\mcitedefaultseppunct}\relax
\EndOfBibitem
\bibitem[Zhang \latin{et~al.}(2018)Zhang, Han, Wang, Car, and E]{Zhang2018PRL}
Zhang,~L.; Han,~J.; Wang,~H.; Car,~R.; E,~W. Deep Potential Molecular Dynamics:
  A Scalable Model with the Accuracy of Quantum Mechanics. \emph{Phys. Rev.
  Lett.} \textbf{2018}, \emph{120}, 143001\relax
\mciteBstWouldAddEndPuncttrue
\mciteSetBstMidEndSepPunct{\mcitedefaultmidpunct}
{\mcitedefaultendpunct}{\mcitedefaultseppunct}\relax
\EndOfBibitem
\bibitem[Chan \latin{et~al.}(2019)Chan, Narayanan, Cherukara, Sen, Sasikumar,
  Gray, Chan, and Sankaranarayanan]{Chan2019JPCC}
Chan,~H.; Narayanan,~B.; Cherukara,~M.~J.; Sen,~F.~G.; Sasikumar,~K.;
  Gray,~S.~K.; Chan,~M. K.~Y.; Sankaranarayanan,~S. K. R.~S. Machine Learning
  Classical Interatomic Potentials for Molecular Dynamics from First-Principles
  Training Data. \emph{J. Phys. Chem. C} \textbf{2019}, \emph{123},
  6941--6957\relax
\mciteBstWouldAddEndPuncttrue
\mciteSetBstMidEndSepPunct{\mcitedefaultmidpunct}
{\mcitedefaultendpunct}{\mcitedefaultseppunct}\relax
\EndOfBibitem
\bibitem[Faber \latin{et~al.}(2018)Faber, Christensen, Huang, and von
  Lilienfeld]{Faber2018JCP}
Faber,~F.~A.; Christensen,~A.~S.; Huang,~B.; von Lilienfeld,~O.~A. Alchemical
  and Structural Distribution Based Representation for Universal Quantum
  Machine Learning. \emph{J. Chem. Phys.} \textbf{2018}, \emph{148},
  241717\relax
\mciteBstWouldAddEndPuncttrue
\mciteSetBstMidEndSepPunct{\mcitedefaultmidpunct}
{\mcitedefaultendpunct}{\mcitedefaultseppunct}\relax
\EndOfBibitem
\bibitem[Wang and Yang(2019)Wang, and Yang]{Wang2019JCTC}
Wang,~H.; Yang,~W. Toward Building Protein Force Fields by Residue-Based
  Systematic Molecular Fragmentation and Neural Network. \emph{J. Chem. Theory
  Comput.} \textbf{2019}, \emph{15}, 1409--1417\relax
\mciteBstWouldAddEndPuncttrue
\mciteSetBstMidEndSepPunct{\mcitedefaultmidpunct}
{\mcitedefaultendpunct}{\mcitedefaultseppunct}\relax
\EndOfBibitem
\bibitem[Gerrits \latin{et~al.}(2019)Gerrits, Shakouri, Behler, and
  Kroes]{Gerrits2019JPCL}
Gerrits,~N.; Shakouri,~K.; Behler,~J.; Kroes,~G.-J. Accurate Probabilities for
  Highly Activated Reaction of Polyatomic Molecules on Surfaces Using a
  High-Dimensional Neural Network Potential: CHD$_3$ + Cu(111). \emph{J. Phys.
  Chem. Lett.} \textbf{2019}, \emph{10}, 1763--1768\relax
\mciteBstWouldAddEndPuncttrue
\mciteSetBstMidEndSepPunct{\mcitedefaultmidpunct}
{\mcitedefaultendpunct}{\mcitedefaultseppunct}\relax
\EndOfBibitem
\bibitem[Chmiela \latin{et~al.}(2019)Chmiela, Sauceda, Poltavsky, Müller, and
  Tkatchenko]{Chmiela2019CPC}
Chmiela,~S.; Sauceda,~H.~E.; Poltavsky,~I.; Müller,~K.-R.; Tkatchenko,~A.
  sGDML: Constructing Accurate and Data Efficient Molecular Force Fields using
  Machine Learning. \emph{Comput. Phys. Commun.} \textbf{2019}, \emph{240}, 38
  -- 45\relax
\mciteBstWouldAddEndPuncttrue
\mciteSetBstMidEndSepPunct{\mcitedefaultmidpunct}
{\mcitedefaultendpunct}{\mcitedefaultseppunct}\relax
\EndOfBibitem
\bibitem[Carleo \latin{et~al.}(2019)Carleo, Cirac, Cranmer, Daudet, Schuld,
  Tishby, Vogt-Maranto, and Zdeborov{\'a}]{Carleo2019RMP}
Carleo,~G.; Cirac,~I.; Cranmer,~K.; Daudet,~L.; Schuld,~M.; Tishby,~N.;
  Vogt-Maranto,~L.; Zdeborov{\'a},~L. Machine Learning and the Physical
  Sciences. \emph{Rev. Mod. Phys.} \textbf{2019}, \emph{91}, 045002\relax
\mciteBstWouldAddEndPuncttrue
\mciteSetBstMidEndSepPunct{\mcitedefaultmidpunct}
{\mcitedefaultendpunct}{\mcitedefaultseppunct}\relax
\EndOfBibitem
\bibitem[Krems(2019)]{Krems2019PCCP}
Krems,~R.~V. Bayesian Machine Learning for Quantum Molecular Dynamics.
  \emph{Phys. Chem. Chem. Phys.} \textbf{2019}, \emph{21}, 13392--13410\relax
\mciteBstWouldAddEndPuncttrue
\mciteSetBstMidEndSepPunct{\mcitedefaultmidpunct}
{\mcitedefaultendpunct}{\mcitedefaultseppunct}\relax
\EndOfBibitem
\bibitem[Deringer \latin{et~al.}(2019)Deringer, Caro, and
  Cs\'{a}nyi]{Deringer2019AM}
Deringer,~V.~L.; Caro,~M.~A.; Cs\'{a}nyi,~G. Machine Learning Interatomic
  Potentials as Emerging Tools for Materials Science. \emph{Adv. Mat.}
  \textbf{2019}, \emph{31}, 1902765\relax
\mciteBstWouldAddEndPuncttrue
\mciteSetBstMidEndSepPunct{\mcitedefaultmidpunct}
{\mcitedefaultendpunct}{\mcitedefaultseppunct}\relax
\EndOfBibitem
\bibitem[Ward \latin{et~al.}(2019)Ward, Blaiszik, Foster, Assary, Narayanan,
  and Curtiss]{Ward2019MRSC}
Ward,~L.; Blaiszik,~B.; Foster,~I.; Assary,~R.~S.; Narayanan,~B.; Curtiss,~L.
  Machine Learning Prediction of Accurate Atomization Energies of Organic
  Molecules from Low-Fidelity Quantum Chemical Calculations. \emph{MRS Commun.}
  \textbf{2019}, \emph{9}, 891–899\relax
\mciteBstWouldAddEndPuncttrue
\mciteSetBstMidEndSepPunct{\mcitedefaultmidpunct}
{\mcitedefaultendpunct}{\mcitedefaultseppunct}\relax
\EndOfBibitem
\bibitem[No\'{e} \latin{et~al.}(2020)No\'{e}, Tkatchenko, M\"{u}ller, and
  Clementi]{Noe2020ARPC}
No\'{e},~F.; Tkatchenko,~A.; M\"{u}ller,~K.-R.; Clementi,~C. Machine Learning
  for Molecular Simulation. \emph{Annu. Rev. Phys. Chem.} \textbf{2020},
  \emph{71}, 361--390\relax
\mciteBstWouldAddEndPuncttrue
\mciteSetBstMidEndSepPunct{\mcitedefaultmidpunct}
{\mcitedefaultendpunct}{\mcitedefaultseppunct}\relax
\EndOfBibitem
\bibitem[Alborzpour \latin{et~al.}(2016)Alborzpour, Tew, and
  Habershon]{Alborzpour2016JCP}
Alborzpour,~J.~P.; Tew,~D.~P.; Habershon,~S. Efficient and Accurate Evaluation
  of Potential Energy Matrix Elements for Quantum Dynamics using Gaussian
  Process Regression. \emph{J. Chem. Phys.} \textbf{2016}, \emph{145},
  174112\relax
\mciteBstWouldAddEndPuncttrue
\mciteSetBstMidEndSepPunct{\mcitedefaultmidpunct}
{\mcitedefaultendpunct}{\mcitedefaultseppunct}\relax
\EndOfBibitem
\bibitem[Cheng \latin{et~al.}(2020)Cheng, Zhao, Ma, Li, and
  Li]{Chen2020JPCA_Dynamics}
Cheng,~Z.; Zhao,~D.; Ma,~J.; Li,~W.; Li,~S. An On-the-Fly Approach to Construct
  Generalized Energy-Based Fragmentation Machine Learning Force Fields of
  Complex Systems. \emph{J. Phys. Chem. A} \textbf{2020}, \emph{124},
  5007--5014\relax
\mciteBstWouldAddEndPuncttrue
\mciteSetBstMidEndSepPunct{\mcitedefaultmidpunct}
{\mcitedefaultendpunct}{\mcitedefaultseppunct}\relax
\EndOfBibitem
\bibitem[Behler \latin{et~al.}(2008)Behler, Reuter, and
  Scheffler]{Behler2008PRB}
Behler,~J.; Reuter,~K.; Scheffler,~M. Nonadiabatic Effects in the Dissociation
  of Oxygen Molecules at the {Al}(111) Surface. \emph{Phys. Rev. B}
  \textbf{2008}, \emph{77}, 115421\relax
\mciteBstWouldAddEndPuncttrue
\mciteSetBstMidEndSepPunct{\mcitedefaultmidpunct}
{\mcitedefaultendpunct}{\mcitedefaultseppunct}\relax
\EndOfBibitem
\bibitem[Hu \latin{et~al.}(2018)Hu, Xie, Li, Li, and Lan]{Hu2018JPCL}
Hu,~D.; Xie,~Y.; Li,~X.; Li,~L.; Lan,~Z. Inclusion of Machine Learning Kernel
  Ridge Regression Potential Energy Surfaces in On-the-Fly Nonadiabatic
  Molecular Dynamics Simulation. \emph{J. Phys. Chem. Lett.} \textbf{2018},
  \emph{9}, 2725--2732\relax
\mciteBstWouldAddEndPuncttrue
\mciteSetBstMidEndSepPunct{\mcitedefaultmidpunct}
{\mcitedefaultendpunct}{\mcitedefaultseppunct}\relax
\EndOfBibitem
\bibitem[Dral \latin{et~al.}(2018)Dral, Barbatti, and Thiel]{Dral2018JPCL}
Dral,~P.~O.; Barbatti,~M.; Thiel,~W. Nonadiabatic Excited-State Dynamics with
  Machine Learning. \emph{J. Phys. Chem. Lett.} \textbf{2018}, \emph{9},
  5660--5663\relax
\mciteBstWouldAddEndPuncttrue
\mciteSetBstMidEndSepPunct{\mcitedefaultmidpunct}
{\mcitedefaultendpunct}{\mcitedefaultseppunct}\relax
\EndOfBibitem
\bibitem[Chen \latin{et~al.}(2018)Chen, Liu, Fang, Dral, and Cui]{Chen2018JPCL}
Chen,~W.-K.; Liu,~X.-Y.; Fang,~W.-H.; Dral,~P.~O.; Cui,~G. Deep Learning for
  Nonadiabatic Excited-State Dynamics. \emph{J. Phys. Chem. Lett.}
  \textbf{2018}, \emph{9}, 6702--6708\relax
\mciteBstWouldAddEndPuncttrue
\mciteSetBstMidEndSepPunct{\mcitedefaultmidpunct}
{\mcitedefaultendpunct}{\mcitedefaultseppunct}\relax
\EndOfBibitem
\bibitem[Williams and Eisfeld(2018)Williams, and Eisfeld]{Williams2018JCP}
Williams,~D. M.~G.; Eisfeld,~W. {Neural Network Diabatization: A New Ansatz for
  Accurate High-Dimensional Coupled Potential Energy Surfaces}. \emph{J. Chem.
  Phys.} \textbf{2018}, \emph{149}, 204106\relax
\mciteBstWouldAddEndPuncttrue
\mciteSetBstMidEndSepPunct{\mcitedefaultmidpunct}
{\mcitedefaultendpunct}{\mcitedefaultseppunct}\relax
\EndOfBibitem
\bibitem[Xie \latin{et~al.}(2018)Xie, Zhu, Yarkony, and Guo]{Xie2018JCP}
Xie,~C.; Zhu,~X.; Yarkony,~D.~R.; Guo,~H. Permutation Invariant Polynomial
  Neural Network Approach to Fitting Potential Energy Surfaces. {IV}. Coupled
  Diabatic Potential Energy Matrices. \emph{J. Chem. Phys.} \textbf{2018},
  \emph{149}, 144107\relax
\mciteBstWouldAddEndPuncttrue
\mciteSetBstMidEndSepPunct{\mcitedefaultmidpunct}
{\mcitedefaultendpunct}{\mcitedefaultseppunct}\relax
\EndOfBibitem
\bibitem[Guan \latin{et~al.}(2019)Guan, Zhang, Guo, and Yarkony]{Guan2019PCCP}
Guan,~Y.; Zhang,~D.~H.; Guo,~H.; Yarkony,~D.~R. Representation of Coupled
  Adiabatic Potential Energy Surfaces using Neural Network Based Quasi-Diabatic
  {Hamiltonian}s: {1,2 $^2$A'} States of {LiFH}. \emph{Phys. Chem. Chem. Phys.}
  \textbf{2019}, 10.1039/C8CP06598E\relax
\mciteBstWouldAddEndPuncttrue
\mciteSetBstMidEndSepPunct{\mcitedefaultmidpunct}
{\mcitedefaultendpunct}{\mcitedefaultseppunct}\relax
\EndOfBibitem
\bibitem[Richings and Habershon(2018)Richings, and Habershon]{Richings2018JCP}
Richings,~G.~W.; Habershon,~S. {MCTDH} on-the-Fly: Efficient Grid-Based Quantum
  Dynamics without Pre-Computed Potential Energy Surfaces. \emph{J. Chem.
  Phys.} \textbf{2018}, \emph{148}, 134116\relax
\mciteBstWouldAddEndPuncttrue
\mciteSetBstMidEndSepPunct{\mcitedefaultmidpunct}
{\mcitedefaultendpunct}{\mcitedefaultseppunct}\relax
\EndOfBibitem
\bibitem[Richings \latin{et~al.}(2019)Richings, Robertson, and
  Habershon]{Richings2019JCTC}
Richings,~G.~W.; Robertson,~C.; Habershon,~S. Improved on-the-Fly {MCTDH}
  Simulations with Many-Body-Potential Tensor Decomposition and Projection
  Diabatization. \emph{J. Chem. Theory Comput.} \textbf{2019}, \emph{15},
  857--870\relax
\mciteBstWouldAddEndPuncttrue
\mciteSetBstMidEndSepPunct{\mcitedefaultmidpunct}
{\mcitedefaultendpunct}{\mcitedefaultseppunct}\relax
\EndOfBibitem
\bibitem[Guan \latin{et~al.}(2019)Guan, Guo, and Yarkony]{Guan2019JCP}
Guan,~Y.; Guo,~H.; Yarkony,~D.~R. Neural Network Based Quasi-Diabatic
  {Hamiltonian}s with Symmetry Adaptation and a Correct Description of Conical
  Intersections. \emph{J. Chem. Phys.} \textbf{2019}, \emph{150}, 214101\relax
\mciteBstWouldAddEndPuncttrue
\mciteSetBstMidEndSepPunct{\mcitedefaultmidpunct}
{\mcitedefaultendpunct}{\mcitedefaultseppunct}\relax
\EndOfBibitem
\bibitem[Wang \latin{et~al.}(2019)Wang, Xie, Guo, and Yarkony]{Wang2019JPCA}
Wang,~Y.; Xie,~C.; Guo,~H.; Yarkony,~D.~R. A Quasi-Diabatic Representation of
  the 1,2$^1$A States of Methylamine. \emph{J. Phys. Chem. A} \textbf{2019},
  \emph{123}, 5231--5241\relax
\mciteBstWouldAddEndPuncttrue
\mciteSetBstMidEndSepPunct{\mcitedefaultmidpunct}
{\mcitedefaultendpunct}{\mcitedefaultseppunct}\relax
\EndOfBibitem
\bibitem[Richings and Habershon(2017)Richings, and Habershon]{Richings2017CPL}
Richings,~G.~W.; Habershon,~S. Direct Grid-Based Quantum Dynamics on Propagated
  Diabatic Potential Energy Surfaces. \emph{Chem. Phys. Lett.} \textbf{2017},
  \emph{683}, 228 -- 233\relax
\mciteBstWouldAddEndPuncttrue
\mciteSetBstMidEndSepPunct{\mcitedefaultmidpunct}
{\mcitedefaultendpunct}{\mcitedefaultseppunct}\relax
\EndOfBibitem
\bibitem[Netzloff \latin{et~al.}(2006)Netzloff, collins, and
  Gordon]{Netzloff2006JCP}
Netzloff,~H.~M.; collins,~M.~A.; Gordon,~M.~S. {Growing Multiconfigurational
  Potential Energy Surfaces with Applications to X+H$_2$ (X=C,N,O) Reactions}.
  \emph{J. Chem. Phys.} \textbf{2006}, \emph{124}, 154104\relax
\mciteBstWouldAddEndPuncttrue
\mciteSetBstMidEndSepPunct{\mcitedefaultmidpunct}
{\mcitedefaultendpunct}{\mcitedefaultseppunct}\relax
\EndOfBibitem
\bibitem[Bettens and Collins(1999)Bettens, and Collins]{Bettens1999JCP}
Bettens,~R. P.~A.; Collins,~M.~A. Learning to Interpolate Molecular Potential
  Energy Surfaces with Confidence: A Bayesian Approach. \emph{J. Chem. Phys.}
  \textbf{1999}, \emph{111}, 816--826\relax
\mciteBstWouldAddEndPuncttrue
\mciteSetBstMidEndSepPunct{\mcitedefaultmidpunct}
{\mcitedefaultendpunct}{\mcitedefaultseppunct}\relax
\EndOfBibitem
\bibitem[Ghosh \latin{et~al.}(2019)Ghosh, Stuke, Todorovi\'{c}, J{\o}rgensen,
  Schmidt, Vehtari, and Rinke]{Ghosh2019AS}
Ghosh,~K.; Stuke,~A.; Todorovi\'{c},~M.; J{\o}rgensen,~P.~B.; Schmidt,~M.~N.;
  Vehtari,~A.; Rinke,~P. Deep Learning Spectroscopy: Neural Networks for
  Molecular Excitation Spectra. \emph{Adv. Sci.} \textbf{2019}, \emph{6},
  1801367\relax
\mciteBstWouldAddEndPuncttrue
\mciteSetBstMidEndSepPunct{\mcitedefaultmidpunct}
{\mcitedefaultendpunct}{\mcitedefaultseppunct}\relax
\EndOfBibitem
\bibitem[Kananenka \latin{et~al.}(2019)Kananenka, Yao, Corcelli, and
  Skinner]{Kananenka2019JCTC}
Kananenka,~A.~A.; Yao,~K.; Corcelli,~S.~A.; Skinner,~J.~L. Machine Learning for
  Vibrational Spectroscopic Maps. \emph{J. Chem. Theory Comput.} \textbf{2019},
  \emph{15}, 6850--6858\relax
\mciteBstWouldAddEndPuncttrue
\mciteSetBstMidEndSepPunct{\mcitedefaultmidpunct}
{\mcitedefaultendpunct}{\mcitedefaultseppunct}\relax
\EndOfBibitem
\bibitem[Roch \latin{et~al.}(2020)Roch, Saikin, Häse, Friederich, Goldsmith,
  Le\'{o}n, and Aspuru-Guzik]{Roch2020ACSN}
Roch,~L.~M.; Saikin,~S.~K.; Häse,~F.; Friederich,~P.; Goldsmith,~R.~H.;
  Le\'{o}n,~S.; Aspuru-Guzik,~A. From Absorption Spectra to Charge Transfer in
  Nanoaggregates of Oligomers with Machine Learning. \emph{ACS Nano}
  \textbf{2020}, \emph{in press}, doi:10.1021/acsnano.0c00384\relax
\mciteBstWouldAddEndPuncttrue
\mciteSetBstMidEndSepPunct{\mcitedefaultmidpunct}
{\mcitedefaultendpunct}{\mcitedefaultseppunct}\relax
\EndOfBibitem
\bibitem[Rankine \latin{et~al.}(2020)Rankine, Madkhali, and
  Penfold]{Rankine2020JPCA}
Rankine,~C.~D.; Madkhali,~M. M.~M.; Penfold,~T.~J. {A Deep Neural Network for
  the Rapid Prediction of X-Ray Absorption Spectra}. \emph{J. Phys. Chem. A}
  \textbf{2020}, \emph{124}, 4263--4270\relax
\mciteBstWouldAddEndPuncttrue
\mciteSetBstMidEndSepPunct{\mcitedefaultmidpunct}
{\mcitedefaultendpunct}{\mcitedefaultseppunct}\relax
\EndOfBibitem
\bibitem[Pereira \latin{et~al.}(2017)Pereira, Xiao, Latino, Wu, Zhang, and
  Aires-de Sousa]{Pereira2017JCIM}
Pereira,~F.; Xiao,~K.; Latino,~D. A. R.~S.; Wu,~C.; Zhang,~Q.; Aires-de
  Sousa,~J. Machine Learning Methods to Predict Density Functional Theory B3LYP
  Energies of HOMO and LUMO Orbitals. \emph{J. Chem. Inf. Model.}
  \textbf{2017}, \emph{57}, 11--21\relax
\mciteBstWouldAddEndPuncttrue
\mciteSetBstMidEndSepPunct{\mcitedefaultmidpunct}
{\mcitedefaultendpunct}{\mcitedefaultseppunct}\relax
\EndOfBibitem
\bibitem[Isayev \latin{et~al.}(2017)Isayev, Oses, Toher, Gossett, Curtarolo,
  and Tropsha]{Isayev2017NC}
Isayev,~O.; Oses,~c.; Toher,~c.; Gossett,~E.; Curtarolo,~S.; Tropsha,~A.
  Universal Fragment Descriptors for Predicting Properties of Inorganic
  Crystals. \emph{Nat. Commun.} \textbf{2017}, \emph{8}, 15679\relax
\mciteBstWouldAddEndPuncttrue
\mciteSetBstMidEndSepPunct{\mcitedefaultmidpunct}
{\mcitedefaultendpunct}{\mcitedefaultseppunct}\relax
\EndOfBibitem
\bibitem[Pronobis \latin{et~al.}(2018)Pronobis, Sch\"{u}tt, Tkatchenko, and
  M\"{u}ller]{Pronobis2018EPJB}
Pronobis,~W.; Sch\"{u}tt,~K.~R.; Tkatchenko,~A.; M\"{u}ller,~K.-R. Capturing
  Intensive and Extensive DFT/TDDFT Molecular Properties with Machine Learning.
  \emph{Eur. Phys. J. B} \textbf{2018}, \emph{91}, 178\relax
\mciteBstWouldAddEndPuncttrue
\mciteSetBstMidEndSepPunct{\mcitedefaultmidpunct}
{\mcitedefaultendpunct}{\mcitedefaultseppunct}\relax
\EndOfBibitem
\bibitem[Stuke \latin{et~al.}(2019)Stuke, Todorovi\'{c}, Rupp, Kunkel, Ghosh,
  Himanen, and Rinke]{Stuke2019JCP}
Stuke,~A.; Todorovi\'{c},~M.; Rupp,~M.; Kunkel,~C.; Ghosh,~K.; Himanen,~L.;
  Rinke,~P. Chemical Diversity in Molecular Orbital Energy Predictions with
  Kernel Ridge Regression. \emph{J. Chem. Phys.} \textbf{2019}, \emph{150},
  204121\relax
\mciteBstWouldAddEndPuncttrue
\mciteSetBstMidEndSepPunct{\mcitedefaultmidpunct}
{\mcitedefaultendpunct}{\mcitedefaultseppunct}\relax
\EndOfBibitem
\bibitem[H\"{a}se \latin{et~al.}(2019)H\"{a}se, Fdez.~Galv\'{a}n, Aspuru-Guzik,
  Lindh, and Vacher]{Haese2019CS}
H\"{a}se,~F.; Fdez.~Galv\'{a}n,~I.; Aspuru-Guzik,~A.; Lindh,~R.; Vacher,~M. How
  Machine Learning can Assist the Interpretation of \emph{Ab Initio} Molecular
  Dynamics Simulations and Conceptual Understanding of Chemistry. \emph{Chem.
  Sci.} \textbf{2019}, \emph{10}, 2298--2307\relax
\mciteBstWouldAddEndPuncttrue
\mciteSetBstMidEndSepPunct{\mcitedefaultmidpunct}
{\mcitedefaultendpunct}{\mcitedefaultseppunct}\relax
\EndOfBibitem
\bibitem[{H\"{a}se, Florian and Kreisbeck, Christoph and Aspuru-Guzik,
  Al\'{a}n}(2017)]{Haese2017CS}
{H\"{a}se, Florian and Kreisbeck, Christoph and Aspuru-Guzik, Al\'{a}n},
  Machine Learning for Quantum Dynamics: Deep Learning of Excitation Energy
  Transfer Properties. \emph{Chem. Sci.} \textbf{2017}, \emph{8},
  8419--8426\relax
\mciteBstWouldAddEndPuncttrue
\mciteSetBstMidEndSepPunct{\mcitedefaultmidpunct}
{\mcitedefaultendpunct}{\mcitedefaultseppunct}\relax
\EndOfBibitem
\bibitem[H\"ase \latin{et~al.}(2016)H\"ase, Valleau, Pyzer-Knapp, and
  Aspuru-Guzik]{Haese2016CS}
H\"ase,~F.; Valleau,~S.; Pyzer-Knapp,~E.; Aspuru-Guzik,~A. Machine Learning
  Exciton Dynamics. \emph{Chem. Sci.} \textbf{2016}, \emph{7}, 5139--5147\relax
\mciteBstWouldAddEndPuncttrue
\mciteSetBstMidEndSepPunct{\mcitedefaultmidpunct}
{\mcitedefaultendpunct}{\mcitedefaultseppunct}\relax
\EndOfBibitem
\bibitem[O’Boyle \latin{et~al.}(2011)O’Boyle, Campbell, and
  Hutchison]{Boyle2011JPCC}
O’Boyle,~N.~M.; Campbell,~C.~M.; Hutchison,~G.~R. Computational Design and
  Selection of Optimal Organic Photovoltaic Materials. \emph{J. Phys. Chem. C}
  \textbf{2011}, \emph{115}, 16200--16210\relax
\mciteBstWouldAddEndPuncttrue
\mciteSetBstMidEndSepPunct{\mcitedefaultmidpunct}
{\mcitedefaultendpunct}{\mcitedefaultseppunct}\relax
\EndOfBibitem
\bibitem[Teunissen \latin{et~al.}(2017)Teunissen, De~Proft, and
  De~Vleeschouwer]{Teunissen2017JCTC}
Teunissen,~J.~L.; De~Proft,~F.; De~Vleeschouwer,~F. Tuning the HOMO–LUMO
  Energy Gap of Small Diamondoids Using Inverse Molecular Design. \emph{J.
  Chem. Theory Comput.} \textbf{2017}, \emph{13}, 1351--1365\relax
\mciteBstWouldAddEndPuncttrue
\mciteSetBstMidEndSepPunct{\mcitedefaultmidpunct}
{\mcitedefaultendpunct}{\mcitedefaultseppunct}\relax
\EndOfBibitem
\bibitem[Liu \latin{et~al.}(2018)Liu, Tan, Khoram, and Yu]{Liu2018ACSP}
Liu,~D.; Tan,~Y.; Khoram,~E.; Yu,~Z. Training Deep Neural Networks for the
  Inverse Design of Nanophotonic Structures. \emph{ACS Photonics}
  \textbf{2018}, \emph{5}, 1365--1369\relax
\mciteBstWouldAddEndPuncttrue
\mciteSetBstMidEndSepPunct{\mcitedefaultmidpunct}
{\mcitedefaultendpunct}{\mcitedefaultseppunct}\relax
\EndOfBibitem
\bibitem[Elton \latin{et~al.}(2019)Elton, Boukouvalas, Fuge, and
  Chung]{Elton2019MSDE}
Elton,~D.~C.; Boukouvalas,~Z.; Fuge,~M.~D.; Chung,~P.~W. {Deep Learning for
  Molecular Design -- A Review of the State of the Art}. \emph{Mol. Syst. Des.
  Eng.} \textbf{2019}, \emph{4}, 828--849\relax
\mciteBstWouldAddEndPuncttrue
\mciteSetBstMidEndSepPunct{\mcitedefaultmidpunct}
{\mcitedefaultendpunct}{\mcitedefaultseppunct}\relax
\EndOfBibitem
\bibitem[Sanchez-Lengeling and Aspuru-Guzik(2018)Sanchez-Lengeling, and
  Aspuru-Guzik]{Sanchez-Lengeling2018S}
Sanchez-Lengeling,~B.; Aspuru-Guzik,~A. Inverse Molecular Design using Machine
  Learning: Generative Models for Matter Engineering. \emph{Science}
  \textbf{2018}, \emph{361}, 360--365\relax
\mciteBstWouldAddEndPuncttrue
\mciteSetBstMidEndSepPunct{\mcitedefaultmidpunct}
{\mcitedefaultendpunct}{\mcitedefaultseppunct}\relax
\EndOfBibitem
\bibitem[Goldsmith \latin{et~al.}(2018)Goldsmith, Esterhuizen, Liu, Bartel, and
  Sutton]{Goldsmith2018AJ}
Goldsmith,~B.~R.; Esterhuizen,~J.; Liu,~J.-X.; Bartel,~C.~J.; Sutton,~C.
  Machine Learning for Heterogeneous Catalyst Design and Discovery. \emph{AIChE
  J.} \textbf{2018}, \emph{64}, 2311--2323\relax
\mciteBstWouldAddEndPuncttrue
\mciteSetBstMidEndSepPunct{\mcitedefaultmidpunct}
{\mcitedefaultendpunct}{\mcitedefaultseppunct}\relax
\EndOfBibitem
\bibitem[Davies \latin{et~al.}(2018)Davies, Butler, Isayev, and
  Walsh]{Davies2018FD}
Davies,~D.~W.; Butler,~K.~T.; Isayev,~O.; Walsh,~A. Materials Discovery by
  Chemical Analogy: Role of Oxidation States in Structure Prediction.
  \emph{Faraday Discuss.} \textbf{2018}, \emph{211}, 553--568\relax
\mciteBstWouldAddEndPuncttrue
\mciteSetBstMidEndSepPunct{\mcitedefaultmidpunct}
{\mcitedefaultendpunct}{\mcitedefaultseppunct}\relax
\EndOfBibitem
\bibitem[{Anatole von Lilienfeld} \latin{et~al.}(2020){Anatole von Lilienfeld},
  {M{\"u}ller}, and {Tkatchenko}]{Lilienfeld2020NRC}
{Anatole von Lilienfeld},~O.; {M{\"u}ller},~K.-R.; {Tkatchenko},~A. {Exploring
  Chemical compound Space with Quantum-Based Machine Learning}. \emph{Nat. Rev.
  Chem.} \textbf{2020}, \relax
\mciteBstWouldAddEndPunctfalse
\mciteSetBstMidEndSepPunct{\mcitedefaultmidpunct}
{}{\mcitedefaultseppunct}\relax
\EndOfBibitem
\bibitem[Freeze \latin{et~al.}(2019)Freeze, Kelly, and Batista]{Freeze2019CR}
Freeze,~J.~G.; Kelly,~H.~R.; Batista,~V.~S. Search for Catalysts by Inverse
  Design: Artificial Intelligence, Mountain Climbers, and Alchemists.
  \emph{Chemical Reviews} \textbf{2019}, \emph{119}, 6595--6612\relax
\mciteBstWouldAddEndPuncttrue
\mciteSetBstMidEndSepPunct{\mcitedefaultmidpunct}
{\mcitedefaultendpunct}{\mcitedefaultseppunct}\relax
\EndOfBibitem
\bibitem[Lee(2020)]{Lee2020OE}
Lee,~M.-H. Robust Random Forest Based Non-Fullerene Organic Solar Cells
  Efficiency Prediction. \emph{Org. Electron.} \textbf{2020}, \emph{76},
  105465\relax
\mciteBstWouldAddEndPuncttrue
\mciteSetBstMidEndSepPunct{\mcitedefaultmidpunct}
{\mcitedefaultendpunct}{\mcitedefaultseppunct}\relax
\EndOfBibitem
\bibitem[Cartwright(2020)]{Cartwright2020}
Cartwright,~H.~M., Ed. \emph{Machine Learning in Chemistry}; Theoretical and
  Computational Chemistry Series; The Royal Society of Chemistry, 2020\relax
\mciteBstWouldAddEndPuncttrue
\mciteSetBstMidEndSepPunct{\mcitedefaultmidpunct}
{\mcitedefaultendpunct}{\mcitedefaultseppunct}\relax
\EndOfBibitem
\bibitem[Gastegger and Marquetand(2020)Gastegger, and
  Marquetand]{Gastegger2020}
Gastegger,~M.; Marquetand,~P. In \emph{Machine Learning Meets Quantum Physics};
  Sch{\"u}tt,~K.~T., Chmiela,~S., von Lilienfeld,~O.~A., Tkatchenko,~A.,
  Tsuda,~K., M{\"u}ller,~K.-R., Eds.; Springer International Publishing: Cham,
  2020; pp 233--252\relax
\mciteBstWouldAddEndPuncttrue
\mciteSetBstMidEndSepPunct{\mcitedefaultmidpunct}
{\mcitedefaultendpunct}{\mcitedefaultseppunct}\relax
\EndOfBibitem
\bibitem[Sch\"{u}tt \latin{et~al.}(2020)Sch\"{u}tt, Chmiela, von Lilienfeld,
  Tkatchenko, Tsuda, and M\"{u}ller]{Schuett2020}
Sch\"{u}tt,~K.~T., Chmiela,~S., von Lilienfeld,~O.~A., Tkatchenko,~A.,
  Tsuda,~K., M\"{u}ller,~K.-R., Eds. \emph{Machine Learning Meets Quantum
  Physics}; Springer International Publishing, 2020\relax
\mciteBstWouldAddEndPuncttrue
\mciteSetBstMidEndSepPunct{\mcitedefaultmidpunct}
{\mcitedefaultendpunct}{\mcitedefaultseppunct}\relax
\EndOfBibitem
\bibitem[Park \latin{et~al.}(2020)Park, Al-Saadon, MacLeod, Shiozaki, and
  Vlaisavljevich]{Park2020CR}
Park,~J.~W.; Al-Saadon,~R.; MacLeod,~M.~K.; Shiozaki,~T.; Vlaisavljevich,~B.
  {Multireference Electron Correlation Methods: Journeys along Potential Energy
  Surfaces}. \emph{Chem. Rev.} \textbf{2020}, \emph{in press}, null\relax
\mciteBstWouldAddEndPuncttrue
\mciteSetBstMidEndSepPunct{\mcitedefaultmidpunct}
{\mcitedefaultendpunct}{\mcitedefaultseppunct}\relax
\EndOfBibitem
\bibitem[Akimov and Prezhdo(2015)Akimov, and Prezhdo]{Akimov2015CR}
Akimov,~A.~V.; Prezhdo,~O.~V. {Large-Scale Computations in Chemistry: A Bird's
  Eye View of a Vibrant Field}. \emph{Chem. Rev.} \textbf{2015}, \emph{115},
  5797--5890\relax
\mciteBstWouldAddEndPuncttrue
\mciteSetBstMidEndSepPunct{\mcitedefaultmidpunct}
{\mcitedefaultendpunct}{\mcitedefaultseppunct}\relax
\EndOfBibitem
\bibitem[Frutos \latin{et~al.}(2007)Frutos, Andruni{\'o}w, Santoro, Ferr{\'e},
  and Olivucci]{Frutos2007PNAS}
Frutos,~L.~M.; Andruni{\'o}w,~T.; Santoro,~F.; Ferr{\'e},~N.; Olivucci,~M.
  Tracking the Excited-State Time Evolution of the Visual Pigment with
  Multiconfigurational Quantum Chemistry. \emph{Proceedings of the National
  Academy of Sciences} \textbf{2007}, \emph{104}, 7764--7769\relax
\mciteBstWouldAddEndPuncttrue
\mciteSetBstMidEndSepPunct{\mcitedefaultmidpunct}
{\mcitedefaultendpunct}{\mcitedefaultseppunct}\relax
\EndOfBibitem
\bibitem[Menger \latin{et~al.}(2018)Menger, Plasser, Mennucci, and
  Gonz\'{a}lez]{Menger2018JCTC}
Menger,~M. F. S.~J.; Plasser,~F.; Mennucci,~B.; Gonz\'{a}lez,~L. Surface
  Hopping within an Exciton Picture. An Electrostatic Embedding Scheme.
  \emph{J. Chem. Theory Comput.} \textbf{2018}, \emph{14}, 6139--6148\relax
\mciteBstWouldAddEndPuncttrue
\mciteSetBstMidEndSepPunct{\mcitedefaultmidpunct}
{\mcitedefaultendpunct}{\mcitedefaultseppunct}\relax
\EndOfBibitem
\bibitem[Dou and Subotnik(2020)Dou, and Subotnik]{Dou2020JPCA}
Dou,~W.; Subotnik,~J.~E. Nonadiabatic Molecular Dynamics at Metal Surfaces.
  \emph{J. Phys. Chem. A} \textbf{2020}, \emph{124}, 757--771\relax
\mciteBstWouldAddEndPuncttrue
\mciteSetBstMidEndSepPunct{\mcitedefaultmidpunct}
{\mcitedefaultendpunct}{\mcitedefaultseppunct}\relax
\EndOfBibitem
\bibitem[Dou \latin{et~al.}(2015)Dou, Nitzan, and Subotnik]{Dou2015JCP}
Dou,~W.; Nitzan,~A.; Subotnik,~J.~E. Frictional Effects Near a Metal Surface.
  \emph{J. Chem. Phys.} \textbf{2015}, \emph{143}, 054103\relax
\mciteBstWouldAddEndPuncttrue
\mciteSetBstMidEndSepPunct{\mcitedefaultmidpunct}
{\mcitedefaultendpunct}{\mcitedefaultseppunct}\relax
\EndOfBibitem
\bibitem[Tavernelli(2006)]{Tavernelli2006PRB}
Tavernelli,~I. Electronic Density Response of Liquid Water using Time-Dependent
  Density Functional Theory. \emph{Phys. Rev. B} \textbf{2006}, \emph{73},
  094204\relax
\mciteBstWouldAddEndPuncttrue
\mciteSetBstMidEndSepPunct{\mcitedefaultmidpunct}
{\mcitedefaultendpunct}{\mcitedefaultseppunct}\relax
\EndOfBibitem
\bibitem[Sch\"utt \latin{et~al.}(2014)Sch\"utt, Glawe, Brockherde, Sanna,
  M\"uller, and Gross]{Schuett2014PRB}
Sch\"utt,~K.~T.; Glawe,~H.; Brockherde,~F.; Sanna,~A.; M\"uller,~K.~R.;
  Gross,~E. K.~U. How to Represent Crystal Structures for Machine Learning:
  Towards Fast Prediction of Electronic Properties. \emph{Phys. Rev. B}
  \textbf{2014}, \emph{89}, 205118\relax
\mciteBstWouldAddEndPuncttrue
\mciteSetBstMidEndSepPunct{\mcitedefaultmidpunct}
{\mcitedefaultendpunct}{\mcitedefaultseppunct}\relax
\EndOfBibitem
\bibitem[Lee \latin{et~al.}(2016)Lee, Seko, Shitara, Nakayama, and
  Tanaka]{Lee2016PRB}
Lee,~J.; Seko,~A.; Shitara,~K.; Nakayama,~K.; Tanaka,~I. Prediction Model of
  Band Gap for Inorganic Compounds by Combination of Density Functional Theory
  Calculations and Machine Learning Techniques. \emph{Phys. Rev. B}
  \textbf{2016}, \emph{93}, 115104\relax
\mciteBstWouldAddEndPuncttrue
\mciteSetBstMidEndSepPunct{\mcitedefaultmidpunct}
{\mcitedefaultendpunct}{\mcitedefaultseppunct}\relax
\EndOfBibitem
\bibitem[Zhuo \latin{et~al.}(2018)Zhuo, Mansouri~Tehrani, and
  Brgoch]{Zhuo2018JPCL}
Zhuo,~Y.; Mansouri~Tehrani,~A.; Brgoch,~J. Predicting the Band Gaps of
  Inorganic Solids by Machine Learning. \emph{J. Phys. Chem. Lett.}
  \textbf{2018}, \emph{9}, 1668--1673\relax
\mciteBstWouldAddEndPuncttrue
\mciteSetBstMidEndSepPunct{\mcitedefaultmidpunct}
{\mcitedefaultendpunct}{\mcitedefaultseppunct}\relax
\EndOfBibitem
\bibitem[Pilania \latin{et~al.}(2017)Pilania, Gubernatis, and
  Lookman]{Pilania2017CMS}
Pilania,~G.; Gubernatis,~J.; Lookman,~T. Multi-Fidelity Machine Learning Models
  for Accurate Bandgap Predictions of Solids. \emph{Comput. Mat. Sci.}
  \textbf{2017}, \emph{129}, 156 -- 163\relax
\mciteBstWouldAddEndPuncttrue
\mciteSetBstMidEndSepPunct{\mcitedefaultmidpunct}
{\mcitedefaultendpunct}{\mcitedefaultseppunct}\relax
\EndOfBibitem
\bibitem[Spiering \latin{et~al.}(2019)Spiering, Shakouri, Behler, Kroes, and
  Meyer]{Spiering2019JPCL}
Spiering,~P.; Shakouri,~K.; Behler,~J.; Kroes,~G.-J.; Meyer,~J.
  Orbital-Dependent Electronic Friction Significantly Affects the Description
  of Reactive Scattering of {N}$_2$ from {Ru}(0001). \emph{J. Phys. Chem.
  Lett.} \textbf{2019}, \emph{10}, 2957--2962\relax
\mciteBstWouldAddEndPuncttrue
\mciteSetBstMidEndSepPunct{\mcitedefaultmidpunct}
{\mcitedefaultendpunct}{\mcitedefaultseppunct}\relax
\EndOfBibitem
\bibitem[Zhang \latin{et~al.}(2020)Zhang, Maurer, and Jiang]{Zhang2020JPCC}
Zhang,~Y.; Maurer,~R.~J.; Jiang,~B. Symmetry-Adapted High Dimensional Neural
  Network Representation of Electronic Friction Tensor of Adsorbates on Metals.
  \emph{J. Phys. Chem. C} \textbf{2020}, \emph{124}, 186--195\relax
\mciteBstWouldAddEndPuncttrue
\mciteSetBstMidEndSepPunct{\mcitedefaultmidpunct}
{\mcitedefaultendpunct}{\mcitedefaultseppunct}\relax
\EndOfBibitem
\bibitem[Zhang \latin{et~al.}(2019)Zhang, Maurer, Guo, and Jiang]{Zhang2019CS}
Zhang,~Y.; Maurer,~R.~J.; Guo,~H.; Jiang,~B. Hot-Electron Effects during
  Reactive Scattering of {H}$_2$ from {A}g(111): The Interplay between
  Mode-Specific Electronic Friction and the Potential Energy Landscape.
  \emph{Chem. Sci.} \textbf{2019}, \emph{10}, 1089--1097\relax
\mciteBstWouldAddEndPuncttrue
\mciteSetBstMidEndSepPunct{\mcitedefaultmidpunct}
{\mcitedefaultendpunct}{\mcitedefaultseppunct}\relax
\EndOfBibitem
\bibitem[Head-Gordon and Tully(1995)Head-Gordon, and Tully]{Head-Gordon1995JCP}
Head-Gordon,~M.; Tully,~J.~C. Molecular Dynamics with Electronic Frictions.
  \emph{J. Chem. Phys.} \textbf{1995}, \emph{103}, 10137--10145\relax
\mciteBstWouldAddEndPuncttrue
\mciteSetBstMidEndSepPunct{\mcitedefaultmidpunct}
{\mcitedefaultendpunct}{\mcitedefaultseppunct}\relax
\EndOfBibitem
\bibitem[Douglas-Gallardo \latin{et~al.}(2019)Douglas-Gallardo, Berdakin,
  Frauenheim, and S{{\'a}}nchez]{Douglas-Gallardo2019N}
Douglas-Gallardo,~O.~A.; Berdakin,~M.; Frauenheim,~T.; S{{\'a}}nchez,~C.~G.
  Plasmon-Induced Hot-Carrier Generation Differences in Gold and Silver
  Nanoclusters. \emph{Nanoscale} \textbf{2019}, \emph{11}, 8604--8615\relax
\mciteBstWouldAddEndPuncttrue
\mciteSetBstMidEndSepPunct{\mcitedefaultmidpunct}
{\mcitedefaultendpunct}{\mcitedefaultseppunct}\relax
\EndOfBibitem
\bibitem[Yin \latin{et~al.}(2019)Yin, Zhang, and Jiang]{Yin2019JPCL}
Yin,~R.; Zhang,~Y.; Jiang,~B. {Strong Vibrational Relaxation of NO Scattered
  from Au(111): Importance of the Adiabatic Potential Energy Surface}. \emph{J.
  Phys. Chem. Lett.} \textbf{2019}, \emph{10}, 5969--5974\relax
\mciteBstWouldAddEndPuncttrue
\mciteSetBstMidEndSepPunct{\mcitedefaultmidpunct}
{\mcitedefaultendpunct}{\mcitedefaultseppunct}\relax
\EndOfBibitem
\bibitem[Rittmeyer \latin{et~al.}(2018)Rittmeyer, Bukas, and
  Reuter]{Rittmeyer2018APX}
Rittmeyer,~S.~P.; Bukas,~V.~J.; Reuter,~K. Energy dissipation at metal
  surfaces. \emph{Adv. Phys-X} \textbf{2018}, \emph{3}, 1381574\relax
\mciteBstWouldAddEndPuncttrue
\mciteSetBstMidEndSepPunct{\mcitedefaultmidpunct}
{\mcitedefaultendpunct}{\mcitedefaultseppunct}\relax
\EndOfBibitem
\bibitem[Therrien \latin{et~al.}(2019)Therrien, Kale, Yuan, Zhang, Halas, and
  Christopher]{Therrien2019FD}
Therrien,~A.~J.; Kale,~M.~J.; Yuan,~L.; Zhang,~C.; Halas,~N.~J.;
  Christopher,~P. Impact of Chemical Interface Damping on Surface Plasmon
  Dephasing. \emph{Faraday Discuss.} \textbf{2019}, \emph{214}, 59--72\relax
\mciteBstWouldAddEndPuncttrue
\mciteSetBstMidEndSepPunct{\mcitedefaultmidpunct}
{\mcitedefaultendpunct}{\mcitedefaultseppunct}\relax
\EndOfBibitem
\bibitem[Wodtke \latin{et~al.}(2004)Wodtke, Tully, and
  Auerbach]{wodtke2004IRPC}
Wodtke,~A.~M.; Tully,~J.~C.; Auerbach,~D.~J. Electronically Non-Adiabatic
  Interactions of Molecules at Metal Surfaces: Can we Trust the
  Born–Oppenheimer Approximation for Surface Chemistry? \emph{Int. Rev. Phys.
  Chem.} \textbf{2004}, \emph{23}, 513--539\relax
\mciteBstWouldAddEndPuncttrue
\mciteSetBstMidEndSepPunct{\mcitedefaultmidpunct}
{\mcitedefaultendpunct}{\mcitedefaultseppunct}\relax
\EndOfBibitem
\bibitem[Park \latin{et~al.}(2019)Park, Krüger, Borodin, Kitsopoulos, and
  Wodtke]{Park2019RPP}
Park,~G.~B.; Krüger,~B.~C.; Borodin,~D.; Kitsopoulos,~T.~N.; Wodtke,~A.~M.
  Fundamental Mechanisms for Molecular Energy Conversion and Chemical Reactions
  at Surfaces. \emph{Rep. Prog. Phys.} \textbf{2019}, \emph{82}, 096401\relax
\mciteBstWouldAddEndPuncttrue
\mciteSetBstMidEndSepPunct{\mcitedefaultmidpunct}
{\mcitedefaultendpunct}{\mcitedefaultseppunct}\relax
\EndOfBibitem
\bibitem[Jiang and Guo(2019)Jiang, and Guo]{Jiang2019JCP}
Jiang,~B.; Guo,~H. Dynamics in Reactions on Metal Surfaces: A Theoretical
  Perspective. \emph{J. Chem. Phys.} \textbf{2019}, \emph{150}, 180901\relax
\mciteBstWouldAddEndPuncttrue
\mciteSetBstMidEndSepPunct{\mcitedefaultmidpunct}
{\mcitedefaultendpunct}{\mcitedefaultseppunct}\relax
\EndOfBibitem
\bibitem[Shenvi \latin{et~al.}(2009)Shenvi, Roy, and Tully]{Shenvi2009JCP}
Shenvi,~N.; Roy,~S.; Tully,~J.~C. Nonadiabatic Dynamics at Metal Surfaces:
  Independent-Electron Surface Hopping. \emph{J. Chem. Phys.} \textbf{2009},
  \emph{130}, 174107\relax
\mciteBstWouldAddEndPuncttrue
\mciteSetBstMidEndSepPunct{\mcitedefaultmidpunct}
{\mcitedefaultendpunct}{\mcitedefaultseppunct}\relax
\EndOfBibitem
\bibitem[Shenvi \latin{et~al.}(2009)Shenvi, Roy, and Tully]{Shenvi2009S}
Shenvi,~N.; Roy,~S.; Tully,~J.~C. {Dynamical Steering and Electronic Excitation
  in NO Scattering from a Gold Surface}. \emph{Science} \textbf{2009},
  \emph{326}, 829--832\relax
\mciteBstWouldAddEndPuncttrue
\mciteSetBstMidEndSepPunct{\mcitedefaultmidpunct}
{\mcitedefaultendpunct}{\mcitedefaultseppunct}\relax
\EndOfBibitem
\bibitem[Dou \latin{et~al.}(2018)Dou, Schinabeck, Thoss, and
  Subotnik]{Dou2018JCP}
Dou,~W.; Schinabeck,~C.; Thoss,~M.; Subotnik,~J.~E. A broadened classical
  master equation approach for treating electron-nuclear coupling in
  non-equilibrium transport. \emph{J. Chem. Phys.} \textbf{2018}, \emph{148},
  102317\relax
\mciteBstWouldAddEndPuncttrue
\mciteSetBstMidEndSepPunct{\mcitedefaultmidpunct}
{\mcitedefaultendpunct}{\mcitedefaultseppunct}\relax
\EndOfBibitem
\bibitem[Jiang \latin{et~al.}(2020)Jiang, Li, and Guo]{Jiang2020JPCL}
Jiang,~B.; Li,~J.; Guo,~H. {High-Fidelity Potential Energy Surfaces for Gas
  Phase and Gas-Surface Scattering Processes from Machine Learning}. \emph{J.
  Phys. Chem. Lett.} \textbf{2020}, \emph{11}, 5120–5131\relax
\mciteBstWouldAddEndPuncttrue
\mciteSetBstMidEndSepPunct{\mcitedefaultmidpunct}
{\mcitedefaultendpunct}{\mcitedefaultseppunct}\relax
\EndOfBibitem
\bibitem[Buhrke and Hildebrandt(2020)Buhrke, and Hildebrandt]{Buhrke2019CR}
Buhrke,~D.; Hildebrandt,~P. Probing Structure and Reaction Dynamics of Proteins
  Using Time-Resolved Resonance Raman Spectroscopy. \emph{Chem. Rev.}
  \textbf{2020}, \emph{120}, 3577--3630\relax
\mciteBstWouldAddEndPuncttrue
\mciteSetBstMidEndSepPunct{\mcitedefaultmidpunct}
{\mcitedefaultendpunct}{\mcitedefaultseppunct}\relax
\EndOfBibitem
\bibitem[Raimbault \latin{et~al.}(2019)Raimbault, Grisafi, Ceriotti, and
  Rossi]{Raimbault2019NJP}
Raimbault,~N.; Grisafi,~A.; Ceriotti,~M.; Rossi,~M. Using Gaussian Process
  Regression to Simulate the Vibrational Raman Spectra of Molecular Crystals.
  \emph{New J. Phys.} \textbf{2019}, \emph{21}, 105001\relax
\mciteBstWouldAddEndPuncttrue
\mciteSetBstMidEndSepPunct{\mcitedefaultmidpunct}
{\mcitedefaultendpunct}{\mcitedefaultseppunct}\relax
\EndOfBibitem
\bibitem[Hu \latin{et~al.}(2019)Hu, Ye, Zhang, Li, Zhang, Luo, Mukamel, and
  Jiang]{Hu2019JPCL}
Hu,~W.; Ye,~S.; Zhang,~Y.; Li,~T.; Zhang,~G.; Luo,~Y.; Mukamel,~S.; Jiang,~J.
  Machine Learning Protocol for Surface-Enhanced Raman Spectroscopy. \emph{J.
  Phys. Chem. Lett.} \textbf{2019}, \emph{10}, 6026--6031\relax
\mciteBstWouldAddEndPuncttrue
\mciteSetBstMidEndSepPunct{\mcitedefaultmidpunct}
{\mcitedefaultendpunct}{\mcitedefaultseppunct}\relax
\EndOfBibitem
\bibitem[Lussier \latin{et~al.}(2020)Lussier, Thibault, Charron, Wallace, and
  Masson]{Lussier2020TRAC}
Lussier,~F.; Thibault,~V.; Charron,~B.; Wallace,~G.~Q.; Masson,~J.-F. Deep
  Learning and Artificial Intelligence Methods for Raman and Surface-Enhanced
  Raman Scattering. \emph{TrAC, Trends Anal. Chem.} \textbf{2020}, \emph{124},
  115796\relax
\mciteBstWouldAddEndPuncttrue
\mciteSetBstMidEndSepPunct{\mcitedefaultmidpunct}
{\mcitedefaultendpunct}{\mcitedefaultseppunct}\relax
\EndOfBibitem
\bibitem[Fu and Hopkins(2018)Fu, and Hopkins]{Fu2018JPCA}
Fu,~W.; Hopkins,~W.~S. Applying Machine Learning to Vibrational Spectroscopy.
  \emph{J. Phys. Chem. A} \textbf{2018}, \emph{122}, 167--171\relax
\mciteBstWouldAddEndPuncttrue
\mciteSetBstMidEndSepPunct{\mcitedefaultmidpunct}
{\mcitedefaultendpunct}{\mcitedefaultseppunct}\relax
\EndOfBibitem
\bibitem[Aires-de Sousa \latin{et~al.}(2002)Aires-de Sousa, Hemmer, and
  Gasteiger]{Aires-de-Sousa2002AC}
Aires-de Sousa,~J.; Hemmer,~M.~C.; Gasteiger,~J. {Prediction of $^1$H NMR
  Chemical Shifts Using Neural Networks}. \emph{Anal. Chem.} \textbf{2002},
  \emph{74}, 80--90\relax
\mciteBstWouldAddEndPuncttrue
\mciteSetBstMidEndSepPunct{\mcitedefaultmidpunct}
{\mcitedefaultendpunct}{\mcitedefaultseppunct}\relax
\EndOfBibitem
\bibitem[Taguchi \latin{et~al.}(2019)Taguchi, Evans, Dikanov, and
  Griffin]{Taguchi2019JPCL}
Taguchi,~A.~T.; Evans,~E.~D.; Dikanov,~S.~A.; Griffin,~R.~G. Convolutional
  Neural Network Analysis of Two-Dimensional Hyperfine Sublevel Correlation
  Electron Paramagnetic Resonance Spectra. \emph{J. Phys. Chem. Lett.}
  \textbf{2019}, \emph{10}, 1115--1119\relax
\mciteBstWouldAddEndPuncttrue
\mciteSetBstMidEndSepPunct{\mcitedefaultmidpunct}
{\mcitedefaultendpunct}{\mcitedefaultseppunct}\relax
\EndOfBibitem
\bibitem[Cobas(2020)]{Cobas2020MRC}
Cobas,~C. NMR Signal Processing, Prediction, and Structure Verification with
  Machine Learning Techniques. \emph{Magn. Reson. Chem.} \textbf{2020},
  \emph{58}, 512--519\relax
\mciteBstWouldAddEndPuncttrue
\mciteSetBstMidEndSepPunct{\mcitedefaultmidpunct}
{\mcitedefaultendpunct}{\mcitedefaultseppunct}\relax
\EndOfBibitem
\bibitem[{Salomon-Ferrer} \latin{et~al.}(2013){Salomon-Ferrer}, Case, and
  Walker]{Salomon-Ferrer2013WCMS}
{Salomon-Ferrer},~R.; Case,~D.~A.; Walker,~R.~C. An overview of the Amber
  biomolecular simulation package. \emph{WIREs Computational Molecular Science}
  \textbf{2013}, \emph{3}, 198--210\relax
\mciteBstWouldAddEndPuncttrue
\mciteSetBstMidEndSepPunct{\mcitedefaultmidpunct}
{\mcitedefaultendpunct}{\mcitedefaultseppunct}\relax
\EndOfBibitem
\bibitem[{B. r. Brooks, et al.}(2009)]{Brooks2009JCC}
{B. r. Brooks, et al.}, {CHARMM: The Biomolecular Simulation Program}. \emph{J.
  Comput. Chem.} \textbf{2009}, \emph{30}, 1545--1614\relax
\mciteBstWouldAddEndPuncttrue
\mciteSetBstMidEndSepPunct{\mcitedefaultmidpunct}
{\mcitedefaultendpunct}{\mcitedefaultseppunct}\relax
\EndOfBibitem
\bibitem[Eichenberger \latin{et~al.}(2011)Eichenberger, Allison, Dolenc,
  Geerke, Horta, Meier, Oostenbrink, Schmid, Steiner, Wang, and van
  Gunsteren]{Eichenberger2011JCTC}
Eichenberger,~A.~P.; Allison,~J.~R.; Dolenc,~J.; Geerke,~D.~P.; Horta,~B.
  A.~C.; Meier,~K.; Oostenbrink,~C.; Schmid,~N.; Steiner,~D.; Wang,~D.; van
  Gunsteren,~W.~F. {GROMOS++ Software for the Analysis of Biomolecular
  Simulation Trajectories}. \emph{J. Chem. Theory Comput.} \textbf{2011},
  \emph{7}, 3379--3390\relax
\mciteBstWouldAddEndPuncttrue
\mciteSetBstMidEndSepPunct{\mcitedefaultmidpunct}
{\mcitedefaultendpunct}{\mcitedefaultseppunct}\relax
\EndOfBibitem
\bibitem[Reif \latin{et~al.}(2012)Reif, Hünenberger, and
  Oostenbrink]{Reif2012JCTC}
Reif,~M.~M.; Hünenberger,~P.~H.; Oostenbrink,~C. {New Interaction Parameters
  for Charged Amino Acid Side Chains in the GROMOS Force Field}. \emph{J. Chem.
  Theory Comput.} \textbf{2012}, \emph{8}, 3705--3723\relax
\mciteBstWouldAddEndPuncttrue
\mciteSetBstMidEndSepPunct{\mcitedefaultmidpunct}
{\mcitedefaultendpunct}{\mcitedefaultseppunct}\relax
\EndOfBibitem
\bibitem[Perthold \latin{et~al.}(2020)Perthold, Petrov, and
  Oostenbrink]{Perthold2020JCIM}
Perthold,~J.~W.; Petrov,~D.; Oostenbrink,~C. Towards Automated Free Energy
  Calculation with Accelerated Enveloping Distribution Sampling (A-EDS).
  \emph{J. Chem. Inf. Model.} \textbf{2020}, \emph{in press},
  doi:10.1021/acs.jcim.0c00456\relax
\mciteBstWouldAddEndPuncttrue
\mciteSetBstMidEndSepPunct{\mcitedefaultmidpunct}
{\mcitedefaultendpunct}{\mcitedefaultseppunct}\relax
\EndOfBibitem
\bibitem[\"{O}hlknecht \latin{et~al.}(2020)\"{O}hlknecht, Lier, Petrov, Fuchs,
  and Oostenbrink]{Oelknecht2020JCC}
\"{O}hlknecht,~C.; Lier,~B.; Petrov,~D.; Fuchs,~J.; Oostenbrink,~C. Correcting
  Electrostatic Artifacts due to Net-Charge Changes in the Calculation of
  Ligand Binding Free Energies. \emph{J. Comput. Chem.} \textbf{2020},
  \emph{41}, 986--999\relax
\mciteBstWouldAddEndPuncttrue
\mciteSetBstMidEndSepPunct{\mcitedefaultmidpunct}
{\mcitedefaultendpunct}{\mcitedefaultseppunct}\relax
\EndOfBibitem
\bibitem[Michlits \latin{et~al.}(2020)Michlits, Lier, Pfanzagl,
  Djinovi\'{c}-Carugo, Furtm\"{u}ller, Oostenbrink, Obinger, and
  Hofbauer]{Michlits2020ACSC}
Michlits,~H.; Lier,~B.; Pfanzagl,~V.; Djinovi\'{c}-Carugo,~K.;
  Furtm\"{u}ller,~P.~G.; Oostenbrink,~C.; Obinger,~C.; Hofbauer,~S.
  Actinobacterial Coproheme Decarboxylases Use Histidine as a Distal Base to
  Promote Compound I Formation. \emph{ACS Catal.} \textbf{2020}, \emph{10},
  5405--5418\relax
\mciteBstWouldAddEndPuncttrue
\mciteSetBstMidEndSepPunct{\mcitedefaultmidpunct}
{\mcitedefaultendpunct}{\mcitedefaultseppunct}\relax
\EndOfBibitem
\bibitem[Brunk and Rothlisberger(2015)Brunk, and Rothlisberger]{Brunk2015CR}
Brunk,~E.; Rothlisberger,~U. {Mixed Quantum Mechanical/Molecular Mechanical
  Molecular Dynamics Simulations of Biological Systems in Ground and
  Electronically Excited States}. \emph{Chem. Rev.} \textbf{2015}, \emph{115},
  6217--6263\relax
\mciteBstWouldAddEndPuncttrue
\mciteSetBstMidEndSepPunct{\mcitedefaultmidpunct}
{\mcitedefaultendpunct}{\mcitedefaultseppunct}\relax
\EndOfBibitem
\bibitem[Bedrov \latin{et~al.}(2019)Bedrov, Piquemal, Borodin, MacKerell, Roux,
  and Schr\"{o}der]{Bedrov2019CR}
Bedrov,~D.; Piquemal,~J.-P.; Borodin,~O.; MacKerell,~A.~D.; Roux,~B.;
  Schr\"{o}der,~C. Molecular Dynamics Simulations of Ionic Liquids and
  Electrolytes Using Polarizable Force Fields. \emph{Chem. Rev.} \textbf{2019},
  \emph{119}, 7940--7995\relax
\mciteBstWouldAddEndPuncttrue
\mciteSetBstMidEndSepPunct{\mcitedefaultmidpunct}
{\mcitedefaultendpunct}{\mcitedefaultseppunct}\relax
\EndOfBibitem
\bibitem[Sosso \latin{et~al.}(2016)Sosso, Chen, Cox, Fitzner, Pedevilla, Zen,
  and Michaelides]{Sosso2016CR}
Sosso,~G.~C.; Chen,~J.; Cox,~S.~J.; Fitzner,~M.; Pedevilla,~P.; Zen,~A.;
  Michaelides,~A. Crystal Nucleation in Liquids: Open Questions and Future
  Challenges in Molecular Dynamics Simulations. \emph{Chem. Rev.}
  \textbf{2016}, \emph{116}, 7078--7116\relax
\mciteBstWouldAddEndPuncttrue
\mciteSetBstMidEndSepPunct{\mcitedefaultmidpunct}
{\mcitedefaultendpunct}{\mcitedefaultseppunct}\relax
\EndOfBibitem
\bibitem[Venable \latin{et~al.}(2019)Venable, Kr\"{a}mer, and
  Pastor]{Venable2019CR}
Venable,~R.~M.; Kr\"{a}mer,~A.; Pastor,~R.~W. Molecular Dynamics Simulations of
  Membrane Permeability. \emph{Chem. Rev.} \textbf{2019}, \emph{119},
  5954--5997\relax
\mciteBstWouldAddEndPuncttrue
\mciteSetBstMidEndSepPunct{\mcitedefaultmidpunct}
{\mcitedefaultendpunct}{\mcitedefaultseppunct}\relax
\EndOfBibitem
\bibitem[Marrink \latin{et~al.}(2019)Marrink, Corradi, Souza, Ingólfsson,
  Tieleman, and Sansom]{Marrink2019CR}
Marrink,~S.~J.; Corradi,~V.; Souza,~P.~C.; Ingólfsson,~H.~I.; Tieleman,~D.~P.;
  Sansom,~M.~S. Computational Modeling of Realistic Cell Membranes.
  \emph{Chemical Reviews} \textbf{2019}, \emph{119}, 6184--6226\relax
\mciteBstWouldAddEndPuncttrue
\mciteSetBstMidEndSepPunct{\mcitedefaultmidpunct}
{\mcitedefaultendpunct}{\mcitedefaultseppunct}\relax
\EndOfBibitem
\bibitem[G.(2013)]{Groenhof2013}
G.,~G. In \emph{Biomolecular Simulations. Methods in Molecular Biology (Methods
  and Protocols)}; Monticelli,~L., Salonen,~E., Eds.; Humana Press, Totowa, NJ,
  2013; Vol. 924\relax
\mciteBstWouldAddEndPuncttrue
\mciteSetBstMidEndSepPunct{\mcitedefaultmidpunct}
{\mcitedefaultendpunct}{\mcitedefaultseppunct}\relax
\EndOfBibitem
\bibitem[{Thomas P. Senftle, et al.}(2016)]{Senftle2016npjCM}
{Thomas P. Senftle, et al.}, The ReaxFF Reactive Force-Field: Development,
  Applications and Future Directions. \emph{npj Comput. Mater.} \textbf{2016},
  \emph{2}\relax
\mciteBstWouldAddEndPuncttrue
\mciteSetBstMidEndSepPunct{\mcitedefaultmidpunct}
{\mcitedefaultendpunct}{\mcitedefaultseppunct}\relax
\EndOfBibitem
\bibitem[Sauceda \latin{et~al.}(2020)Sauceda, Chmiela, Poltavsky, M{\"u}ller,
  and Tkatchenko]{Sauceda2020}
Sauceda,~H.~E.; Chmiela,~S.; Poltavsky,~I.; M{\"u}ller,~K.-R.; Tkatchenko,~A.
  In \emph{Machine Learning Meets Quantum Physics}; Sch{\"u}tt,~K.~T.,
  Chmiela,~S., von Lilienfeld,~O.~A., Tkatchenko,~A., Tsuda,~K.,
  M{\"u}ller,~K.-R., Eds.; Springer International Publishing: Cham, 2020; pp
  277--307\relax
\mciteBstWouldAddEndPuncttrue
\mciteSetBstMidEndSepPunct{\mcitedefaultmidpunct}
{\mcitedefaultendpunct}{\mcitedefaultseppunct}\relax
\EndOfBibitem
\bibitem[No{\'e}(2020)]{Noe2020}
No{\'e},~F. In \emph{Machine Learning Meets Quantum Physics};
  Sch{\"u}tt,~K.~T., Chmiela,~S., von Lilienfeld,~O.~A., Tkatchenko,~A.,
  Tsuda,~K., M{\"u}ller,~K.-R., Eds.; Springer International Publishing: Cham,
  2020; pp 331--372\relax
\mciteBstWouldAddEndPuncttrue
\mciteSetBstMidEndSepPunct{\mcitedefaultmidpunct}
{\mcitedefaultendpunct}{\mcitedefaultseppunct}\relax
\EndOfBibitem
\bibitem[Glielmo \latin{et~al.}(2020)Glielmo, Zeni, Fekete, and
  De~Vita]{Glielmo2020}
Glielmo,~A.; Zeni,~C.; Fekete,~{\'A}.; De~Vita,~A. In \emph{Machine Learning
  Meets Quantum Physics}; Sch{\"u}tt,~K.~T., Chmiela,~S., von
  Lilienfeld,~O.~A., Tkatchenko,~A., Tsuda,~K., M{\"u}ller,~K.-R., Eds.;
  Springer International Publishing: Cham, 2020; pp 67--98\relax
\mciteBstWouldAddEndPuncttrue
\mciteSetBstMidEndSepPunct{\mcitedefaultmidpunct}
{\mcitedefaultendpunct}{\mcitedefaultseppunct}\relax
\EndOfBibitem
\bibitem[Abbott \latin{et~al.}(2019)Abbott, Turney, Zhang, Smith, Altarawy, and
  Schaefer]{Abbott2019JCTC}
Abbott,~A.~S.; Turney,~J.~M.; Zhang,~B.; Smith,~D. G.~A.; Altarawy,~D.;
  Schaefer,~H.~F. PES-Learn: An Open-Source Software Package for the Automated
  Generation of Machine Learning Models of Molecular Potential Energy Surfaces.
  \emph{J. Chem. Theory Comput.} \textbf{2019}, \emph{15}, 4386--4398\relax
\mciteBstWouldAddEndPuncttrue
\mciteSetBstMidEndSepPunct{\mcitedefaultmidpunct}
{\mcitedefaultendpunct}{\mcitedefaultseppunct}\relax
\EndOfBibitem
\bibitem[Hellstr{\"o}m and Behler(2020)Hellstr{\"o}m, and
  Behler]{Hellstroem2020}
Hellstr{\"o}m,~M.; Behler,~J. In \emph{Machine Learning Meets Quantum Physics};
  Sch{\"u}tt,~K.~T., Chmiela,~S., von Lilienfeld,~O.~A., Tkatchenko,~A.,
  Tsuda,~K., M{\"u}ller,~K.-R., Eds.; Springer International Publishing: Cham,
  2020; pp 253--275\relax
\mciteBstWouldAddEndPuncttrue
\mciteSetBstMidEndSepPunct{\mcitedefaultmidpunct}
{\mcitedefaultendpunct}{\mcitedefaultseppunct}\relax
\EndOfBibitem
\bibitem[Vargas-Hern{\'a}ndez and Krems(2020)Vargas-Hern{\'a}ndez, and
  Krems]{Vargas-Hernandez2020}
Vargas-Hern{\'a}ndez,~R.~A.; Krems,~R.~V. In \emph{Machine Learning Meets
  Quantum Physics}; Sch{\"u}tt,~K.~T., Chmiela,~S., von Lilienfeld,~O.~A.,
  Tkatchenko,~A., Tsuda,~K., M{\"u}ller,~K.-R., Eds.; Springer International
  Publishing: Cham, 2020; pp 171--194\relax
\mciteBstWouldAddEndPuncttrue
\mciteSetBstMidEndSepPunct{\mcitedefaultmidpunct}
{\mcitedefaultendpunct}{\mcitedefaultseppunct}\relax
\EndOfBibitem
\bibitem[K{\"o}ppel \latin{et~al.}(2004)K{\"o}ppel, Domcke, and
  Cederbaum]{Koeppel04}
K{\"o}ppel,~H.; Domcke,~W.; Cederbaum,~L.~S. \emph{in: Conical Intersections
  (W. Domcke, D. R. Yarkony, H. K{\"o}ppel, Eds.)}; World Scientific: New York,
  2004\relax
\mciteBstWouldAddEndPuncttrue
\mciteSetBstMidEndSepPunct{\mcitedefaultmidpunct}
{\mcitedefaultendpunct}{\mcitedefaultseppunct}\relax
\EndOfBibitem
\bibitem[Plasser \latin{et~al.}(2019)Plasser, Gómez, Menger, Mai, and
  Gonz\'{a}lez]{Plasser2019PCCP}
Plasser,~F.; Gómez,~S.; Menger,~M. F. S.~J.; Mai,~S.; Gonz\'{a}lez,~L. Highly
  Efficient Surface Hopping Dynamics using a Linear Vibronic Coupling Model.
  \emph{Phys. Chem. Chem. Phys.} \textbf{2019}, \emph{21}, 57--69\relax
\mciteBstWouldAddEndPuncttrue
\mciteSetBstMidEndSepPunct{\mcitedefaultmidpunct}
{\mcitedefaultendpunct}{\mcitedefaultseppunct}\relax
\EndOfBibitem
\bibitem[He \latin{et~al.}(2008)He, Tan, Zheng, and Prasad]{He2008CR}
He,~G.~S.; Tan,~L.-S.; Zheng,~Q.; Prasad,~P.~N. {Multiphoton Absorbing
  Materials: Molecular Designs, Characterizations, and Applications}.
  \emph{Chem. Rev.} \textbf{2008}, \emph{108}, 1245--1330\relax
\mciteBstWouldAddEndPuncttrue
\mciteSetBstMidEndSepPunct{\mcitedefaultmidpunct}
{\mcitedefaultendpunct}{\mcitedefaultseppunct}\relax
\EndOfBibitem
\bibitem[Marquetand \latin{et~al.}(2014)Marquetand, Weinacht, Rozgonyi,
  Gonz\'{a}lez-Vazquez, Geißler, and Gonz\'{a}lez]{Marquetand2014}
Marquetand,~P.; Weinacht,~T.; Rozgonyi,~T.; Gonz\'{a}lez-Vazquez,~J.;
  Geißler,~D.; Gonz\'{a}lez,~L. In \emph{Advances in Multiphoton Processes and
  Spectroscopy}; Fujimura,~Y., Ed.; World Scientific, Singapore, 2014; Vol.~21;
  pp 1--54\relax
\mciteBstWouldAddEndPuncttrue
\mciteSetBstMidEndSepPunct{\mcitedefaultmidpunct}
{\mcitedefaultendpunct}{\mcitedefaultseppunct}\relax
\EndOfBibitem
\bibitem[Tagliamonti \latin{et~al.}(2016)Tagliamonti, S\'andor, Zhao, Rozgonyi,
  Marquetand, and Weinacht]{Tagliamonti2016PRA}
Tagliamonti,~V.; S\'andor,~P.; Zhao,~A.; Rozgonyi,~T.; Marquetand,~P.;
  Weinacht,~T. Nonadiabatic Dynamics and Multiphoton Resonances in Strong-Field
  Molecular Ionization with Few-Cycle Laser Pulses. \emph{Phys. Rev. A}
  \textbf{2016}, \emph{93}, 051401\relax
\mciteBstWouldAddEndPuncttrue
\mciteSetBstMidEndSepPunct{\mcitedefaultmidpunct}
{\mcitedefaultendpunct}{\mcitedefaultseppunct}\relax
\EndOfBibitem
\bibitem[M.~Wollenhaupt and Baumert(2007)M.~Wollenhaupt, and
  Baumert]{Wollenhaupt2007}
M.~Wollenhaupt,~A.~A.; Baumert,~T. In \emph{Springer Handbook of Lasers and
  Optics}; Tr\"{a}ger,~F., Ed.; Springer Science and Business Media, LLC New
  York, 2007; Chapter 12, pp 937--983\relax
\mciteBstWouldAddEndPuncttrue
\mciteSetBstMidEndSepPunct{\mcitedefaultmidpunct}
{\mcitedefaultendpunct}{\mcitedefaultseppunct}\relax
\EndOfBibitem
\bibitem[Hilborn(1982)]{Hilborn1982AJP}
Hilborn,~R.~C. Einstein Coefficients, Cross Sections, f Values, Dipole Moments,
  and All That. \emph{Am. J. Phys.} \textbf{1982}, \emph{50}, 982--986\relax
\mciteBstWouldAddEndPuncttrue
\mciteSetBstMidEndSepPunct{\mcitedefaultmidpunct}
{\mcitedefaultendpunct}{\mcitedefaultseppunct}\relax
\EndOfBibitem
\bibitem[Andrews(2014)]{Andrews2014}
Andrews,~D.~L. \emph{Molecular Photophysics and Spectroscopy}; 2053-2571;
  Morgan \& Claypool Publishers, 2014; pp 9--1 to 9--4\relax
\mciteBstWouldAddEndPuncttrue
\mciteSetBstMidEndSepPunct{\mcitedefaultmidpunct}
{\mcitedefaultendpunct}{\mcitedefaultseppunct}\relax
\EndOfBibitem
\bibitem[Silva \latin{et~al.}(2007)Silva, Ediz, Yaron, and
  Armitage]{Silva2007JACS}
Silva,~G.~L.; Ediz,~V.; Yaron,~D.; Armitage,~B.~A. {Experimental and
  Computational Investigation of Unsymmetrical Cyanine Dyes: Understanding
  Torsionally Fluorogenic Dyes}. \emph{J. Am. Chem. Soc.} \textbf{2007},
  \emph{129}, 5710--5718\relax
\mciteBstWouldAddEndPuncttrue
\mciteSetBstMidEndSepPunct{\mcitedefaultmidpunct}
{\mcitedefaultendpunct}{\mcitedefaultseppunct}\relax
\EndOfBibitem
\bibitem[Hartschuh \latin{et~al.}(2003)Hartschuh, Pedrosa, Novotny, and
  Krauss]{Hartschuh2003S}
Hartschuh,~A.; Pedrosa,~H.~N.; Novotny,~L.; Krauss,~T.~D. Simultaneous
  Fluorescence and Raman Scattering from Single Carbon Nanotubes.
  \emph{Science} \textbf{2003}, \emph{301}, 1354--1356\relax
\mciteBstWouldAddEndPuncttrue
\mciteSetBstMidEndSepPunct{\mcitedefaultmidpunct}
{\mcitedefaultendpunct}{\mcitedefaultseppunct}\relax
\EndOfBibitem
\bibitem[Terenziani \latin{et~al.}(2008)Terenziani, Katan, Badaeva, Tretiak,
  and Blanchard-Desce]{Terenziani2008AM}
Terenziani,~F.; Katan,~C.; Badaeva,~E.; Tretiak,~S.; Blanchard-Desce,~M.
  {Enhanced Two-Photon Absorption of Organic Chromophores: Theoretical and
  Experimental Assessments}. \emph{Adv. Mater.} \textbf{2008}, \emph{20},
  4641--4678\relax
\mciteBstWouldAddEndPuncttrue
\mciteSetBstMidEndSepPunct{\mcitedefaultmidpunct}
{\mcitedefaultendpunct}{\mcitedefaultseppunct}\relax
\EndOfBibitem
\bibitem[Richings and Habershon(2020)Richings, and Habershon]{Richings2020JCP}
Richings,~G.~W.; Habershon,~S. A New Diabatization Scheme for Direct Quantum
  Dynamics: Procrustes Diabatization. \emph{J. Chem. Phys.} \textbf{2020},
  \emph{152}, 154108\relax
\mciteBstWouldAddEndPuncttrue
\mciteSetBstMidEndSepPunct{\mcitedefaultmidpunct}
{\mcitedefaultendpunct}{\mcitedefaultseppunct}\relax
\EndOfBibitem
\bibitem[Tannor(2006)]{Tannor2006}
Tannor,~D. \emph{Introduction to Quantum Mechanics: A Time-Dependent
  Perspective}; University Science Books: Sausalito, 2006\relax
\mciteBstWouldAddEndPuncttrue
\mciteSetBstMidEndSepPunct{\mcitedefaultmidpunct}
{\mcitedefaultendpunct}{\mcitedefaultseppunct}\relax
\EndOfBibitem
\bibitem[Weinacht and Pearson(2019)Weinacht, and Pearson]{Weinacht2019}
Weinacht,~T.; Pearson,~B. \emph{Time-Resolved Spectroscopy: An Experimental
  Perspective}; CRC Press: New York, 2019\relax
\mciteBstWouldAddEndPuncttrue
\mciteSetBstMidEndSepPunct{\mcitedefaultmidpunct}
{\mcitedefaultendpunct}{\mcitedefaultseppunct}\relax
\EndOfBibitem
\bibitem[Mai \latin{et~al.}(2018)Mai, Marquetand, and
  Gonz{\'a}lez]{Mai2018WCMS}
Mai,~S.; Marquetand,~P.; Gonz{\'a}lez,~L. {Nonadiabatic Dynamics: The SHARC
  Approach}. \emph{WIREs Comput. Mol. Sci.} \textbf{2018}, \emph{8},
  e1370\relax
\mciteBstWouldAddEndPuncttrue
\mciteSetBstMidEndSepPunct{\mcitedefaultmidpunct}
{\mcitedefaultendpunct}{\mcitedefaultseppunct}\relax
\EndOfBibitem
\bibitem[Yonehara \latin{et~al.}(2012)Yonehara, Hanasaki, and
  Takatsuka]{Yonehara2012CR}
Yonehara,~T.; Hanasaki,~K.; Takatsuka,~K. Fundamental Approaches to
  Nonadiabaticity: Toward a Chemical Theory beyond the Born--Oppenheimer
  Paradigm. \emph{Chem. Rev.} \textbf{2012}, \emph{112}, 499--542\relax
\mciteBstWouldAddEndPuncttrue
\mciteSetBstMidEndSepPunct{\mcitedefaultmidpunct}
{\mcitedefaultendpunct}{\mcitedefaultseppunct}\relax
\EndOfBibitem
\bibitem[Casida and Huix-Rotllant(2012)Casida, and
  Huix-Rotllant]{Casida2012ARPC}
Casida,~M.; Huix-Rotllant,~M. Progress in Time-Dependent Density-Functional
  Theory. \emph{Annu. Rev. Phys. Chem.} \textbf{2012}, \emph{63},
  287--323\relax
\mciteBstWouldAddEndPuncttrue
\mciteSetBstMidEndSepPunct{\mcitedefaultmidpunct}
{\mcitedefaultendpunct}{\mcitedefaultseppunct}\relax
\EndOfBibitem
\bibitem[Maitra(2016)]{Maitra2016JCP}
Maitra,~N.~T. Perspective: Fundamental Aspects of Time-Dependent Density
  Functional Theory. \emph{J. Chem. Phys.} \textbf{2016}, \emph{144},
  220901\relax
\mciteBstWouldAddEndPuncttrue
\mciteSetBstMidEndSepPunct{\mcitedefaultmidpunct}
{\mcitedefaultendpunct}{\mcitedefaultseppunct}\relax
\EndOfBibitem
\bibitem[Szalay \latin{et~al.}(2012)Szalay, M\"uller, Gidofalvi, Lischka, and
  Shepard]{Szalay2012CR}
Szalay,~P.~G.; M\"uller,~T.; Gidofalvi,~G.; Lischka,~H.; Shepard,~R.
  Multiconfiguration Self-Consistent Field and Multireference Configuration
  Interaction Methods and Applications. \emph{Chem. Rev.} \textbf{2012},
  \emph{112}, 108--181\relax
\mciteBstWouldAddEndPuncttrue
\mciteSetBstMidEndSepPunct{\mcitedefaultmidpunct}
{\mcitedefaultendpunct}{\mcitedefaultseppunct}\relax
\EndOfBibitem
\bibitem[Helgaker \latin{et~al.}(2014)Helgaker, J{\o}rgensen, and
  Olsen]{Helgaker2014}
Helgaker,~T.; J{\o}rgensen,~P.; Olsen,~J. \emph{Molecular
  Electronic‐Structure Theory}; John Wiley \& Sons, Ltd, 2014\relax
\mciteBstWouldAddEndPuncttrue
\mciteSetBstMidEndSepPunct{\mcitedefaultmidpunct}
{\mcitedefaultendpunct}{\mcitedefaultseppunct}\relax
\EndOfBibitem
\bibitem[Roos \latin{et~al.}(2016)Roos, Lindh, Malmqvist, Veryazov, and
  Widmark]{Roos2016}
Roos,~B.~O.; Lindh,~R.; Malmqvist,~P.~{\AA}.; Veryazov,~V.; Widmark,~P.
  \emph{Multiconfigurational Quantum Chemistry}; John Wiley \& Sons, Ltd,
  2016\relax
\mciteBstWouldAddEndPuncttrue
\mciteSetBstMidEndSepPunct{\mcitedefaultmidpunct}
{\mcitedefaultendpunct}{\mcitedefaultseppunct}\relax
\EndOfBibitem
\bibitem[Born and Oppenheimer(1927)Born, and Oppenheimer]{Born1927AP}
Born,~M.; Oppenheimer,~R. Zur Quantentheorie der Molekeln. \emph{Ann. Phys.}
  \textbf{1927}, \emph{389}, 457--484\relax
\mciteBstWouldAddEndPuncttrue
\mciteSetBstMidEndSepPunct{\mcitedefaultmidpunct}
{\mcitedefaultendpunct}{\mcitedefaultseppunct}\relax
\EndOfBibitem
\bibitem[Kohn(1999)]{Kohn1999RMP}
Kohn,~W. {Nobel Lecture: Electronic Structure of Matter -- Wave Functions and
  Density Functionals}. \emph{Rev. Mod. Phys.} \textbf{1999}, \emph{71},
  1253--1266\relax
\mciteBstWouldAddEndPuncttrue
\mciteSetBstMidEndSepPunct{\mcitedefaultmidpunct}
{\mcitedefaultendpunct}{\mcitedefaultseppunct}\relax
\EndOfBibitem
\bibitem[Schr\"odinger(1926)]{Schroedinger1926PR}
Schr\"odinger,~E. {An Undulatory Theory of the Mechanics of Atoms and
  Molecules}. \emph{Phys. Rev.} \textbf{1926}, \emph{28}, 1049--1070\relax
\mciteBstWouldAddEndPuncttrue
\mciteSetBstMidEndSepPunct{\mcitedefaultmidpunct}
{\mcitedefaultendpunct}{\mcitedefaultseppunct}\relax
\EndOfBibitem
\bibitem[Sch()]{Schroedinger1933}
{Erwin-Schr\"{o}dinger -- Nobel Lecture}.
  \href{https://www.nobelprize.org/prizes/physics/1933/schrodinger/lecture/}{https://www.nobelprize.org/prizes/phy-sics/1933/schrodinger/lecture/}\relax
\mciteBstWouldAddEndPuncttrue
\mciteSetBstMidEndSepPunct{\mcitedefaultmidpunct}
{\mcitedefaultendpunct}{\mcitedefaultseppunct}\relax
\EndOfBibitem
\bibitem[Yu \latin{et~al.}(2016)Yu, Li, and Truhlar]{Yu2016JCP}
Yu,~H.~S.; Li,~S.~L.; Truhlar,~D.~G. Perspective: {Kohn-Sham} Density
  Functional Theory Descending a Staircase. \emph{J. Chem. Phys.}
  \textbf{2016}, \emph{145}, 130901\relax
\mciteBstWouldAddEndPuncttrue
\mciteSetBstMidEndSepPunct{\mcitedefaultmidpunct}
{\mcitedefaultendpunct}{\mcitedefaultseppunct}\relax
\EndOfBibitem
\bibitem[Maurer \latin{et~al.}(2019)Maurer, Freysoldt, Reilly, Brandenburg,
  Hofmann, Björkman, Lebègue, and Tkatchenko]{Maurer2019ARMR}
Maurer,~R.~J.; Freysoldt,~C.; Reilly,~A.~M.; Brandenburg,~J.~G.;
  Hofmann,~O.~T.; Björkman,~T.; Lebègue,~S.; Tkatchenko,~A. Advances in
  Density-Functional Calculations for Materials Modeling. \emph{Annual Review
  of Materials Research} \textbf{2019}, \emph{49}, 1--30\relax
\mciteBstWouldAddEndPuncttrue
\mciteSetBstMidEndSepPunct{\mcitedefaultmidpunct}
{\mcitedefaultendpunct}{\mcitedefaultseppunct}\relax
\EndOfBibitem
\bibitem[Benavides-Riveros \latin{et~al.}(2017)Benavides-Riveros, Lathiotakis,
  and Marques]{Benavides-Riveros2017PCCP}
Benavides-Riveros,~C.~L.; Lathiotakis,~N.~N.; Marques,~M. A.~L. Towards a
  Formal Definition of Static and Dynamic Electronic Correlations. \emph{Phys.
  Chem. Chem. Phys.} \textbf{2017}, \emph{19}, 12655--12664\relax
\mciteBstWouldAddEndPuncttrue
\mciteSetBstMidEndSepPunct{\mcitedefaultmidpunct}
{\mcitedefaultendpunct}{\mcitedefaultseppunct}\relax
\EndOfBibitem
\bibitem[Szabo and Ostlund(2012)Szabo, and Ostlund]{Szabo2012}
Szabo,~A.; Ostlund,~N. \emph{{Modern Quantum Chemistry: Introduction to
  Advanced Electronic Structure Theory}}; Dover Books on Chemistry; Dover
  Publications, 2012\relax
\mciteBstWouldAddEndPuncttrue
\mciteSetBstMidEndSepPunct{\mcitedefaultmidpunct}
{\mcitedefaultendpunct}{\mcitedefaultseppunct}\relax
\EndOfBibitem
\bibitem[Helgaker \latin{et~al.}(2014)Helgaker, J{\o}rgensen, and
  Olsen]{Helgaker2014HF}
Helgaker,~T.; J{\o}rgensen,~P.; Olsen,~J. \emph{Molecular
  Electronic‐Structure Theory}; John Wiley \& Sons, Ltd, 2014; Chapter 10, pp
  433--522\relax
\mciteBstWouldAddEndPuncttrue
\mciteSetBstMidEndSepPunct{\mcitedefaultmidpunct}
{\mcitedefaultendpunct}{\mcitedefaultseppunct}\relax
\EndOfBibitem
\bibitem[Helgaker \latin{et~al.}(2014)Helgaker, J{\o}rgensen, and
  Olsen]{Helgaker2014CI}
Helgaker,~T.; J{\o}rgensen,~P.; Olsen,~J. \emph{Molecular
  Electronic‐Structure Theory}; John Wiley \& Sons, Ltd, 2014; Chapter 11, pp
  523--597\relax
\mciteBstWouldAddEndPuncttrue
\mciteSetBstMidEndSepPunct{\mcitedefaultmidpunct}
{\mcitedefaultendpunct}{\mcitedefaultseppunct}\relax
\EndOfBibitem
\bibitem[Dreuw and Wormit(2015)Dreuw, and Wormit]{Dreuw2015WIREs}
Dreuw,~A.; Wormit,~M. The algebraic diagrammatic construction scheme for the
  polarization propagator for the calculation of excited states. \emph{WIREs
  Comput. Mol. Sci.} \textbf{2015}, \emph{5}, 82--95\relax
\mciteBstWouldAddEndPuncttrue
\mciteSetBstMidEndSepPunct{\mcitedefaultmidpunct}
{\mcitedefaultendpunct}{\mcitedefaultseppunct}\relax
\EndOfBibitem
\bibitem[{von Niessen} \latin{et~al.}(1984){von Niessen}, Schirmer, and
  Cederbaum]{Niessen1984}
{von Niessen},~W.; Schirmer,~J.; Cederbaum,~L. Computational Methods for the
  One-Particle Green's Function. \emph{Comp. Phys. Rep.} \textbf{1984},
  \emph{1}, 57 -- 125\relax
\mciteBstWouldAddEndPuncttrue
\mciteSetBstMidEndSepPunct{\mcitedefaultmidpunct}
{\mcitedefaultendpunct}{\mcitedefaultseppunct}\relax
\EndOfBibitem
\bibitem[Linderberg and \"{O}hrn(2005)Linderberg, and \"{O}hrn]{Lindenberg2005}
Linderberg,~J.; \"{O}hrn,~Y. \emph{{Propagators in Quantum Chemistry}}; John
  Wiley \& Sons, Ltd, 2005; Chapter 2, pp 3--6\relax
\mciteBstWouldAddEndPuncttrue
\mciteSetBstMidEndSepPunct{\mcitedefaultmidpunct}
{\mcitedefaultendpunct}{\mcitedefaultseppunct}\relax
\EndOfBibitem
\bibitem[Melin \latin{et~al.}(2005)Melin, Ayers, and Ortiz]{Melin2005JCS}
Melin,~J.; Ayers,~P.; Ortiz,~J. The Electron-Propagator Approach to Conceptual
  Density-Functional Theory. \emph{J. Chem. Sci.} \textbf{2005}, \emph{117},
  387--400\relax
\mciteBstWouldAddEndPuncttrue
\mciteSetBstMidEndSepPunct{\mcitedefaultmidpunct}
{\mcitedefaultendpunct}{\mcitedefaultseppunct}\relax
\EndOfBibitem
\bibitem[Corzo and Ortiz(2017)Corzo, and Ortiz]{Corzo2017}
Corzo,~H.~H.; Ortiz,~J.~V. In \emph{L{\"{o}}wdin Volume}; Sabin,~J.~R.,
  Br{\"{a}}ndas,~E.~J., Eds.; Advances in Quantum Chemistry; Academic Press,
  2017; Vol.~74; pp 267 -- 298\relax
\mciteBstWouldAddEndPuncttrue
\mciteSetBstMidEndSepPunct{\mcitedefaultmidpunct}
{\mcitedefaultendpunct}{\mcitedefaultseppunct}\relax
\EndOfBibitem
\bibitem[M\"{o}ller and Plesset(1934)M\"{o}ller, and Plesset]{Moeller1934PR}
M\"{o}ller,~C.; Plesset,~M.~S. {Note on an Approximation Treatment for
  Many-Electron Systems}. \emph{Phys. Rev.} \textbf{1934}, \emph{46},
  618--622\relax
\mciteBstWouldAddEndPuncttrue
\mciteSetBstMidEndSepPunct{\mcitedefaultmidpunct}
{\mcitedefaultendpunct}{\mcitedefaultseppunct}\relax
\EndOfBibitem
\bibitem[Bartlett(1981)]{Bartlett1981ARPC}
Bartlett,~R.~J. {Many-Body Perturbation Theory and Coupled Cluster Theory for
  Electron Correlation in Molecules}. \emph{Annu. Rev. Phys. Chem.}
  \textbf{1981}, \emph{32}, 359--401\relax
\mciteBstWouldAddEndPuncttrue
\mciteSetBstMidEndSepPunct{\mcitedefaultmidpunct}
{\mcitedefaultendpunct}{\mcitedefaultseppunct}\relax
\EndOfBibitem
\bibitem[Helgaker \latin{et~al.}(2014)Helgaker, J{\o}rgensen, and
  Olsen]{Helgaker2014CC}
Helgaker,~T.; J{\o}rgensen,~P.; Olsen,~J. \emph{Molecular
  Electronic‐Structure Theory}; John Wiley \& Sons, Ltd, 2014; Chapter 13, pp
  648--723\relax
\mciteBstWouldAddEndPuncttrue
\mciteSetBstMidEndSepPunct{\mcitedefaultmidpunct}
{\mcitedefaultendpunct}{\mcitedefaultseppunct}\relax
\EndOfBibitem
\bibitem[Izs\'{a}k(2020)]{Izsak2020WCMS}
Izs\'{a}k,~R. Single-Reference Coupled Cluster Methods for Computing Excitation
  Energies in Large Molecules: The Efficiency and Accuracy of Approximations.
  \emph{WIREs Comput. Mol. Sci.} \textbf{2020}, \emph{10}, e1445\relax
\mciteBstWouldAddEndPuncttrue
\mciteSetBstMidEndSepPunct{\mcitedefaultmidpunct}
{\mcitedefaultendpunct}{\mcitedefaultseppunct}\relax
\EndOfBibitem
\bibitem[Krylov(2008)]{Krylov2008ARPC}
Krylov,~A.~I. Equation-of-Motion Coupled-Cluster Methods for Open-Shell and
  Electronically Excited Species: The Hitchhiker's Guide to Fock Space.
  \emph{Annu. Rev. Phys. Chem.} \textbf{2008}, \emph{59}, 433--462\relax
\mciteBstWouldAddEndPuncttrue
\mciteSetBstMidEndSepPunct{\mcitedefaultmidpunct}
{\mcitedefaultendpunct}{\mcitedefaultseppunct}\relax
\EndOfBibitem
\bibitem[Parrill and Lipkowitz(2018)Parrill, and Lipkowitz]{Parrill2018}
Parrill,~A.; Lipkowitz,~K. \emph{Reviews in Computational Chemistry, Volume
  31}; Reviews in Computational Chemistry; Wiley, 2018\relax
\mciteBstWouldAddEndPuncttrue
\mciteSetBstMidEndSepPunct{\mcitedefaultmidpunct}
{\mcitedefaultendpunct}{\mcitedefaultseppunct}\relax
\EndOfBibitem
\bibitem[Pacifici~L.(2013)]{Pacifici2013}
Pacifici~L.,~L.~A.,~Verdicchio~M. In \emph{Computational Science and Its
  Applications – ICCSA 2013}; {B. Murgante, \emph{et. al}},, Ed.; Springer,
  Berlin, Heidelberg, 2013; Vol. 7971\relax
\mciteBstWouldAddEndPuncttrue
\mciteSetBstMidEndSepPunct{\mcitedefaultmidpunct}
{\mcitedefaultendpunct}{\mcitedefaultseppunct}\relax
\EndOfBibitem
\bibitem[Helgaker \latin{et~al.}(2014)Helgaker, J{\o}rgensen, and
  Olsen]{Helgaker2014MCSCF}
Helgaker,~T.; J{\o}rgensen,~P.; Olsen,~J. \emph{Molecular
  Electronic‐Structure Theory}; John Wiley \& Sons, Ltd, 2014; Chapter 12, pp
  598--647\relax
\mciteBstWouldAddEndPuncttrue
\mciteSetBstMidEndSepPunct{\mcitedefaultmidpunct}
{\mcitedefaultendpunct}{\mcitedefaultseppunct}\relax
\EndOfBibitem
\bibitem[Roos \latin{et~al.}(1980)Roos, Taylor, and Siegbahn]{Roos1980CP}
Roos,~B.~O.; Taylor,~P.~R.; Siegbahn,~P.~E. A Complete Active Space SCF Method
  (CASSCF) using a Density Matrix Formulated Super-CI Approach. \emph{Chem.
  Phys.} \textbf{1980}, \emph{48}, 157--173\relax
\mciteBstWouldAddEndPuncttrue
\mciteSetBstMidEndSepPunct{\mcitedefaultmidpunct}
{\mcitedefaultendpunct}{\mcitedefaultseppunct}\relax
\EndOfBibitem
\bibitem[Roos and Siegbahn(1980)Roos, and Siegbahn]{Roos1980IJQC}
Roos,~B.~O.; Siegbahn,~P. E.~M. {A Direct CI Method with a Multiconfigurational
  Reference State}. \emph{Int. J. Quantum Chem.} \textbf{1980}, \emph{17},
  485--500\relax
\mciteBstWouldAddEndPuncttrue
\mciteSetBstMidEndSepPunct{\mcitedefaultmidpunct}
{\mcitedefaultendpunct}{\mcitedefaultseppunct}\relax
\EndOfBibitem
\bibitem[Lischka \latin{et~al.}(2004)Lischka, Dallos, Szalay, Yarkony, and
  Shepard]{Lischka2004JCP}
Lischka,~H.; Dallos,~M.; Szalay,~P.~G.; Yarkony,~D.~R.; Shepard,~R. Analytic
  Evaluation of Nonadiabatic coupling Terms at the {MR-CI} Level. {I.}
  {Formalism}. \emph{J. Chem. Phys.} \textbf{2004}, \emph{120},
  7322--7329\relax
\mciteBstWouldAddEndPuncttrue
\mciteSetBstMidEndSepPunct{\mcitedefaultmidpunct}
{\mcitedefaultendpunct}{\mcitedefaultseppunct}\relax
\EndOfBibitem
\bibitem[Lischka \latin{et~al.}(2020)Lischka, Shepard, M\"{u}ller, Szalay,
  Pitzer, Aquino, Ara\'{u}jo~do Nascimento, Barbatti, Belcher, Blaudeau,
  Borges, Brozell, Carter, Das, Gidofalvi, Gonz\'{a}lez, Hase, Kedziora,
  Kertesz, Kossoski, Machado, Matsika, do~Monte, Nachtigallov\'{a}, Nieman,
  Oppel, Parish, Plasser, Spada, Stahlberg, Ventura, Yarkony, and
  Zhang]{Lischka2020JCP}
Lischka,~H. \latin{et~al.}  The Generality of the GUGA MRCI Approach in
  COLUMBUS for Treating Complex Quantum Chemistry. \emph{J. Chem. Phys.}
  \textbf{2020}, \emph{152}, 134110\relax
\mciteBstWouldAddEndPuncttrue
\mciteSetBstMidEndSepPunct{\mcitedefaultmidpunct}
{\mcitedefaultendpunct}{\mcitedefaultseppunct}\relax
\EndOfBibitem
\bibitem[Andersson \latin{et~al.}(1990)Andersson, Malmqvist, Roos, Sadlej, and
  Wolinski]{Andersson1990JPC}
Andersson,~K.; Malmqvist,~P.~A.; Roos,~B.~O.; Sadlej,~A.~J.; Wolinski,~K.
  {Second-Order Perturbation Theory with a CASSCF Reference Function}. \emph{J.
  Phys. Chem.} \textbf{1990}, \emph{94}, 5483--5488\relax
\mciteBstWouldAddEndPuncttrue
\mciteSetBstMidEndSepPunct{\mcitedefaultmidpunct}
{\mcitedefaultendpunct}{\mcitedefaultseppunct}\relax
\EndOfBibitem
\bibitem[Andersson \latin{et~al.}(1992)Andersson, Malmqvist, and
  Roos]{Andersson1992JCP}
Andersson,~K.; Malmqvist,~P.; Roos,~B.~O. Second‐Order Perturbation Theory
  with a Complete Active Space Self‐Consistent Field Reference Function.
  \emph{J. Phys. Chem.} \textbf{1992}, \emph{96}, 1218--1226\relax
\mciteBstWouldAddEndPuncttrue
\mciteSetBstMidEndSepPunct{\mcitedefaultmidpunct}
{\mcitedefaultendpunct}{\mcitedefaultseppunct}\relax
\EndOfBibitem
\bibitem[Finley \latin{et~al.}(1998)Finley, Malmqvist, Roos, and
  Serrano-Andr{\'e}s]{Finley1998CPL}
Finley,~J.; Malmqvist,~P.-A.; Roos,~B.~O.; Serrano-Andr{\'e}s,~L. The
  Multi-State \{CASPT2\} Method. \emph{Chem. Phys. Lett.} \textbf{1998},
  \emph{288}, 299 -- 306\relax
\mciteBstWouldAddEndPuncttrue
\mciteSetBstMidEndSepPunct{\mcitedefaultmidpunct}
{\mcitedefaultendpunct}{\mcitedefaultseppunct}\relax
\EndOfBibitem
\bibitem[Angeli \latin{et~al.}(2001)Angeli, Cimiraglia, Evangelisti, Leininger,
  and Malrieu]{Angeli2001JCP}
Angeli,~C.; Cimiraglia,~R.; Evangelisti,~S.; Leininger,~T.; Malrieu,~J.-P.
  Introduction of N-Electron Valence States for Multireference Perturbation
  Theory. \emph{J. Chem. Phys.} \textbf{2001}, \emph{114}, 10252--10264\relax
\mciteBstWouldAddEndPuncttrue
\mciteSetBstMidEndSepPunct{\mcitedefaultmidpunct}
{\mcitedefaultendpunct}{\mcitedefaultseppunct}\relax
\EndOfBibitem
\bibitem[Roemelt \latin{et~al.}(2016)Roemelt, Guo, and Chan]{Roemelt2016JCP}
Roemelt,~M.; Guo,~S.; Chan,~G. K.-L. A Projected Approximation to Strongly
  Contracted N-Electron Valence Perturbation Theory for DMRG Wavefunctions.
  \emph{J. Chem. Phys.} \textbf{2016}, \emph{144}, 204113\relax
\mciteBstWouldAddEndPuncttrue
\mciteSetBstMidEndSepPunct{\mcitedefaultmidpunct}
{\mcitedefaultendpunct}{\mcitedefaultseppunct}\relax
\EndOfBibitem
\bibitem[Guo \latin{et~al.}(2017)Guo, Sivalingam, Valeev, and
  Neese]{Guo2017JCP}
Guo,~Y.; Sivalingam,~K.; Valeev,~E.~F.; Neese,~F. Explicitly Correlated
  N-Electron Valence State Perturbation Theory (NEVPT2-F12). \emph{J. Chem.
  Phys.} \textbf{2017}, \emph{147}, 064110\relax
\mciteBstWouldAddEndPuncttrue
\mciteSetBstMidEndSepPunct{\mcitedefaultmidpunct}
{\mcitedefaultendpunct}{\mcitedefaultseppunct}\relax
\EndOfBibitem
\bibitem[Maitra \latin{et~al.}(2012)Maitra, Sinha, and
  Mukherjee]{Maitra2012JCP}
Maitra,~R.; Sinha,~D.; Mukherjee,~D. Unitary Group Adapted State-Specific
  Multi-Reference Coupled Cluster Theory: Formulation and Pilot Numerical
  Applications. \emph{J. Chem. Phys.} \textbf{2012}, \emph{137}, 024105\relax
\mciteBstWouldAddEndPuncttrue
\mciteSetBstMidEndSepPunct{\mcitedefaultmidpunct}
{\mcitedefaultendpunct}{\mcitedefaultseppunct}\relax
\EndOfBibitem
\bibitem[M{\'{a}\c{s}}ik and Huba{\c{c}}(1998)M{\'{a}\c{s}}ik, and
  Huba{\c{c}}]{Masik1998}
M{\'{a}\c{s}}ik,~J.; Huba{\c{c}},~I. In \emph{{Multireference Brillouin-Wigner
  Coupled-Cluster Theory. Single-Root Approach.}}; Sabin,~J.~R., Zerner,~M.~C.,
  Br{\"{a}}ndas,~E., Wilson,~S., Maruani,~J., Smeyers,~Y., Grout,~P.,
  McWeeny,~R., Eds.; Advances in Quantum Chemistry; Academic Press, 1998;
  Vol.~31; pp 75 -- 104\relax
\mciteBstWouldAddEndPuncttrue
\mciteSetBstMidEndSepPunct{\mcitedefaultmidpunct}
{\mcitedefaultendpunct}{\mcitedefaultseppunct}\relax
\EndOfBibitem
\bibitem[Musia{\l} \latin{et~al.}(2011)Musia{\l}, Perera, and
  Bartlett]{Musial2011JCP}
Musia{\l},~M.; Perera,~A.; Bartlett,~R.~J. Multireference Coupled-Cluster
  Theory: The Easy Way. \emph{J. Chem. Phys.} \textbf{2011}, \emph{134},
  114108\relax
\mciteBstWouldAddEndPuncttrue
\mciteSetBstMidEndSepPunct{\mcitedefaultmidpunct}
{\mcitedefaultendpunct}{\mcitedefaultseppunct}\relax
\EndOfBibitem
\bibitem[Evangelista(2018)]{Evangelista2018JCP}
Evangelista,~F.~A. {Perspective: Multireference Coupled Cluster Theories of
  Dynamical Electron Correlation}. \emph{J. Chem. Phys.} \textbf{2018},
  \emph{149}, 030901\relax
\mciteBstWouldAddEndPuncttrue
\mciteSetBstMidEndSepPunct{\mcitedefaultmidpunct}
{\mcitedefaultendpunct}{\mcitedefaultseppunct}\relax
\EndOfBibitem
\bibitem[Fdez.~Galv\'an \latin{et~al.}(2019)Fdez.~Galv\'an, Vacher, Alavi,
  Angeli, Aquilante, Autschbach, Bao, Bokarev, Bogdanov, Carlson, Chibotaru,
  Creutzberg, Dattani, Delcey, Dong, Dreuw, Freitag, Frutos, Gagliardi,
  Gendron, Giussani, Gonz\'alez, Grell, Guo, Hoyer, Johansson, Keller, Knecht,
  Kova{\v{c}}evi\'c, K\"allman, Li~Manni, Lundberg, Ma, Mai, Malhado,
  Malmqvist, Marquetand, Mewes, Norell, Olivucci, Oppel, Phung, Pierloot,
  Plasser, Reiher, Sand, Schapiro, Sharma, Stein, S{\o}rensen, Truhlar, Ugandi,
  Ungur, Valentini, Vancoillie, Veryazov, Weser, Weso{\l}owski, Widmark,
  Wouters, Zech, Zobel, and Lindh]{Fdez-Galvan2019JCTC}
Fdez.~Galv\'an,~I. \latin{et~al.}  OpenMolcas: From Source Code to Insight.
  \emph{J. Chem. Theory Comput.} \textbf{2019}, \emph{15}, 5925--5964\relax
\mciteBstWouldAddEndPuncttrue
\mciteSetBstMidEndSepPunct{\mcitedefaultmidpunct}
{\mcitedefaultendpunct}{\mcitedefaultseppunct}\relax
\EndOfBibitem
\bibitem[Roos \latin{et~al.}(2004)Roos, Lindh, Malmqvist, Veryazov, and
  Widmark]{Roos2004JPCA}
Roos,~B.; Lindh,~R.; Malmqvist,~P.-{\AA}.; Veryazov,~V.; Widmark,~P.-O. Main
  Group Atoms and Dimers Studied with a new Relativistic ANO Basis Set.
  \emph{J. Phys. Chem. A} \textbf{2004}, \emph{108}, 2851--2858\relax
\mciteBstWouldAddEndPuncttrue
\mciteSetBstMidEndSepPunct{\mcitedefaultmidpunct}
{\mcitedefaultendpunct}{\mcitedefaultseppunct}\relax
\EndOfBibitem
\bibitem[Vogiatzis \latin{et~al.}(2017)Vogiatzis, Ma, Olsen, Gagliardi, and
  de~Jong]{Vogiatzis2017JCP}
Vogiatzis,~K.~D.; Ma,~D.; Olsen,~J.; Gagliardi,~L.; de~Jong,~W.~A. Pushing
  Configuration-Interaction to the limit: Towards Massively Parallel MCSCF
  Calculations. \emph{J. Chem. Phys.} \textbf{2017}, \emph{147}, 184111\relax
\mciteBstWouldAddEndPuncttrue
\mciteSetBstMidEndSepPunct{\mcitedefaultmidpunct}
{\mcitedefaultendpunct}{\mcitedefaultseppunct}\relax
\EndOfBibitem
\bibitem[Kato and Baba(1995)Kato, and Baba]{Kato1995CR}
Kato,~H.; Baba,~M. Dynamics of Excited Molecules: Predissociation. \emph{Chem.
  Rev.} \textbf{1995}, \emph{95}, 2311--2349\relax
\mciteBstWouldAddEndPuncttrue
\mciteSetBstMidEndSepPunct{\mcitedefaultmidpunct}
{\mcitedefaultendpunct}{\mcitedefaultseppunct}\relax
\EndOfBibitem
\bibitem[Merer and Mulliken(1969)Merer, and Mulliken]{Merer1969CR}
Merer,~A.~J.; Mulliken,~R.~S. Ultraviolet Spectra and Excited States of
  Ethylene and its Alkyl Derivatives. \emph{Chem. Rev.} \textbf{1969},
  \emph{69}, 639--656\relax
\mciteBstWouldAddEndPuncttrue
\mciteSetBstMidEndSepPunct{\mcitedefaultmidpunct}
{\mcitedefaultendpunct}{\mcitedefaultseppunct}\relax
\EndOfBibitem
\bibitem[Ashfold and Langford(1999)Ashfold, and Langford]{Ashfold1999}
Ashfold,~M. N.~R.; Langford,~S.~R. In \emph{The Role of Rydberg States in
  Spectroscopy and Photochemistry: Low and High Rydberg States};
  S{\'a}ndorfy,~C., Ed.; Springer Netherlands: Dordrecht, 1999; pp 23--56\relax
\mciteBstWouldAddEndPuncttrue
\mciteSetBstMidEndSepPunct{\mcitedefaultmidpunct}
{\mcitedefaultendpunct}{\mcitedefaultseppunct}\relax
\EndOfBibitem
\bibitem[Merkt(1997)]{Merkt1997ARPC}
Merkt,~F. {Molecules in High Rydberg States}. \emph{Annu. Rev. Phys. Chem.}
  \textbf{1997}, \emph{48}, 675--709\relax
\mciteBstWouldAddEndPuncttrue
\mciteSetBstMidEndSepPunct{\mcitedefaultmidpunct}
{\mcitedefaultendpunct}{\mcitedefaultseppunct}\relax
\EndOfBibitem
\bibitem[Stein and Reiher(2016)Stein, and Reiher]{Stein2016JCTC}
Stein,~C.~J.; Reiher,~M. Automated Selection of Active Orbital Spaces. \emph{J.
  Chem. Theory Comput.} \textbf{2016}, \emph{12}, 1760--1771\relax
\mciteBstWouldAddEndPuncttrue
\mciteSetBstMidEndSepPunct{\mcitedefaultmidpunct}
{\mcitedefaultendpunct}{\mcitedefaultseppunct}\relax
\EndOfBibitem
\bibitem[Stein and Reiher(2017)Stein, and Reiher]{Stein2017MP}
Stein,~C.~J.; Reiher,~M. Measuring Multi-Configurational Character by Orbital
  Entanglement. \emph{Mol. Phys.} \textbf{2017}, \emph{115}, 2110--2119\relax
\mciteBstWouldAddEndPuncttrue
\mciteSetBstMidEndSepPunct{\mcitedefaultmidpunct}
{\mcitedefaultendpunct}{\mcitedefaultseppunct}\relax
\EndOfBibitem
\bibitem[Stein and Reiher(2017)Stein, and Reiher]{Stein2017CHIMIA}
Stein,~C.~J.; Reiher,~M. {Automated Identification of Relevant Frontier
  Orbitals for Chemical Compounds and Processes}. \emph{CHIMIA} \textbf{2017},
  \emph{71}, 170--176\relax
\mciteBstWouldAddEndPuncttrue
\mciteSetBstMidEndSepPunct{\mcitedefaultmidpunct}
{\mcitedefaultendpunct}{\mcitedefaultseppunct}\relax
\EndOfBibitem
\bibitem[Chan and Van~Voorhis(2005)Chan, and Van~Voorhis]{Chan2006JCP}
Chan,~G. K.-L.; Van~Voorhis,~T. Density-Matrix Renormalization-Group Algorithms
  with Nonorthogonal Orbitals and Non-Hermitian Operators, and Applications to
  Polyenes. \emph{J. Chem. Phys.} \textbf{2005}, \emph{122}, 204101\relax
\mciteBstWouldAddEndPuncttrue
\mciteSetBstMidEndSepPunct{\mcitedefaultmidpunct}
{\mcitedefaultendpunct}{\mcitedefaultseppunct}\relax
\EndOfBibitem
\bibitem[Zgid and Nooijen(2008)Zgid, and Nooijen]{Zgid2008JCP}
Zgid,~D.; Nooijen,~M. The Density Matrix Renormalization Group Self-Consistent
  Field Method: Orbital Optimization with the Density Matrix Renormalization
  Group Method in the Active Space. \emph{J. Chem. Phys.} \textbf{2008},
  \emph{128}, 144116\relax
\mciteBstWouldAddEndPuncttrue
\mciteSetBstMidEndSepPunct{\mcitedefaultmidpunct}
{\mcitedefaultendpunct}{\mcitedefaultseppunct}\relax
\EndOfBibitem
\bibitem[Keller \latin{et~al.}(2015)Keller, Dolfi, Troyer, and
  Reiher]{Keller2015JCP}
Keller,~S.; Dolfi,~M.; Troyer,~M.; Reiher,~M. An Efficient Matrix Product
  Operator Representation of the Quantum Chemical Hamiltonian. \emph{J. Chem.
  Phys.} \textbf{2015}, \emph{143}, 244118\relax
\mciteBstWouldAddEndPuncttrue
\mciteSetBstMidEndSepPunct{\mcitedefaultmidpunct}
{\mcitedefaultendpunct}{\mcitedefaultseppunct}\relax
\EndOfBibitem
\bibitem[Knecht \latin{et~al.}(2016)Knecht, Keller, Autschbach, and
  Reiher]{Knecht2016JCTC}
Knecht,~S.; Keller,~S.; Autschbach,~J.; Reiher,~M. {A Nonorthogonal
  State-Interaction Approach for Matrix Product State Wave Functions}. \emph{J.
  Chem. Theory Comput.} \textbf{2016}, \emph{12}, 5881--5894\relax
\mciteBstWouldAddEndPuncttrue
\mciteSetBstMidEndSepPunct{\mcitedefaultmidpunct}
{\mcitedefaultendpunct}{\mcitedefaultseppunct}\relax
\EndOfBibitem
\bibitem[Freitag \latin{et~al.}(2017)Freitag, Knecht, Angeli, and
  Reiher]{Freitag2017JCTC}
Freitag,~L.; Knecht,~S.; Angeli,~C.; Reiher,~M. {Multireference Perturbation
  Theory with Cholesky Decomposition for the Density Matrix Renormalization
  Group}. \emph{J. Chem. Theory Comput.} \textbf{2017}, \emph{13},
  451--459\relax
\mciteBstWouldAddEndPuncttrue
\mciteSetBstMidEndSepPunct{\mcitedefaultmidpunct}
{\mcitedefaultendpunct}{\mcitedefaultseppunct}\relax
\EndOfBibitem
\bibitem[Freitag \latin{et~al.}(2019)Freitag, Ma, Baiardi, Knecht, and
  Reiher]{Freitag2019JCTC}
Freitag,~L.; Ma,~Y.; Baiardi,~A.; Knecht,~S.; Reiher,~M. {Approximate
  Analytical Gradients and Nonadiabatic Couplings for the State-Average Density
  Matrix Renormalization Group Self-Consistent-Field Method}. \emph{J. Chem.
  Theory Comput.} \textbf{2019}, \emph{15}, 6724--6737\relax
\mciteBstWouldAddEndPuncttrue
\mciteSetBstMidEndSepPunct{\mcitedefaultmidpunct}
{\mcitedefaultendpunct}{\mcitedefaultseppunct}\relax
\EndOfBibitem
\bibitem[Hohenberg and Kohn(1964)Hohenberg, and Kohn]{Hohenberg1964PR}
Hohenberg,~P.; Kohn,~W. Inhomogeneous Electron Gas. \emph{Phys. Rev.}
  \textbf{1964}, \emph{136}, B864--B871\relax
\mciteBstWouldAddEndPuncttrue
\mciteSetBstMidEndSepPunct{\mcitedefaultmidpunct}
{\mcitedefaultendpunct}{\mcitedefaultseppunct}\relax
\EndOfBibitem
\bibitem[Kohn and Sham(1965)Kohn, and Sham]{Kohn1965PR}
Kohn,~W.; Sham,~L.~J. Self-Consistent Equations Including Exchange and
  Correlation Effects. \emph{Phys. Rev.} \textbf{1965}, \emph{140},
  A1133--A1138\relax
\mciteBstWouldAddEndPuncttrue
\mciteSetBstMidEndSepPunct{\mcitedefaultmidpunct}
{\mcitedefaultendpunct}{\mcitedefaultseppunct}\relax
\EndOfBibitem
\bibitem[Casida(2009)]{Casida2009JMST}
Casida,~M.~E. Time-Dependent Density-Functional Theory for Molecules and
  Molecular Solids. \emph{J. Mol. Struc.-Theochem} \textbf{2009}, \emph{914}, 3
  -- 18\relax
\mciteBstWouldAddEndPuncttrue
\mciteSetBstMidEndSepPunct{\mcitedefaultmidpunct}
{\mcitedefaultendpunct}{\mcitedefaultseppunct}\relax
\EndOfBibitem
\bibitem[Runge and Gross(1984)Runge, and Gross]{Runge1984PRL}
Runge,~E.; Gross,~E. K.~U. Density-Functional Theory for Time-Dependent
  Systems. \emph{Phys. Rev. Lett.} \textbf{1984}, \emph{52}, 997--1000\relax
\mciteBstWouldAddEndPuncttrue
\mciteSetBstMidEndSepPunct{\mcitedefaultmidpunct}
{\mcitedefaultendpunct}{\mcitedefaultseppunct}\relax
\EndOfBibitem
\bibitem[Zangwill and Soven(1980)Zangwill, and Soven]{Zangwill1980PRA}
Zangwill,~A.; Soven,~P. Density-Functional Approach to Local-Field Effects in
  Finite Systems: Photoabsorption in the Rare Gases. \emph{Phys. Rev. A}
  \textbf{1980}, \emph{21}, 1561--1572\relax
\mciteBstWouldAddEndPuncttrue
\mciteSetBstMidEndSepPunct{\mcitedefaultmidpunct}
{\mcitedefaultendpunct}{\mcitedefaultseppunct}\relax
\EndOfBibitem
\bibitem[Chong(1995)]{Chong1995}
Chong,~D.~P. \emph{{Recent Advances in Density Functional Methods}}; World
  Scientific, 1995\relax
\mciteBstWouldAddEndPuncttrue
\mciteSetBstMidEndSepPunct{\mcitedefaultmidpunct}
{\mcitedefaultendpunct}{\mcitedefaultseppunct}\relax
\EndOfBibitem
\bibitem[Tamm(1945)]{Tamm1945JP}
Tamm,~I. Relativistic Interaction of Elementary Particles. \emph{J. Phys.
  (Moscow)} \textbf{1945}, \emph{9}, 449\relax
\mciteBstWouldAddEndPuncttrue
\mciteSetBstMidEndSepPunct{\mcitedefaultmidpunct}
{\mcitedefaultendpunct}{\mcitedefaultseppunct}\relax
\EndOfBibitem
\bibitem[Dancoff(1950)]{Dancoff1950PR}
Dancoff,~S.~M. Non-Adiabatic Meson Theory of Nuclear Forces. \emph{Phys. Rev.}
  \textbf{1950}, \emph{78}, 382--385\relax
\mciteBstWouldAddEndPuncttrue
\mciteSetBstMidEndSepPunct{\mcitedefaultmidpunct}
{\mcitedefaultendpunct}{\mcitedefaultseppunct}\relax
\EndOfBibitem
\bibitem[Hirata and Head-Gordon(1999)Hirata, and Head-Gordon]{Hirata1999CPL}
Hirata,~S.; Head-Gordon,~M. Time-Dependent Density Functional Theory within the
  Tamm–Dancoff Approximation. \emph{Chem. Phys. Lett.} \textbf{1999},
  \emph{314}, 291 -- 299\relax
\mciteBstWouldAddEndPuncttrue
\mciteSetBstMidEndSepPunct{\mcitedefaultmidpunct}
{\mcitedefaultendpunct}{\mcitedefaultseppunct}\relax
\EndOfBibitem
\bibitem[Cordova \latin{et~al.}(2007)Cordova, Doriol, Ipatov, Casida, Filippi,
  and Vela]{Cordova2007JCP}
Cordova,~F.; Doriol,~L.~J.; Ipatov,~A.; Casida,~M.~E.; Filippi,~C.; Vela,~A.
  Troubleshooting Time-Dependent Density-Functional Theory for Photochemical
  Applications: Oxirane. \emph{J. Chem. Phys.} \textbf{2007}, \emph{127},
  164111\relax
\mciteBstWouldAddEndPuncttrue
\mciteSetBstMidEndSepPunct{\mcitedefaultmidpunct}
{\mcitedefaultendpunct}{\mcitedefaultseppunct}\relax
\EndOfBibitem
\bibitem[Goerigk and Casanova-Pa\'ez(2020)Goerigk, and
  Casanova-Pa\'ez]{Goerigk2020AJC}
Goerigk,~L.; Casanova-Pa\'ez,~M. The Trip to the Density Functional Theory Zoo
  Continues: Making a Case for Time-Dependent Double Hybrids for Excited-State
  Problems. \emph{Aust. J. Chem.} \textbf{2020}, \emph{in press},
  DOI:10.1071/CH20093\relax
\mciteBstWouldAddEndPuncttrue
\mciteSetBstMidEndSepPunct{\mcitedefaultmidpunct}
{\mcitedefaultendpunct}{\mcitedefaultseppunct}\relax
\EndOfBibitem
\bibitem[Worth and Cederbaum(2004)Worth, and Cederbaum]{Worth2004ARPC}
Worth,~G.~A.; Cederbaum,~L.~S. Beyond Born-Oppenheimer: Molecular Dynamics
  Through a conical Intersection. \emph{Annu. Rev. Phys. Chem.} \textbf{2004},
  \emph{55}, 127--158\relax
\mciteBstWouldAddEndPuncttrue
\mciteSetBstMidEndSepPunct{\mcitedefaultmidpunct}
{\mcitedefaultendpunct}{\mcitedefaultseppunct}\relax
\EndOfBibitem
\bibitem[Doltsinis(2006)]{Doltsinis2006NIC}
Doltsinis,~N.~L. \emph{Molecular Dynamics Beyond the Born-Oppenheimer
  Approximation: Mixed Quantum-Classical Approaches}; NIC Series; John von
  Neuman Institut for computing, 2006; Vol.~31; pp 389--409\relax
\mciteBstWouldAddEndPuncttrue
\mciteSetBstMidEndSepPunct{\mcitedefaultmidpunct}
{\mcitedefaultendpunct}{\mcitedefaultseppunct}\relax
\EndOfBibitem
\bibitem[Jacquemin and Adamo(2016)Jacquemin, and Adamo]{Jacquemin2016}
Jacquemin,~D.; Adamo,~C. In \emph{Density-Functional Methods for Excited
  States}; Ferr{\'e},~N., Filatov,~M., Huix-Rotllant,~M., Eds.; Springer
  International Publishing: Cham, 2016; pp 347--375\relax
\mciteBstWouldAddEndPuncttrue
\mciteSetBstMidEndSepPunct{\mcitedefaultmidpunct}
{\mcitedefaultendpunct}{\mcitedefaultseppunct}\relax
\EndOfBibitem
\bibitem[Li \latin{et~al.}(2014)Li, Marenich, Xu, and Truhlar]{Li2014JPCL}
Li,~S.~L.; Marenich,~A.~V.; Xu,~X.; Truhlar,~D.~G. Configuration
  Interaction-Corrected Tamm–Dancoff Approximation: A Time-Dependent Density
  Functional Method with the Correct Dimensionality of Conical Intersections.
  \emph{J. Phys. Chem. Lett.} \textbf{2014}, \emph{5}, 322--328\relax
\mciteBstWouldAddEndPuncttrue
\mciteSetBstMidEndSepPunct{\mcitedefaultmidpunct}
{\mcitedefaultendpunct}{\mcitedefaultseppunct}\relax
\EndOfBibitem
\bibitem[Bannwarth \latin{et~al.}(2020)Bannwarth, Yu, Hohenstein, and
  Mart\'{i}nez]{Bannwarth2020chemrxiv}
Bannwarth,~C.; Yu,~J.~K.; Hohenstein,~E.~G.; Mart\'{i}nez,~T.~J. {Hole-Hole
  Tamm-Dancoff-Approximated Density Functional Theory: A Highly Efficient
  Electronic Structure Method Incorporating Dynamic and Static Correlation}.
  \emph{ChemRxiv} \textbf{2020}, \relax
\mciteBstWouldAddEndPunctfalse
\mciteSetBstMidEndSepPunct{\mcitedefaultmidpunct}
{}{\mcitedefaultseppunct}\relax
\EndOfBibitem
\bibitem[Maitra \latin{et~al.}(2004)Maitra, Zhang, Cave, and
  Burke]{Maitra2004JCP}
Maitra,~N.~T.; Zhang,~F.; Cave,~R.~J.; Burke,~K. Double Excitations within
  Time-Dependent Density Functional Theory Linear Response. \emph{J. Chem.
  Phys.} \textbf{2004}, \emph{120}, 5932--5937\relax
\mciteBstWouldAddEndPuncttrue
\mciteSetBstMidEndSepPunct{\mcitedefaultmidpunct}
{\mcitedefaultendpunct}{\mcitedefaultseppunct}\relax
\EndOfBibitem
\bibitem[Elliott \latin{et~al.}(2011)Elliott, Goldson, Canahui, and
  Maitra]{Elliot2011CP}
Elliott,~P.; Goldson,~S.; Canahui,~C.; Maitra,~N.~T. Perspectives on
  Double-Excitations in TDDFT. \emph{Chem. Phys.} \textbf{2011}, \emph{391},
  110 -- 119\relax
\mciteBstWouldAddEndPuncttrue
\mciteSetBstMidEndSepPunct{\mcitedefaultmidpunct}
{\mcitedefaultendpunct}{\mcitedefaultseppunct}\relax
\EndOfBibitem
\bibitem[Katriel \latin{et~al.}(2001)Katriel, Zahariev, and
  Burke]{Katriel2001IJQC}
Katriel,~J.; Zahariev,~F.; Burke,~K. Symmetry and Degeneracy in Density
  Functional Theory. \emph{Int. J. Quantum Chem.} \textbf{2001}, \emph{85},
  432--435\relax
\mciteBstWouldAddEndPuncttrue
\mciteSetBstMidEndSepPunct{\mcitedefaultmidpunct}
{\mcitedefaultendpunct}{\mcitedefaultseppunct}\relax
\EndOfBibitem
\bibitem[Shao \latin{et~al.}(2003)Shao, Head-Gordon, and Krylov]{Shao2003JCP}
Shao,~Y.; Head-Gordon,~M.; Krylov,~A.~I. The Spin–Flip Approach within
  Time-Dependent Density Functional Theory: Theory and Applications to
  Diradicals. \emph{J. Chem. Phys.} \textbf{2003}, \emph{118}, 4807--4818\relax
\mciteBstWouldAddEndPuncttrue
\mciteSetBstMidEndSepPunct{\mcitedefaultmidpunct}
{\mcitedefaultendpunct}{\mcitedefaultseppunct}\relax
\EndOfBibitem
\bibitem[Gavnholt \latin{et~al.}(2008)Gavnholt, Olsen, Engelund, and
  Schi\o{}tz]{Gavnholt2008PRB}
Gavnholt,~J.; Olsen,~T.; Engelund,~M.; Schi\o{}tz,~J. $\ensuremath{\Delta}$
  Self-Consistent Field Method to obtain Potential Energy Surfaces of Excited
  Molecules on Surfaces. \emph{Phys. Rev. B} \textbf{2008}, \emph{78},
  075441\relax
\mciteBstWouldAddEndPuncttrue
\mciteSetBstMidEndSepPunct{\mcitedefaultmidpunct}
{\mcitedefaultendpunct}{\mcitedefaultseppunct}\relax
\EndOfBibitem
\bibitem[Maurer and Reuter(2011)Maurer, and Reuter]{Maurer2011JCP}
Maurer,~R.~J.; Reuter,~K. Assessing Computationally Efficient Isomerization
  Dynamics: $\Delta$-SCF Density-Functional Theory Study of Azobenzene
  Molecular Switching. \emph{J. Chem. Phys.} \textbf{2011}, \emph{135},
  224303\relax
\mciteBstWouldAddEndPuncttrue
\mciteSetBstMidEndSepPunct{\mcitedefaultmidpunct}
{\mcitedefaultendpunct}{\mcitedefaultseppunct}\relax
\EndOfBibitem
\bibitem[Maurer and Reuter(2013)Maurer, and Reuter]{Maurer2013JCP}
Maurer,~R.~J.; Reuter,~K. Excited-State Potential-Energy Surfaces of
  Metal-Adsorbed Organic Molecules from Linear Expansion
  {$\Delta$}-Self-Consistent Field Density-Functional Theory
  ({$\Delta$}{SCF-DFT}). \emph{J. Chem. Phys.} \textbf{2013}, \emph{139},
  014708\relax
\mciteBstWouldAddEndPuncttrue
\mciteSetBstMidEndSepPunct{\mcitedefaultmidpunct}
{\mcitedefaultendpunct}{\mcitedefaultseppunct}\relax
\EndOfBibitem
\bibitem[Chai and Head-Gordon(2008)Chai, and Head-Gordon]{Chai2008JCP}
Chai,~J.-D.; Head-Gordon,~M. Systematic Optimization of Long-Range Corrected
  Hybrid Density Functionals. \emph{J. Chem. Phys.} \textbf{2008}, \emph{128},
  084106\relax
\mciteBstWouldAddEndPuncttrue
\mciteSetBstMidEndSepPunct{\mcitedefaultmidpunct}
{\mcitedefaultendpunct}{\mcitedefaultseppunct}\relax
\EndOfBibitem
\bibitem[Tozer and Handy(2000)Tozer, and Handy]{Tozer2000PCCP}
Tozer,~D.~J.; Handy,~N.~C. On the Determination of Excitation Energies using
  Density Functional Theory. \emph{Phys. Chem. Chem. Phys.} \textbf{2000},
  \emph{2}, 2117--2121\relax
\mciteBstWouldAddEndPuncttrue
\mciteSetBstMidEndSepPunct{\mcitedefaultmidpunct}
{\mcitedefaultendpunct}{\mcitedefaultseppunct}\relax
\EndOfBibitem
\bibitem[Ramakrishnan \latin{et~al.}(2015)Ramakrishnan, Hartmann, Tapavicza,
  and von Lilienfeld]{Ramakrishnan2015JCP}
Ramakrishnan,~R.; Hartmann,~M.; Tapavicza,~E.; von Lilienfeld,~O.~A. Electronic
  Spectra from {TDDFT} and Machine Learning in Chemical Space. \emph{J. Chem.
  Phys.} \textbf{2015}, \emph{143}, 084111\relax
\mciteBstWouldAddEndPuncttrue
\mciteSetBstMidEndSepPunct{\mcitedefaultmidpunct}
{\mcitedefaultendpunct}{\mcitedefaultseppunct}\relax
\EndOfBibitem
\bibitem[Dral \latin{et~al.}(2020)Dral, Owens, Dral, and
  Cs\'{a}nyi]{Dral2020JCP}
Dral,~P.~O.; Owens,~A.; Dral,~A.; Cs\'{a}nyi,~G. Hierarchical Machine Learning
  of Potential Energy Surfaces. \emph{J. Chem. Phys.} \textbf{2020},
  \emph{152}, 204110\relax
\mciteBstWouldAddEndPuncttrue
\mciteSetBstMidEndSepPunct{\mcitedefaultmidpunct}
{\mcitedefaultendpunct}{\mcitedefaultseppunct}\relax
\EndOfBibitem
\bibitem[Smith \latin{et~al.}(2019)Smith, Nebgen, Zubatyuk, Lubbers, Devereuz,
  Barros, Tretiak, Isayev, and Roitberg]{Smith2019NC}
Smith,~J.~S.; Nebgen,~B.~T.; Zubatyuk,~R.; Lubbers,~N.; Devereuz,~C.;
  Barros,~K.; Tretiak,~S.; Isayev,~O.; Roitberg,~A.~E. Approaching Coupled
  Cluster Accuracy with a General-Purpose Neural Network Potential Through
  Transfer Learning. \emph{Nat. Commun.} \textbf{2019}, \emph{10}\relax
\mciteBstWouldAddEndPuncttrue
\mciteSetBstMidEndSepPunct{\mcitedefaultmidpunct}
{\mcitedefaultendpunct}{\mcitedefaultseppunct}\relax
\EndOfBibitem
\bibitem[Abedi \latin{et~al.}(2010)Abedi, Maitra, and Gross]{Abedi2010PRL}
Abedi,~A.; Maitra,~N.~T.; Gross,~E. K.~U. Exact Factorization of the
  Time-Dependent Electron-Nuclear Wave Function. \emph{Phys. Rev. Lett.}
  \textbf{2010}, \emph{105}, 123002\relax
\mciteBstWouldAddEndPuncttrue
\mciteSetBstMidEndSepPunct{\mcitedefaultmidpunct}
{\mcitedefaultendpunct}{\mcitedefaultseppunct}\relax
\EndOfBibitem
\bibitem[Thachuk \latin{et~al.}(1996)Thachuk, Ivanov, and
  Wardlaw]{Thachuk1996JCP}
Thachuk,~M.; Ivanov,~M.~Y.; Wardlaw,~D.~M. A Semiclassical Approach to
  Intense‐Field Above‐Threshold Dissociation in the Long Wavelength Limit.
  \emph{J. Chem. Phys.} \textbf{1996}, \emph{105}, 4094--4104\relax
\mciteBstWouldAddEndPuncttrue
\mciteSetBstMidEndSepPunct{\mcitedefaultmidpunct}
{\mcitedefaultendpunct}{\mcitedefaultseppunct}\relax
\EndOfBibitem
\bibitem[Mitri\ifmmode~\acute{c}\else \'{c}\fi{}
  \latin{et~al.}(2009)Mitri\ifmmode~\acute{c}\else \'{c}\fi{}, Petersen, and
  Bona\ifmmode \check{c}\else \v{c}\fi{}i\ifmmode \acute{c}\else~\'{c}\fi{}
  Kouteck\'y]{Mitric2009PRA}
Mitri\ifmmode~\acute{c}\else \'{c}\fi{},~R.; Petersen,~J.; Bona\ifmmode
  \check{c}\else \v{c}\fi{}i\ifmmode \acute{c}\else~\'{c}\fi{} Kouteck\'y,~V.
  Laser-Field-Induced Surface-Hopping Method for the Simulation and Control of
  Ultrafast Photodynamics. \emph{Phys. Rev. A} \textbf{2009}, \emph{79},
  053416\relax
\mciteBstWouldAddEndPuncttrue
\mciteSetBstMidEndSepPunct{\mcitedefaultmidpunct}
{\mcitedefaultendpunct}{\mcitedefaultseppunct}\relax
\EndOfBibitem
\bibitem[Mitri\'{c} \latin{et~al.}(2011)Mitri\'{c}, Petersen, Wohlgemuth,
  Werner, and Bona\c{c}i\'{c}-Kouteck\'{y}]{Mitric2011PCCP}
Mitri\'{c},~R.; Petersen,~J.; Wohlgemuth,~M.; Werner,~U.;
  Bona\c{c}i\'{c}-Kouteck\'{y},~V. Field-Induced Surface Hopping Method for
  Probing Transition State Nonadiabatic Dynamics of Ag{$_3$}. \emph{Phys. Chem.
  Chem. Phys.} \textbf{2011}, \emph{13}, 8690--8696\relax
\mciteBstWouldAddEndPuncttrue
\mciteSetBstMidEndSepPunct{\mcitedefaultmidpunct}
{\mcitedefaultendpunct}{\mcitedefaultseppunct}\relax
\EndOfBibitem
\bibitem[Granucci \latin{et~al.}(2012)Granucci, Persico, and
  Spighi]{Granucci2012JCP}
Granucci,~G.; Persico,~M.; Spighi,~G. Surface Hopping Trajectory Simulations
  with Spin-Orbit and Dynamical Couplings. \emph{J. Chem. Phys.} \textbf{2012},
  \emph{137}, 22A501\relax
\mciteBstWouldAddEndPuncttrue
\mciteSetBstMidEndSepPunct{\mcitedefaultmidpunct}
{\mcitedefaultendpunct}{\mcitedefaultseppunct}\relax
\EndOfBibitem
\bibitem[Baer(2002)]{Baer2002PR}
Baer,~M. Introduction to the Theory of Electronic Non-Adiabatic Coupling Terms
  in Molecular Systems. \emph{Phys. Rep.} \textbf{2002}, \emph{358},
  75--142\relax
\mciteBstWouldAddEndPuncttrue
\mciteSetBstMidEndSepPunct{\mcitedefaultmidpunct}
{\mcitedefaultendpunct}{\mcitedefaultseppunct}\relax
\EndOfBibitem
\bibitem[Kryachko and Yarkony(2000)Kryachko, and Yarkony]{Kryachko2000IJQC}
Kryachko,~E.~S.; Yarkony,~D.~R. {Diabatic Bases and Molecular Properties}.
  \emph{Int. J. Quantum Chem.} \textbf{2000}, \emph{76}, 235--243\relax
\mciteBstWouldAddEndPuncttrue
\mciteSetBstMidEndSepPunct{\mcitedefaultmidpunct}
{\mcitedefaultendpunct}{\mcitedefaultseppunct}\relax
\EndOfBibitem
\bibitem[Marian(2012)]{Marian2012WIRE}
Marian,~C.~M. {Spin–Orbit Coupling and Intersystem Crossing in Molecules}.
  \emph{WIREs Comput. Mol. Sci.} \textbf{2012}, \emph{2}, 187--203\relax
\mciteBstWouldAddEndPuncttrue
\mciteSetBstMidEndSepPunct{\mcitedefaultmidpunct}
{\mcitedefaultendpunct}{\mcitedefaultseppunct}\relax
\EndOfBibitem
\bibitem[K.~G.~Dyall(2007)]{Dyall2007}
K.~G.~Dyall,~K.~F. \emph{{ Introduction to Relativistic Quantum Chemistry}};
  Oxford University Press, 2007\relax
\mciteBstWouldAddEndPuncttrue
\mciteSetBstMidEndSepPunct{\mcitedefaultmidpunct}
{\mcitedefaultendpunct}{\mcitedefaultseppunct}\relax
\EndOfBibitem
\bibitem[M.~Reiher(2009)]{Reiher2009}
M.~Reiher,~A.~W. \emph{{Relativistic Quantum Chemistry}}; Wiley VCH Verlag
  Weinheim, 2009\relax
\mciteBstWouldAddEndPuncttrue
\mciteSetBstMidEndSepPunct{\mcitedefaultmidpunct}
{\mcitedefaultendpunct}{\mcitedefaultseppunct}\relax
\EndOfBibitem
\bibitem[H.~A.~Bethe(1957)]{Bethe1957}
H.~A.~Bethe,~E. E.~S. \emph{{Quantum Mechanics of One- and Two-Electron
  Atoms}}; Springer, Berlin, 1957\relax
\mciteBstWouldAddEndPuncttrue
\mciteSetBstMidEndSepPunct{\mcitedefaultmidpunct}
{\mcitedefaultendpunct}{\mcitedefaultseppunct}\relax
\EndOfBibitem
\bibitem[Pyykko(1988)]{Pyykko1988CR}
Pyykko,~P. {Relativistic Effects in Structural Chemistry}. \emph{Chem. Rev.}
  \textbf{1988}, \emph{88}, 563--594\relax
\mciteBstWouldAddEndPuncttrue
\mciteSetBstMidEndSepPunct{\mcitedefaultmidpunct}
{\mcitedefaultendpunct}{\mcitedefaultseppunct}\relax
\EndOfBibitem
\bibitem[Neese \latin{et~al.}(2007)Neese, Petrenko, Ganyushin, and
  Olbrich]{Neese2007CCR}
Neese,~F.; Petrenko,~T.; Ganyushin,~D.; Olbrich,~G. {Advanced Aspects of
  {\emph{Ab Initio}} Theoretical Optical Spectroscopy of Transition Metal
  complexes: Multiplets, spin-orbit coupling and resonance Raman intensities}.
  \emph{Coord. Chem. Rev.} \textbf{2007}, \emph{251}, 288 -- 327\relax
\mciteBstWouldAddEndPuncttrue
\mciteSetBstMidEndSepPunct{\mcitedefaultmidpunct}
{\mcitedefaultendpunct}{\mcitedefaultseppunct}\relax
\EndOfBibitem
\bibitem[Neese(2005)]{Neese2005JCP}
Neese,~F. Efficient and Accurate Approximations to the Molecular Spin-Orbit
  Coupling Operator and their Use in Molecular G-Tensor Calculations. \emph{J.
  Chem. Phys.} \textbf{2005}, \emph{122}, 034107\relax
\mciteBstWouldAddEndPuncttrue
\mciteSetBstMidEndSepPunct{\mcitedefaultmidpunct}
{\mcitedefaultendpunct}{\mcitedefaultseppunct}\relax
\EndOfBibitem
\bibitem[Richter \latin{et~al.}(2011)Richter, Marquetand,
  Gonz\'{a}lez-V\'{a}zquez, Sola, and Gonz\'{a}lez]{Richter2011JCTC}
Richter,~M.; Marquetand,~P.; Gonz\'{a}lez-V\'{a}zquez,~J.; Sola,~I.;
  Gonz\'{a}lez,~L. {SHARC:} {Ab} Initio Molecular Dynamics with Surface Hopping
  in the Adiabatic Representation Including Arbitrary couplings. \emph{J. Chem.
  Theory Comput.} \textbf{2011}, \emph{7}, 1253--1258\relax
\mciteBstWouldAddEndPuncttrue
\mciteSetBstMidEndSepPunct{\mcitedefaultmidpunct}
{\mcitedefaultendpunct}{\mcitedefaultseppunct}\relax
\EndOfBibitem
\bibitem[Mai \latin{et~al.}(2015)Mai, Marquetand, and
  Gonz\'{a}lez]{Mai2015IJQC}
Mai,~S.; Marquetand,~P.; Gonz\'{a}lez,~L. A General Method to Describe
  Intersystem Crossing Dynamics in Trajectory Surface Hopping. \emph{Int. J.
  Quantum Chem.} \textbf{2015}, \emph{115}, 1215--1231\relax
\mciteBstWouldAddEndPuncttrue
\mciteSetBstMidEndSepPunct{\mcitedefaultmidpunct}
{\mcitedefaultendpunct}{\mcitedefaultseppunct}\relax
\EndOfBibitem
\bibitem[Mai \latin{et~al.}(2018)Mai, Plasser, Marquetand, and
  González]{Mai2018}
Mai,~S.; Plasser,~F.; Marquetand,~P.; González,~L. \emph{Attosecond Molecular
  Dynamics}; The Royal Society of Chemistry, 2018; pp 348--385\relax
\mciteBstWouldAddEndPuncttrue
\mciteSetBstMidEndSepPunct{\mcitedefaultmidpunct}
{\mcitedefaultendpunct}{\mcitedefaultseppunct}\relax
\EndOfBibitem
\bibitem[K\"{o}ppel \latin{et~al.}(2001)K\"{o}ppel, Gronki, and
  Mahapatra]{Koeppel2001JCP}
K\"{o}ppel,~H.; Gronki,~J.; Mahapatra,~S. Construction Scheme for Regularized
  Diabatic States. \emph{J. Chem. Phys.} \textbf{2001}, \emph{115},
  2377--2388\relax
\mciteBstWouldAddEndPuncttrue
\mciteSetBstMidEndSepPunct{\mcitedefaultmidpunct}
{\mcitedefaultendpunct}{\mcitedefaultseppunct}\relax
\EndOfBibitem
\bibitem[Richings and Worth(2015)Richings, and Worth]{Richings2015JPCA}
Richings,~G.~W.; Worth,~G.~A. A Practical Diabatisation Scheme for Use with the
  Direct-Dynamics Variational Multi-configuration Gaussian Method. \emph{J.
  Phys. Chem. A} \textbf{2015}, \emph{119}, 12457--12470\relax
\mciteBstWouldAddEndPuncttrue
\mciteSetBstMidEndSepPunct{\mcitedefaultmidpunct}
{\mcitedefaultendpunct}{\mcitedefaultseppunct}\relax
\EndOfBibitem
\bibitem[Accomasso \latin{et~al.}(2019)Accomasso, Persico, and
  Granucci]{Accomasso2019CPC}
Accomasso,~D.; Persico,~M.; Granucci,~G. Diabatization by Localization in the
  Framework of Configuration Interaction Based on Floating Occupation Molecular
  Orbitals ({FOMO-CI}). \emph{ChemPhotoChem} \textbf{2019}, \emph{3},
  933--944\relax
\mciteBstWouldAddEndPuncttrue
\mciteSetBstMidEndSepPunct{\mcitedefaultmidpunct}
{\mcitedefaultendpunct}{\mcitedefaultseppunct}\relax
\EndOfBibitem
\bibitem[Lenzen and Manthe(2017)Lenzen, and Manthe]{Lenzen2017JCP}
Lenzen,~T.; Manthe,~U. Neural Network Based Coupled Diabatic Potential Energy
  Surfaces for Reactive Scattering. \emph{J. Chem. Phys.} \textbf{2017},
  \emph{147}, 084105\relax
\mciteBstWouldAddEndPuncttrue
\mciteSetBstMidEndSepPunct{\mcitedefaultmidpunct}
{\mcitedefaultendpunct}{\mcitedefaultseppunct}\relax
\EndOfBibitem
\bibitem[Subotnik \latin{et~al.}(2008)Subotnik, Yeganeh, Cave, and
  Ratner]{Subotnik2008JCP}
Subotnik,~J.~E.; Yeganeh,~S.; Cave,~R.~J.; Ratner,~M.~A. {Constructing Diabatic
  States from Adiabatic States: Extending Generalized Mulliken–Hush to
  Multiple Charge Centers with Boys Localization}. \emph{J. Chem. Phys.}
  \textbf{2008}, \emph{129}, 244101\relax
\mciteBstWouldAddEndPuncttrue
\mciteSetBstMidEndSepPunct{\mcitedefaultmidpunct}
{\mcitedefaultendpunct}{\mcitedefaultseppunct}\relax
\EndOfBibitem
\bibitem[Hoyer \latin{et~al.}(2016)Hoyer, Parker, Gagliardi, and
  Truhlar]{Hoyer2016JCP}
Hoyer,~C.~E.; Parker,~K.; Gagliardi,~L.; Truhlar,~D.~G. The {DQ} and {DQ\O}
  Electronic Structure Diabatization Methods: Validation for General
  Applications. \emph{J. Chem. Phys.} \textbf{2016}, \emph{144}, 194101\relax
\mciteBstWouldAddEndPuncttrue
\mciteSetBstMidEndSepPunct{\mcitedefaultmidpunct}
{\mcitedefaultendpunct}{\mcitedefaultseppunct}\relax
\EndOfBibitem
\bibitem[Wittenbrink \latin{et~al.}(2013)Wittenbrink, Ndome, and
  Eisfeld]{Wittenbrink2013JPCA}
Wittenbrink,~N.; Ndome,~H.; Eisfeld,~W. Toward Spin–Orbit coupled Diabatic
  Potential Energy Surfaces for Methyl Iodide Using Effective Relativistic
  Coupling by Asymptotic Representation. \emph{J. Phys. Chem. A} \textbf{2013},
  \emph{117}, 7408--7420\relax
\mciteBstWouldAddEndPuncttrue
\mciteSetBstMidEndSepPunct{\mcitedefaultmidpunct}
{\mcitedefaultendpunct}{\mcitedefaultseppunct}\relax
\EndOfBibitem
\bibitem[Varga \latin{et~al.}(2018)Varga, Parker, and Truhlar]{Varga2018PCCP}
Varga,~Z.; Parker,~K.~A.; Truhlar,~D.~G. Direct Diabatization Based on
  Nonadiabatic Couplings: The N/D Method. \emph{Phys. Chem. Chem. Phys.}
  \textbf{2018}, \emph{20}, 26643--26659\relax
\mciteBstWouldAddEndPuncttrue
\mciteSetBstMidEndSepPunct{\mcitedefaultmidpunct}
{\mcitedefaultendpunct}{\mcitedefaultseppunct}\relax
\EndOfBibitem
\bibitem[Nakamura and Truhlar(2002)Nakamura, and Truhlar]{Nakamura2002JCP}
Nakamura,~H.; Truhlar,~D.~G. Direct Diabatization of Electronic States by the
  Fourfold Way. II. Dynamical Correlation and Rearrangement Processes. \emph{J.
  Chem. Phys.} \textbf{2002}, \emph{117}, 5576--5593\relax
\mciteBstWouldAddEndPuncttrue
\mciteSetBstMidEndSepPunct{\mcitedefaultmidpunct}
{\mcitedefaultendpunct}{\mcitedefaultseppunct}\relax
\EndOfBibitem
\bibitem[Cave and Stanton(2014)Cave, and Stanton]{Cave2014JCP}
Cave,~R.~J.; Stanton,~J.~F. Block Diagonalization of the Equation-of-Motion
  Coupled Cluster Effective Hamiltonian: Treatment of Diabatic Potential
  Constants and Triple Excitations. \emph{J. Chem. Phys.} \textbf{2014},
  \emph{140}, 214112\relax
\mciteBstWouldAddEndPuncttrue
\mciteSetBstMidEndSepPunct{\mcitedefaultmidpunct}
{\mcitedefaultendpunct}{\mcitedefaultseppunct}\relax
\EndOfBibitem
\bibitem[Venghaus and Eisfeld(2016)Venghaus, and Eisfeld]{Venghaus2016JCP}
Venghaus,~F.; Eisfeld,~W. Block-Diagonalization as a Tool for the Robust
  Diabatization of High-Dimensional Potential Energy Surfaces. \emph{J. Chem.
  Phys.} \textbf{2016}, \emph{144}, 114110\relax
\mciteBstWouldAddEndPuncttrue
\mciteSetBstMidEndSepPunct{\mcitedefaultmidpunct}
{\mcitedefaultendpunct}{\mcitedefaultseppunct}\relax
\EndOfBibitem
\bibitem[Robertson \latin{et~al.}(2019)Robertson, Gonz\'{a}lez-V\'{a}zquez,
  corral, D\'{i}az-Tendero, and D\'{i}az]{Robertson2019JCC}
Robertson,~C.; Gonz\'{a}lez-V\'{a}zquez,~J.; corral,~I.; D\'{i}az-Tendero,~S.;
  D\'{i}az,~C. Nonadiabatic Scattering of {NO} off {A}u3 Clusters: A Simple and
  Robust Diabatic State Manifold Generation Method for Multiconfigurational
  Wavefunctions. \emph{J. Comput. Chem.} \textbf{2019}, \emph{40},
  794--810\relax
\mciteBstWouldAddEndPuncttrue
\mciteSetBstMidEndSepPunct{\mcitedefaultmidpunct}
{\mcitedefaultendpunct}{\mcitedefaultseppunct}\relax
\EndOfBibitem
\bibitem[Li \latin{et~al.}(2013)Li, Jiang, and Guo]{Li2013JCP}
Li,~J.; Jiang,~B.; Guo,~H. Permutation Invariant Polynomial Neural Network
  Approach to Fitting Potential Energy Surfaces. {II}. {Four}-Atom Systems.
  \emph{J. Chem. Phys.} \textbf{2013}, \emph{139}, 204103\relax
\mciteBstWouldAddEndPuncttrue
\mciteSetBstMidEndSepPunct{\mcitedefaultmidpunct}
{\mcitedefaultendpunct}{\mcitedefaultseppunct}\relax
\EndOfBibitem
\bibitem[Jiang and Guo(2013)Jiang, and Guo]{Jiang2013JCP}
Jiang,~B.; Guo,~H. Permutation Invariant Polynomial Neural Network Approach to
  Fitting Potential Energy Surfaces. \emph{J. Chem. Phys.} \textbf{2013},
  \emph{139}, 054112\relax
\mciteBstWouldAddEndPuncttrue
\mciteSetBstMidEndSepPunct{\mcitedefaultmidpunct}
{\mcitedefaultendpunct}{\mcitedefaultseppunct}\relax
\EndOfBibitem
\bibitem[Jiang and Guo(2014)Jiang, and Guo]{Jiang2014JCP}
Jiang,~B.; Guo,~H. Permutation Invariant Polynomial Neural Network Approach to
  Fitting Potential Energy Surfaces. {III}. Molecule-Surface Interactions.
  \emph{J. Chem. Phys.} \textbf{2014}, \emph{141}, 034109\relax
\mciteBstWouldAddEndPuncttrue
\mciteSetBstMidEndSepPunct{\mcitedefaultmidpunct}
{\mcitedefaultendpunct}{\mcitedefaultseppunct}\relax
\EndOfBibitem
\bibitem[Jiang \latin{et~al.}(2016)Jiang, Li, and Guo]{Jiang2016IRPC}
Jiang,~B.; Li,~J.; Guo,~H. Potential Energy Surfaces from High Fidelity Fitting
  of \emph{Ab Initio} Points: The Permutation Invariant Polynomial - Neural
  Network Approach. \emph{Int. Rev. Phys. Chem.} \textbf{2016}, \emph{35},
  479--506\relax
\mciteBstWouldAddEndPuncttrue
\mciteSetBstMidEndSepPunct{\mcitedefaultmidpunct}
{\mcitedefaultendpunct}{\mcitedefaultseppunct}\relax
\EndOfBibitem
\bibitem[Mai \latin{et~al.}(2014)Mai, M.~Richter, Marquetand, and
  Gonz\'alez]{Mai2014}
Mai,~S.; M.~Richter,~M.; Marquetand,~P.; Gonz\'alez,~L. Excitation of
  Nucleobases from a Computational Perspective II: Dynamics. \textbf{2014},
  \relax
\mciteBstWouldAddEndPunctfalse
\mciteSetBstMidEndSepPunct{\mcitedefaultmidpunct}
{}{\mcitedefaultseppunct}\relax
\EndOfBibitem
\bibitem[Liu(2020)]{Liu2020JCP}
Liu,~W. Essentials of Relativistic Quantum Chemistry. \emph{J. Chem. Phys.}
  \textbf{2020}, \emph{152}, 180901\relax
\mciteBstWouldAddEndPuncttrue
\mciteSetBstMidEndSepPunct{\mcitedefaultmidpunct}
{\mcitedefaultendpunct}{\mcitedefaultseppunct}\relax
\EndOfBibitem
\bibitem[Horton \latin{et~al.}(2019)Horton, Liu, Forbes, Makhija, Lausten,
  Stolow, Hockett, Marquetand, Rozgonyi, and Weinacht]{Horton2019JCP}
Horton,~S.~L.; Liu,~Y.; Forbes,~R.; Makhija,~V.; Lausten,~R.; Stolow,~A.;
  Hockett,~P.; Marquetand,~P.; Rozgonyi,~T.; Weinacht,~T. {Excited state
  dynamics of CH$_2$I$_2$ and CH$_2$BrI studied with UV pump VUV probe
  photoelectron spectroscopy}. \emph{J. Chem. Phys.} \textbf{2019}, \emph{150},
  174201\relax
\mciteBstWouldAddEndPuncttrue
\mciteSetBstMidEndSepPunct{\mcitedefaultmidpunct}
{\mcitedefaultendpunct}{\mcitedefaultseppunct}\relax
\EndOfBibitem
\bibitem[Horton \latin{et~al.}(2018)Horton, Liu, Chakraborty, Marquetand,
  Rozgonyi, Matsika, and Weinacht]{Horton2018PRA}
Horton,~S.~L.; Liu,~Y.; Chakraborty,~P.; Marquetand,~P.; Rozgonyi,~T.;
  Matsika,~S.; Weinacht,~T. Strong-Field- Versus Weak-Field-Ionization
  Pump-Probe Spectroscopy. \emph{Phys. Rev. A} \textbf{2018}, \emph{98},
  053416\relax
\mciteBstWouldAddEndPuncttrue
\mciteSetBstMidEndSepPunct{\mcitedefaultmidpunct}
{\mcitedefaultendpunct}{\mcitedefaultseppunct}\relax
\EndOfBibitem
\bibitem[Sussman \latin{et~al.}(2006)Sussman, Townsend, Ivanov, and
  Stolow]{Sussman2006S}
Sussman,~B.~J.; Townsend,~D.; Ivanov,~M.~Y.; Stolow,~A. Dynamic Stark Control
  of Photochemical Processes. \emph{Science} \textbf{2006}, \emph{314},
  278--281\relax
\mciteBstWouldAddEndPuncttrue
\mciteSetBstMidEndSepPunct{\mcitedefaultmidpunct}
{\mcitedefaultendpunct}{\mcitedefaultseppunct}\relax
\EndOfBibitem
\bibitem[Marquetand \latin{et~al.}(2011)Marquetand, Richter,
  {Gonz\'alez-V\'azquez}, Sola, and Gonz\'alez]{Marquetand2011FD}
Marquetand,~P.; Richter,~M.; {Gonz\'alez-V\'azquez},~J.; Sola,~I.;
  Gonz\'alez,~L. Nonadiabatic Ab Initio Molecular Dynamics Including Spin-Orbit
  Coupling and Laser Fields. \emph{Faraday Discuss.} \textbf{2011}, \emph{153},
  261--273\relax
\mciteBstWouldAddEndPuncttrue
\mciteSetBstMidEndSepPunct{\mcitedefaultmidpunct}
{\mcitedefaultendpunct}{\mcitedefaultseppunct}\relax
\EndOfBibitem
\bibitem[Bajo \latin{et~al.}(2012)Bajo, {Gonz\'alez-V\'azquez}, Sola,
  Santamaria, Richter, Marquetand, and Gonz\'alez]{Bajo2012JPCA}
Bajo,~J.~J.; {Gonz\'alez-V\'azquez},~J.; Sola,~I.; Santamaria,~J.; Richter,~M.;
  Marquetand,~P.; Gonz\'alez,~L. Mixed Quantum-Classical Dynamics in the
  Adiabatic Representation to Simulate Molecules Driven by Strong Laser Pulses.
  \emph{J. Phys. Chem. A} \textbf{2012}, \emph{116}, 2800--2807\relax
\mciteBstWouldAddEndPuncttrue
\mciteSetBstMidEndSepPunct{\mcitedefaultmidpunct}
{\mcitedefaultendpunct}{\mcitedefaultseppunct}\relax
\EndOfBibitem
\bibitem[K{\"o}ppel \latin{et~al.}(1984)K{\"o}ppel, Domcke, and
  Cederbaum]{Koeppel84ACP}
K{\"o}ppel,~H.; Domcke,~W.; Cederbaum,~L.~S. Multimode Molecular Dynamics
  Beyond the Born-Oppenheimer Approximation. \emph{Adv. Chem. Phys.}
  \textbf{1984}, \emph{57}, 59--246\relax
\mciteBstWouldAddEndPuncttrue
\mciteSetBstMidEndSepPunct{\mcitedefaultmidpunct}
{\mcitedefaultendpunct}{\mcitedefaultseppunct}\relax
\EndOfBibitem
\bibitem[Ben-Nun and Mart\'{i}nez(2002)Ben-Nun, and Mart\'{i}nez]{Ben-Nun2002}
Ben-Nun,~M.; Mart\'{i}nez,~T.~J. \emph{Advances in Chemical Physics}; John
  Wiley \& Sons, Ltd, 2002; pp 439--512\relax
\mciteBstWouldAddEndPuncttrue
\mciteSetBstMidEndSepPunct{\mcitedefaultmidpunct}
{\mcitedefaultendpunct}{\mcitedefaultseppunct}\relax
\EndOfBibitem
\bibitem[Beck \latin{et~al.}(2000)Beck, Jäckle, Worth, and Meyer]{Beck2000PR}
Beck,~M.; Jäckle,~A.; Worth,~G.; Meyer,~H.-D. The Multiconfiguration
  Time-Dependent Hartree (MCTDH) Method: A Highly Efficient Algorithm for
  Propagating Wavepackets. \emph{Phys. Rep.} \textbf{2000}, \emph{324},
  1--105\relax
\mciteBstWouldAddEndPuncttrue
\mciteSetBstMidEndSepPunct{\mcitedefaultmidpunct}
{\mcitedefaultendpunct}{\mcitedefaultseppunct}\relax
\EndOfBibitem
\bibitem[Yeager and J{\o}rgensen(1979)Yeager, and J{\o}rgensen]{Yeager1979CPL}
Yeager,~D.~L.; J{\o}rgensen,~P. A Multiconfigurational Time-Dependent
  Hartree-Fock Approach. \emph{Chem. Phys. Lett.} \textbf{1979}, \emph{65},
  77--80\relax
\mciteBstWouldAddEndPuncttrue
\mciteSetBstMidEndSepPunct{\mcitedefaultmidpunct}
{\mcitedefaultendpunct}{\mcitedefaultseppunct}\relax
\EndOfBibitem
\bibitem[Manthe(2017)]{Manthe2017JPCM}
Manthe,~U. Wavepacket Dynamics and the Multi-Configurational Time-Dependent
  Hartree Approach. \emph{J. Phys.: Condens. Matter} \textbf{2017}, \emph{29},
  253001\relax
\mciteBstWouldAddEndPuncttrue
\mciteSetBstMidEndSepPunct{\mcitedefaultmidpunct}
{\mcitedefaultendpunct}{\mcitedefaultseppunct}\relax
\EndOfBibitem
\bibitem[Eng \latin{et~al.}(2015)Eng, Gourlaouen, Gindensperger, and
  Daniel]{Eng2015ACR}
Eng,~J.; Gourlaouen,~C.; Gindensperger,~E.; Daniel,~C. Spin-Vibronic Quantum
  Dynamics for Ultrafast Excited-State Processes. \emph{Acc. Chem. Res.}
  \textbf{2015}, \emph{48}, 809--817\relax
\mciteBstWouldAddEndPuncttrue
\mciteSetBstMidEndSepPunct{\mcitedefaultmidpunct}
{\mcitedefaultendpunct}{\mcitedefaultseppunct}\relax
\EndOfBibitem
\bibitem[G{\'o}mez \latin{et~al.}(2019)G{\'o}mez, Heindl, Szabadi, and
  Gonz{\'a}lez]{Gomez2019JPCA}
G{\'o}mez,~S.; Heindl,~M.; Szabadi,~A.; Gonz{\'a}lez,~L. From Surface Hopping
  to Quantum Dynamics and Back. Finding Essential Electronic and Nuclear
  Degrees of Freedom and Optimal Surface Hopping Parameters. \emph{J. Phys.
  Chem. A} \textbf{2019}, \emph{123}, 8321--8332\relax
\mciteBstWouldAddEndPuncttrue
\mciteSetBstMidEndSepPunct{\mcitedefaultmidpunct}
{\mcitedefaultendpunct}{\mcitedefaultseppunct}\relax
\EndOfBibitem
\bibitem[Ischtwan and Collins(1994)Ischtwan, and Collins]{Ischtwan1994JCP}
Ischtwan,~J.; Collins,~M.~A. Molecular Potential Energy Surfaces by
  Interpolation. \emph{J. Chem. Phys.} \textbf{1994}, \emph{100},
  8080--8088\relax
\mciteBstWouldAddEndPuncttrue
\mciteSetBstMidEndSepPunct{\mcitedefaultmidpunct}
{\mcitedefaultendpunct}{\mcitedefaultseppunct}\relax
\EndOfBibitem
\bibitem[Evenhuis and collins(2004)Evenhuis, and collins]{Evenhuis2004JCP}
Evenhuis,~C.~R.; collins,~M.~A. Interpolation of Diabatic Potential Energy
  Surfaces. \emph{J. Chem. Phys.} \textbf{2004}, \emph{121}, 2515--2527\relax
\mciteBstWouldAddEndPuncttrue
\mciteSetBstMidEndSepPunct{\mcitedefaultmidpunct}
{\mcitedefaultendpunct}{\mcitedefaultseppunct}\relax
\EndOfBibitem
\bibitem[Evenhuis and Mart{\'i}nez(2011)Evenhuis, and
  Mart{\'i}nez]{Evenhuis2011JCP}
Evenhuis,~C.; Mart{\'i}nez,~T.~J. A Scheme to Interpolate Potential Energy
  Surfaces and Derivative Coupling Vectors without Performing a Global
  Diabatization. \emph{J. Chem. Phys.} \textbf{2011}, \emph{135}, 224110\relax
\mciteBstWouldAddEndPuncttrue
\mciteSetBstMidEndSepPunct{\mcitedefaultmidpunct}
{\mcitedefaultendpunct}{\mcitedefaultseppunct}\relax
\EndOfBibitem
\bibitem[Mukherjee \latin{et~al.}(2013)Mukherjee, Bandyopadhyay, Paul, and
  Adhikari]{Mukherjee2013JPCA}
Mukherjee,~S.; Bandyopadhyay,~S.; Paul,~A.~K.; Adhikari,~S. {Construction of
  Diabatic {Hamiltonian} Matrix from \emph{Ab Initio} Calculated Molecular
  Symmetry Adapted Nonadiabatic Coupling Terms and Nuclear Dynamics for the
  Excited States of Na$_3$ Cluster}. \emph{J. Phys. Chem. A} \textbf{2013},
  \emph{117}, 3475--3495\relax
\mciteBstWouldAddEndPuncttrue
\mciteSetBstMidEndSepPunct{\mcitedefaultmidpunct}
{\mcitedefaultendpunct}{\mcitedefaultseppunct}\relax
\EndOfBibitem
\bibitem[Worth \latin{et~al.}(2008)Worth, Robb, and Lasorne]{Worth2008MP}
Worth,~G.; Robb,~M.; Lasorne,~B. Solving the Time-Dependent Schr\"{o}dinger
  Equation for Nuclear Motion in One Step: Direct Dynamics of Non-Adiabatic
  Systems. \emph{Mol. Phys.} \textbf{2008}, \emph{106}, 2077--2091\relax
\mciteBstWouldAddEndPuncttrue
\mciteSetBstMidEndSepPunct{\mcitedefaultmidpunct}
{\mcitedefaultendpunct}{\mcitedefaultseppunct}\relax
\EndOfBibitem
\bibitem[Persico and Granucci(2014)Persico, and Granucci]{Persico2014TCA}
Persico,~M.; Granucci,~G. An Overview of Nonadiabatic Dynamics Simulations
  Methods, with Focus on the Direct Approach Versus the Fitting of Potential
  Energy Surfaces. \emph{Theor. Chem. Acc.} \textbf{2014}, \emph{133},
  1526\relax
\mciteBstWouldAddEndPuncttrue
\mciteSetBstMidEndSepPunct{\mcitedefaultmidpunct}
{\mcitedefaultendpunct}{\mcitedefaultseppunct}\relax
\EndOfBibitem
\bibitem[Komarova \latin{et~al.}(2018)Komarova, Remacle, and
  Levine]{Komarova2018PCL}
Komarova,~K.~G.; Remacle,~F.; Levine,~R. On the Fly Quantum Dynamics of
  Electronic and Nuclear Wave Packets. \emph{Chem. Phys. Lett.} \textbf{2018},
  \emph{699}, 155 -- 161\relax
\mciteBstWouldAddEndPuncttrue
\mciteSetBstMidEndSepPunct{\mcitedefaultmidpunct}
{\mcitedefaultendpunct}{\mcitedefaultseppunct}\relax
\EndOfBibitem
\bibitem[Lasorne \latin{et~al.}(2007)Lasorne, Robb, and Worth]{Lasorne2007PCCP}
Lasorne,~B.; Robb,~M.~A.; Worth,~G.~A. Direct Quantum Dynamics using
  Variational Multi-Configuration Gaussian Wavepackets. Implementation Details
  and Test Case. \emph{Phys. Chem. Chem. Phys.} \textbf{2007}, \emph{9},
  3210--3227\relax
\mciteBstWouldAddEndPuncttrue
\mciteSetBstMidEndSepPunct{\mcitedefaultmidpunct}
{\mcitedefaultendpunct}{\mcitedefaultseppunct}\relax
\EndOfBibitem
\bibitem[Ben-Nun and Mart\'{i}nez(2000)Ben-Nun, and
  Mart\'{i}nez]{Ben-Nun2000CP}
Ben-Nun,~M.; Mart\'{i}nez,~T.~J. Photodynamics of Ethylene: \emph{Ab Initio}
  Studies of Conical Intersections. \emph{Chem. Phys.} \textbf{2000},
  \emph{259}, 237 -- 248\relax
\mciteBstWouldAddEndPuncttrue
\mciteSetBstMidEndSepPunct{\mcitedefaultmidpunct}
{\mcitedefaultendpunct}{\mcitedefaultseppunct}\relax
\EndOfBibitem
\bibitem[Curchod \latin{et~al.}(2016)Curchod, Rauer, Marquetand, Gonz\'{a}lez,
  and Mart\'{i}nez]{Curchod2016JCP}
Curchod,~B. F.~E.; Rauer,~C.; Marquetand,~P.; Gonz\'{a}lez,~L.;
  Mart\'{i}nez,~T.~J. Communication: GAIMS—Generalized \emph{Ab Initio}
  Multiple Spawning for Both Internal Conversion and Intersystem Crossing
  Processes. \emph{J. Chem. Phys.} \textbf{2016}, \emph{144}, 101102\relax
\mciteBstWouldAddEndPuncttrue
\mciteSetBstMidEndSepPunct{\mcitedefaultmidpunct}
{\mcitedefaultendpunct}{\mcitedefaultseppunct}\relax
\EndOfBibitem
\bibitem[Mignolet and Curchod(2018)Mignolet, and Curchod]{Mignolet2018JCP}
Mignolet,~B.; Curchod,~B. F.~E. A Walk Through the Approximations of {\emph{Ab
  Initio}} Multiple Spawning. \emph{J. Chem. Phys.} \textbf{2018}, \emph{148},
  134110\relax
\mciteBstWouldAddEndPuncttrue
\mciteSetBstMidEndSepPunct{\mcitedefaultmidpunct}
{\mcitedefaultendpunct}{\mcitedefaultseppunct}\relax
\EndOfBibitem
\bibitem[Freixas \latin{et~al.}(2018)Freixas, Fernandez-Alberti, Makhov,
  Tretiak, and Shalashilin]{Freixas2018PCCP}
Freixas,~V.~M.; Fernandez-Alberti,~S.; Makhov,~D.~V.; Tretiak,~S.;
  Shalashilin,~D. An {\emph{Ab Initio}} Multiple Cloning Approach for the
  Simulation of Photoinduced Dynamics in Conjugated Molecules. \emph{Phys.
  Chem. Chem. Phys.} \textbf{2018}, \emph{20}, 17762--17772\relax
\mciteBstWouldAddEndPuncttrue
\mciteSetBstMidEndSepPunct{\mcitedefaultmidpunct}
{\mcitedefaultendpunct}{\mcitedefaultseppunct}\relax
\EndOfBibitem
\bibitem[{Tomislav Begu\c{s}i\'c and Aur\'elien Patoz and Miroslav \c{S}ulc and
  Ji\c{r}\'i Van\'i\c{c}ek}(2018)]{Begusic2018CP}
{Tomislav Begu\c{s}i\'c and Aur\'elien Patoz and Miroslav \c{S}ulc and
  Ji\c{r}\'i Van\'i\c{c}ek}, On-the-Fly {\emph{Ab Initio}} Three Thawed
  Gaussians Approximation: A Semiclassical Approach to Herzberg-Teller Spectra.
  \emph{Chem. Phys.} \textbf{2018}, \emph{515}, 152 -- 163\relax
\mciteBstWouldAddEndPuncttrue
\mciteSetBstMidEndSepPunct{\mcitedefaultmidpunct}
{\mcitedefaultendpunct}{\mcitedefaultseppunct}\relax
\EndOfBibitem
\bibitem[Markland and Ceriotti(2018)Markland, and Ceriotti]{Markland2018NRC}
Markland,~T.; Ceriotti,~M. Nuclear Quantum Effects Enter the Mainstream.
  \emph{Nat. Rev. Chem.} \textbf{2018}, \emph{2}\relax
\mciteBstWouldAddEndPuncttrue
\mciteSetBstMidEndSepPunct{\mcitedefaultmidpunct}
{\mcitedefaultendpunct}{\mcitedefaultseppunct}\relax
\EndOfBibitem
\bibitem[Miller(1970)]{Miller1970JCP}
Miller,~W.~H. Classical S Matrix: Numerical Application to Inelastic
  Collisions. \emph{J. Chem. Phys.} \textbf{1970}, \emph{53}, 3578--3587\relax
\mciteBstWouldAddEndPuncttrue
\mciteSetBstMidEndSepPunct{\mcitedefaultmidpunct}
{\mcitedefaultendpunct}{\mcitedefaultseppunct}\relax
\EndOfBibitem
\bibitem[Ceotto \latin{et~al.}(2009)Ceotto, Atahan, Shim, Tantardini, and
  Aspuru-Guzik]{Ceotto2009PCCP}
Ceotto,~M.; Atahan,~S.; Shim,~S.; Tantardini,~G.~F.; Aspuru-Guzik,~A.
  First-Principles Semiclassical Initial Value Representation Molecular
  Dynamics. \emph{Phys. Chem. Chem. Phys.} \textbf{2009}, \emph{11},
  3861--3867\relax
\mciteBstWouldAddEndPuncttrue
\mciteSetBstMidEndSepPunct{\mcitedefaultmidpunct}
{\mcitedefaultendpunct}{\mcitedefaultseppunct}\relax
\EndOfBibitem
\bibitem[Nakamura \latin{et~al.}(2016)Nakamura, Nanbu, Teranishi, and
  Ohta]{Nakamura2016PCCP}
Nakamura,~H.; Nanbu,~S.; Teranishi,~Y.; Ohta,~A. Development of Semiclassical
  Molecular Dynamics Simulation Method. \emph{Phys. Chem. Chem. Phys.}
  \textbf{2016}, \emph{18}, 11972--11985\relax
\mciteBstWouldAddEndPuncttrue
\mciteSetBstMidEndSepPunct{\mcitedefaultmidpunct}
{\mcitedefaultendpunct}{\mcitedefaultseppunct}\relax
\EndOfBibitem
\bibitem[Gao \latin{et~al.}(2020)Gao, Saller, Liu, Kelly, Richardson, and
  Geva]{Gao2020JCTC}
Gao,~X.; Saller,~M. A.~C.; Liu,~Y.; Kelly,~A.; Richardson,~J.~O.; Geva,~E.
  Benchmarking Quasiclassical Mapping Hamiltonian Methods for Simulating
  Electronically Nonadiabatic Molecular Dynamics. \emph{J. Chem. Theory
  Comput.} \textbf{2020}, \emph{16}, 2883--2895\relax
\mciteBstWouldAddEndPuncttrue
\mciteSetBstMidEndSepPunct{\mcitedefaultmidpunct}
{\mcitedefaultendpunct}{\mcitedefaultseppunct}\relax
\EndOfBibitem
\bibitem[Ceriotti \latin{et~al.}(2014)Ceriotti, More, and
  Manolopoulos]{Ceriotti2014CPC}
Ceriotti,~M.; More,~J.; Manolopoulos,~D.~E. i-PI: A Python Interface for
  {\emph{Ab Initio}} Path Integral Molecular Dynamics Simulations.
  \emph{Comput. Phys. Commun.} \textbf{2014}, \emph{185}, 1019 -- 1026\relax
\mciteBstWouldAddEndPuncttrue
\mciteSetBstMidEndSepPunct{\mcitedefaultmidpunct}
{\mcitedefaultendpunct}{\mcitedefaultseppunct}\relax
\EndOfBibitem
\bibitem[Kapil \latin{et~al.}(2019)Kapil, Rossi, Marsalek, Petraglia, Litman,
  Spura, Cheng, Cuzzocrea, Mei{\ss}ner, Wilkins, Helfrecht, Juda, Bienvenue,
  Fang, Kessler, Poltavsky, Vandenbrande, Wieme, Corminboeuf, K\"uhne,
  Manolopoulos, Markland, Richardson, Tkatchenko, Tribello, {Van Speybroeck},
  and Ceriotti]{Kapil2019CPC}
Kapil,~V. \latin{et~al.}  i-PI 2.0: A Universal Force Engine for Advanced
  Molecular Simulations. \emph{Comput. Phys. Commun.} \textbf{2019},
  \emph{236}, 214--223\relax
\mciteBstWouldAddEndPuncttrue
\mciteSetBstMidEndSepPunct{\mcitedefaultmidpunct}
{\mcitedefaultendpunct}{\mcitedefaultseppunct}\relax
\EndOfBibitem
\bibitem[Thoss \latin{et~al.}(2000)Thoss, Miller, and Stock]{Thoss2000JCP}
Thoss,~M.; Miller,~W.~H.; Stock,~G. Semiclassical Description of Nonadiabatic
  Quantum Dynamics: Application to the S$_1$–S$_2$ Conical Intersection in
  Pyrazine. \emph{J. Chem. Phys.} \textbf{2000}, \emph{112}, 10282--10292\relax
\mciteBstWouldAddEndPuncttrue
\mciteSetBstMidEndSepPunct{\mcitedefaultmidpunct}
{\mcitedefaultendpunct}{\mcitedefaultseppunct}\relax
\EndOfBibitem
\bibitem[Lee \latin{et~al.}(2016)Lee, Huo, and Coker]{Lee2016ARPC}
Lee,~M.~K.; Huo,~P.; Coker,~D.~F. Semiclassical Path Integral Dynamics:
  Photosynthetic Energy Transfer with Realistic Environment Interactions.
  \emph{Annu. Rev. Phys. Chem.} \textbf{2016}, \emph{67}, 639--668\relax
\mciteBstWouldAddEndPuncttrue
\mciteSetBstMidEndSepPunct{\mcitedefaultmidpunct}
{\mcitedefaultendpunct}{\mcitedefaultseppunct}\relax
\EndOfBibitem
\bibitem[Stock and Thoss(1997)Stock, and Thoss]{Stock1997PRL}
Stock,~G.; Thoss,~M. Semiclassical Description of Nonadiabatic Quantum
  Dynamics. \emph{Phys. Rev. Lett.} \textbf{1997}, \emph{78}, 578--581\relax
\mciteBstWouldAddEndPuncttrue
\mciteSetBstMidEndSepPunct{\mcitedefaultmidpunct}
{\mcitedefaultendpunct}{\mcitedefaultseppunct}\relax
\EndOfBibitem
\bibitem[Westermayr and Marquetand(2020)Westermayr, and
  Marquetand]{Westermayr2020MLST_Perspective}
Westermayr,~J.; Marquetand,~P. Machine Learning and Excited-State Molecular
  Dynamics. \emph{Mach. Learn.: Sci. Technol.} \textbf{2020}, \emph{in press},
  doi:10.1088/2632--2153/ab9c3e\relax
\mciteBstWouldAddEndPuncttrue
\mciteSetBstMidEndSepPunct{\mcitedefaultmidpunct}
{\mcitedefaultendpunct}{\mcitedefaultseppunct}\relax
\EndOfBibitem
\bibitem[Weinreich \latin{et~al.}(2020)Weinreich, R\"{o}mer, Paleico, and
  Behler]{Weinreich2020JPCC}
Weinreich,~J.; R\"{o}mer,~A.; Paleico,~M.~L.; Behler,~J. Properties of
  $\alpha$-Brass Nanoparticles. 1. Neural Network Potential Energy Surface.
  \emph{J. Phys. Chem. C} \textbf{2020}, \emph{in press}\relax
\mciteBstWouldAddEndPuncttrue
\mciteSetBstMidEndSepPunct{\mcitedefaultmidpunct}
{\mcitedefaultendpunct}{\mcitedefaultseppunct}\relax
\EndOfBibitem
\bibitem[Lin \latin{et~al.}(2020)Lin, Zhang, Zhao, and Jiang]{Lin2020JCP}
Lin,~Q.; Zhang,~Y.; Zhao,~B.; Jiang,~B. Automatically Growing Global Reactive
  Neural Network Potential Energy Surfaces: A Trajectory-Free Active Learning
  Strategy. \emph{J. Chem. Phys.} \textbf{2020}, \emph{152}, 154104\relax
\mciteBstWouldAddEndPuncttrue
\mciteSetBstMidEndSepPunct{\mcitedefaultmidpunct}
{\mcitedefaultendpunct}{\mcitedefaultseppunct}\relax
\EndOfBibitem
\bibitem[Tully(1990)]{Tully1990JCP}
Tully,~J.~C. Molecular Dynamics with Electronic Transitions. \emph{J. Chem.
  Phys.} \textbf{1990}, \emph{93}, 1061--1071\relax
\mciteBstWouldAddEndPuncttrue
\mciteSetBstMidEndSepPunct{\mcitedefaultmidpunct}
{\mcitedefaultendpunct}{\mcitedefaultseppunct}\relax
\EndOfBibitem
\bibitem[Tully(1991)]{Tully1991IJQC}
Tully,~J.~C. Nonadiabatic Molecular Dynamics. \emph{Int. J. Quantum Chem.}
  \textbf{1991}, \emph{40}, 299--309\relax
\mciteBstWouldAddEndPuncttrue
\mciteSetBstMidEndSepPunct{\mcitedefaultmidpunct}
{\mcitedefaultendpunct}{\mcitedefaultseppunct}\relax
\EndOfBibitem
\bibitem[C.~Tully(1998)]{Tully1998FD}
C.~Tully,~J. Mixed Quantum–Classical Dynamics. \emph{Faraday Discuss.}
  \textbf{1998}, \emph{110}, 407--419\relax
\mciteBstWouldAddEndPuncttrue
\mciteSetBstMidEndSepPunct{\mcitedefaultmidpunct}
{\mcitedefaultendpunct}{\mcitedefaultseppunct}\relax
\EndOfBibitem
\bibitem[S.~Mai(2020)]{Mai2020}
S.~Mai,~L.~G.,~P.~Marquetand In \emph{Quantum Chemistry and Dynamics of Excited
  States: Methods and Applications}; Gonz\'alez,~L., Lindh,~R., Eds.; Wiley,
  2020; in press\relax
\mciteBstWouldAddEndPuncttrue
\mciteSetBstMidEndSepPunct{\mcitedefaultmidpunct}
{\mcitedefaultendpunct}{\mcitedefaultseppunct}\relax
\EndOfBibitem
\bibitem[Zener(1932)]{Zener1932}
Zener,~C. Non-Adiabatic Crossing of Energy Levels. \emph{Proc. Roy. Soc. Lond.
  A} \textbf{1932}, \emph{137}, 696--701\relax
\mciteBstWouldAddEndPuncttrue
\mciteSetBstMidEndSepPunct{\mcitedefaultmidpunct}
{\mcitedefaultendpunct}{\mcitedefaultseppunct}\relax
\EndOfBibitem
\bibitem[Wittig(2005)]{Wittig2005JPCB}
Wittig,~C. {The Landau-Zener Formula}. \emph{J. Phys. Chem. B} \textbf{2005},
  \emph{109}, 8428--8430\relax
\mciteBstWouldAddEndPuncttrue
\mciteSetBstMidEndSepPunct{\mcitedefaultmidpunct}
{\mcitedefaultendpunct}{\mcitedefaultseppunct}\relax
\EndOfBibitem
\bibitem[Zhu \latin{et~al.}(2001)Zhu, Kamisaka, and Nakamura]{Zhu2001JCP}
Zhu,~C.; Kamisaka,~H.; Nakamura,~H. Significant Improvement of the Trajectory
  Surface Hopping Method by the Zhu–Nakamura Theory. \emph{J. Chem. Phys}
  \textbf{2001}, \emph{115}, 11036--11039\relax
\mciteBstWouldAddEndPuncttrue
\mciteSetBstMidEndSepPunct{\mcitedefaultmidpunct}
{\mcitedefaultendpunct}{\mcitedefaultseppunct}\relax
\EndOfBibitem
\bibitem[Zhu \latin{et~al.}(2002)Zhu, Kamisaka, and Nakamura]{Zhu2002JCP}
Zhu,~C.; Kamisaka,~H.; Nakamura,~H. {New Implementation of the Trajectory
  Surface Hopping Method with Use of the Zhu–Nakamura Theory. II. Application
  to the Charge Transfer Processes in the 3D DH$_2$+ System}. \emph{J. Chem.
  Phys.} \textbf{2002}, \emph{116}, 3234--3247\relax
\mciteBstWouldAddEndPuncttrue
\mciteSetBstMidEndSepPunct{\mcitedefaultmidpunct}
{\mcitedefaultendpunct}{\mcitedefaultseppunct}\relax
\EndOfBibitem
\bibitem[Oloyede \latin{et~al.}(2006)Oloyede, Mil’nikov, and
  Nakamura]{Oloyede2006JCP}
Oloyede,~P.; Mil’nikov,~G.; Nakamura,~H. Generalized Trajectory Surface
  Hopping Method Based on the Zhu-Nakamura Theory. \emph{J. Chem. Phys.}
  \textbf{2006}, \emph{124}, 144110\relax
\mciteBstWouldAddEndPuncttrue
\mciteSetBstMidEndSepPunct{\mcitedefaultmidpunct}
{\mcitedefaultendpunct}{\mcitedefaultseppunct}\relax
\EndOfBibitem
\bibitem[Ishida \latin{et~al.}(2017)Ishida, Nanbu, and
  Nakamura]{Ishida2017IRPC}
Ishida,~T.; Nanbu,~S.; Nakamura,~H. Clarification of Nonadiabatic Chemical
  Dynamics by the Zhu-Nakamura Theory of Nonadiabatic Transition: From
  Tri-Atomic Systems to Reactions in Solutions. \emph{Int. Rev. Phys. Chem.}
  \textbf{2017}, \emph{36}, 229--286\relax
\mciteBstWouldAddEndPuncttrue
\mciteSetBstMidEndSepPunct{\mcitedefaultmidpunct}
{\mcitedefaultendpunct}{\mcitedefaultseppunct}\relax
\EndOfBibitem
\bibitem[Zhu \latin{et~al.}(1995)Zhu, Kleiman, Li, Lu, Trentelman, and
  Gordon]{Zhu1995PRL}
Zhu,~L.; Kleiman,~V.; Li,~X.; Lu,~S.~P.; Trentelman,~K.; Gordon,~R.~J.
  Ultrafast coherent control and Destruction of Excitons in Quantum Wells.
  \emph{Phys. Rev. Lett.} \textbf{1995}, \emph{75}, 2598--2601\relax
\mciteBstWouldAddEndPuncttrue
\mciteSetBstMidEndSepPunct{\mcitedefaultmidpunct}
{\mcitedefaultendpunct}{\mcitedefaultseppunct}\relax
\EndOfBibitem
\bibitem[Granucci and Persico(2007)Granucci, and Persico]{Granucci2007JCP}
Granucci,~G.; Persico,~M. Critical Appraisal of the Fewest Switching Algorithm
  for Surface Hopping. \emph{J. Chem. Phys.} \textbf{2007}, \emph{126},
  134114\relax
\mciteBstWouldAddEndPuncttrue
\mciteSetBstMidEndSepPunct{\mcitedefaultmidpunct}
{\mcitedefaultendpunct}{\mcitedefaultseppunct}\relax
\EndOfBibitem
\bibitem[Fabiano \latin{et~al.}(2008)Fabiano, Keal, and Thiel]{Fabiano2008CP}
Fabiano,~E.; Keal,~T.; Thiel,~W. Implementation of Surface Hopping Molecular
  Dynamics using Semiempirical Methods. \emph{Chem. Phys.} \textbf{2008},
  \emph{349}, 334 -- 347\relax
\mciteBstWouldAddEndPuncttrue
\mciteSetBstMidEndSepPunct{\mcitedefaultmidpunct}
{\mcitedefaultendpunct}{\mcitedefaultseppunct}\relax
\EndOfBibitem
\bibitem[Malhado \latin{et~al.}(2014)Malhado, Bearpark, and
  Hynes]{Malhado2014FC}
Malhado,~J.~P.; Bearpark,~M.~J.; Hynes,~J.~T. Non-Adiabatic Dynamics Close to
  Conical Intersections and the Surface Hopping Perspective. \emph{Front.
  Chem.} \textbf{2014}, \emph{2}, 97\relax
\mciteBstWouldAddEndPuncttrue
\mciteSetBstMidEndSepPunct{\mcitedefaultmidpunct}
{\mcitedefaultendpunct}{\mcitedefaultseppunct}\relax
\EndOfBibitem
\bibitem[Wang \latin{et~al.}(2016)Wang, Akimov, and Prezhdo]{Wang2016JPCL}
Wang,~L.; Akimov,~A.; Prezhdo,~O.~V. Recent Progress in Surface Hopping:
  2011-2015. \emph{J. Phys. Chem. Lett.} \textbf{2016}, \emph{7},
  2100--2112\relax
\mciteBstWouldAddEndPuncttrue
\mciteSetBstMidEndSepPunct{\mcitedefaultmidpunct}
{\mcitedefaultendpunct}{\mcitedefaultseppunct}\relax
\EndOfBibitem
\bibitem[Subotnik and Rhee(2015)Subotnik, and Rhee]{Subotnik2015JPCA}
Subotnik,~J.~E.; Rhee,~Y.~M. On Surface Hopping and Time-Reversal. \emph{J.
  Phys. Chem. A} \textbf{2015}, \emph{119}, 990--995\relax
\mciteBstWouldAddEndPuncttrue
\mciteSetBstMidEndSepPunct{\mcitedefaultmidpunct}
{\mcitedefaultendpunct}{\mcitedefaultseppunct}\relax
\EndOfBibitem
\bibitem[Hammes‐Schiffer and Tully(1994)Hammes‐Schiffer, and
  Tully]{Hammes-Schiffer1994JCP}
Hammes‐Schiffer,~S.; Tully,~J.~C. Proton Transfer in Solution: Molecular
  Dynamics with Quantum Transitions. \emph{J. Chem. Phys.} \textbf{1994},
  \emph{101}, 4657--4667\relax
\mciteBstWouldAddEndPuncttrue
\mciteSetBstMidEndSepPunct{\mcitedefaultmidpunct}
{\mcitedefaultendpunct}{\mcitedefaultseppunct}\relax
\EndOfBibitem
\bibitem[Sawada \latin{et~al.}(1985)Sawada, Nitzan, and Metiu]{Sawada1985PRB}
Sawada,~S.-I.; Nitzan,~A.; Metiu,~H. Mean-Trajectory Approximation for Charge-
  and Energy-Transfer Processes at Surfaces. \emph{Phys. Rev. B} \textbf{1985},
  \emph{32}, 851--867\relax
\mciteBstWouldAddEndPuncttrue
\mciteSetBstMidEndSepPunct{\mcitedefaultmidpunct}
{\mcitedefaultendpunct}{\mcitedefaultseppunct}\relax
\EndOfBibitem
\bibitem[Li \latin{et~al.}(2005)Li, Tully, Schlegel, and Frisch]{Li2005JCP}
Li,~X.; Tully,~J.~C.; Schlegel,~H.~B.; Frisch,~M.~J. {\emph{Ab Initio}}
  Ehrenfest Dynamics. \emph{J. Chem. Phys.} \textbf{2005}, \emph{123},
  084106\relax
\mciteBstWouldAddEndPuncttrue
\mciteSetBstMidEndSepPunct{\mcitedefaultmidpunct}
{\mcitedefaultendpunct}{\mcitedefaultseppunct}\relax
\EndOfBibitem
\bibitem[Mai \latin{et~al.}(2016)Mai, Marquetand, and
  Gonz{\'a}lez]{Mai2016JPCL}
Mai,~S.; Marquetand,~P.; Gonz{\'a}lez,~L. Intersystem Crossing Pathways in the
  Noncanonical Nucleobase 2-Thiouracil: A Time-Dependent Picture. \emph{J.
  Phys. Chem. Lett.} \textbf{2016}, \emph{7}, 1978--1983\relax
\mciteBstWouldAddEndPuncttrue
\mciteSetBstMidEndSepPunct{\mcitedefaultmidpunct}
{\mcitedefaultendpunct}{\mcitedefaultseppunct}\relax
\EndOfBibitem
\bibitem[Mai \latin{et~al.}(2017)Mai, Richter, Marquetand, and
  Gonz{\'a}lez]{Mai2017CP}
Mai,~S.; Richter,~M.; Marquetand,~P.; Gonz{\'a}lez,~L. The DNA Nucleobase
  Thymine in Motion -- Intersystem Crossing Simulated with Surface Hopping.
  \emph{Chem. Phys.} \textbf{2017}, \emph{482}, 9 -- 15\relax
\mciteBstWouldAddEndPuncttrue
\mciteSetBstMidEndSepPunct{\mcitedefaultmidpunct}
{\mcitedefaultendpunct}{\mcitedefaultseppunct}\relax
\EndOfBibitem
\bibitem[Raghunathan~Ramakrishnan and von
  Lilienfeld(2014)Raghunathan~Ramakrishnan, and von
  Lilienfeld]{Ramakrishnan2014SD}
Raghunathan~Ramakrishnan,~M.~R.,~Pavlo O.~Dral; von Lilienfeld,~O.~A. Quantum
  Chemistry Structures and Properties of 134 Kilo Molecules. \emph{Sci. Data}
  \textbf{2014}, \emph{1}\relax
\mciteBstWouldAddEndPuncttrue
\mciteSetBstMidEndSepPunct{\mcitedefaultmidpunct}
{\mcitedefaultendpunct}{\mcitedefaultseppunct}\relax
\EndOfBibitem
\bibitem[Artrith \latin{et~al.}(2011)Artrith, Morawietz, and
  Behler]{Artrith2011PRB}
Artrith,~N.; Morawietz,~T.; Behler,~J. High-Dimensional Neural-Network
  Potentials for Multicomponent Systems: {Applications} to Zinc Oxide.
  \emph{Phys. Rev. B} \textbf{2011}, \emph{83}, 153101\relax
\mciteBstWouldAddEndPuncttrue
\mciteSetBstMidEndSepPunct{\mcitedefaultmidpunct}
{\mcitedefaultendpunct}{\mcitedefaultseppunct}\relax
\EndOfBibitem
\bibitem[Huang and von Lilienfeld(2016)Huang, and von Lilienfeld]{Huang2016JCP}
Huang,~B.; von Lilienfeld,~O.~A. Communication: Understanding Molecular
  Representations in Machine Learning: The Role of Uniqueness and Target
  Similarity. \emph{J. Chem. Phys.} \textbf{2016}, \emph{145}, 161102\relax
\mciteBstWouldAddEndPuncttrue
\mciteSetBstMidEndSepPunct{\mcitedefaultmidpunct}
{\mcitedefaultendpunct}{\mcitedefaultseppunct}\relax
\EndOfBibitem
\bibitem[Yao \latin{et~al.}(2018)Yao, Herr, Toth, Mckintyre, and
  Parkhill]{Yao2018CS}
Yao,~K.; Herr,~J.~E.; Toth,~D.~W.; Mckintyre,~R.; Parkhill,~J. The
  {TensorMol-0.1} Model Chemistry: A Neural Network Augmented with Long-Range
  Physics. \emph{Chem. Sci.} \textbf{2018}, \emph{9}, 2261--2269\relax
\mciteBstWouldAddEndPuncttrue
\mciteSetBstMidEndSepPunct{\mcitedefaultmidpunct}
{\mcitedefaultendpunct}{\mcitedefaultseppunct}\relax
\EndOfBibitem
\bibitem[Sch\"{u}tt \latin{et~al.}(2018)Sch\"{u}tt, Sauceda, Kindermans,
  Tkatchenko, and M\"{u}ller]{Schuett2018JCP}
Sch\"{u}tt,~K.~T.; Sauceda,~H.~E.; Kindermans,~P.-J.; Tkatchenko,~A.;
  M\"{u}ller,~K.-R. {SchNet} -- A Deep Learning Architecture for Molecules and
  Materials. \emph{J. Chem. Phys.} \textbf{2018}, \emph{148}, 241722\relax
\mciteBstWouldAddEndPuncttrue
\mciteSetBstMidEndSepPunct{\mcitedefaultmidpunct}
{\mcitedefaultendpunct}{\mcitedefaultseppunct}\relax
\EndOfBibitem
\bibitem[Nebgen \latin{et~al.}(2018)Nebgen, Lubbers, Smith, Sifain, Lokhov,
  Isayev, Roitberg, Barros, and Tretiak]{Nebgen2018JCTC}
Nebgen,~B.; Lubbers,~N.; Smith,~J.~S.; Sifain,~A.~E.; Lokhov,~A.; Isayev,~O.;
  Roitberg,~A.~E.; Barros,~K.; Tretiak,~S. Transferable Dynamic Molecular
  Charge Assignment Using Deep Neural Networks. \emph{J. Chem. Theory Comput.}
  \textbf{2018}, \emph{14}, 4687--4698\relax
\mciteBstWouldAddEndPuncttrue
\mciteSetBstMidEndSepPunct{\mcitedefaultmidpunct}
{\mcitedefaultendpunct}{\mcitedefaultseppunct}\relax
\EndOfBibitem
\bibitem[Sifain \latin{et~al.}(2018)Sifain, Lubbers, Nebgen, Smith, Lokhov,
  Isayev, Roitberg, Barros, and Tretiak]{Sifain2018JPCL}
Sifain,~A.~E.; Lubbers,~N.; Nebgen,~B.~T.; Smith,~J.~S.; Lokhov,~A.~Y.;
  Isayev,~O.; Roitberg,~A.~E.; Barros,~K.; Tretiak,~S. Discovering a
  Transferable Charge Assignment Model Using Machine Learning. \emph{J. Phys.
  Chem. Lett.} \textbf{2018}, \emph{9}, 4495--4501\relax
\mciteBstWouldAddEndPuncttrue
\mciteSetBstMidEndSepPunct{\mcitedefaultmidpunct}
{\mcitedefaultendpunct}{\mcitedefaultseppunct}\relax
\EndOfBibitem
\bibitem[Sch{\"u}tt \latin{et~al.}(2019)Sch{\"u}tt, Gastegger, Tkatchenko, and
  M{\"u}ller]{Schuett2019}
Sch{\"u}tt,~K.~T.; Gastegger,~M.; Tkatchenko,~A.; M{\"u}ller,~K.-R.
  \emph{Explainable AI: Interpreting, Explaining and Visualizing Deep
  Learning}; Springer International Publishing, 2019; pp 311--330\relax
\mciteBstWouldAddEndPuncttrue
\mciteSetBstMidEndSepPunct{\mcitedefaultmidpunct}
{\mcitedefaultendpunct}{\mcitedefaultseppunct}\relax
\EndOfBibitem
\bibitem[Sch\"utt \latin{et~al.}(2019)Sch\"utt, Kessel, Gastegger, Nicoli,
  Tkatchenko, and M\"uller]{Schuett2019JCTC}
Sch\"utt,~K.~T.; Kessel,~P.; Gastegger,~M.; Nicoli,~K.~A.; Tkatchenko,~A.;
  M\"uller,~K.-R. SchNetPack: A Deep Learning Toolbox For Atomistic Systems.
  \emph{J. Chem. Theory Comput.} \textbf{2019}, \emph{15}, 448--455\relax
\mciteBstWouldAddEndPuncttrue
\mciteSetBstMidEndSepPunct{\mcitedefaultmidpunct}
{\mcitedefaultendpunct}{\mcitedefaultseppunct}\relax
\EndOfBibitem
\bibitem[Christensen \latin{et~al.}(2019)Christensen, Faber, and von
  Lilienfeld]{Christensen2019JCP}
Christensen,~A.~S.; Faber,~F.~A.; von Lilienfeld,~O.~A. Operators in Quantum
  Machine Learning: Response Properties in Chemical Space. \emph{J. Chem.
  Phys.} \textbf{2019}, \emph{150}, 064105\relax
\mciteBstWouldAddEndPuncttrue
\mciteSetBstMidEndSepPunct{\mcitedefaultmidpunct}
{\mcitedefaultendpunct}{\mcitedefaultseppunct}\relax
\EndOfBibitem
\bibitem[Veit \latin{et~al.}(2020)Veit, Wilkins, Yang, Jr., and
  Ceriotti]{Veit2020arXiv}
Veit,~M.; Wilkins,~D.~M.; Yang,~Y.; Jr.,~R. A.~D.; Ceriotti,~M. Predicting
  Molecular Dipole Moments by Combining Atomic Partial Charges and Atomic
  Dipoles. \emph{arXiv} \textbf{2020}, \emph{2003.12437}\relax
\mciteBstWouldAddEndPuncttrue
\mciteSetBstMidEndSepPunct{\mcitedefaultmidpunct}
{\mcitedefaultendpunct}{\mcitedefaultseppunct}\relax
\EndOfBibitem
\bibitem[Gastegger and Marquetand(2018)Gastegger, and
  Marquetand]{Gastegger2018arXiv}
Gastegger,~M.; Marquetand,~P. Molecular dynamics with neural-network
  potentials. \emph{arXiv:1812.07676 [physics.chem-ph]} \textbf{2018}, \relax
\mciteBstWouldAddEndPunctfalse
\mciteSetBstMidEndSepPunct{\mcitedefaultmidpunct}
{}{\mcitedefaultseppunct}\relax
\EndOfBibitem
\bibitem[Thomas \latin{et~al.}(2013)Thomas, Brehm, Fligg, V{\"o}hringer, and
  Kirchner]{Thomas2013PCCP}
Thomas,~M.; Brehm,~M.; Fligg,~R.; V{\"o}hringer,~P.; Kirchner,~B. Computing
  Vibrational Spectra from {\emph{Ab Initio}} Molecular Dynamics. \emph{Phys.
  Chem. Chem. Phys.} \textbf{2013}, \emph{15}, 6608--6622\relax
\mciteBstWouldAddEndPuncttrue
\mciteSetBstMidEndSepPunct{\mcitedefaultmidpunct}
{\mcitedefaultendpunct}{\mcitedefaultseppunct}\relax
\EndOfBibitem
\bibitem[Wilke \latin{et~al.}(2016)Wilke, Wilke, Meerts, and
  Schmitt]{Wilke2016JCP}
Wilke,~J.; Wilke,~M.; Meerts,~W.~L.; Schmitt,~M. {Determination of Ground and
  Excited State Dipole Moments via Electronic Stark Spectroscopy:
  5-Methoxyindole}. \emph{J. Chem. Phys.} \textbf{2016}, \emph{144},
  044201\relax
\mciteBstWouldAddEndPuncttrue
\mciteSetBstMidEndSepPunct{\mcitedefaultmidpunct}
{\mcitedefaultendpunct}{\mcitedefaultseppunct}\relax
\EndOfBibitem
\bibitem[Tennyson(2016)]{Tennyson2016JCP}
Tennyson,~J. Perspective: Accurate Ro-Vibrational Calculations on Small
  Molecules. \emph{J. Chem. Phys.} \textbf{2016}, \emph{145}, 120901\relax
\mciteBstWouldAddEndPuncttrue
\mciteSetBstMidEndSepPunct{\mcitedefaultmidpunct}
{\mcitedefaultendpunct}{\mcitedefaultseppunct}\relax
\EndOfBibitem
\bibitem[Marquetand \latin{et~al.}(2016)Marquetand, Nogueira, Mai, Plasser, and
  Gonz\'alez]{Marquetand2017M}
Marquetand,~P.; Nogueira,~J.; Mai,~S.; Plasser,~F.; Gonz\'alez,~L. Challenges
  in Simulating Light-Induced Processes in DNA. \emph{Molecules} \textbf{2016},
  \emph{22}, 49\relax
\mciteBstWouldAddEndPuncttrue
\mciteSetBstMidEndSepPunct{\mcitedefaultmidpunct}
{\mcitedefaultendpunct}{\mcitedefaultseppunct}\relax
\EndOfBibitem
\bibitem[Nogueira and Gonz\'{a}lez(2018)Nogueira, and
  Gonz\'{a}lez]{Nogueira2018ARPC}
Nogueira,~J.~J.; Gonz\'{a}lez,~L. {Computational Photophysics in the Presence
  of an Environment}. \emph{Annu. Rev. Phys. Chem.} \textbf{2018}, \emph{69},
  473--497\relax
\mciteBstWouldAddEndPuncttrue
\mciteSetBstMidEndSepPunct{\mcitedefaultmidpunct}
{\mcitedefaultendpunct}{\mcitedefaultseppunct}\relax
\EndOfBibitem
\bibitem[Barbatti \latin{et~al.}(2007)Barbatti, Granucci, Persico, Ruckenbauer,
  Vazdar, {Eckert-Maksi\'c}, and Lischka]{Barbatti2007JPPA}
Barbatti,~M.; Granucci,~G.; Persico,~M.; Ruckenbauer,~M.; Vazdar,~M.;
  {Eckert-Maksi\'c},~M.; Lischka,~H. The on-the-Fly Surface-Hopping Program
  System {Newton-X:} Application to {\emph{Ab Initio}} Simulation of the
  Nonadiabatic Photodynamics of Benchmark Systems. \emph{J. Photochem.
  Photobiol. A} \textbf{2007}, \emph{190}, 228--240\relax
\mciteBstWouldAddEndPuncttrue
\mciteSetBstMidEndSepPunct{\mcitedefaultmidpunct}
{\mcitedefaultendpunct}{\mcitedefaultseppunct}\relax
\EndOfBibitem
\bibitem[Dral(2020)]{Dral2020JPCL}
Dral,~P.~O. Quantum Chemistry in the Age of Machine Learning. \emph{J. Phys.
  Chem. Lett.} \textbf{2020}, \emph{11}, 2336--2347\relax
\mciteBstWouldAddEndPuncttrue
\mciteSetBstMidEndSepPunct{\mcitedefaultmidpunct}
{\mcitedefaultendpunct}{\mcitedefaultseppunct}\relax
\EndOfBibitem
\bibitem[Chen \latin{et~al.}(2019)Chen, Fang, and Cui]{Chen2019JPCL}
Chen,~W.-K.; Fang,~W.-H.; Cui,~G. Integrating Machine Learning with the
  Multilayer Energy-Based Fragment Method for Excited States of Large Systems.
  \emph{J. Phys. Chem. Lett.} \textbf{2019}, \emph{10}, 7836--7841\relax
\mciteBstWouldAddEndPuncttrue
\mciteSetBstMidEndSepPunct{\mcitedefaultmidpunct}
{\mcitedefaultendpunct}{\mcitedefaultseppunct}\relax
\EndOfBibitem
\bibitem[Chen \latin{et~al.}(2020)Chen, Zhang, Jiang, Fang, and
  Cui]{Chen2020JPCA_Hessian}
Chen,~W.-K.; Zhang,~Y.; Jiang,~B.; Fang,~W.-H.; Cui,~G. Efficient Construction
  of Excited-State Hessian Matrices with Machine Learning Accelerated
  Multilayer Energy-Based Fragment Method. \emph{The Journal of Physical
  Chemistry A} \textbf{2020}, \emph{in press},
  DOI:10.1021/acs.jpca.0c04117\relax
\mciteBstWouldAddEndPuncttrue
\mciteSetBstMidEndSepPunct{\mcitedefaultmidpunct}
{\mcitedefaultendpunct}{\mcitedefaultseppunct}\relax
\EndOfBibitem
\bibitem[Behler \latin{et~al.}(2008)Behler, Marto{{\'a}}k, Donadio, and
  Parrinello]{Behler2008PRL}
Behler,~J.; Marto{{\'a}}k,~R.; Donadio,~D.; Parrinello,~M. Metadynamics
  Simulations of the High-Pressure Phases of Silicon Employing a
  High-Dimensional Neural Network Potential. \emph{Phys. Rev. Lett.}
  \textbf{2008}, \emph{100}, 185501\relax
\mciteBstWouldAddEndPuncttrue
\mciteSetBstMidEndSepPunct{\mcitedefaultmidpunct}
{\mcitedefaultendpunct}{\mcitedefaultseppunct}\relax
\EndOfBibitem
\bibitem[Gastegger \latin{et~al.}(2018)Gastegger, Schwiedrzik, Bittermann,
  Berzsenyi, and Marquetand]{Gastegger2018JCP}
Gastegger,~M.; Schwiedrzik,~L.; Bittermann,~M.; Berzsenyi,~F.; Marquetand,~P.
  wACSF -- Weighted Atom-Centered Symmetry Functions as Descriptors in Machine
  Learning Potentials. \emph{J. Chem. Phys.} \textbf{2018}, \emph{148},
  241709\relax
\mciteBstWouldAddEndPuncttrue
\mciteSetBstMidEndSepPunct{\mcitedefaultmidpunct}
{\mcitedefaultendpunct}{\mcitedefaultseppunct}\relax
\EndOfBibitem
\bibitem[Chen \latin{et~al.}(2020)Chen, Zuehlsdorff, Morawietz, Isborn, and
  Markland]{Chen2020arXiv}
Chen,~M.~S.; Zuehlsdorff,~T.~J.; Morawietz,~T.; Isborn,~C.~M.; Markland,~T.~E.
  Exploiting Machine Learning to Efficiently Predict Multidimensional Optical
  Spectra in Complex Environments. \emph{arXiv} \textbf{2020},
  \emph{2005.09776}\relax
\mciteBstWouldAddEndPuncttrue
\mciteSetBstMidEndSepPunct{\mcitedefaultmidpunct}
{\mcitedefaultendpunct}{\mcitedefaultseppunct}\relax
\EndOfBibitem
\bibitem[Koch and Zhang(2014)Koch, and Zhang]{Koch2014JCP}
Koch,~W.; Zhang,~D.~H. {Communication: Separable Potential Energy Surfaces from
  Multiplicative Artificial Neural Networks}. \emph{J. Chem. Phys.}
  \textbf{2014}, \emph{141}, 021101\relax
\mciteBstWouldAddEndPuncttrue
\mciteSetBstMidEndSepPunct{\mcitedefaultmidpunct}
{\mcitedefaultendpunct}{\mcitedefaultseppunct}\relax
\EndOfBibitem
\bibitem[He \latin{et~al.}(2016)He, Yuan, Li, and Chen]{He2016SR}
He,~D.; Yuan,~J.; Li,~H.; Chen,~M. {Global Diabatic Potential Energy Surfaces
  and Quantum Dynamical Studies for the Li(2p) + H$_2(X^1\Sigma^+_g)
  \rightarrow$ LiH($X^1\Sigma^+)$ + H Reaction}. \emph{Sci. Rep.}
  \textbf{2016}, \emph{6}\relax
\mciteBstWouldAddEndPuncttrue
\mciteSetBstMidEndSepPunct{\mcitedefaultmidpunct}
{\mcitedefaultendpunct}{\mcitedefaultseppunct}\relax
\EndOfBibitem
\bibitem[Guan \latin{et~al.}(2017)Guan, Fu, and Zhang]{Guan2017JCP}
Guan,~Y.; Fu,~B.; Zhang,~D.~H. Construction of Diabatic Energy Surfaces for
  LiFH with Artificial Neural Networks. \emph{J. Chem. Phys.} \textbf{2017},
  \emph{147}, 224307\relax
\mciteBstWouldAddEndPuncttrue
\mciteSetBstMidEndSepPunct{\mcitedefaultmidpunct}
{\mcitedefaultendpunct}{\mcitedefaultseppunct}\relax
\EndOfBibitem
\bibitem[Wang \latin{et~al.}(2018)Wang, Yang, Yuan, and Chen]{Wang2018SR}
Wang,~S.; Yang,~Z.; Yuan,~J.; Chen,~M. {New Diabatic Potential Energy Surfaces
  of the NaH$_2$ System and Dynamics Studies for the Na(3p) + H$_2$
  $\rightarrow$ NaH + H Reaction}. \emph{Sci. Rep.} \textbf{2018},
  \emph{8}\relax
\mciteBstWouldAddEndPuncttrue
\mciteSetBstMidEndSepPunct{\mcitedefaultmidpunct}
{\mcitedefaultendpunct}{\mcitedefaultseppunct}\relax
\EndOfBibitem
\bibitem[Yuan \latin{et~al.}(2018)Yuan, He, Wang, Chen, and Han]{Yuan2018PCCP}
Yuan,~J.; He,~D.; Wang,~S.; Chen,~M.; Han,~K. {Diabatic Potential Energy
  Surfaces of MgH$_2^+$ and Dynamic Studies for the Mg$^+$(3p) +H$_2$
  $\rightarrow$ MgH$^+$ + H Reaction}. \emph{Phys. Chem. Chem. Phys.}
  \textbf{2018}, \emph{20}, 6638--6647\relax
\mciteBstWouldAddEndPuncttrue
\mciteSetBstMidEndSepPunct{\mcitedefaultmidpunct}
{\mcitedefaultendpunct}{\mcitedefaultseppunct}\relax
\EndOfBibitem
\bibitem[Yin \latin{et~al.}(2019)Yin, Guan, Fu, and Zhang]{Yin2019PCCP}
Yin,~Z.; Guan,~Y.; Fu,~B.; Zhang,~D.~H. {Two-State Diabatic Potential Energy
  Surfaces of ClH$_2$ Based on Nonadiabatic Couplings with Neural Networks}.
  \emph{Phys. Chem. Chem. Phys.} \textbf{2019}, \emph{21}, 20372--20383\relax
\mciteBstWouldAddEndPuncttrue
\mciteSetBstMidEndSepPunct{\mcitedefaultmidpunct}
{\mcitedefaultendpunct}{\mcitedefaultseppunct}\relax
\EndOfBibitem
\bibitem[Akimov(2018)]{Akimov2018JPCL}
Akimov,~A.~V. A Simple Phase correction Makes a Big Difference in Nonadiabatic
  Molecular Dynamics. \emph{J. Phys. Chem. Lett.} \textbf{2018}, \emph{9},
  6096--6102\relax
\mciteBstWouldAddEndPuncttrue
\mciteSetBstMidEndSepPunct{\mcitedefaultmidpunct}
{\mcitedefaultendpunct}{\mcitedefaultseppunct}\relax
\EndOfBibitem
\bibitem[Shu \latin{et~al.}(2019)Shu, Kryven, Sampaio~de Oliveira-Filho, Zhang,
  Song, Li, Meana-Pa$\tilde{n}$eda, Fu, Bowman, and Truhlar]{Shu2019JCP}
Shu,~Y.; Kryven,~J.; Sampaio~de Oliveira-Filho,~A.~G.; Zhang,~L.; Song,~G.-L.;
  Li,~S.~L.; Meana-Pa$\tilde{n}$eda,~R.; Fu,~B.; Bowman,~J.~M.; Truhlar,~D.~G.
  {Direct Diabatization and Analytic Representation of Coupled Potential Energy
  Surfaces and Couplings for the Reactive Quenching of the Excited $^2\Sigma^+$
  State of {OH} by Molecular Hydrogen}. \emph{J. Chem. Phys.} \textbf{2019},
  \emph{151}, 104311\relax
\mciteBstWouldAddEndPuncttrue
\mciteSetBstMidEndSepPunct{\mcitedefaultmidpunct}
{\mcitedefaultendpunct}{\mcitedefaultseppunct}\relax
\EndOfBibitem
\bibitem[Yarkony(1996)]{Yarkony1996JCPa}
Yarkony,~D.~R. {On the Consequences of Nonremovable Derivative Couplings. I.
  The Geometric Phase and Quasidiabatic States: A Numerical Study}. \emph{J.
  Chem. Phys.} \textbf{1996}, \emph{105}, 10456--10461\relax
\mciteBstWouldAddEndPuncttrue
\mciteSetBstMidEndSepPunct{\mcitedefaultmidpunct}
{\mcitedefaultendpunct}{\mcitedefaultseppunct}\relax
\EndOfBibitem
\bibitem[Yarkony(1996)]{Yarkony1996JCPb}
Yarkony,~D.~R. {On the Role of Conical Intersections in Photodissociation. V.
  Conical Intersections and the Geometric Phase in the Photodissociation of
  Methyl Mercaptan}. \emph{J. Chem. Phys.} \textbf{1996}, \emph{104},
  7866--7881\relax
\mciteBstWouldAddEndPuncttrue
\mciteSetBstMidEndSepPunct{\mcitedefaultmidpunct}
{\mcitedefaultendpunct}{\mcitedefaultseppunct}\relax
\EndOfBibitem
\bibitem[Ryabinkin \latin{et~al.}(2014)Ryabinkin, Joubert-Doriol, and
  Izmaylov]{Ryabinkin2014JCP}
Ryabinkin,~I.~G.; Joubert-Doriol,~L.; Izmaylov,~A.~F. When Do We Need to
  Account for the Geometric Phase in Excited State Dynamics? \emph{J. Chem.
  Phys.} \textbf{2014}, \emph{140}, 214116\relax
\mciteBstWouldAddEndPuncttrue
\mciteSetBstMidEndSepPunct{\mcitedefaultmidpunct}
{\mcitedefaultendpunct}{\mcitedefaultseppunct}\relax
\EndOfBibitem
\bibitem[Gherib \latin{et~al.}(2015)Gherib, Ryabinkin, and
  Izmaylov]{Gherib2015JCTC}
Gherib,~R.; Ryabinkin,~I.~G.; Izmaylov,~A.~F. {Why Do Mixed Quantum-Classical
  Methods Describe Short-Time Dynamics through Conical Intersections So Well?
  Analysis of Geometric Phase Effects}. \emph{J. Chem. Theory Comput.}
  \textbf{2015}, \emph{11}, 1375--1382\relax
\mciteBstWouldAddEndPuncttrue
\mciteSetBstMidEndSepPunct{\mcitedefaultmidpunct}
{\mcitedefaultendpunct}{\mcitedefaultseppunct}\relax
\EndOfBibitem
\bibitem[Ryabinkin \latin{et~al.}(2017)Ryabinkin, Joubert-Doriol, and
  Izmaylov]{Ryabinkin2017ACR}
Ryabinkin,~I.~G.; Joubert-Doriol,~L.; Izmaylov,~A.~F. Geometric Phase Effects
  in Nonadiabatic Dynamics Near Conical Intersections. \emph{Acc. Chem. Res.}
  \textbf{2017}, \emph{50}, 1785--1793\relax
\mciteBstWouldAddEndPuncttrue
\mciteSetBstMidEndSepPunct{\mcitedefaultmidpunct}
{\mcitedefaultendpunct}{\mcitedefaultseppunct}\relax
\EndOfBibitem
\bibitem[Plasser \latin{et~al.}(2016)Plasser, Ruckenbauer, Mai, Oppel,
  Marquetand, and Gonz{\'a}lez]{Plasser2016JCTC}
Plasser,~F.; Ruckenbauer,~M.; Mai,~S.; Oppel,~M.; Marquetand,~P.;
  Gonz{\'a}lez,~L. Efficient and Flexible Computation of Many-Electron Wave
  Function Overlaps. \emph{J. Chem. Theory Comput.} \textbf{2016}, \emph{12},
  1207\relax
\mciteBstWouldAddEndPuncttrue
\mciteSetBstMidEndSepPunct{\mcitedefaultmidpunct}
{\mcitedefaultendpunct}{\mcitedefaultseppunct}\relax
\EndOfBibitem
\bibitem[Zhu and Yarkony(2010)Zhu, and Yarkony]{Zhu2010JCP}
Zhu,~X.; Yarkony,~D.~R. Toward Eliminating the Electronic Structure Bottleneck
  in Nonadiabatic Dynamics on the Fly: An Algorithm to Fit Nonlocal,
  Quasidiabatic, Coupled Electronic State Hamiltonians Based on \emph{Ab
  Initio} Electronic Structure Data. \emph{J. Chem. Phys.} \textbf{2010},
  \emph{132}, 104101\relax
\mciteBstWouldAddEndPuncttrue
\mciteSetBstMidEndSepPunct{\mcitedefaultmidpunct}
{\mcitedefaultendpunct}{\mcitedefaultseppunct}\relax
\EndOfBibitem
\bibitem[Zhu and Yarkony(2012)Zhu, and Yarkony]{Zhu2012JCP}
Zhu,~X.; Yarkony,~D.~R. Quasi-Diabatic Representations of Adiabatic Potential
  Energy Surfaces Coupled by Conical Intersections including Bond Breaking: A
  More General Construction Procedure and an Analysis of the Diabatic
  Representation. \emph{J. Chem. Phys.} \textbf{2012}, \emph{137}, 22A511\relax
\mciteBstWouldAddEndPuncttrue
\mciteSetBstMidEndSepPunct{\mcitedefaultmidpunct}
{\mcitedefaultendpunct}{\mcitedefaultseppunct}\relax
\EndOfBibitem
\bibitem[Zhu and Yarkony(2012)Zhu, and Yarkony]{Zhu2012JCPb}
Zhu,~X.; Yarkony,~D.~R. On the Representation of Coupled Adiabatic Potential
  Energy Surfaces using Quasi-Diabatic Hamiltonians: A Distributed Origins
  Expansion Approach. \emph{J. Chem. Phys.} \textbf{2012}, \emph{136},
  174110\relax
\mciteBstWouldAddEndPuncttrue
\mciteSetBstMidEndSepPunct{\mcitedefaultmidpunct}
{\mcitedefaultendpunct}{\mcitedefaultseppunct}\relax
\EndOfBibitem
\bibitem[Rasmussen(2004)]{Rasmussen2004}
Rasmussen,~C.~E. In \emph{Advanced Lectures on Machine Learning: ML Summer
  Schools 2003, Canberra, Australia, February 2 - 14, 2003, T{\"u}bingen,
  Germany, August 4 - 16, 2003, Revised Lectures}; Bousquet,~O., von
  Luxburg,~U., R{\"a}tsch,~G., Eds.; Springer Berlin Heidelberg: Berlin,
  Heidelberg, 2004; pp 63--71\relax
\mciteBstWouldAddEndPuncttrue
\mciteSetBstMidEndSepPunct{\mcitedefaultmidpunct}
{\mcitedefaultendpunct}{\mcitedefaultseppunct}\relax
\EndOfBibitem
\bibitem[Mai \latin{et~al.}(2018)Mai, Richter, Ruckenbauer, Oppel, Marquetand,
  and Gonz{\'a}lez]{sharc-md2}
Mai,~S.; Richter,~M.; Ruckenbauer,~M.; Oppel,~M.; Marquetand,~P.;
  Gonz{\'a}lez,~L. {SHARC2.0: Surface Hopping Including ARbitrary Couplings --
  Program Package for Non-Adiabatic Dynamics}. sharc-md.org, 2018\relax
\mciteBstWouldAddEndPuncttrue
\mciteSetBstMidEndSepPunct{\mcitedefaultmidpunct}
{\mcitedefaultendpunct}{\mcitedefaultseppunct}\relax
\EndOfBibitem
\bibitem[Beard \latin{et~al.}(2019)Beard, Sivaraman, V\'{a}zquez-Mayagoitia,
  Vishwanath, and Cole]{Beard2019SD}
Beard,~E.~J.; Sivaraman,~G.; V\'{a}zquez-Mayagoitia,~A.; Vishwanath,~V.;
  Cole,~J.~M. {Comparative Dataset of Experimental and Computational Attributes
  of UV/Vis Absorption Spectra}. \emph{Sci. Data} \textbf{2019}, \emph{6}\relax
\mciteBstWouldAddEndPuncttrue
\mciteSetBstMidEndSepPunct{\mcitedefaultmidpunct}
{\mcitedefaultendpunct}{\mcitedefaultseppunct}\relax
\EndOfBibitem
\bibitem[Barbatti \latin{et~al.}(2005)Barbatti, Ruckenbauer, and
  Lischka]{Barbatti2005JCP}
Barbatti,~M.; Ruckenbauer,~M.; Lischka,~H. The Photodynamics of Ethylene: A
  Surface-Hopping Study on Structural Aspects. \emph{J. Chem. Phys.}
  \textbf{2005}, \emph{122}, 174307\relax
\mciteBstWouldAddEndPuncttrue
\mciteSetBstMidEndSepPunct{\mcitedefaultmidpunct}
{\mcitedefaultendpunct}{\mcitedefaultseppunct}\relax
\EndOfBibitem
\bibitem[Tapavicza \latin{et~al.}(2007)Tapavicza, Tavernelli, and
  Rothlisberger]{Tapavicza2007PRL}
Tapavicza,~E.; Tavernelli,~I.; Rothlisberger,~U. {Trajectory Surface Hopping
  within Linear Response Time-Dependent Density-Functional Theory}. \emph{Phys.
  Rev. Lett.} \textbf{2007}, \emph{98}, 023001\relax
\mciteBstWouldAddEndPuncttrue
\mciteSetBstMidEndSepPunct{\mcitedefaultmidpunct}
{\mcitedefaultendpunct}{\mcitedefaultseppunct}\relax
\EndOfBibitem
\bibitem[Tavernelli \latin{et~al.}(2009)Tavernelli, Tapavicza, and
  Rothlisberger]{Tavernelli2009JCP}
Tavernelli,~I.; Tapavicza,~E.; Rothlisberger,~U. Nonadiabatic Coupling Vectors
  within Linear Response Time-Dependent Density Functional Theory. \emph{J.
  Chem. Phys.} \textbf{2009}, \emph{130}, 124107\relax
\mciteBstWouldAddEndPuncttrue
\mciteSetBstMidEndSepPunct{\mcitedefaultmidpunct}
{\mcitedefaultendpunct}{\mcitedefaultseppunct}\relax
\EndOfBibitem
\bibitem[Tavernelli \latin{et~al.}(2009)Tavernelli, Tapavicza, and
  Rothlisberger]{Tavernelli2009JMS}
Tavernelli,~I.; Tapavicza,~E.; Rothlisberger,~U. Non-Adiabatic Dynamics using
  Time-Dependent Density Functional Theory: Assessing the Coupling Strengths.
  \emph{J. Mol. Struct.: THEOCHEM} \textbf{2009}, \emph{914}, 22 -- 29\relax
\mciteBstWouldAddEndPuncttrue
\mciteSetBstMidEndSepPunct{\mcitedefaultmidpunct}
{\mcitedefaultendpunct}{\mcitedefaultseppunct}\relax
\EndOfBibitem
\bibitem[Barbatti \latin{et~al.}(2006)Barbatti, Aquino, and
  Lischka]{Barbatti2006MP}
Barbatti,~M.; Aquino,~A. J.~A.; Lischka,~H. Ultrafast Two-Step Process in the
  Non-Adiabatic Relaxation of the {CH$_{2}$NH$_{2}$} Molecule. \emph{Mol.
  Phys.} \textbf{2006}, \emph{104}, 1053--1060\relax
\mciteBstWouldAddEndPuncttrue
\mciteSetBstMidEndSepPunct{\mcitedefaultmidpunct}
{\mcitedefaultendpunct}{\mcitedefaultseppunct}\relax
\EndOfBibitem
\bibitem[Tao \latin{et~al.}(2011)Tao, Allison, Wright, Stooke, Khurmi, van
  Tilborg, Liu, Falcone, Belkacem, and Martinez]{Tao2011JCP}
Tao,~H.; Allison,~T.~K.; Wright,~T.~W.; Stooke,~A.~M.; Khurmi,~C.; van
  Tilborg,~J.; Liu,~Y.; Falcone,~R.~W.; Belkacem,~A.; Martinez,~T.~J. Ultrafast
  internal conversion in ethylene. I. The excited state lifetime. \emph{J.
  Chem. Phys.} \textbf{2011}, \emph{134}, 244306\relax
\mciteBstWouldAddEndPuncttrue
\mciteSetBstMidEndSepPunct{\mcitedefaultmidpunct}
{\mcitedefaultendpunct}{\mcitedefaultseppunct}\relax
\EndOfBibitem
\bibitem[Allison \latin{et~al.}(2012)Allison, Tao, Glover, Wright, Stooke,
  Khurmi, van Tilborg, Liu, Falcone, Mart\'{i}nez, and
  Belkacem]{Allison2012JCP}
Allison,~T.~K.; Tao,~H.; Glover,~W.~J.; Wright,~T.~W.; Stooke,~A.~M.;
  Khurmi,~C.; van Tilborg,~J.; Liu,~Y.; Falcone,~R.~W.; Mart\'{i}nez,~T.~J.;
  Belkacem,~A. Ultrafast internal conversion in ethylene. {II.} Mechanisms and
  pathways for quenching and hydrogen elimination. \emph{J. Chem. Phys.}
  \textbf{2012}, \emph{136}, 124317\relax
\mciteBstWouldAddEndPuncttrue
\mciteSetBstMidEndSepPunct{\mcitedefaultmidpunct}
{\mcitedefaultendpunct}{\mcitedefaultseppunct}\relax
\EndOfBibitem
\bibitem[Mori \latin{et~al.}(2012)Mori, Glover, Schuurman, and
  Martinez]{Mori2012JPCA}
Mori,~T.; Glover,~W.~J.; Schuurman,~M.~S.; Martinez,~T.~J. Role of Rydberg
  States in the Photochemical Dynamics of Ethylene. \emph{J. Phys. Chem. A}
  \textbf{2012}, \emph{116}, 2808--2818\relax
\mciteBstWouldAddEndPuncttrue
\mciteSetBstMidEndSepPunct{\mcitedefaultmidpunct}
{\mcitedefaultendpunct}{\mcitedefaultseppunct}\relax
\EndOfBibitem
\bibitem[Sellner \latin{et~al.}(2013)Sellner, Barbatti, M\"{u}ller, Domcke, and
  Lischka]{Sellner2013MP}
Sellner,~B.; Barbatti,~M.; M\"{u}ller,~T.; Domcke,~W.; Lischka,~H. Ultrafast
  Non-Adiabatic Dynamics of Ethylene including Rydberg States. \emph{Mol.
  Phys.} \textbf{2013}, \emph{111}, 2439--2450\relax
\mciteBstWouldAddEndPuncttrue
\mciteSetBstMidEndSepPunct{\mcitedefaultmidpunct}
{\mcitedefaultendpunct}{\mcitedefaultseppunct}\relax
\EndOfBibitem
\bibitem[Barbatti \latin{et~al.}(2012)Barbatti, Lan, Crespo-Otero, Szymczak,
  Lischka, and Thiel]{Barbatti2012JCP}
Barbatti,~M.; Lan,~Z.; Crespo-Otero,~R.; Szymczak,~J.~J.; Lischka,~H.;
  Thiel,~W. Critical Appraisal of Excited State Nonadiabatic Dynamics
  Simulations of 9H-Adenine. \emph{J. Chem. Phys.} \textbf{2012}, \emph{117},
  22A503\relax
\mciteBstWouldAddEndPuncttrue
\mciteSetBstMidEndSepPunct{\mcitedefaultmidpunct}
{\mcitedefaultendpunct}{\mcitedefaultseppunct}\relax
\EndOfBibitem
\bibitem[Hollas \latin{et~al.}(2018)Hollas, \v{S}i\v{s}t{\'i}k, Hohenstein,
  Mart{\'i}nez, and Slav{\'i}\v{c}ek]{Hollas2018JCTC}
Hollas,~D.; \v{S}i\v{s}t{\'i}k,~L.; Hohenstein,~E.~G.; Mart{\'i}nez,~T.~J.;
  Slav{\'i}\v{c}ek,~P. Nonadiabatic {\emph{Ab Initio}} Molecular Dynamics with
  the Floating Occupation Molecular Orbital-Complete Active Space Configuration
  Interaction Method. \emph{J. Chem. Theory Comput.} \textbf{2018}, \emph{14},
  339--350\relax
\mciteBstWouldAddEndPuncttrue
\mciteSetBstMidEndSepPunct{\mcitedefaultmidpunct}
{\mcitedefaultendpunct}{\mcitedefaultseppunct}\relax
\EndOfBibitem
\bibitem[Botu and Ramprasad(2015)Botu, and Ramprasad]{Botu2015IJQC}
Botu,~V.; Ramprasad,~R. Adaptive Machine Learning Framework to Accelerate
  \emph{Ab Initio} Molecular Dynamics. \emph{Int. J. Quant. Chem.}
  \textbf{2015}, \emph{115}, 1074--1083\relax
\mciteBstWouldAddEndPuncttrue
\mciteSetBstMidEndSepPunct{\mcitedefaultmidpunct}
{\mcitedefaultendpunct}{\mcitedefaultseppunct}\relax
\EndOfBibitem
\bibitem[Behler(2015)]{Behler2015IJQC}
Behler,~J. Constructing High-Dimensional Neural Network Potentials: A Tutorial
  Review. \emph{Int. J. Quantum Chem.} \textbf{2015}, \emph{115},
  1032--1050\relax
\mciteBstWouldAddEndPuncttrue
\mciteSetBstMidEndSepPunct{\mcitedefaultmidpunct}
{\mcitedefaultendpunct}{\mcitedefaultseppunct}\relax
\EndOfBibitem
\bibitem[Ceriotti \latin{et~al.}(2013)Ceriotti, Tribello, and
  Parrinello]{Ceriotti2013JCTC}
Ceriotti,~M.; Tribello,~G.~A.; Parrinello,~M. Demonstrating the Transferability
  and the Descriptive Power of Sketch-Map. \emph{J. Chem. Theory Comput.}
  \textbf{2013}, \emph{9}, 1521--1532\relax
\mciteBstWouldAddEndPuncttrue
\mciteSetBstMidEndSepPunct{\mcitedefaultmidpunct}
{\mcitedefaultendpunct}{\mcitedefaultseppunct}\relax
\EndOfBibitem
\bibitem[Dral \latin{et~al.}(2017)Dral, Owens, Yurchenko, and
  Thiel]{Dral2017JCP}
Dral,~P.~O.; Owens,~A.; Yurchenko,~S.~N.; Thiel,~W. Structure-Based Sampling
  and Self-Correcting Machine Learning for Accurate Calculations of Potential
  Energy Surfaces and Vibrational Levels. \emph{J. Chem. Phys.} \textbf{2017},
  \emph{146}, 244108\relax
\mciteBstWouldAddEndPuncttrue
\mciteSetBstMidEndSepPunct{\mcitedefaultmidpunct}
{\mcitedefaultendpunct}{\mcitedefaultseppunct}\relax
\EndOfBibitem
\bibitem[Sobol' \latin{et~al.}(2011)Sobol', Asotsky, Kreinin, and
  Kucherenko]{Sobol2011W}
Sobol',~I.~M.; Asotsky,~D.; Kreinin,~A.; Kucherenko,~S. Construction and
  Comparison of High-Dimensional Sobol' Generators. \emph{Wilmott}
  \textbf{2011}, \emph{2011}, 64--79\relax
\mciteBstWouldAddEndPuncttrue
\mciteSetBstMidEndSepPunct{\mcitedefaultmidpunct}
{\mcitedefaultendpunct}{\mcitedefaultseppunct}\relax
\EndOfBibitem
\bibitem[Uteva \latin{et~al.}(2018)Uteva, Graham, Wilkinson, and
  Wheatley]{Uteva2018JCP}
Uteva,~E.; Graham,~R.~S.; Wilkinson,~R.~D.; Wheatley,~R.~J. Active Learning in
  Gaussian Process Interpolation of Potential Energy Surfaces. \emph{J. Chem.
  Phys.} \textbf{2018}, \emph{149}, 174114\relax
\mciteBstWouldAddEndPuncttrue
\mciteSetBstMidEndSepPunct{\mcitedefaultmidpunct}
{\mcitedefaultendpunct}{\mcitedefaultseppunct}\relax
\EndOfBibitem
\bibitem[Hansen \latin{et~al.}(2013)Hansen, Montavon, Biegler, Fazli, Rupp,
  Scheffler, von Lilienfeld, Tkatchenko, and M\"{u}ller]{Hansen2013JCTC}
Hansen,~K.; Montavon,~G.; Biegler,~F.; Fazli,~S.; Rupp,~M.; Scheffler,~M.; von
  Lilienfeld,~O.~A.; Tkatchenko,~A.; M\"{u}ller,~K.-R. Assessment and
  Validation of Machine Learning Methods for Predicting Molecular Atomization
  Energies. \emph{J. Chem. Theory Comput.} \textbf{2013}, \emph{9},
  3404--3419\relax
\mciteBstWouldAddEndPuncttrue
\mciteSetBstMidEndSepPunct{\mcitedefaultmidpunct}
{\mcitedefaultendpunct}{\mcitedefaultseppunct}\relax
\EndOfBibitem
\bibitem[Chmiela \latin{et~al.}(2017)Chmiela, Tkatchenko, Sauceda, Poltavsky,
  Sch{\"u}tt, and M{\"u}ller]{Chmiela2017SA}
Chmiela,~S.; Tkatchenko,~A.; Sauceda,~H.~E.; Poltavsky,~I.; Sch{\"u}tt,~K.~T.;
  M{\"u}ller,~K.-R. Machine Learning of Accurate Energy-Conserving Molecular
  Force Fields. \emph{Sci. Adv.} \textbf{2017}, \emph{3}\relax
\mciteBstWouldAddEndPuncttrue
\mciteSetBstMidEndSepPunct{\mcitedefaultmidpunct}
{\mcitedefaultendpunct}{\mcitedefaultseppunct}\relax
\EndOfBibitem
\bibitem[Christensen \latin{et~al.}(2020)Christensen, Bratholm, Faber, and
  Anatole~von Lilienfeld]{Christensen2020JCP}
Christensen,~A.~S.; Bratholm,~L.~A.; Faber,~F.~A.; Anatole~von Lilienfeld,~O.
  FCHL Revisited: Faster and More Accurate Quantum Machine Learning. \emph{J.
  Chem. Phys.} \textbf{2020}, \emph{152}, 044107\relax
\mciteBstWouldAddEndPuncttrue
\mciteSetBstMidEndSepPunct{\mcitedefaultmidpunct}
{\mcitedefaultendpunct}{\mcitedefaultseppunct}\relax
\EndOfBibitem
\bibitem[Kim \latin{et~al.}(2019)Kim, Park, and Choi]{Kim2019SD}
Kim,~H.; Park,~J.; Choi,~S. Energy Refinement and Analysis of Structures in the
  QM9 Database via a Highly Accurate Quantum Chemical Method. \emph{Sci. Data}
  \textbf{2019}, \emph{6}\relax
\mciteBstWouldAddEndPuncttrue
\mciteSetBstMidEndSepPunct{\mcitedefaultmidpunct}
{\mcitedefaultendpunct}{\mcitedefaultseppunct}\relax
\EndOfBibitem
\bibitem[Glavatskikh \latin{et~al.}(2019)Glavatskikh, Leguy, Hunault, Cauchi,
  and Da~Mota]{Glavatskikh2019JC}
Glavatskikh,~M.; Leguy,~J.; Hunault,~G.; Cauchi,~T.; Da~Mota,~B. Dataset's
  Chemical Diversity Limits the Generalizability of Machine Learning
  Predictions. \emph{J. Cheminform.} \textbf{2019}, \emph{11}\relax
\mciteBstWouldAddEndPuncttrue
\mciteSetBstMidEndSepPunct{\mcitedefaultmidpunct}
{\mcitedefaultendpunct}{\mcitedefaultseppunct}\relax
\EndOfBibitem
\bibitem[kag()]{kaggle1}
https://www.kaggle.com/c/champs-scalar-coupling/, 2020-05-01\relax
\mciteBstWouldAddEndPuncttrue
\mciteSetBstMidEndSepPunct{\mcitedefaultmidpunct}
{\mcitedefaultendpunct}{\mcitedefaultseppunct}\relax
\EndOfBibitem
\bibitem[{von Lilienfeld}(2018)]{vonLilienfeld2018twitter}
{von Lilienfeld},~O.~A. The QM9 challenge.
  https://twitter.com/ProfvLilienfeld/status/1073179005854121984, 2018\relax
\mciteBstWouldAddEndPuncttrue
\mciteSetBstMidEndSepPunct{\mcitedefaultmidpunct}
{\mcitedefaultendpunct}{\mcitedefaultseppunct}\relax
\EndOfBibitem
\bibitem[Fink \latin{et~al.}(2005)Fink, Bruggesser, and Reymond]{Fink2005ACIE}
Fink,~T.; Bruggesser,~H.; Reymond,~J.-L. Virtual Exploration of the
  Small-Molecule Chemical Universe Below 160 Daltons. \emph{Angew. Chem., Int.
  Ed.} \textbf{2005}, \emph{44}, 1504--1508\relax
\mciteBstWouldAddEndPuncttrue
\mciteSetBstMidEndSepPunct{\mcitedefaultmidpunct}
{\mcitedefaultendpunct}{\mcitedefaultseppunct}\relax
\EndOfBibitem
\bibitem[Fink and Reymond(2007)Fink, and Reymond]{Fink2007JCIM}
Fink,~T.; Reymond,~J.-L. {Virtual Exploration of the Chemical Universe up to 11
  Atoms of C, N, O, F: Assembly of 26.4 Million Structures (110.9 Million
  Stereoisomers) and Analysis for New Ring Systems, Stereochemistry,
  Physicochemical Properties, Compound Classes, and Drug Discovery}. \emph{J.
  Chem. Inf. Model.} \textbf{2007}, \emph{47}, 342--353\relax
\mciteBstWouldAddEndPuncttrue
\mciteSetBstMidEndSepPunct{\mcitedefaultmidpunct}
{\mcitedefaultendpunct}{\mcitedefaultseppunct}\relax
\EndOfBibitem
\bibitem[Blum and Reymond(2009)Blum, and Reymond]{Blum2009JACS}
Blum,~L.~C.; Reymond,~J.-L. 970 Million Druglike Small Molecules for Virtual
  Screening in the Chemical Universe Database {GDB-13}. \emph{J. Am. Chem.
  Soc.} \textbf{2009}, \emph{131}, 8732\relax
\mciteBstWouldAddEndPuncttrue
\mciteSetBstMidEndSepPunct{\mcitedefaultmidpunct}
{\mcitedefaultendpunct}{\mcitedefaultseppunct}\relax
\EndOfBibitem
\bibitem[Montavon \latin{et~al.}(2013)Montavon, Rupp, Gobre,
  Vazquez-Mayagoitia, Hansen, Tkatchenko, M{\"u}ller, and von
  Lilienfeld]{Montavon2013NJP}
Montavon,~G.; Rupp,~M.; Gobre,~V.; Vazquez-Mayagoitia,~A.; Hansen,~K.;
  Tkatchenko,~A.; M{\"u}ller,~K.-R.; von Lilienfeld,~O.~A. Machine Learning of
  Molecular Electronic Properties in Chemical Compound Space. \emph{New J.
  Phys.} \textbf{2013}, \emph{15}, 095003\relax
\mciteBstWouldAddEndPuncttrue
\mciteSetBstMidEndSepPunct{\mcitedefaultmidpunct}
{\mcitedefaultendpunct}{\mcitedefaultseppunct}\relax
\EndOfBibitem
\bibitem[Ruddigkeit \latin{et~al.}(2012)Ruddigkeit, van Deursen, Blum, and
  Reymond]{Ruddigkeit2012JCIM}
Ruddigkeit,~L.; van Deursen,~R.; Blum,~L.~C.; Reymond,~J.-L. Enumeration of 166
  Billion Organic Small Molecules in the Chemical Universe Database GDB-17.
  \emph{J. Chem. Inf. Model.} \textbf{2012}, \emph{52}, 2864--2875\relax
\mciteBstWouldAddEndPuncttrue
\mciteSetBstMidEndSepPunct{\mcitedefaultmidpunct}
{\mcitedefaultendpunct}{\mcitedefaultseppunct}\relax
\EndOfBibitem
\bibitem[Ramakrishnan \latin{et~al.}(2015)Ramakrishnan, Dral, Rupp, and von
  Lilienfeld]{Ramakrishnan2015JCTC}
Ramakrishnan,~R.; Dral,~P.~O.; Rupp,~M.; von Lilienfeld,~O.~A. Big Data Meets
  Quantum Chemistry Approximations: The $\Delta$-Machine Learning Approach.
  \emph{J. Chem. Theory Comput.} \textbf{2015}, \emph{11}, 2087--2096\relax
\mciteBstWouldAddEndPuncttrue
\mciteSetBstMidEndSepPunct{\mcitedefaultmidpunct}
{\mcitedefaultendpunct}{\mcitedefaultseppunct}\relax
\EndOfBibitem
\bibitem[Lee \latin{et~al.}(1988)Lee, Yang, and Parr]{Lee1988PRB}
Lee,~C.; Yang,~W.; Parr,~R.~G. Development of the Colle-Salvetti
  Correlation-Energy Formula into a Functional of the Electron Density.
  \emph{Phys. Rev. B} \textbf{1988}, \emph{37}, 785--789\relax
\mciteBstWouldAddEndPuncttrue
\mciteSetBstMidEndSepPunct{\mcitedefaultmidpunct}
{\mcitedefaultendpunct}{\mcitedefaultseppunct}\relax
\EndOfBibitem
\bibitem[Becke(1988)]{Becke1988PRA}
Becke,~A.~D. Density-Functional Exchange-Energy Approximation with Correct
  Asymptotic Behavior. \emph{Phys. Rev. A} \textbf{1988}, \emph{38},
  3098--3100\relax
\mciteBstWouldAddEndPuncttrue
\mciteSetBstMidEndSepPunct{\mcitedefaultmidpunct}
{\mcitedefaultendpunct}{\mcitedefaultseppunct}\relax
\EndOfBibitem
\bibitem[Stuke \latin{et~al.}(2020)Stuke, Kunkel, Golze, Todorovi\'{c},
  Margraf, Reuter, Rinke, and Oberhofer]{Stuke2020SD}
Stuke,~A.; Kunkel,~C.; Golze,~D.; Todorovi\'{c},~M.; Margraf,~J.~T.;
  Reuter,~K.; Rinke,~P.; Oberhofer,~H. Atomic Structures and Orbital Energies
  of 61,489 Crystal-Forming Organic Molecules. \emph{Sci. Data} \textbf{2020},
  \emph{7}\relax
\mciteBstWouldAddEndPuncttrue
\mciteSetBstMidEndSepPunct{\mcitedefaultmidpunct}
{\mcitedefaultendpunct}{\mcitedefaultseppunct}\relax
\EndOfBibitem
\bibitem[Perdew \latin{et~al.}(1996)Perdew, Burke, and
  Ernzerhof]{Perdew1996PRL}
Perdew,~J.~P.; Burke,~K.; Ernzerhof,~M. Generalized Gradient Approximation Made
  Simple. \emph{Phys. Rev. Lett.} \textbf{1996}, \emph{77}, 3865--3868\relax
\mciteBstWouldAddEndPuncttrue
\mciteSetBstMidEndSepPunct{\mcitedefaultmidpunct}
{\mcitedefaultendpunct}{\mcitedefaultseppunct}\relax
\EndOfBibitem
\bibitem[Nakata and Shimazaki(2017)Nakata, and Shimazaki]{Nakata2017JCIM}
Nakata,~M.; Shimazaki,~T. {PubChemQC Project: A Large-Scale First-Principles
  Electronic Structure Database for Data-Driven Chemistry}. \emph{J. Chem. Inf.
  Model.} \textbf{2017}, \emph{57}, 1300--1308\relax
\mciteBstWouldAddEndPuncttrue
\mciteSetBstMidEndSepPunct{\mcitedefaultmidpunct}
{\mcitedefaultendpunct}{\mcitedefaultseppunct}\relax
\EndOfBibitem
\bibitem[Kolb \latin{et~al.}(2016)Kolb, Zhao, Li, Jiang, and Guo]{Kolb2016JCP}
Kolb,~B.; Zhao,~B.; Li,~J.; Jiang,~B.; Guo,~H. Permutation Invariant Potential
  Energy Surfaces for Polyatomic Reactions using Atomistic Neural Networks.
  \emph{J. Chem. Phys.} \textbf{2016}, \emph{144}, 224103\relax
\mciteBstWouldAddEndPuncttrue
\mciteSetBstMidEndSepPunct{\mcitedefaultmidpunct}
{\mcitedefaultendpunct}{\mcitedefaultseppunct}\relax
\EndOfBibitem
\bibitem[Bernstein \latin{et~al.}(2019)Bernstein, Cs\'{a}nyi, and
  Deringer]{Bernstein2019npjCM}
Bernstein,~N.; Cs\'{a}nyi,~G.; Deringer,~V.~L. De Novo Exploration and
  Self-Guided Learning of Potential-Energy Surfaces. \emph{npj Comput. Mater.}
  \textbf{2019}, \emph{5}\relax
\mciteBstWouldAddEndPuncttrue
\mciteSetBstMidEndSepPunct{\mcitedefaultmidpunct}
{\mcitedefaultendpunct}{\mcitedefaultseppunct}\relax
\EndOfBibitem
\bibitem[Yao \latin{et~al.}(2020)Yao, Sanchez-Lengeling, Bobbitt, Bucior,
  Kumar, Collins, Burns, Woo, Farha, Snurr, and Aspuru-Guzik]{Yao2020arXiv}
Yao,~Z.; Sanchez-Lengeling,~B.; Bobbitt,~N.~S.; Bucior,~B.~J.; Kumar,~S. G.~H.;
  Collins,~S.~P.; Burns,~T.; Woo,~T.~K.; Farha,~O.; Snurr,~R.~Q.;
  Aspuru-Guzik,~A. {Inverse Design of Nanoporous Crystalline Reticular
  Materials with Deep Generative Models}. \emph{ChemRxiv} \textbf{2020},
  \emph{DOI:10.26434/chemrxiv.12186681.v1}\relax
\mciteBstWouldAddEndPuncttrue
\mciteSetBstMidEndSepPunct{\mcitedefaultmidpunct}
{\mcitedefaultendpunct}{\mcitedefaultseppunct}\relax
\EndOfBibitem
\bibitem[Krenn \latin{et~al.}(2019)Krenn, H{\"{a}}se, Nigam, Friederich, and
  Aspuru{-}Guzik]{Krenn2019arXiv}
Krenn,~M.; H{\"{a}}se,~F.; Nigam,~A.; Friederich,~P.; Aspuru{-}Guzik,~A.
  {SELFIES:} A Robust Representation of Semantically Constrained Graphs with an
  Example Application in Chemistry. \emph{arXiv} \textbf{2019},
  \emph{abs/1905.13741}\relax
\mciteBstWouldAddEndPuncttrue
\mciteSetBstMidEndSepPunct{\mcitedefaultmidpunct}
{\mcitedefaultendpunct}{\mcitedefaultseppunct}\relax
\EndOfBibitem
\bibitem[Gebauer \latin{et~al.}(2019)Gebauer, Gastegger, and
  Sch\"{u}tt]{Gebauer2019arXiv}
Gebauer,~N. W.~A.; Gastegger,~M.; Sch\"{u}tt,~K.~T. Symmetry-Adapted Generation
  of 3D Point Sets for the Targeted Discovery of Molecules. \emph{arXiv}
  \textbf{2019}, \emph{1906.00957}\relax
\mciteBstWouldAddEndPuncttrue
\mciteSetBstMidEndSepPunct{\mcitedefaultmidpunct}
{\mcitedefaultendpunct}{\mcitedefaultseppunct}\relax
\EndOfBibitem
\bibitem[Sch{\"u}tt \latin{et~al.}(2017)Sch{\"u}tt, Arbabzadah, Chmiela,
  M{\"u}ller, and Tkatchenko]{Schuett2017NC}
Sch{\"u}tt,~K.~T.; Arbabzadah,~F.; Chmiela,~S.; M{\"u}ller,~K.~R.;
  Tkatchenko,~A. Quantum-Chemical Insights from Deep Tensor Neural Networks.
  \emph{Nat. Commun.} \textbf{2017}, \emph{8}, 13890 EP --\relax
\mciteBstWouldAddEndPuncttrue
\mciteSetBstMidEndSepPunct{\mcitedefaultmidpunct}
{\mcitedefaultendpunct}{\mcitedefaultseppunct}\relax
\EndOfBibitem
\bibitem[Ye \latin{et~al.}(2019)Ye, Hu, Li, Zhang, Zhong, Zhang, Luo, Mukamel,
  and Jiang]{Ye2019PNAS}
Ye,~S.; Hu,~W.; Li,~X.; Zhang,~J.; Zhong,~K.; Zhang,~G.; Luo,~Y.; Mukamel,~S.;
  Jiang,~J. A Neural Network Protocol for Electronic Excitations of
  N-Methylacetamide. \emph{Proc. Natl. Acad. Sci.} \textbf{2019}, \emph{116},
  11612--11617\relax
\mciteBstWouldAddEndPuncttrue
\mciteSetBstMidEndSepPunct{\mcitedefaultmidpunct}
{\mcitedefaultendpunct}{\mcitedefaultseppunct}\relax
\EndOfBibitem
\bibitem[Jorgensen \latin{et~al.}(1996)Jorgensen, Maxwell, and
  Tirado-Rives]{Jorgensen1996JACS}
Jorgensen,~W.~L.; Maxwell,~D.~S.; Tirado-Rives,~J. Development and Testing of
  the OPLS All-Atom Force Field on Conformational Energetics and Properties of
  Organic Liquids. \emph{J. Am. Chem. Soc.} \textbf{1996}, \emph{118},
  11225--11236\relax
\mciteBstWouldAddEndPuncttrue
\mciteSetBstMidEndSepPunct{\mcitedefaultmidpunct}
{\mcitedefaultendpunct}{\mcitedefaultseppunct}\relax
\EndOfBibitem
\bibitem[Van Der~Spoel \latin{et~al.}(2005)Van Der~Spoel, Lindahl, Hess,
  Groenhof, Mark, and Berendsen]{VanDerSpoel2005JCC}
Van Der~Spoel,~D.; Lindahl,~E.; Hess,~B.; Groenhof,~G.; Mark,~A.~E.;
  Berendsen,~H. J.~C. GROMACS: Fast, Flexible, and Free. \emph{J. Comp. Chem.}
  \textbf{2005}, \emph{26}, 1701--1718\relax
\mciteBstWouldAddEndPuncttrue
\mciteSetBstMidEndSepPunct{\mcitedefaultmidpunct}
{\mcitedefaultendpunct}{\mcitedefaultseppunct}\relax
\EndOfBibitem
\bibitem[Ceriotti \latin{et~al.}(2011)Ceriotti, Tribello, and
  Parrinello]{Ceriotti2011PNAS}
Ceriotti,~M.; Tribello,~G.~A.; Parrinello,~M. Simplifying the Representation of
  Complex Free-Energy Landscapes using Sketch-Map. \emph{Proc. Natl. Acad. Sci}
  \textbf{2011}, \emph{108}, 13023--13028\relax
\mciteBstWouldAddEndPuncttrue
\mciteSetBstMidEndSepPunct{\mcitedefaultmidpunct}
{\mcitedefaultendpunct}{\mcitedefaultseppunct}\relax
\EndOfBibitem
\bibitem[Tribello \latin{et~al.}(2012)Tribello, Ceriotti, and
  Parrinello]{Tribello2012PNAS}
Tribello,~G.~A.; Ceriotti,~M.; Parrinello,~M. Using Sketch-Map Coordinates to
  Analyze and Bias Molecular Dynamics Simulations. \emph{Proc. Natl. Acad. Sci}
  \textbf{2012}, \emph{109}, 5196--5201\relax
\mciteBstWouldAddEndPuncttrue
\mciteSetBstMidEndSepPunct{\mcitedefaultmidpunct}
{\mcitedefaultendpunct}{\mcitedefaultseppunct}\relax
\EndOfBibitem
\bibitem[Seung \latin{et~al.}(1992)Seung, Opper, and Sompolinsky]{Seung1992}
Seung,~H.~S.; Opper,~M.; Sompolinsky,~H. Query by Committee. Proceedings of the
  Fifth Annual Workshop on Computational Learning Theory. New York, NY, USA,
  1992; p 287–294\relax
\mciteBstWouldAddEndPuncttrue
\mciteSetBstMidEndSepPunct{\mcitedefaultmidpunct}
{\mcitedefaultendpunct}{\mcitedefaultseppunct}\relax
\EndOfBibitem
\bibitem[Collins(2002)]{Collins2002TCA}
Collins,~M. Molecular Potential-Energy Surfaces for Chemical Reaction Dynamics.
  \emph{Theor. Chem. Acc.} \textbf{2002}, \emph{108}, 313–324\relax
\mciteBstWouldAddEndPuncttrue
\mciteSetBstMidEndSepPunct{\mcitedefaultmidpunct}
{\mcitedefaultendpunct}{\mcitedefaultseppunct}\relax
\EndOfBibitem
\bibitem[Godsi \latin{et~al.}(2010)Godsi, Collins, and Peskin]{Godsi2010JCP}
Godsi,~O.; Collins,~M.~A.; Peskin,~U. Quantum Grow—A Quantum Dynamics
  Sampling Approach for Growing Potential Energy Surfaces and Nonadiabatic
  Couplings. \emph{J. Chem. Phys.} \textbf{2010}, \emph{132}, 124106\relax
\mciteBstWouldAddEndPuncttrue
\mciteSetBstMidEndSepPunct{\mcitedefaultmidpunct}
{\mcitedefaultendpunct}{\mcitedefaultseppunct}\relax
\EndOfBibitem
\bibitem[Dawes \latin{et~al.}(2007)Dawes, Thompson, Guo, Wagner, and
  Minkoff]{Dawes2007JCP}
Dawes,~R.; Thompson,~D.~L.; Guo,~Y.; Wagner,~A.~F.; Minkoff,~M. Interpolating
  Moving Least-Squares Methods for Fitting Potential Energy Surfaces: Computing
  High-Density Potential Energy Surface Data from Low-Density \emph{Ab Initio}
  Data Points. \emph{J. Chem. Phys.} \textbf{2007}, \emph{126}, 184108\relax
\mciteBstWouldAddEndPuncttrue
\mciteSetBstMidEndSepPunct{\mcitedefaultmidpunct}
{\mcitedefaultendpunct}{\mcitedefaultseppunct}\relax
\EndOfBibitem
\bibitem[Dawes \latin{et~al.}(2008)Dawes, Thompson, Wagner, and
  Minkoff]{Dawes2008JCP}
Dawes,~R.; Thompson,~D.~L.; Wagner,~A.~F.; Minkoff,~M. Interpolating Moving
  Least-Squares Methods for Fitting Potential Energy Surfaces: A Strategy for
  Efficient Automatic Data Point Placement in High Dimensions. \emph{J. Chem.
  Phys.} \textbf{2008}, \emph{128}, 084107\relax
\mciteBstWouldAddEndPuncttrue
\mciteSetBstMidEndSepPunct{\mcitedefaultmidpunct}
{\mcitedefaultendpunct}{\mcitedefaultseppunct}\relax
\EndOfBibitem
\bibitem[Braams and Bowman(2009)Braams, and Bowman]{Braams2009IRPC}
Braams,~B.~J.; Bowman,~J.~M. Permutationally Invariant Potential Energy
  Surfaces in High Dimensionality. \emph{Int. Rev. Phys. Chem.} \textbf{2009},
  \emph{28}, 577--606\relax
\mciteBstWouldAddEndPuncttrue
\mciteSetBstMidEndSepPunct{\mcitedefaultmidpunct}
{\mcitedefaultendpunct}{\mcitedefaultseppunct}\relax
\EndOfBibitem
\bibitem[Qu \latin{et~al.}(2018)Qu, Yu, and Bowman]{Qu2018ARPC}
Qu,~C.; Yu,~Q.; Bowman,~J.~M. {Permutationally Invariant Potential Energy
  Surfaces}. \emph{Annu. Rev. Phys. Chem.} \textbf{2018}, \emph{69},
  151--175\relax
\mciteBstWouldAddEndPuncttrue
\mciteSetBstMidEndSepPunct{\mcitedefaultmidpunct}
{\mcitedefaultendpunct}{\mcitedefaultseppunct}\relax
\EndOfBibitem
\bibitem[Lorenz \latin{et~al.}(2004)Lorenz, Groß, and
  Scheffler]{Lorenz2004CPL}
Lorenz,~S.; Groß,~A.; Scheffler,~M. Representing High-Dimensional
  Potential-Energy Surfaces for Reactions at Surfaces by Neural Networks.
  \emph{Chem. Phys. Lett.} \textbf{2004}, \emph{395}, 210 -- 215\relax
\mciteBstWouldAddEndPuncttrue
\mciteSetBstMidEndSepPunct{\mcitedefaultmidpunct}
{\mcitedefaultendpunct}{\mcitedefaultseppunct}\relax
\EndOfBibitem
\bibitem[Raff \latin{et~al.}(2005)Raff, Malshe, Hagan, Doughan, Rockley, and
  Komanduri]{Raff2005JCP}
Raff,~L.~M.; Malshe,~M.; Hagan,~M.; Doughan,~D.~I.; Rockley,~M.~G.;
  Komanduri,~R. \emph{Ab Initio} Potential-Energy Surfaces for Complex,
  Multichannel Systems using Modified Novelty Sampling and Feedforward Neural
  Networks. \emph{J. Chem. Phys.} \textbf{2005}, \emph{122}, 084104\relax
\mciteBstWouldAddEndPuncttrue
\mciteSetBstMidEndSepPunct{\mcitedefaultmidpunct}
{\mcitedefaultendpunct}{\mcitedefaultseppunct}\relax
\EndOfBibitem
\bibitem[Behler and Parrinello(2007)Behler, and Parrinello]{Behler2007PRL}
Behler,~J.; Parrinello,~M. Generalized {Neural}-{Network} {Representation} of
  {High}-{Dimensional} {Potential}-{Energy} {Surfaces}. \emph{Phys. Rev. Lett.}
  \textbf{2007}, \emph{98}, 146401\relax
\mciteBstWouldAddEndPuncttrue
\mciteSetBstMidEndSepPunct{\mcitedefaultmidpunct}
{\mcitedefaultendpunct}{\mcitedefaultseppunct}\relax
\EndOfBibitem
\bibitem[Chen \latin{et~al.}(2013)Chen, Xu, Xu, and Zhang]{Chen2013JCP}
Chen,~J.; Xu,~X.; Xu,~X.; Zhang,~D.~H. A Global Potential Energy Surface for
  the {H}$_2$ + {OH} $\leftrightarrow$ {H}$_2${O + H} Reaction using Neural
  Networks. \emph{J. Chem. Phys.} \textbf{2013}, \emph{138}, 154301\relax
\mciteBstWouldAddEndPuncttrue
\mciteSetBstMidEndSepPunct{\mcitedefaultmidpunct}
{\mcitedefaultendpunct}{\mcitedefaultseppunct}\relax
\EndOfBibitem
\bibitem[Jiang and Guo(2015)Jiang, and Guo]{Jiang2015PRL}
Jiang,~B.; Guo,~H. Dynamics of Water Dissociative Chemisorption on Ni(111):
  Effects of Impact Sites and Incident Angles. \emph{Phys. Rev. Lett.}
  \textbf{2015}, \emph{114}, 166101\relax
\mciteBstWouldAddEndPuncttrue
\mciteSetBstMidEndSepPunct{\mcitedefaultmidpunct}
{\mcitedefaultendpunct}{\mcitedefaultseppunct}\relax
\EndOfBibitem
\bibitem[Shen \latin{et~al.}(2015)Shen, Chen, Zhang, Shao, and
  Zhang]{Shen2015JCP}
Shen,~X.; Chen,~J.; Zhang,~Z.; Shao,~K.; Zhang,~D.~H. {Methane Dissociation on
  Ni(111): A Fifteen-Dimensional Potential Energy Surface using Neural Network
  Method}. \emph{J. Chem. Phys.} \textbf{2015}, \emph{143}, 144701\relax
\mciteBstWouldAddEndPuncttrue
\mciteSetBstMidEndSepPunct{\mcitedefaultmidpunct}
{\mcitedefaultendpunct}{\mcitedefaultseppunct}\relax
\EndOfBibitem
\bibitem[Shao \latin{et~al.}(2016)Shao, Chen, Zhao, and Zhang]{Shao2016JCP}
Shao,~K.; Chen,~J.; Zhao,~Z.; Zhang,~D.~H. Communication: Fitting Potential
  Energy Surfaces with Fundamental Invariant Neural Network. \emph{J. Chem.
  Phys.} \textbf{2016}, \emph{145}, 071101\relax
\mciteBstWouldAddEndPuncttrue
\mciteSetBstMidEndSepPunct{\mcitedefaultmidpunct}
{\mcitedefaultendpunct}{\mcitedefaultseppunct}\relax
\EndOfBibitem
\bibitem[Cui and Krems(2016)Cui, and Krems]{Cui2016JPB}
Cui,~J.; Krems,~R.~V. Efficient Non-Parametric Fitting of Potential Energy
  Surfaces for Polyatomic Molecules with Gaussian Processes. \emph{J. Phys. B:
  At., Mol. Opt. Phys.} \textbf{2016}, \emph{49}, 224001\relax
\mciteBstWouldAddEndPuncttrue
\mciteSetBstMidEndSepPunct{\mcitedefaultmidpunct}
{\mcitedefaultendpunct}{\mcitedefaultseppunct}\relax
\EndOfBibitem
\bibitem[Kolb \latin{et~al.}(2017)Kolb, Marshall, Zhao, Jiang, and
  Guo]{Kolb2017JPCA}
Kolb,~B.; Marshall,~P.; Zhao,~B.; Jiang,~B.; Guo,~H. Representing Global
  Reactive Potential Energy Surfaces Using Gaussian Processes. \emph{J. Phys.
  Chem. A} \textbf{2017}, \emph{121}, 2552--2557\relax
\mciteBstWouldAddEndPuncttrue
\mciteSetBstMidEndSepPunct{\mcitedefaultmidpunct}
{\mcitedefaultendpunct}{\mcitedefaultseppunct}\relax
\EndOfBibitem
\bibitem[Kolb \latin{et~al.}(2017)Kolb, Luo, Zhou, Jiang, and
  Guo]{Kolb2017JPCL}
Kolb,~B.; Luo,~X.; Zhou,~X.; Jiang,~B.; Guo,~H. {High-Dimensional Atomistic
  Neural Network Potentials for Molecule–Surface Interactions: HCl Scattering
  from Au(111)}. \emph{J. Phys. Chem. Lett.} \textbf{2017}, \emph{8},
  666--672\relax
\mciteBstWouldAddEndPuncttrue
\mciteSetBstMidEndSepPunct{\mcitedefaultmidpunct}
{\mcitedefaultendpunct}{\mcitedefaultseppunct}\relax
\EndOfBibitem
\bibitem[Huang \latin{et~al.}(2017)Huang, Shang, Zhang, and Liu]{Huang2017CS}
Huang,~S.-D.; Shang,~C.; Zhang,~X.-J.; Liu,~Z.-P. Material Discovery by
  Combining Stochastic Surface Walking Global Optimization with a Neural
  Network. \emph{Chem. Sci.} \textbf{2017}, \emph{8}, 6327--6337\relax
\mciteBstWouldAddEndPuncttrue
\mciteSetBstMidEndSepPunct{\mcitedefaultmidpunct}
{\mcitedefaultendpunct}{\mcitedefaultseppunct}\relax
\EndOfBibitem
\bibitem[Zhou \latin{et~al.}(2017)Zhou, Nattino, Zhang, Chen, Kroes, Guo, and
  Jiang]{Zhou2017PCCP}
Zhou,~X.; Nattino,~F.; Zhang,~Y.; Chen,~J.; Kroes,~G.-J.; Guo,~H.; Jiang,~B.
  {Dissociative Chemisorption of Methane on Ni(111) using a Chemically Accurate
  Fifteen Dimensional Potential Energy Surface}. \emph{Phys. Chem. Chem. Phys.}
  \textbf{2017}, \emph{19}, 30540--30550\relax
\mciteBstWouldAddEndPuncttrue
\mciteSetBstMidEndSepPunct{\mcitedefaultmidpunct}
{\mcitedefaultendpunct}{\mcitedefaultseppunct}\relax
\EndOfBibitem
\bibitem[Zhang \latin{et~al.}(2019)Zhang, Hu, and Jiang]{Zhang2019JPCL}
Zhang,~Y.; Hu,~C.; Jiang,~B. Embedded Atom Neural Network Potentials: Efficient
  and Accurate Machine Learning with a Physically Inspired Representation.
  \emph{J. Phys. Chem. Lett.} \textbf{2019}, \emph{10}, 4962--4967\relax
\mciteBstWouldAddEndPuncttrue
\mciteSetBstMidEndSepPunct{\mcitedefaultmidpunct}
{\mcitedefaultendpunct}{\mcitedefaultseppunct}\relax
\EndOfBibitem
\bibitem[Wigner(1932)]{Wigner1932PR}
Wigner,~E. On The Quantum Correction for Thermodynamic Equilibrium. \emph{Phys.
  Rev.} \textbf{1932}, \emph{40}, 749--750\relax
\mciteBstWouldAddEndPuncttrue
\mciteSetBstMidEndSepPunct{\mcitedefaultmidpunct}
{\mcitedefaultendpunct}{\mcitedefaultseppunct}\relax
\EndOfBibitem
\bibitem[Bruccoleri and Karplus(1990)Bruccoleri, and Karplus]{Bruccoleri1990B}
Bruccoleri,~R.~E.; Karplus,~M. Conformational Sampling using High-Temperature
  Molecular Dynamics. \emph{Biopolymers} \textbf{1990}, \emph{29},
  1847--1862\relax
\mciteBstWouldAddEndPuncttrue
\mciteSetBstMidEndSepPunct{\mcitedefaultmidpunct}
{\mcitedefaultendpunct}{\mcitedefaultseppunct}\relax
\EndOfBibitem
\bibitem[Maximova \latin{et~al.}(2016)Maximova, Moffatt, Ma, Nussinov, and
  Shehu]{Maximova2016PLOSCB}
Maximova,~T.; Moffatt,~R.; Ma,~B.; Nussinov,~R.; Shehu,~A. Principles and
  Overview of Sampling Methods for Modeling Macromolecular Structure and
  Dynamics. \emph{PLOS computational Biology} \textbf{2016}, \emph{12},
  1--70\relax
\mciteBstWouldAddEndPuncttrue
\mciteSetBstMidEndSepPunct{\mcitedefaultmidpunct}
{\mcitedefaultendpunct}{\mcitedefaultseppunct}\relax
\EndOfBibitem
\bibitem[K\"astner(2011)]{Kaestner2011WIRCMS}
K\"astner,~J. Umbrella Sampling. \emph{Wiley Interdiscip. Rev. Comput. Mol.
  Sci.} \textbf{2011}, \emph{1}, 932--942\relax
\mciteBstWouldAddEndPuncttrue
\mciteSetBstMidEndSepPunct{\mcitedefaultmidpunct}
{\mcitedefaultendpunct}{\mcitedefaultseppunct}\relax
\EndOfBibitem
\bibitem[Tao(2019)]{Tao2019TCA}
Tao,~G. Trajectory-Guided Sampling for Molecular Dynamics Simulation.
  \emph{Theor. Chem. Acc.} \textbf{2019}, \emph{138}, 34\relax
\mciteBstWouldAddEndPuncttrue
\mciteSetBstMidEndSepPunct{\mcitedefaultmidpunct}
{\mcitedefaultendpunct}{\mcitedefaultseppunct}\relax
\EndOfBibitem
\bibitem[Yang \latin{et~al.}(2019)Yang, Shao, Zhang, Yang, and
  Gao]{Yang2019JCP}
Yang,~Y.~I.; Shao,~Q.; Zhang,~J.; Yang,~L.; Gao,~Y.~Q. Enhanced Sampling in
  Molecular Dynamics. \emph{J. Chem. Phys.} \textbf{2019}, \emph{151},
  070902\relax
\mciteBstWouldAddEndPuncttrue
\mciteSetBstMidEndSepPunct{\mcitedefaultmidpunct}
{\mcitedefaultendpunct}{\mcitedefaultseppunct}\relax
\EndOfBibitem
\bibitem[Herr \latin{et~al.}(2018)Herr, Yao, McIntyre, Toth, and
  Parkhill]{Herr2018JCP}
Herr,~J.~E.; Yao,~K.; McIntyre,~R.; Toth,~D.~W.; Parkhill,~J. Metadynamics for
  Training Neural Network Model Chemistries: A Competitive Assessment. \emph{J.
  Chem. Phys.} \textbf{2018}, \emph{148}, 241710\relax
\mciteBstWouldAddEndPuncttrue
\mciteSetBstMidEndSepPunct{\mcitedefaultmidpunct}
{\mcitedefaultendpunct}{\mcitedefaultseppunct}\relax
\EndOfBibitem
\bibitem[Grimme(2019)]{Grimme2019JCTC}
Grimme,~S. Exploration of Chemical compound, conformer, and Reaction Space with
  Meta-Dynamics Simulations Based on Tight-Binding Quantum Chemical
  Calculations. \emph{J. Chem. Theory Comput.} \textbf{2019}, \emph{15},
  2847--2862\relax
\mciteBstWouldAddEndPuncttrue
\mciteSetBstMidEndSepPunct{\mcitedefaultmidpunct}
{\mcitedefaultendpunct}{\mcitedefaultseppunct}\relax
\EndOfBibitem
\bibitem[Smith \latin{et~al.}(2018)Smith, Nebgen, Lubbers, Isayev, and
  Roitberg]{Smith2018JCP}
Smith,~J.~S.; Nebgen,~B.; Lubbers,~N.; Isayev,~O.; Roitberg,~A.~E. Less is
  More: Sampling Chemical Space with Active Learning. \emph{J. Chem. Phys.}
  \textbf{2018}, \emph{148}, 241733\relax
\mciteBstWouldAddEndPuncttrue
\mciteSetBstMidEndSepPunct{\mcitedefaultmidpunct}
{\mcitedefaultendpunct}{\mcitedefaultseppunct}\relax
\EndOfBibitem
\bibitem[Malbon \latin{et~al.}(2020)Malbon, Zhao, Guo, and
  Yarkony]{Malbon2020PCCP}
Malbon,~C.~L.; Zhao,~B.; Guo,~H.; Yarkony,~D.~R. {On the Nonadiabatic
  Collisional Quenching of OH(A) by H$_2$: A Four Coupled Quasi-Diabatic State
  Description}. \emph{Phys. Chem. Chem. Phys.} \textbf{2020}, --\relax
\mciteBstWouldAddEndPuncttrue
\mciteSetBstMidEndSepPunct{\mcitedefaultmidpunct}
{\mcitedefaultendpunct}{\mcitedefaultseppunct}\relax
\EndOfBibitem
\bibitem[Xu \latin{et~al.}(2014)Xu, Chen, and Zhang]{Xu2014CJCP}
Xu,~X.; Chen,~J.; Zhang,~D.~H. Global Potential Energy Surface for the H+CH$_4
  \leftrightarrow$ H$_2$+CH$_3$ Reaction using Neural Networks. \emph{Chin. J.
  Chem. Phys.} \textbf{2014}, \emph{27}, 373--379\relax
\mciteBstWouldAddEndPuncttrue
\mciteSetBstMidEndSepPunct{\mcitedefaultmidpunct}
{\mcitedefaultendpunct}{\mcitedefaultseppunct}\relax
\EndOfBibitem
\bibitem[Li and Guo(2015)Li, and Guo]{Li2015JCP}
Li,~J.; Guo,~H. {Communication: An Accurate Full 15 Dimensional Permutationally
  Invariant Potential Energy Surface for the {OH + CH$_4$ $\rightarrow$ H$_2$O
  + CH$_3$} Reaction}. \emph{J. Chem. Phys.} \textbf{2015}, \emph{143},
  221103\relax
\mciteBstWouldAddEndPuncttrue
\mciteSetBstMidEndSepPunct{\mcitedefaultmidpunct}
{\mcitedefaultendpunct}{\mcitedefaultseppunct}\relax
\EndOfBibitem
\bibitem[Jiang and Guo(2014)Jiang, and Guo]{Jiang2014PCCP}
Jiang,~B.; Guo,~H. {Six-Dimensional Quantum Dynamics for Dissociative
  Chemisorption of H$_2$ and D$_2$ on Ag(111) on a Permutation Invariant
  Potential Energy Surface}. \emph{Phys. Chem. Chem. Phys.} \textbf{2014},
  \emph{16}, 24704--24715\relax
\mciteBstWouldAddEndPuncttrue
\mciteSetBstMidEndSepPunct{\mcitedefaultmidpunct}
{\mcitedefaultendpunct}{\mcitedefaultseppunct}\relax
\EndOfBibitem
\bibitem[Toyoura \latin{et~al.}(2016)Toyoura, Hirano, Seko, Shiga, Kuwabara,
  Karasuyama, Shitara, and Takeuchi]{Toyoura2016PRB}
Toyoura,~K.; Hirano,~D.; Seko,~A.; Shiga,~M.; Kuwabara,~A.; Karasuyama,~M.;
  Shitara,~K.; Takeuchi,~I. Machine-Learning-Based Selective Sampling Procedure
  for Identifying the Low-Energy Region in a Potential Energy Surface: A Case
  Study on Proton Conduction in Oxides. \emph{Phys. Rev. B} \textbf{2016},
  \emph{93}, 054112\relax
\mciteBstWouldAddEndPuncttrue
\mciteSetBstMidEndSepPunct{\mcitedefaultmidpunct}
{\mcitedefaultendpunct}{\mcitedefaultseppunct}\relax
\EndOfBibitem
\bibitem[Guan \latin{et~al.}(2018)Guan, Yang, and Zhang]{Guan2018MP}
Guan,~Y.; Yang,~S.; Zhang,~D.~H. Construction of Reactive Potential Energy
  Surfaces with Gaussian Process Regression: Active Data Selection. \emph{Mol.
  Phys.} \textbf{2018}, \emph{116}, 823--834\relax
\mciteBstWouldAddEndPuncttrue
\mciteSetBstMidEndSepPunct{\mcitedefaultmidpunct}
{\mcitedefaultendpunct}{\mcitedefaultseppunct}\relax
\EndOfBibitem
\bibitem[Vargas-Hern{\'{a}}ndez \latin{et~al.}(2019)Vargas-Hern{\'{a}}ndez,
  Guan, Zhang, and Krems]{Vargas-Hernandez2019NJP}
Vargas-Hern{\'{a}}ndez,~R.~A.; Guan,~Y.; Zhang,~D.~H.; Krems,~R.~V. Bayesian
  Optimization for the Inverse Scattering Problem in Quantum Reaction Dynamics.
  \emph{New J. Phys.} \textbf{2019}, \emph{21}, 022001\relax
\mciteBstWouldAddEndPuncttrue
\mciteSetBstMidEndSepPunct{\mcitedefaultmidpunct}
{\mcitedefaultendpunct}{\mcitedefaultseppunct}\relax
\EndOfBibitem
\bibitem[Todorovi\'{c} \latin{et~al.}(2019)Todorovi\'{c}, Gutmann, Corander,
  and Rinke]{Todorovic2018npjCM}
Todorovi\'{c},~M.; Gutmann,~M.~U.; Corander,~J.; Rinke,~P. Bayesian Inference
  of Atomistic Structure in Functional Materials. \emph{npj Comput. Mater.}
  \textbf{2019}, \emph{5}\relax
\mciteBstWouldAddEndPuncttrue
\mciteSetBstMidEndSepPunct{\mcitedefaultmidpunct}
{\mcitedefaultendpunct}{\mcitedefaultseppunct}\relax
\EndOfBibitem
\bibitem[Butler \latin{et~al.}(2018)Butler, Davies, Cartwright, Isayev, and
  Walsh]{Butler2018N}
Butler,~K.~T.; Davies,~D.~W.; Cartwright,~H.; Isayev,~O.; Walsh,~A. {Machine
  Learning for Molecular and Materials Science}. \emph{Nature} \textbf{2018},
  \emph{559}, 547--555\relax
\mciteBstWouldAddEndPuncttrue
\mciteSetBstMidEndSepPunct{\mcitedefaultmidpunct}
{\mcitedefaultendpunct}{\mcitedefaultseppunct}\relax
\EndOfBibitem
\bibitem[Haghighatlari \latin{et~al.}(2020)Haghighatlari, Li, Heidar-Zadeh,
  Liu, Guan, and {Head-Gordon}]{Haghighatlari2020C}
Haghighatlari,~M.; Li,~J.; Heidar-Zadeh,~F.; Liu,~Y.; Guan,~X.;
  {Head-Gordon},~T. Learning to Make Chemical Predictions: The Interplay of
  Feature Representation, Data, and Machine Learning Methods. \emph{Chem}
  \textbf{2020}, \relax
\mciteBstWouldAddEndPunctfalse
\mciteSetBstMidEndSepPunct{\mcitedefaultmidpunct}
{}{\mcitedefaultseppunct}\relax
\EndOfBibitem
\bibitem[{B}ishop(2006)]{Bishop2006}
{B}ishop,~C.~M. \emph{{P}attern {R}ecognition and {M}achine {L}earning}, 1st
  ed.; {S}pringer: New York, 2006\relax
\mciteBstWouldAddEndPuncttrue
\mciteSetBstMidEndSepPunct{\mcitedefaultmidpunct}
{\mcitedefaultendpunct}{\mcitedefaultseppunct}\relax
\EndOfBibitem
\bibitem[Halama(2019)]{Halama2019BJC}
Halama,~N. Machine Learning for Tissue Diagnostics in Oncology: Brave New
  World. \emph{Br. J. Cancer} \textbf{2019}, \emph{121}, 431–433\relax
\mciteBstWouldAddEndPuncttrue
\mciteSetBstMidEndSepPunct{\mcitedefaultmidpunct}
{\mcitedefaultendpunct}{\mcitedefaultseppunct}\relax
\EndOfBibitem
\bibitem[Bychkov \latin{et~al.}(2018)Bychkov, Linder, Turkki, Nordling,
  Kovanen, Verrill, Walliander, Lundin, Haglund, and Lundin]{Bychkov2018SR}
Bychkov,~D.; Linder,~N.; Turkki,~R.; Nordling,~S.; Kovanen,~P.~E.; Verrill,~C.;
  Walliander,~M.; Lundin,~M.; Haglund,~C.; Lundin,~J. Deep Learning Based
  Tissue Analysis Predicts Outcome in Colorectal Cancer. \emph{Sci. Rep.}
  \textbf{2018}, \emph{8}, 3395\relax
\mciteBstWouldAddEndPuncttrue
\mciteSetBstMidEndSepPunct{\mcitedefaultmidpunct}
{\mcitedefaultendpunct}{\mcitedefaultseppunct}\relax
\EndOfBibitem
\bibitem[G{\'o}mez-Meire \latin{et~al.}(2014)G{\'o}mez-Meire, Campos,
  Falqu{\'e}, D{\'i}az, and Fdez-Riverola]{Gomez2014FRI}
G{\'o}mez-Meire,~S.; Campos,~C.; Falqu{\'e},~E.; D{\'i}az,~F.;
  Fdez-Riverola,~F. Assuring the Authenticity of Northwest Spain White Wine
  Varieties using Machine Learning Techniques. \emph{Food Res. Int.}
  \textbf{2014}, \emph{60}, 230 -- 240\relax
\mciteBstWouldAddEndPuncttrue
\mciteSetBstMidEndSepPunct{\mcitedefaultmidpunct}
{\mcitedefaultendpunct}{\mcitedefaultseppunct}\relax
\EndOfBibitem
\bibitem[Watanabe \latin{et~al.}(2020)Watanabe, Murata, Ogawa, Vavricka, Kondo,
  Ogino, and Araki]{Watanabe2020JCIM}
Watanabe,~N.; Murata,~M.; Ogawa,~T.; Vavricka,~C.~J.; Kondo,~A.; Ogino,~C.;
  Araki,~M. Exploration and Evaluation of Machine Learning-Based Models for
  Predicting Enzymatic Reactions. \emph{J. Chem. Inf. Model.} \textbf{2020},
  \emph{60}, 1833--1843\relax
\mciteBstWouldAddEndPuncttrue
\mciteSetBstMidEndSepPunct{\mcitedefaultmidpunct}
{\mcitedefaultendpunct}{\mcitedefaultseppunct}\relax
\EndOfBibitem
\bibitem[Chen and Guestrin(2016)Chen, and Guestrin]{Chen2016}
Chen,~T.; Guestrin,~C. XGBoost: A Scalable Tree Boosting System. Proceedings of
  the 22nd ACM SIGKDD International Conference on Knowledge Discovery and Data
  Mining. New York, NY, USA, 2016; p 785–794\relax
\mciteBstWouldAddEndPuncttrue
\mciteSetBstMidEndSepPunct{\mcitedefaultmidpunct}
{\mcitedefaultendpunct}{\mcitedefaultseppunct}\relax
\EndOfBibitem
\bibitem[Ahneman \latin{et~al.}(2018)Ahneman, Estrada, Lin, Dreher, and
  Doyle]{Ahneman2018S}
Ahneman,~D.~T.; Estrada,~J.~G.; Lin,~S.; Dreher,~S.~D.; Doyle,~A.~G. Predicting
  Reaction Performance in C{\textendash}N Cross-Coupling using Machine
  Learning. \emph{Science} \textbf{2018}, \emph{360}, 186--190\relax
\mciteBstWouldAddEndPuncttrue
\mciteSetBstMidEndSepPunct{\mcitedefaultmidpunct}
{\mcitedefaultendpunct}{\mcitedefaultseppunct}\relax
\EndOfBibitem
\bibitem[Atahan-Evrenk and Atalay(2019)Atahan-Evrenk, and
  Atalay]{Atahan-Evrenk2019JPCA}
Atahan-Evrenk,~S.; Atalay,~F.~B. Prediction of Intramolecular Reorganization
  Energy Using Machine Learning. \emph{J. Phys. Chem. A} \textbf{2019},
  \emph{123}, 7855--7863\relax
\mciteBstWouldAddEndPuncttrue
\mciteSetBstMidEndSepPunct{\mcitedefaultmidpunct}
{\mcitedefaultendpunct}{\mcitedefaultseppunct}\relax
\EndOfBibitem
\bibitem[Hofmann \latin{et~al.}(2008)Hofmann, Schölkopf, and
  Smola]{Hofmann2008}
Hofmann,~T.; Schölkopf,~B.; Smola,~A.~J. Kernel Methods in Machine Learning.
  \emph{Ann. Statist.} \textbf{2008}, \emph{36}, 1171--1220\relax
\mciteBstWouldAddEndPuncttrue
\mciteSetBstMidEndSepPunct{\mcitedefaultmidpunct}
{\mcitedefaultendpunct}{\mcitedefaultseppunct}\relax
\EndOfBibitem
\bibitem[Raschka and Mirjalili(2019)Raschka, and Mirjalili]{Raschka2019}
Raschka,~S.; Mirjalili,~V. \emph{Python Machine Learning}, 3rd ed.; Packt
  Publishing, 2019\relax
\mciteBstWouldAddEndPuncttrue
\mciteSetBstMidEndSepPunct{\mcitedefaultmidpunct}
{\mcitedefaultendpunct}{\mcitedefaultseppunct}\relax
\EndOfBibitem
\bibitem[Xue \latin{et~al.}(2020)Xue, Barbatti, and Dral]{Xue2020chemrxiv}
Xue,~B.-X.; Barbatti,~M.; Dral,~P.~O. {Machine Learning for Absorption Cross
  Sections}. \emph{{ChemRxiv}} \textbf{2020},
  \emph{{DOI:10.26434/chemrxiv.12594191.v1}}\relax
\mciteBstWouldAddEndPuncttrue
\mciteSetBstMidEndSepPunct{\mcitedefaultmidpunct}
{\mcitedefaultendpunct}{\mcitedefaultseppunct}\relax
\EndOfBibitem
\bibitem[Ramakrishnan and von Lilienfeld(2017)Ramakrishnan, and von
  Lilienfeld]{Ramakrishnan2015}
Ramakrishnan,~R.; von Lilienfeld,~O.~A. \emph{Reviews in Computational
  Chemistry}; John Wiley \& Sons, Ltd, 2017; Chapter 5, pp 225--256\relax
\mciteBstWouldAddEndPuncttrue
\mciteSetBstMidEndSepPunct{\mcitedefaultmidpunct}
{\mcitedefaultendpunct}{\mcitedefaultseppunct}\relax
\EndOfBibitem
\bibitem[Bart{{\'o}}k \latin{et~al.}(2013)Bart{{\'o}}k, Kondor, and
  Cs{{\'a}}nyi]{Bartok2013PRB}
Bart{{\'o}}k,~A.~P.; Kondor,~R.; Cs{{\'a}}nyi,~G. On Representing Chemical
  Environments. \emph{Phys. Rev. B} \textbf{2013}, \emph{87}, 184115\relax
\mciteBstWouldAddEndPuncttrue
\mciteSetBstMidEndSepPunct{\mcitedefaultmidpunct}
{\mcitedefaultendpunct}{\mcitedefaultseppunct}\relax
\EndOfBibitem
\bibitem[Glorot and Bengio(2010)Glorot, and Bengio]{Glorot2010}
Glorot,~X.; Bengio,~Y. Understanding the Difficulty of Training Deep
  Feedforward Neural Networks. Proceedings of the Thirteenth International
  conference on Artificial Intelligence and Statistics. Chia Laguna Resort,
  Sardinia, Italy, 2010; pp 249--256\relax
\mciteBstWouldAddEndPuncttrue
\mciteSetBstMidEndSepPunct{\mcitedefaultmidpunct}
{\mcitedefaultendpunct}{\mcitedefaultseppunct}\relax
\EndOfBibitem
\bibitem[Duchi \latin{et~al.}(2011)Duchi, Hazan, and Singer]{Duchi2011JMLR}
Duchi,~J.; Hazan,~E.; Singer,~Y. {Adaptive Subgradient Methods for Online
  Learning and Stochastic Optimization}. \emph{J. Mach. Learn. Res.}
  \textbf{2011}, \emph{12}, 2121--2159\relax
\mciteBstWouldAddEndPuncttrue
\mciteSetBstMidEndSepPunct{\mcitedefaultmidpunct}
{\mcitedefaultendpunct}{\mcitedefaultseppunct}\relax
\EndOfBibitem
\bibitem[Kingma and Ba(2014)Kingma, and Ba]{Adam2014}
Kingma,~D.~P.; Ba,~J. Adam: {A} Method for Stochastic Optimization.
  \emph{arXiv} \textbf{2014}, \emph{abs/1412.6980}, 1412.6980\relax
\mciteBstWouldAddEndPuncttrue
\mciteSetBstMidEndSepPunct{\mcitedefaultmidpunct}
{\mcitedefaultendpunct}{\mcitedefaultseppunct}\relax
\EndOfBibitem
\bibitem[{P}uskorius and {F}eldkamp(1991){P}uskorius, and
  {F}eldkamp]{Puskorius1991}
{P}uskorius,~G.~V.; {F}eldkamp,~L.~A. {D}ecoupled extended {Kalman} filter
  training of feedforward layered networks. {IJCNN}-91-{Seattle}
  {International} {Joint} {Conference} on {Neural} {Networks}. 1991; pp
  771--777\relax
\mciteBstWouldAddEndPuncttrue
\mciteSetBstMidEndSepPunct{\mcitedefaultmidpunct}
{\mcitedefaultendpunct}{\mcitedefaultseppunct}\relax
\EndOfBibitem
\bibitem[Singraber \latin{et~al.}(2019)Singraber, Morawietz, Behler, and
  Dellago]{Singraber2019JCTCa}
Singraber,~A.; Morawietz,~T.; Behler,~J.; Dellago,~C. Parallel Multistream
  Training of High-Dimensional Neural Network Potentials. \emph{J. Chem. Theory
  Comput.} \textbf{2019}, \emph{15}, 3075--3092\relax
\mciteBstWouldAddEndPuncttrue
\mciteSetBstMidEndSepPunct{\mcitedefaultmidpunct}
{\mcitedefaultendpunct}{\mcitedefaultseppunct}\relax
\EndOfBibitem
\bibitem[Behler(2011)]{Behler2011JCP}
Behler,~J. Atom-Centered Symmetry Functions for Constructing High-Dimensional
  Neural Network Potentials. \emph{J. Chem. Phys.} \textbf{2011}, \emph{134},
  074106\relax
\mciteBstWouldAddEndPuncttrue
\mciteSetBstMidEndSepPunct{\mcitedefaultmidpunct}
{\mcitedefaultendpunct}{\mcitedefaultseppunct}\relax
\EndOfBibitem
\bibitem[LeCun and Bengio(1995)LeCun, and Bengio]{LeCun1995}
LeCun,~Y.; Bengio,~Y. \emph{The Handbook of Brain Theory and Neural Networks};
  The MIT Press, Cambridge, MA, USA, 1995; pp 255--257\relax
\mciteBstWouldAddEndPuncttrue
\mciteSetBstMidEndSepPunct{\mcitedefaultmidpunct}
{\mcitedefaultendpunct}{\mcitedefaultseppunct}\relax
\EndOfBibitem
\bibitem[Krizhevsky \latin{et~al.}(2012)Krizhevsky, Sutskever, and
  Hinton]{Krizhevsky2012NIPS}
Krizhevsky,~A.; Sutskever,~I.; Hinton,~G.~E. ImageNet Classification with Deep
  Convolutional Neural Networks. \textbf{2012}, 1097--1105\relax
\mciteBstWouldAddEndPuncttrue
\mciteSetBstMidEndSepPunct{\mcitedefaultmidpunct}
{\mcitedefaultendpunct}{\mcitedefaultseppunct}\relax
\EndOfBibitem
\bibitem[Sainath \latin{et~al.}(2015)Sainath, Kingsbury, Saon, Soltau, rahman
  Mohamed, Dahl, and Ramabhadran]{Sainath2015NN}
Sainath,~T.~N.; Kingsbury,~B.; Saon,~G.; Soltau,~H.; rahman Mohamed,~A.;
  Dahl,~G.; Ramabhadran,~B. {Deep Convolutional Neural Networks for Large-scale
  Speech Tasks}. \emph{Neural Networks} \textbf{2015}, \emph{64}, 39 --
  48\relax
\mciteBstWouldAddEndPuncttrue
\mciteSetBstMidEndSepPunct{\mcitedefaultmidpunct}
{\mcitedefaultendpunct}{\mcitedefaultseppunct}\relax
\EndOfBibitem
\bibitem[Gilmer \latin{et~al.}(2017)Gilmer, Schoenholz, Riley, Vinyals, and
  Dahl]{Gilmer2017}
Gilmer,~J.; Schoenholz,~S.~S.; Riley,~P.~F.; Vinyals,~O.; Dahl,~G.~E. Neural
  Message Passing for Quantum Chemistry. Proceedings of the 34th International
  Conference on Machine Learning - Volume 70. 2017; p 1263–1272\relax
\mciteBstWouldAddEndPuncttrue
\mciteSetBstMidEndSepPunct{\mcitedefaultmidpunct}
{\mcitedefaultendpunct}{\mcitedefaultseppunct}\relax
\EndOfBibitem
\bibitem[Sch\"{u}tt(2018)]{phdschuett}
Sch\"{u}tt,~K. Learning Representations of Atomistic Systems with Deep Neural
  Networks. Doctoral Thesis, Technische Universit\"{a}t Berlin, Berlin,
  2018\relax
\mciteBstWouldAddEndPuncttrue
\mciteSetBstMidEndSepPunct{\mcitedefaultmidpunct}
{\mcitedefaultendpunct}{\mcitedefaultseppunct}\relax
\EndOfBibitem
\bibitem[Dral(2019)]{Dral2019JCC}
Dral,~P.~O. MLatom: A Program Package for Quantum Chemical Research Assisted by
  Machine Learning. \emph{J. Comput. Chem.} \textbf{2019}, \emph{40},
  2339--2347\relax
\mciteBstWouldAddEndPuncttrue
\mciteSetBstMidEndSepPunct{\mcitedefaultmidpunct}
{\mcitedefaultendpunct}{\mcitedefaultseppunct}\relax
\EndOfBibitem
\bibitem[Christensen \latin{et~al.}(2017)Christensen, Faber, Huang, Bratholm,
  Tkatchenko, M\"{u}ller, and Lilienfeld]{QML}
Christensen,~A.; Faber,~F.; Huang,~B.; Bratholm,~L.; Tkatchenko,~A.;
  M\"{u}ller,~K.; Lilienfeld,~O. QML: A Python Toolkit for Quantum Machine
  Learning. https://github.com/qmlcode/qml, 2017\relax
\mciteBstWouldAddEndPuncttrue
\mciteSetBstMidEndSepPunct{\mcitedefaultmidpunct}
{\mcitedefaultendpunct}{\mcitedefaultseppunct}\relax
\EndOfBibitem
\bibitem[Hansen \latin{et~al.}(2015)Hansen, Biegler, Ramakrishnan, Pronobis,
  von Lilienfeld, M\"uller, and Tkatchenko]{Hansen2015JPCL}
Hansen,~K.; Biegler,~F.; Ramakrishnan,~R.; Pronobis,~W.; von Lilienfeld,~O.~A.;
  M\"uller,~K.-R.; Tkatchenko,~A. Machine Learning Predictions of Molecular
  Properties: Accurate Many-Body Potentials and Nonlocality in Chemical Space.
  \emph{J. Phys. Chem. Lett.} \textbf{2015}, \emph{6}, 2326--2331\relax
\mciteBstWouldAddEndPuncttrue
\mciteSetBstMidEndSepPunct{\mcitedefaultmidpunct}
{\mcitedefaultendpunct}{\mcitedefaultseppunct}\relax
\EndOfBibitem
\bibitem[Pozdnyakov \latin{et~al.}(2020)Pozdnyakov, Willatt, Bart\'{o}k,
  Ortner, Cs\'{a}nyi, and Ceriotti]{Pozdnyakov2020arXiv}
Pozdnyakov,~S.~N.; Willatt,~M.~J.; Bart\'{o}k,~A.~P.; Ortner,~C.;
  Cs\'{a}nyi,~G.; Ceriotti,~M. On the Completeness of Atomic Structure
  Representations. \emph{arXiv} \textbf{2020}, \emph{2001.11696}\relax
\mciteBstWouldAddEndPuncttrue
\mciteSetBstMidEndSepPunct{\mcitedefaultmidpunct}
{\mcitedefaultendpunct}{\mcitedefaultseppunct}\relax
\EndOfBibitem
\bibitem[\c{C}aylak \latin{et~al.}(2020)\c{C}aylak, von Lilienfeld, and
  Baumeier]{Caylak2020arXiv}
\c{C}aylak,~O.; von Lilienfeld,~O.~A.; Baumeier,~B. Wasserstein Metric for
  Improved QML with Adjacency Matrix Representations. \emph{arXiv}
  \textbf{2020}, \emph{2001.11005}\relax
\mciteBstWouldAddEndPuncttrue
\mciteSetBstMidEndSepPunct{\mcitedefaultmidpunct}
{\mcitedefaultendpunct}{\mcitedefaultseppunct}\relax
\EndOfBibitem
\bibitem[Bowman \latin{et~al.}(2011)Bowman, Czak\'{o}, and Fu]{Bowman2011PCCP}
Bowman,~J.~M.; Czak\'{o},~G.; Fu,~B. High-Dimensional {\emph{Ab Initio}}
  Potential Energy Surfaces for Reaction Dynamics Calculations. \emph{Phys.
  Chem. Chem. Phys.} \textbf{2011}, \emph{13}, 8094--8111\relax
\mciteBstWouldAddEndPuncttrue
\mciteSetBstMidEndSepPunct{\mcitedefaultmidpunct}
{\mcitedefaultendpunct}{\mcitedefaultseppunct}\relax
\EndOfBibitem
\bibitem[Herr \latin{et~al.}(2019)Herr, Koh, Yao, and Parkhill]{Herr2019JCP}
Herr,~J.~E.; Koh,~K.; Yao,~K.; Parkhill,~J. Compressing Physics with an
  Autoencoder: Creating an Atomic Species Representation to Improve Machine
  Learning Models in the Chemical Sciences. \emph{J. Chem. Phys.}
  \textbf{2019}, \emph{151}, 084103\relax
\mciteBstWouldAddEndPuncttrue
\mciteSetBstMidEndSepPunct{\mcitedefaultmidpunct}
{\mcitedefaultendpunct}{\mcitedefaultseppunct}\relax
\EndOfBibitem
\bibitem[Kang \latin{et~al.}(2020)Kang, Seok, and Lee]{Kang2020arXiv}
Kang,~B.; Seok,~C.; Lee,~J. {Prediction of Molecular Electronic Transitions
  using Random Forests}. \emph{ChemRxiv} \textbf{2020},
  \emph{DOI:10.26434/chemrxiv.12482840.v1}\relax
\mciteBstWouldAddEndPuncttrue
\mciteSetBstMidEndSepPunct{\mcitedefaultmidpunct}
{\mcitedefaultendpunct}{\mcitedefaultseppunct}\relax
\EndOfBibitem
\bibitem[Richings and Habershon(2017)Richings, and Habershon]{Richings2017JCTC}
Richings,~G.~W.; Habershon,~S. {Direct Quantum Dynamics Using Grid-Based Wave
  Function Propagation and Machine-Learned Potential Energy Surfaces}. \emph{J.
  Chem. Theory Comput.} \textbf{2017}, \emph{13}, 4012--4024\relax
\mciteBstWouldAddEndPuncttrue
\mciteSetBstMidEndSepPunct{\mcitedefaultmidpunct}
{\mcitedefaultendpunct}{\mcitedefaultseppunct}\relax
\EndOfBibitem
\bibitem[Unke and Meuwly(2019)Unke, and Meuwly]{Unke2019JCTC}
Unke,~O.~T.; Meuwly,~M. {PhysNet: A Neural Network for Predicting Energies,
  Forces, Dipole Moments, and Partial Charges}. \emph{J. Chem. Theory Comput.}
  \textbf{2019}, \emph{15}, 3678--3693\relax
\mciteBstWouldAddEndPuncttrue
\mciteSetBstMidEndSepPunct{\mcitedefaultmidpunct}
{\mcitedefaultendpunct}{\mcitedefaultseppunct}\relax
\EndOfBibitem
\bibitem[Lubbers \latin{et~al.}(2018)Lubbers, Smith, and
  Barros]{Lubbers2018JCP}
Lubbers,~N.; Smith,~J.~S.; Barros,~K. Hierarchical Modeling of Molecular
  Energies using a Deep Neural Network. \emph{J. Chem. Phys.} \textbf{2018},
  \emph{148}, 241715\relax
\mciteBstWouldAddEndPuncttrue
\mciteSetBstMidEndSepPunct{\mcitedefaultmidpunct}
{\mcitedefaultendpunct}{\mcitedefaultseppunct}\relax
\EndOfBibitem
\bibitem[Zheng \latin{et~al.}(2019)Zheng, Gao, and Eisfeld]{Zheng2019PRL}
Zheng,~F.; Gao,~X.; Eisfeld,~A. Excitonic Wave Function Reconstruction from
  Near-Field Spectra Using Machine Learning Techniques. \emph{Phys. Rev. Lett.}
  \textbf{2019}, \emph{123}, 163202\relax
\mciteBstWouldAddEndPuncttrue
\mciteSetBstMidEndSepPunct{\mcitedefaultmidpunct}
{\mcitedefaultendpunct}{\mcitedefaultseppunct}\relax
\EndOfBibitem
\bibitem[Fabrizio \latin{et~al.}(2019)Fabrizio, Grisafi, Meyer, Ceriotti, and
  Corminboeuf]{Fabrizio2019CS}
Fabrizio,~A.; Grisafi,~A.; Meyer,~B.; Ceriotti,~M.; Corminboeuf,~C. Electron
  Density Learning of Non-Covalent Systems. \emph{Chem. Sci.} \textbf{2019},
  \emph{10}, 9424--9432\relax
\mciteBstWouldAddEndPuncttrue
\mciteSetBstMidEndSepPunct{\mcitedefaultmidpunct}
{\mcitedefaultendpunct}{\mcitedefaultseppunct}\relax
\EndOfBibitem
\bibitem[Grisafi \latin{et~al.}(2019)Grisafi, Fabrizio, Meyer, Wilkins,
  Corminboeuf, and Ceriotti]{Grisafi2019ACSCS}
Grisafi,~A.; Fabrizio,~A.; Meyer,~B.; Wilkins,~D.~M.; Corminboeuf,~C.;
  Ceriotti,~M. Transferable Machine-Learning Model of the Electron Density.
  \emph{ACS Cent. Sci.} \textbf{2019}, \emph{5}, 57--64\relax
\mciteBstWouldAddEndPuncttrue
\mciteSetBstMidEndSepPunct{\mcitedefaultmidpunct}
{\mcitedefaultendpunct}{\mcitedefaultseppunct}\relax
\EndOfBibitem
\bibitem[Fabrizio \latin{et~al.}(2020)Fabrizio, Briling, Grisafi, and
  Corminboeuf]{Fabrizio2020CHIMIA}
Fabrizio,~A.; Briling,~K.; Grisafi,~A.; Corminboeuf,~C. Learning (from) the
  Electron Density: Transferability, Conformational and Chemical Diversity.
  \emph{CHIMIA Int. J. Chem.} \textbf{2020}, \emph{74}, 232--236\relax
\mciteBstWouldAddEndPuncttrue
\mciteSetBstMidEndSepPunct{\mcitedefaultmidpunct}
{\mcitedefaultendpunct}{\mcitedefaultseppunct}\relax
\EndOfBibitem
\bibitem[Mai \latin{et~al.}(2018)Mai, Plasser, Dorn, Fumanal, Daniel, and
  Gonz\'alez]{Mai2018CCR}
Mai,~S.; Plasser,~F.; Dorn,~J.; Fumanal,~M.; Daniel,~C.; Gonz\'alez,~L.
  Quantitative Wave Function Analysis for Excited States of Transition Metal
  Complexes. \emph{Coord. Chem. Rev.} \textbf{2018}, \emph{361}, 74 -- 97\relax
\mciteBstWouldAddEndPuncttrue
\mciteSetBstMidEndSepPunct{\mcitedefaultmidpunct}
{\mcitedefaultendpunct}{\mcitedefaultseppunct}\relax
\EndOfBibitem
\bibitem[Mennucci \latin{et~al.}(2009)Mennucci, Cappelli, Guido, Cammi, and
  Tomasi]{Mennucci2009JPCA}
Mennucci,~B.; Cappelli,~C.; Guido,~C.~A.; Cammi,~R.; Tomasi,~J. Structures and
  Properties of Electronically Excited Chromophores in Solution from the
  {Polarizable Continuum Model} Coupled to the {Time-Dependent Density
  Functional Theory}. \emph{J. Phys. Chem. A} \textbf{2009}, \emph{113},
  3009--3020\relax
\mciteBstWouldAddEndPuncttrue
\mciteSetBstMidEndSepPunct{\mcitedefaultmidpunct}
{\mcitedefaultendpunct}{\mcitedefaultseppunct}\relax
\EndOfBibitem
\bibitem[Jasper \latin{et~al.}(2004)Jasper, Kendrick, Mead, and
  Truhlar]{Jasper2004}
Jasper,~A.~W.; Kendrick,~B.~K.; Mead,~C.~A.; Truhlar,~D.~G. \emph{Modern Trends
  in Chemical Reaction Dynamics}; World Scientific, 2004; pp 329--391\relax
\mciteBstWouldAddEndPuncttrue
\mciteSetBstMidEndSepPunct{\mcitedefaultmidpunct}
{\mcitedefaultendpunct}{\mcitedefaultseppunct}\relax
\EndOfBibitem
\bibitem[Yarkony(2004)]{Yarkony2004}
Yarkony,~D.~R. In \emph{conical Intersections}; Domcke,~W., Yarkony,~D.~R.,
  K\"oppel,~H., Eds.; Advanced Series in Physical Chemistry; World Scientific,
  2004; Vol.~15\relax
\mciteBstWouldAddEndPuncttrue
\mciteSetBstMidEndSepPunct{\mcitedefaultmidpunct}
{\mcitedefaultendpunct}{\mcitedefaultseppunct}\relax
\EndOfBibitem
\bibitem[Cupellini \latin{et~al.}(2020)Cupellini, Bondanza, Nottoli, and
  Mennucci]{Cupellini2020BBAB}
Cupellini,~L.; Bondanza,~M.; Nottoli,~M.; Mennucci,~B. Successes \& Challenges
  in the Atomistic Modeling of Light-Harvesting and its Photoregulation.
  \emph{Biochim. Biophys. Acta, Bioenerg.} \textbf{2020}, \emph{1861},
  148049\relax
\mciteBstWouldAddEndPuncttrue
\mciteSetBstMidEndSepPunct{\mcitedefaultmidpunct}
{\mcitedefaultendpunct}{\mcitedefaultseppunct}\relax
\EndOfBibitem
\bibitem[Richings \latin{et~al.}(2019)Richings, Robertson, and
  Habershon]{Richings2019FD}
Richings,~G.~W.; Robertson,~C.; Habershon,~S. Can We Use on-the-Fly Quantum
  Simulations to Connect Molecular Structure and Sunscreen Action?
  \emph{Faraday Discuss.} \textbf{2019}, \emph{216}, 476--493\relax
\mciteBstWouldAddEndPuncttrue
\mciteSetBstMidEndSepPunct{\mcitedefaultmidpunct}
{\mcitedefaultendpunct}{\mcitedefaultseppunct}\relax
\EndOfBibitem
\bibitem[Behler \latin{et~al.}(2007)Behler, Lorenz, and Reuter]{Behler2007JCP}
Behler,~J.; Lorenz,~S.; Reuter,~K. Representing Molecule-Surface Interactions
  with Symmetry-Adapted Neural Networks. \emph{J. Chem. Phys.} \textbf{2007},
  \emph{127}, 014705\relax
\mciteBstWouldAddEndPuncttrue
\mciteSetBstMidEndSepPunct{\mcitedefaultmidpunct}
{\mcitedefaultendpunct}{\mcitedefaultseppunct}\relax
\EndOfBibitem
\bibitem[Behler(2004)]{phdbehler}
Behler,~J. Dissociation of Oxygen Molecules on the Al(111) Surface. Ph.D.\
  thesis, Technical University Berlin, 2004\relax
\mciteBstWouldAddEndPuncttrue
\mciteSetBstMidEndSepPunct{\mcitedefaultmidpunct}
{\mcitedefaultendpunct}{\mcitedefaultseppunct}\relax
\EndOfBibitem
\bibitem[{la Cour Jansen} \latin{et~al.}(1999){la Cour Jansen}, Rettrup, Sarma,
  Snijders, and Palmieri]{Jansen1999IJQC}
{la Cour Jansen},~T.; Rettrup,~S.; Sarma,~C.; Snijders,~J.; Palmieri,~P. On the
  Evaluation of Spin-Orbit Coupling Matrix Elements in a Spin-Adapted Basis.
  \emph{Int. J. Quantum Chem.} \textbf{1999}, 23--27\relax
\mciteBstWouldAddEndPuncttrue
\mciteSetBstMidEndSepPunct{\mcitedefaultmidpunct}
{\mcitedefaultendpunct}{\mcitedefaultseppunct}\relax
\EndOfBibitem
\bibitem[Chen \latin{et~al.}(2019)Chen, Fang, and Cui]{Chen2019PCCP}
Chen,~W.-K.; Fang,~W.-H.; Cui,~G. A Multi-Layer Energy-Based Fragment Method
  for Excited States and Nonadiabatic Dynamics. \emph{Phys. Chem. Chem. Phys.}
  \textbf{2019}, \emph{21}, 22695--22699\relax
\mciteBstWouldAddEndPuncttrue
\mciteSetBstMidEndSepPunct{\mcitedefaultmidpunct}
{\mcitedefaultendpunct}{\mcitedefaultseppunct}\relax
\EndOfBibitem
\bibitem[Christensen and von Lilienfeld(2019)Christensen, and von
  Lilienfeld]{Christensen2019CHIMIA}
Christensen,~A.~S.; von Lilienfeld,~O.~A. Operator Quantum Machine Learning:
  Navigating the Chemical Space of Response Properties. \emph{CHIMIA}
  \textbf{2019}, \emph{73}, 1028--1031\relax
\mciteBstWouldAddEndPuncttrue
\mciteSetBstMidEndSepPunct{\mcitedefaultmidpunct}
{\mcitedefaultendpunct}{\mcitedefaultseppunct}\relax
\EndOfBibitem
\bibitem[Kasha \latin{et~al.}(1965)Kasha, Rawls, and El-Bayoumi]{Kasha1965}
Kasha,~M.; Rawls,~H.~R.; El-Bayoumi,~M.~A. The Exciton Model in Molecular
  Spectroscopy. \emph{Pure and Applied Chemistry} \textbf{1965}, \emph{11}, 371
  -- 392\relax
\mciteBstWouldAddEndPuncttrue
\mciteSetBstMidEndSepPunct{\mcitedefaultmidpunct}
{\mcitedefaultendpunct}{\mcitedefaultseppunct}\relax
\EndOfBibitem
\bibitem[Rogers and Hahn(2010)Rogers, and Hahn]{Rogers2010JCIM}
Rogers,~D.; Hahn,~M. {Extended-Connectivity Fingerprints}. \emph{J. Chem. Inf.
  Model.} \textbf{2010}, \emph{50}, 742--754\relax
\mciteBstWouldAddEndPuncttrue
\mciteSetBstMidEndSepPunct{\mcitedefaultmidpunct}
{\mcitedefaultendpunct}{\mcitedefaultseppunct}\relax
\EndOfBibitem
\bibitem[Durant \latin{et~al.}(2002)Durant, Leland, Henry, and
  Nourse]{Durant2002JCICS}
Durant,~J.~L.; Leland,~B.~A.; Henry,~D.~R.; Nourse,~J.~G. {Reoptimization of
  MDL Keys for Use in Drug Discovery}. \emph{J. Chem. Inf. Comput. Sci.}
  \textbf{2002}, \emph{42}, 1273--1280\relax
\mciteBstWouldAddEndPuncttrue
\mciteSetBstMidEndSepPunct{\mcitedefaultmidpunct}
{\mcitedefaultendpunct}{\mcitedefaultseppunct}\relax
\EndOfBibitem
\bibitem[Landrum(2016)]{Landrum2016}
Landrum,~G. {RDKit: Open-Source Cheminformatics Software}. \textbf{2016},
  \relax
\mciteBstWouldAddEndPunctfalse
\mciteSetBstMidEndSepPunct{\mcitedefaultmidpunct}
{}{\mcitedefaultseppunct}\relax
\EndOfBibitem
\bibitem[Jain \latin{et~al.}(2013)Jain, Ong, Hautier, Chen, Richards, Dacek,
  Cholia, Gunter, Skinner, Ceder, and Persson]{Jain2013APLM}
Jain,~A.; Ong,~S.~P.; Hautier,~G.; Chen,~W.; Richards,~W.~D.; Dacek,~S.;
  Cholia,~S.; Gunter,~D.; Skinner,~D.; Ceder,~G.; Persson,~K.~A. {Commentary:
  The Materials Project: A Materials Genome Approach to Accelerating Materials
  Innovation}. \emph{APL Mater.} \textbf{2013}, \emph{1}, 011002\relax
\mciteBstWouldAddEndPuncttrue
\mciteSetBstMidEndSepPunct{\mcitedefaultmidpunct}
{\mcitedefaultendpunct}{\mcitedefaultseppunct}\relax
\EndOfBibitem
\bibitem[Aarva \latin{et~al.}(2019)Aarva, Deringer, Sainio, Laurila, and
  Caro]{Aarva2019CMb}
Aarva,~A.; Deringer,~V.~L.; Sainio,~S.; Laurila,~T.; Caro,~M.~A. {Understanding
  X-Ray Spectroscopy of Carbonaceous Materials by Combining Experiments,
  Density Functional Theory, and Machine Learning. Part II: Quantitative
  Fitting of Spectra}. \emph{Chem. Mat.} \textbf{2019}, \emph{31},
  9256--9267\relax
\mciteBstWouldAddEndPuncttrue
\mciteSetBstMidEndSepPunct{\mcitedefaultmidpunct}
{\mcitedefaultendpunct}{\mcitedefaultseppunct}\relax
\EndOfBibitem
\bibitem[Bart{\'o}k \latin{et~al.}(2017)Bart{\'o}k, De, Poelking, Bernstein,
  Kermode, Cs{\'a}nyi, and Ceriotti]{Bartok2017SA}
Bart{\'o}k,~A.~P.; De,~S.; Poelking,~C.; Bernstein,~N.; Kermode,~J.~R.;
  Cs{\'a}nyi,~G.; Ceriotti,~M. Machine Learning Unifies the Modeling of
  Materials and Molecules. \emph{Sci. Adv.} \textbf{2017}, \emph{3}\relax
\mciteBstWouldAddEndPuncttrue
\mciteSetBstMidEndSepPunct{\mcitedefaultmidpunct}
{\mcitedefaultendpunct}{\mcitedefaultseppunct}\relax
\EndOfBibitem
\bibitem[Aarva \latin{et~al.}(2019)Aarva, Deringer, Sainio, Laurila, and
  Caro]{Aarva2019CMa}
Aarva,~A.; Deringer,~V.~L.; Sainio,~S.; Laurila,~T.; Caro,~M.~A. {Understanding
  X-ray Spectroscopy of Carbonaceous Materials by Combining Experiments,
  Density Functional Theory, and Machine Learning. Part I: Fingerprint
  Spectra}. \emph{Chem. Mat.} \textbf{2019}, \emph{31}, 9243--9255\relax
\mciteBstWouldAddEndPuncttrue
\mciteSetBstMidEndSepPunct{\mcitedefaultmidpunct}
{\mcitedefaultendpunct}{\mcitedefaultseppunct}\relax
\EndOfBibitem
\bibitem[Deringer \latin{et~al.}(2018)Deringer, Caro, Jana, Aarva, Elliott,
  Laurila, Csányi, and Pastewka]{Deringer2018CM}
Deringer,~V.~L.; Caro,~M.~A.; Jana,~R.; Aarva,~A.; Elliott,~S.~R.; Laurila,~T.;
  Csányi,~G.; Pastewka,~L. Computational Surface Chemistry of Tetrahedral
  Amorphous Carbon by Combining Machine Learning and Density Functional Theory.
  \emph{Chem. Mat.} \textbf{2018}, \emph{30}, 7438--7445\relax
\mciteBstWouldAddEndPuncttrue
\mciteSetBstMidEndSepPunct{\mcitedefaultmidpunct}
{\mcitedefaultendpunct}{\mcitedefaultseppunct}\relax
\EndOfBibitem
\bibitem[Janet and Kulik(2017)Janet, and Kulik]{Janet2017CS}
Janet,~J.~P.; Kulik,~H.~J. Predicting Electronic Structure Properties of
  Transition Metal Complexes with Neural Networks. \emph{Chem. Sci.}
  \textbf{2017}, \emph{8}, 5137--5152\relax
\mciteBstWouldAddEndPuncttrue
\mciteSetBstMidEndSepPunct{\mcitedefaultmidpunct}
{\mcitedefaultendpunct}{\mcitedefaultseppunct}\relax
\EndOfBibitem
\bibitem[Janet \latin{et~al.}(2019)Janet, Duan, Yang, Nandy, and
  Kulik]{Janet2019CS}
Janet,~J.~P.; Duan,~C.; Yang,~T.; Nandy,~A.; Kulik,~H.~J. A Quantitative
  Uncertainty Metric Controls Error in Neural Network-Driven Chemical
  Discovery. \emph{Chem. Sci.} \textbf{2019}, \emph{10}, 7913--7922\relax
\mciteBstWouldAddEndPuncttrue
\mciteSetBstMidEndSepPunct{\mcitedefaultmidpunct}
{\mcitedefaultendpunct}{\mcitedefaultseppunct}\relax
\EndOfBibitem
\bibitem[Janet \latin{et~al.}(2017)Janet, Gani, Steeves, Ioannidis, and
  Kulik]{Janet2017IECR}
Janet,~J.~P.; Gani,~T. Z.~H.; Steeves,~A.~H.; Ioannidis,~E.~I.; Kulik,~H.~J.
  Leveraging Cheminformatics Strategies for Inorganic Discovery: Application to
  Redox Potential Design. \emph{Ind. Eng. Chem. Res.} \textbf{2017}, \emph{56},
  4898--4910\relax
\mciteBstWouldAddEndPuncttrue
\mciteSetBstMidEndSepPunct{\mcitedefaultmidpunct}
{\mcitedefaultendpunct}{\mcitedefaultseppunct}\relax
\EndOfBibitem
\bibitem[Janet \latin{et~al.}(2018)Janet, Chan, and Kulik]{Janet2018JPCL}
Janet,~J.~P.; Chan,~L.; Kulik,~H.~J. {Accelerating Chemical Discovery with
  Machine Learning: Simulated Evolution of Spin Crossover Complexes with an
  Artificial Neural Network}. \emph{J. Phys. Chem. Lett.} \textbf{2018},
  \emph{9}, 1064--1071\relax
\mciteBstWouldAddEndPuncttrue
\mciteSetBstMidEndSepPunct{\mcitedefaultmidpunct}
{\mcitedefaultendpunct}{\mcitedefaultseppunct}\relax
\EndOfBibitem
\bibitem[Janet \latin{et~al.}(2020)Janet, Ramesh, Duan, and
  Kulik]{Janet2020ACSCS}
Janet,~J.~P.; Ramesh,~S.; Duan,~C.; Kulik,~H.~J. {Accurate Multiobjective
  Design in a Space of Millions of Transition Metal Complexes with
  Neural-Network-Driven Efficient Global Optimization}. \emph{ACS Cent. Sci.}
  \textbf{2020}, \emph{6}, 513--524\relax
\mciteBstWouldAddEndPuncttrue
\mciteSetBstMidEndSepPunct{\mcitedefaultmidpunct}
{\mcitedefaultendpunct}{\mcitedefaultseppunct}\relax
\EndOfBibitem
\bibitem[{\DH}or{\dj}evi\'{c} \latin{et~al.}(2018){\DH}or{\dj}evi\'{c},
  Beckwith, Yarema, Yarema, Rosspeintner, Yazdani, Leuthold, Vauthey, and
  Wood]{Dordevic2018ACSP}
{\DH}or{\dj}evi\'{c},~N.; Beckwith,~J.~S.; Yarema,~M.; Yarema,~O.;
  Rosspeintner,~A.; Yazdani,~N.; Leuthold,~J.; Vauthey,~E.; Wood,~V. {Machine
  Learning for Analysis of Time-Resolved Luminescence Data}. \emph{ACS
  Photonics} \textbf{2018}, \emph{5}, 4888--4895\relax
\mciteBstWouldAddEndPuncttrue
\mciteSetBstMidEndSepPunct{\mcitedefaultmidpunct}
{\mcitedefaultendpunct}{\mcitedefaultseppunct}\relax
\EndOfBibitem
\bibitem[Abramavicius \latin{et~al.}(2010)Abramavicius, Jiang, Bulheller,
  Hirst, and Mukamel]{Abramavicius2010JACS}
Abramavicius,~D.; Jiang,~J.; Bulheller,~B.~M.; Hirst,~J.~D.; Mukamel,~S.
  Simulation Study of Chiral Two-Dimensional Ultraviolet Spectroscopy of the
  Protein Backbone. \emph{Journal of the American Chemical Society}
  \textbf{2010}, \emph{132}, 7769--7775\relax
\mciteBstWouldAddEndPuncttrue
\mciteSetBstMidEndSepPunct{\mcitedefaultmidpunct}
{\mcitedefaultendpunct}{\mcitedefaultseppunct}\relax
\EndOfBibitem
\bibitem[Zhang \latin{et~al.}(2011)Zhang, W\"{u}rth, Zhao, Resch-Genger,
  Ernsting, and Sajadi]{Zhang2011RSI}
Zhang,~X.-X.; W\"{u}rth,~C.; Zhao,~L.; Resch-Genger,~U.; Ernsting,~N.~P.;
  Sajadi,~M. Femtosecond Broadband Fluorescence Upconversion Spectroscopy:
  Improved Setup and Photometric Correction. \emph{Rev. Sci. Instrum.}
  \textbf{2011}, \emph{82}, 063108\relax
\mciteBstWouldAddEndPuncttrue
\mciteSetBstMidEndSepPunct{\mcitedefaultmidpunct}
{\mcitedefaultendpunct}{\mcitedefaultseppunct}\relax
\EndOfBibitem
\bibitem[Ahmad \latin{et~al.}(2016)Ahmad, Ahmed, Anwar, Sheraz, and
  Sikorski]{Ahmad2016IJP}
Ahmad,~I.; Ahmed,~S.; Anwar,~Z.; Sheraz,~M.~A.; Sikorski,~M. Photostability and
  Photostabilization of Drugs and Drug Products. \emph{Int. J. Photoenergy}
  \textbf{2016}, \emph{2016}, 1--19\relax
\mciteBstWouldAddEndPuncttrue
\mciteSetBstMidEndSepPunct{\mcitedefaultmidpunct}
{\mcitedefaultendpunct}{\mcitedefaultseppunct}\relax
\EndOfBibitem
\bibitem[Mathew \latin{et~al.}(2014)Mathew, Yella, Gao, Humphry-Baker, Curchod,
  Ashari-Astani, Tavernelli, Rothlisberger, Nazeeruddin, and
  Gr{\"a}tzel]{Mathew2014NC}
Mathew,~S.; Yella,~A.; Gao,~P.; Humphry-Baker,~R.; Curchod,~B. F.~E.;
  Ashari-Astani,~N.; Tavernelli,~I.; Rothlisberger,~U.; Nazeeruddin,~M.~K.;
  Gr{\"a}tzel,~M. Dye-Sensitized Solar Cells with 13{\%} Efficiency Achieved
  Through the Molecular Engineering of Porphyrin Sensitizers. \emph{Nat. Chem.}
  \textbf{2014}, \emph{6}, 242--247\relax
\mciteBstWouldAddEndPuncttrue
\mciteSetBstMidEndSepPunct{\mcitedefaultmidpunct}
{\mcitedefaultendpunct}{\mcitedefaultseppunct}\relax
\EndOfBibitem
\bibitem[Dobson(2004)]{Dobson2004N}
Dobson,~C.~M. Chemical Space and Biology. \emph{Nature} \textbf{2004},
  \emph{432}, 824--828\relax
\mciteBstWouldAddEndPuncttrue
\mciteSetBstMidEndSepPunct{\mcitedefaultmidpunct}
{\mcitedefaultendpunct}{\mcitedefaultseppunct}\relax
\EndOfBibitem
\end{mcitethebibliography}
\providecommand{\latin}[1]{#1}
\makeatletter
\providecommand{\doi}
  {\begingroup\let\do\@makeother\dospecials
  \catcode`\{=1 \catcode`\}=2 \doi@aux}
\providecommand{\doi@aux}[1]{\endgroup\texttt{#1}}
\makeatother
\providecommand*\mcitethebibliography{\thebibliography}
\csname @ifundefined\endcsname{endmcitethebibliography}
  {\let\endmcitethebibliography\endthebibliography}{}

\end{document}